\documentclass[reprint,aps,prc,twocolumn,showpacs,nofootinbib,longbibliography,superscriptaddress]{revtex4-2}  % for review and submission
\usepackage{graphicx}  % needed for figures
\usepackage{dcolumn}   % needed for some tables
\usepackage{bm}        % for math
\usepackage{amssymb}   % for math
\usepackage{amsmath}

\usepackage{pifont} % for xmark

\usepackage{multirow}
\usepackage{array}

\usepackage{comment}

\usepackage[caption=false]{subfig}

\usepackage{hhline}
\usepackage[table]{xcolor}

\newcommand{\D}{\mathrm{d}} % Roman d for integration measure/derivative
\renewcommand\vec[1]{\ensuremath\boldsymbol{#1}} % bold font for vectors
\DeclareMathOperator{\Tr}{Tr}  % Trace operator
\DeclareMathOperator{\sech}{sech}
\definecolor{dgreen}{rgb}{0.01,0.75,0.24}   % dark green color
\definecolor{RowColor}{rgb}{0.88,1,0.9}
\definecolor{RowColor2}{rgb}{0.88,0.9,1}

\usepackage[T1]{fontenc} % for Tanja Duric

\begin{document}

\title{Extended Hubbard model in undoped and doped monolayer and bilayer graphene: Selection rules and organizing principle among competing orders}

\author{Andr$\acute{\mbox{a}}$s L. Szab$\acute{\mbox{o}}$}\email{szabo@pks.mpg.de}
\affiliation{Max-Planck-Institut f\"{u}r Physik komplexer Systeme, N\"{o}thnitzer Str. 38, 01187 Dresden, Germany}

\author{Bitan Roy}\email{bitan.roy@lehigh.edu}
\affiliation{Department of Physics, Lehigh University, Bethlehem, Pennsylvania, 18015, USA}

\date{\today}
\begin{abstract}
Performing a leading-order renormalization group analysis, here we compute the effects of generic local or short-range electronic interactions in monolayer and the Bernal bilayer graphene. Respectively in these two systems gapless chiral quasiparticles display linear and biquadratic band touching, leading to linearly vanishing and constant density of states. Consequently, the former system remains stable for weak enough local interactions, and supports a variety of ordered phases only beyond a critical strength of interactions. By contrast, ordered phases can nucleate for sufficiently weak interactions in bilayer graphene. By tuning the strength of all symmetry allowed local interactions, we construct various cuts of the phase diagram at zero and finite temperature and chemical doping. Typically, at zero doping, insulating phases (such as charge density wave, antiferromagnet, quantum anomalous and spin Hall insulators) prevail at the lowest temperature, while gapless nematic or smectic liquids stabilize at higher temperatures. On the other hand, at finite doping, the lowest temperature ordered phase is occupied by a superconductor. Besides anchoring such an organizing principle among the candidate ordered phases, we also establish a selection rule between them and the interaction channel responsible for the breakdown of linear or biquadratic chiral nodal Fermi liquid. In addition, we also demonstrate the role of the normal state band structure in selecting the pattern of symmetry breaking from a soup of preselected incipient competing orders. As a direct consequence of the selection rule, while an antiferromagnetic phase develops in undoped monolayer and bilayer graphene, the linear (biquadratic) band dispersion favors condensation of a spin-singlet nematic (translational symmetry breaking Kekul\'e) superconductor in doped monolayer (bilayer) graphene, when the on-site Hubbard repulsion dominates in these systems. On the other hand, nearest-neighbor (next-nearest-neighbor) repulsion accommodates charge density wave (quantum spin Hall insulator) and $s+if$ ($s$-wave) pairing at zero and finite chemical doping in both systems, respectively.           
\end{abstract}

\maketitle

\section{Introduction}

Multiband electronic materials constitute a rich landscape harboring a plethora of competing phases, among which spin- and charge density waves, nematicity, unconventional superconductivity are the most prominent and commonly occurring ones, leaving aside the mysterious quantum spin liquids. Indeed, the global phase diagram of strongly correlated electronic materials, such as cuprates, pnictides, iridates, ruthenates, and heavy-fermion compounds, displays a confluence of at least some of these phases. In experiments, these phases can be realized as the temperature, pressure, and chemical doping is tuned in the system. However, owing to the complexity of the band structure, it is often challenging to construct a unifying minimal model for these correlated materials that captures the salient features of their global phase diagram. In this regard emergence of a new class of materials, \emph{semimetals}, featuring band touching at isolated points in the Brillouin zone (BZ), opened a new frontier. Even though semimetals are fascinating phases of matter from the perspective of their topological properties~\cite{RevModPhys.88.035005, RevModPhys.90.015001, 2016NatMa..15.1145B}, here we focus to address the role of strong electronic interactions in such gapless electronic materials. Specifically, we consider two paradigmatic representatives of two-dimensional semimetallic systems, namely monolayer graphene (MLG) and Bernal-stacked bilayer graphene (BLG), where noninteracting itinerant electrons display linear and biquadratic band touchings at the corners of the hexagonal BZ, respectively~\cite{RevModPhys.81.109, katsnelson2020}.

%%%%%%%%%%%%%%%%%%%%%%%%%%%%%%%%%%%%%%%%%%%%%%%%%%%%%%%%%%%%%%%%%%%%%%%%%%%%%%%%%%%%%%%%%%%%
%%%%%%%%%%%%%%%%%%%%%%%%%%%%%%%%%%%%%%%%%%%%%%%%%%%%%%%%%%%%%%%%%%%%%%%%%%%%%%%%%%%%%%%%%%%%
%%%%%%%%%%%%%%%%%%%%%%%%%%%%%%%%%%% FLOWCHART FIGURE: %%%%%%%%%%%%%%%%%%%%%%%%%%%%%%%%%%%%%%
%%%%%%%%%%%%%%%%%%%%%%%%%%%%%%%%%%%%%%%%%%%%%%%%%%%%%%%%%%%%%%%%%%%%%%%%%%%%%%%%%%%%%%%%%%%%
%%%%%%%%%%%%%%%%%%%%%%%%%%%%%%%%%%%%%%%%%%%%%%%%%%%%%%%%%%%%%%%%%%%%%%%%%%%%%%%%%%%%%%%%%%%%
\begin{figure}[t]
\includegraphics[width=1.0\linewidth]{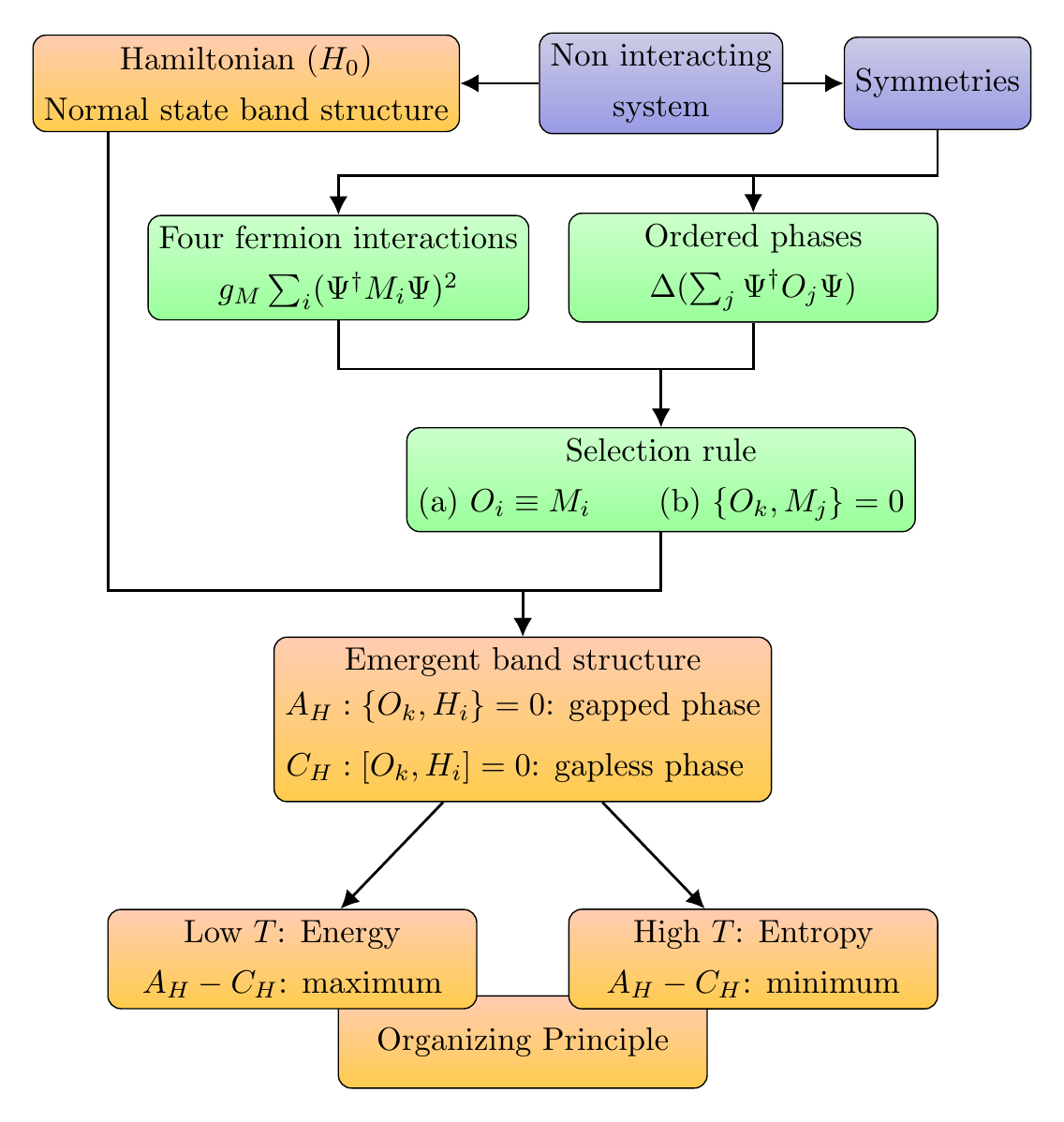}
\caption{A schematic flowchart depicting the selection rules and organizing principle. Consider a noninteracting system, described by the single-particle Hamiltonian $H_0$ containing a set of Hermitian matrices $\{ H_i \}$, a four-fermion interaction with coupling $g_{_M}$ and a fermionic order parameter coupled with the conjugate field $\Delta_O$, all determined by the symmetry of the system. Here $M_i$ and $O_i$ are Hermitian matrices of the same dimension as $H_i$. Then, an ordered phase is boosted by the interaction if (a) $O_i \equiv M_i$ or (b) these two set of matrices maximally anticommute. The selected ordered phases are then arranged along the temperature axis according to the organizing principle, quantified by $A_H$ and $C_H$, respectively counting the pairs of matrices from the sets $\{O_i \}$ and $\{ H_i \}$ that mutually anticommute and commute. Order parameter matrices $O_i$ that maximally anticommute (commute) with $H_i$ yield gapped (gapless) quasiparticles inside the ordered phase and inhabit comparatively lower (higher) temperatures, following the energy-entropy argument. Finite chemical doping promotes superconductivity at the lowest temperature even from repulsive electronic interactions, as the underlying Fermi surface can only be gapped by them, also in agreement with the organizing principle. The nature of the nucleated paired state follows the selection rule.
}~\label{fig:flowchart}
\end{figure}
%%%%%%%%%%%%%%%%%%%%%%%%%%%%%%%%%%%%%%%%%%%%%%%%%%%%%%%%%%%%%%%%%%%%%%%%%%%%%%%%%%%%%%%%%%%%
%%%%%%%%%%%%%%%%%%%%%%%%%%%%%%%%%%%%%%%%%%%%%%%%%%%%%%%%%%%%%%%%%%%%%%%%%%%%%%%%%%%%%%%%%%%%
%%%%%%%%%%%%%%%%%%%%%%%%%%%%%%%%%%%%%%%%%%%%%%%%%%%%%%%%%%%%%%%%%%%%%%%%%%%%%%%%%%%%%%%%%%%%
%%%%%%%%%%%%%%%%%%%%%%%%%%%%%%%%%%%%%%%%%%%%%%%%%%%%%%%%%%%%%%%%%%%%%%%%%%%%%%%%%%%%%%%%%%%%
%%%%%%%%%%%%%%%%%%%%%%%%%%%%%%%%%%%%%%%%%%%%%%%%%%%%%%%%%%%%%%%%%%%%%%%%%%%%%%%%%%%%%%%%%%%%

\begin{figure}[t!]
\includegraphics[width=0.45\linewidth]{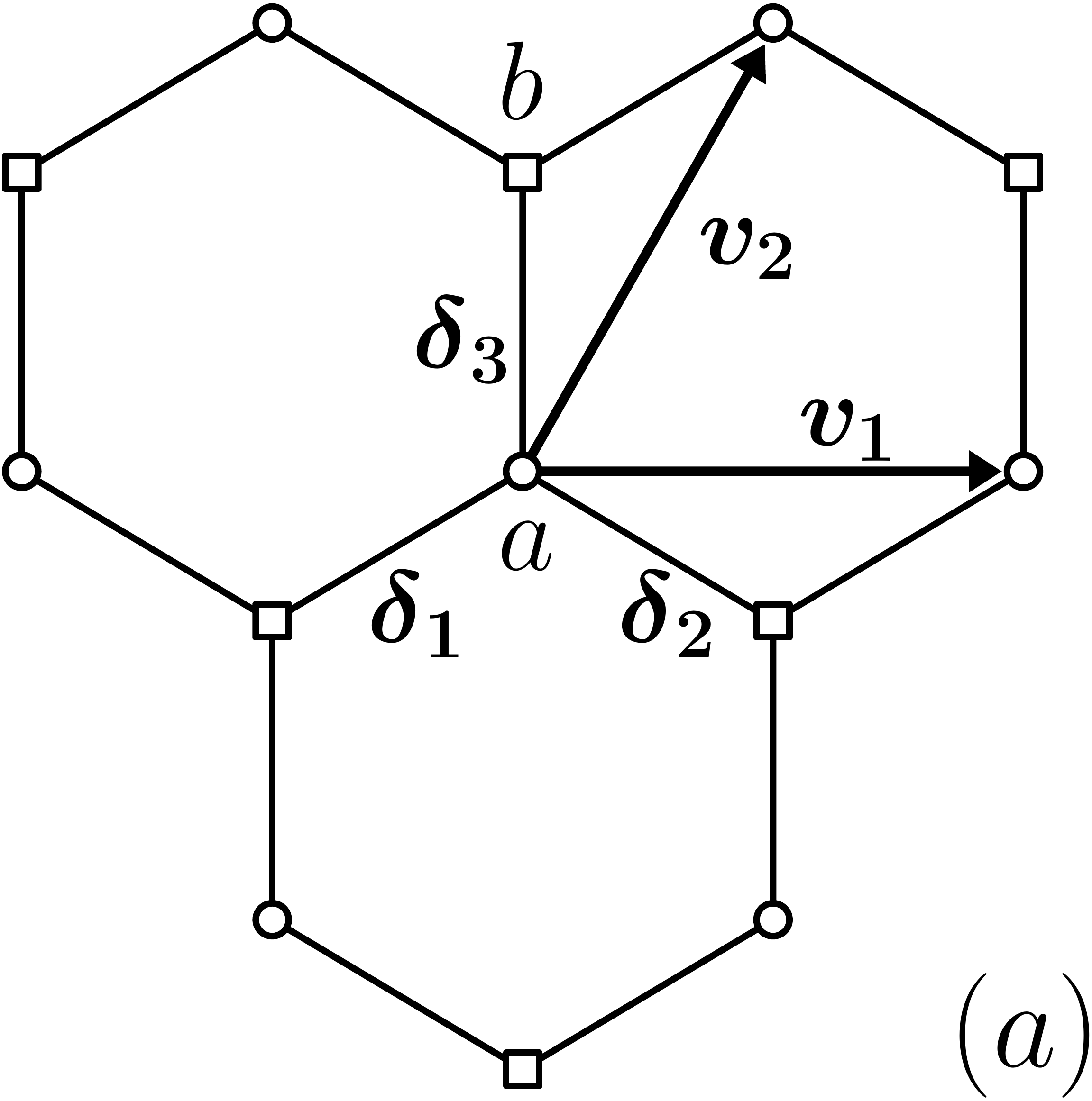}\hspace{0.7cm}
\includegraphics[width=0.45\linewidth]{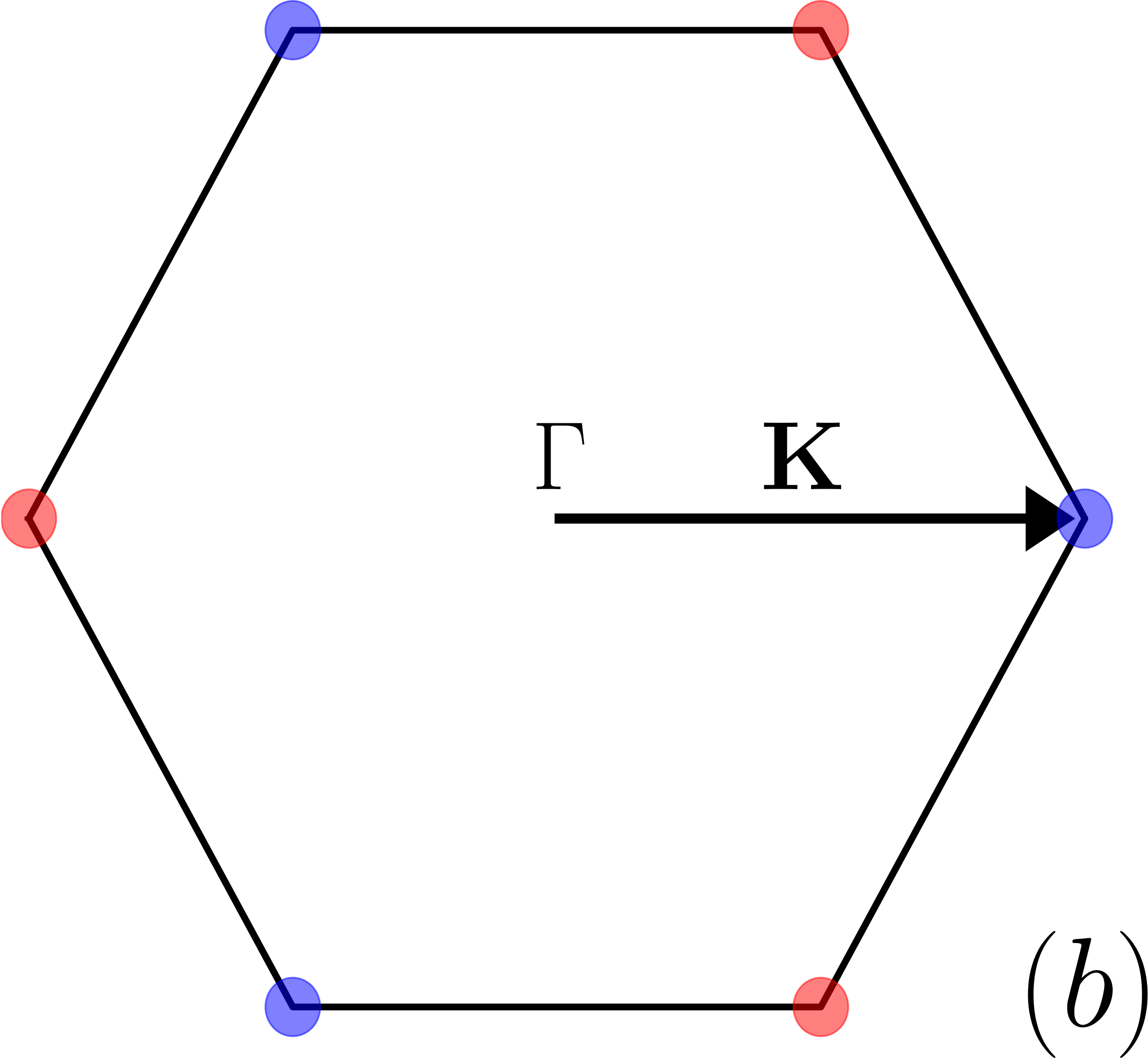} \\
\vspace{0.7cm}
\includegraphics[width=0.45\linewidth]{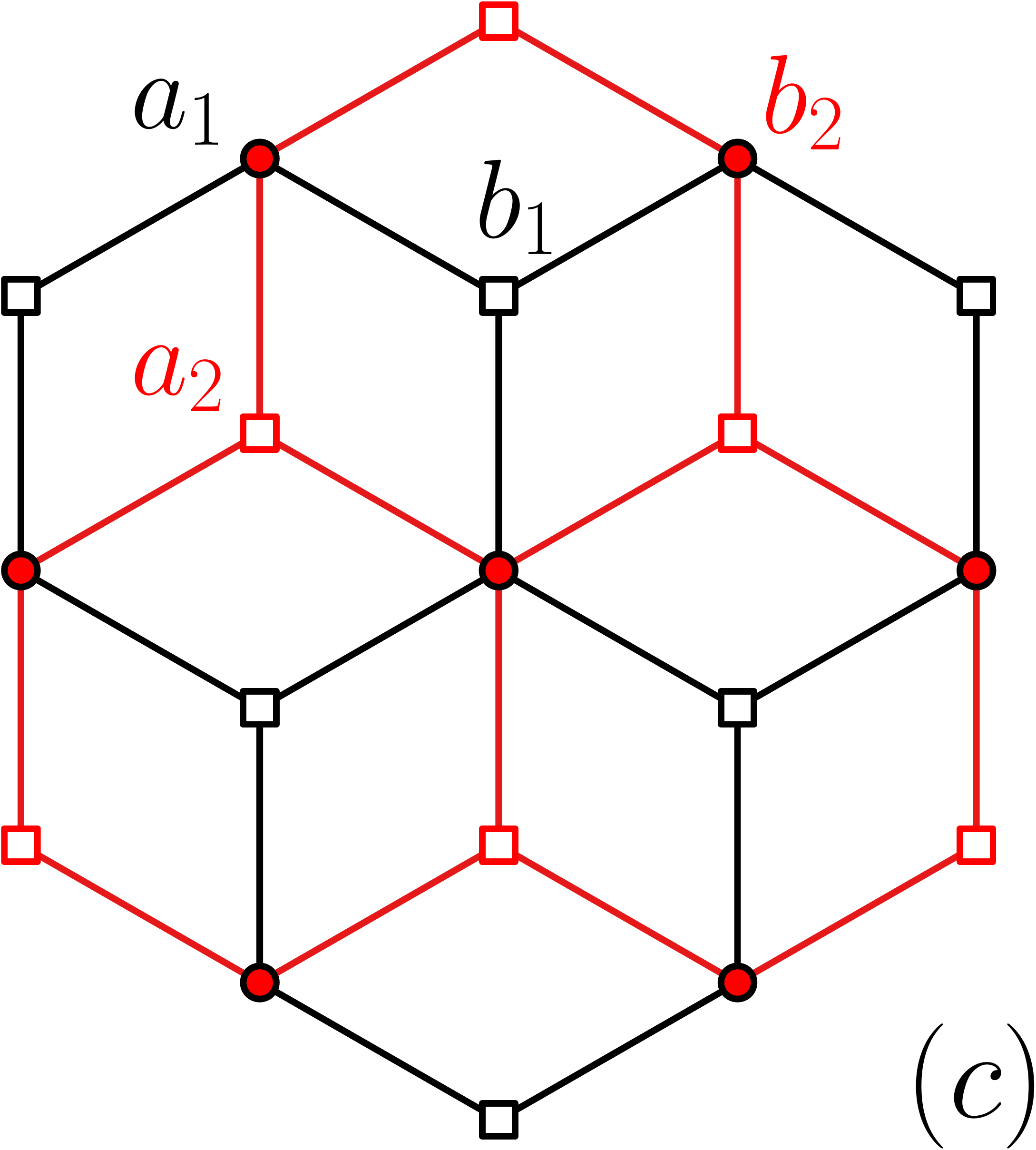}\hspace{0.7cm}
\includegraphics[width=0.45\linewidth]{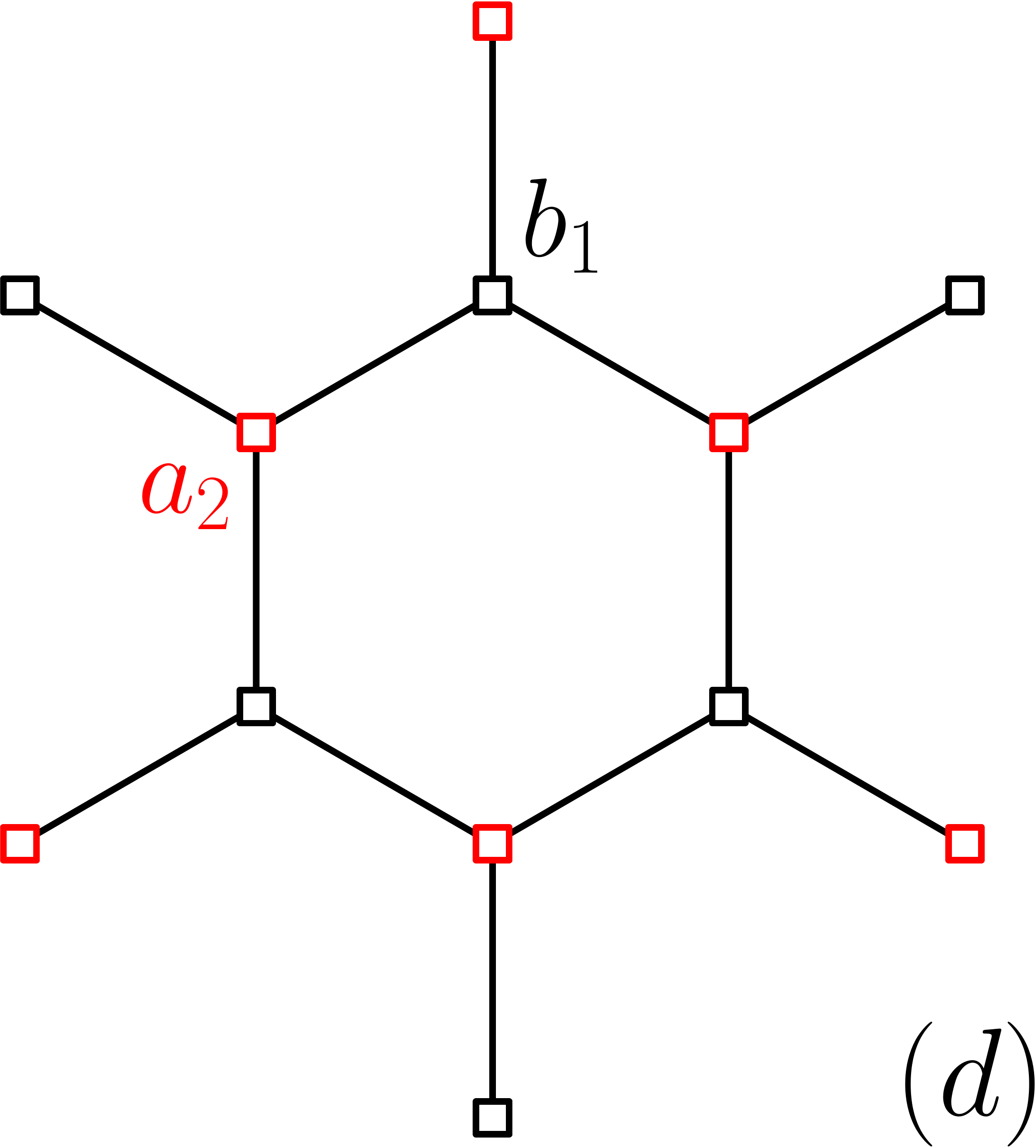}
\caption{(a) Honeycomb monolayer, consisting of two interpenetrating triangular sublattices, A and B. The primitive lattice vectors $\vec{v}_1$ and $\vec{v}_2$ span the $a$ sites (belonging to sublattice A). The $b$ sites (belonging to sublattice B) are generated by $\boldsymbol{\delta}_1$, $\boldsymbol{\delta}_2$ and $\boldsymbol{\delta}_3$. (b) The BZ of MLG, a hexagonal structure rotated by $90$ degrees compared to the real space lattice. Six Dirac points are located at its corners, of which two sets are inequivalent, marked with blue and red circles. The representative valleys are chosen to be at $\vec{K}$ and $-\vec{K}$, respectively. (c) Bernal stacked honeycomb bilayer, with the subscript denoting the layer index $i=1,2$ of the sites. Each $a_1$ and $b_2$ sites overlap. The eigenstates of the high energy split-off bands reside dominantly on these overlapping dimer sites. (d) After integrating out the $a_1$ and $b_2$ sites, the resulting structure is another honeycomb lattice, demonstrating the identical point group symmetry of MLG and BLG.}
~\label{fig:lattices}
\end{figure}

In what follows, restricting ourselves to local or momentum-independent repulsive four-fermions interactions on monolayer and Bernal bilayer honeycomb membrane of carbon atoms, we accomplish the following. (1) We establish a correspondence between a given local four-fermion interaction that ultimately causes the breakdown of either the linear or the biquadratic nodal chiral Fermi liquid and the broken symmetry phase(s) it supports in the ordered state. Throughout this paper, such a correspondence is referred to as the \emph{selection rule}. (2) In addition, we demonstrate an \emph{organizing principle} among competing phases as the temperature or chemical doping is tuned in the system, based on a generalized energy-entropy argument. And finally, (3) we highlight the key role played by the normal state band structure in selecting an ordered phase from a soup of incipient competing orders, otherwise chosen by the selection rule. The whole construction is summarized in Fig.~\ref{fig:flowchart}. Specifically, in this regard, MLG and BLG provide a unique opportunity as these two systems are \emph{indistinguishable} from the point of view of the point or space group symmetry, although they display distinct normal state band structures, see Fig.~\ref{fig:lattices}. It is worthwhile mentioning that recently the author(s) proposed a set of selection rules and organizing principle among competing orders for (a) only the on-site Hubbard repulsion in a two-dimensional slow Dirac liquid in twisted bilayer graphene in Ref.~\cite{PhysRevB.99.121407} and (b) generic repulsive quartic interactions for three-dimensional strong spin-orbit coupled biquadratic effective spin-3/2 Luttinger fermions in Ref.~\cite{PhysRevB.103.165139} that can be relevant for 227 pyrochlore iridates and half-Heuslers, for example. The findings reported in this work are completely in agreement with the ones in these two previous works~\cite{PhysRevB.99.121407, PhysRevB.103.165139}. Therefore, the present discussion besides providing new insights into the global phase diagram of the extended honeycomb Hubbard model in (un)doped MLG and BLG that remained unnoticed despite the existing vast literature on this topic
~\cite{PhysRevB.88.075415,
PhysRevB.82.201408,
PhysRevB.79.085116,
PhysRevLett.97.146401,
PhysRevB.85.245451,
PhysRevLett.112.147002,
PhysRevB.86.075467,
PhysRevB.79.201403,
PhysRevB.82.035429,
PhysRevB.82.153406,
PhysRevB.84.113404,
wang2011nature,
2012PhLA..376..779N,
PhysRevB.86.245431,
PhysRevB.88.045425,
PhysRevX.3.031010,
PhysRevB.89.035103,
PhysRevB.89.165123,
PhysRevB.89.205128,
PhysRevB.90.054521,
2014NJPh...16j3008W,
2014JPCM...26U5601G,
PhysRevB.91.165108,
2015NJPh...17h5003L,
PhysRevB.92.085146,
PhysRevB.92.155137,
PhysRevX.6.011029,
2017JPCM...29d3002C,
PhysRevB.95.235124,
PhysRevB.99.125145,
boyack2020quantum,
PhysRevB.102.245105,
PhysRevLett.98.146801,
PhysRevLett.100.146404,
PhysRevB.102.245140,
PhysRevB.93.155149,
PhysRevB.94.104508,
PhysRevB.94.115105,
PhysRevLett.100.246808,
PhysRevB.78.205431,
PhysRevB.92.085147,
refId0,
PhysRevLett.100.156401,
PhysRevB.72.085123,
PhysRevLett.87.206401,
PhysRevLett.87.246802,
PhysRevLett.61.2015,
PhysRevLett.98.186809,
Sorella_1992,
1997ZPhyB.103..335M,
PhysRevLett.109.126402,
PhysRevB.86.155128,
PhysRevB.81.041402,
PhysRevB.82.115431,
PhysRevLett.104.156803,
PhysRevB.90.205407,
PhysRevB.102.134204}, 
more importantly extends the jurisdiction of the existing selection rule and the organizing principle to a completely different class of semimetallic compounds.
Altogether these analyses suggest a general applicability of the proposed selection rule and organizing principle among competing orders in generic correlated multiband systems, which can be predicted a priory from the symmetry of the system, the noninteracting band structure Hamiltonian, and the (anti)commutation relations among matrices characterizing the four-fermion interaction channels and the order parameters, see Fig.~\ref{fig:flowchart}. Subsequently, these predictions can be anchored by performing an unbiased RG analysis, as we show here and also in Refs.~\cite{PhysRevB.99.121407, PhysRevB.103.165139}.

To briefly and schematically summarize our main findings, here we focus on generic short-range or local, but \emph{repulsive} four-fermion interactions and construct various cuts of the phase diagram for MLG and BLG at zero and finite temperature and chemical doping by performing an unbiased leading-order renormalization group (RG) analysis in the spirit of appropriate $\epsilon$ expansions. Typically, these cuts show excitonic orders at zero chemical doping, among which charge density wave, antiferromagnet, quantum anomalous and spin Hall insulators, and various nematic and smectic liquids are the dominant ones. Often insulators (such as, antiferromagnet) are realized at the lowest temperature, while their nodal counterparts (such as, nematic and smectic orders) nucleate at a relatively higher temperature. On the other hand, as the chemical doping is tuned away from the band touching points some pairing order sets in at the lowest temperature that maximally gaps the underlying Fermi surface, following the spirit of the Kohn-Luttinger mechanism~\cite{PhysRevLett.15.524, doi:10.1142/S0217979292001249}.

\begin{figure}[t!]
\includegraphics[width=0.475\linewidth]{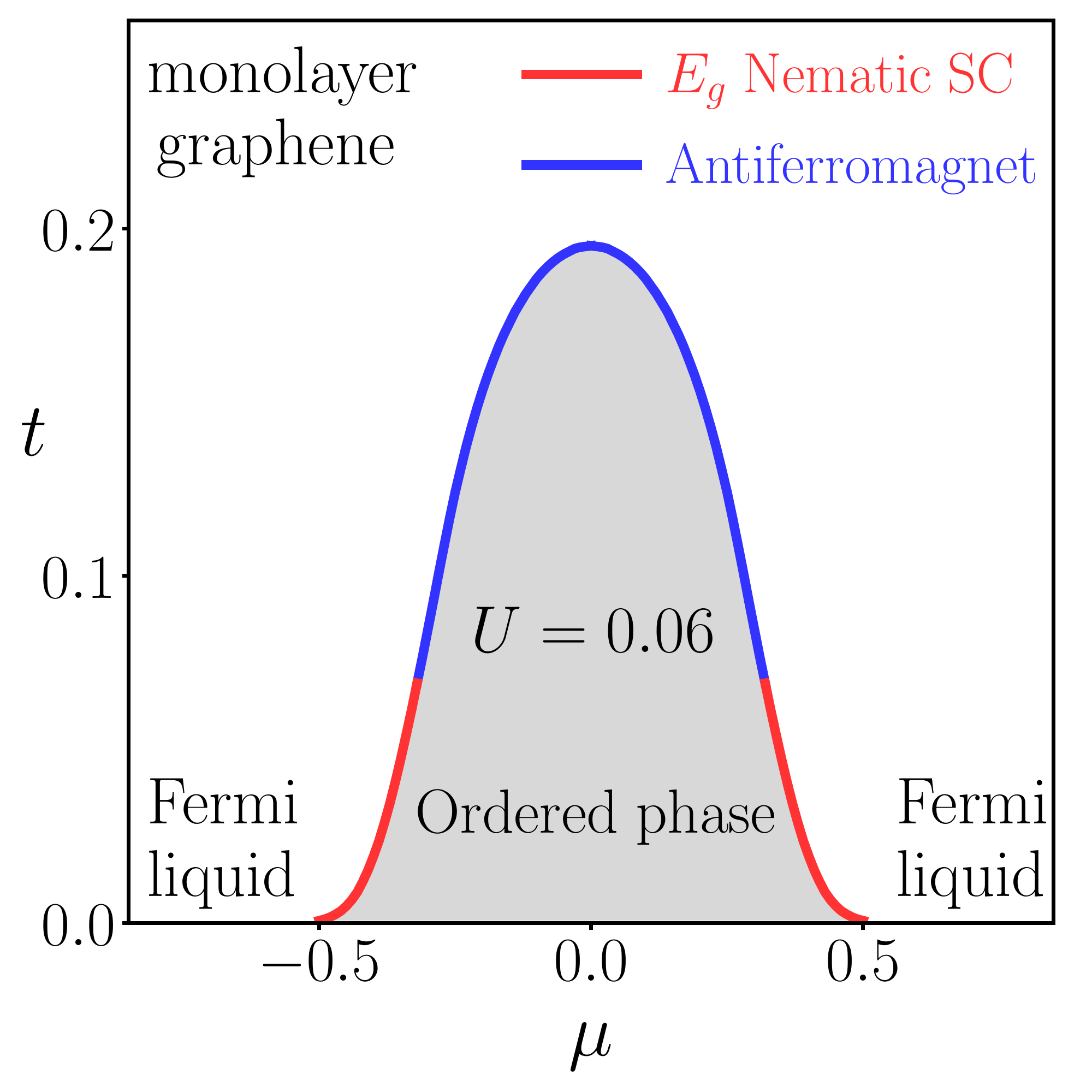}
\includegraphics[width=0.475\linewidth]{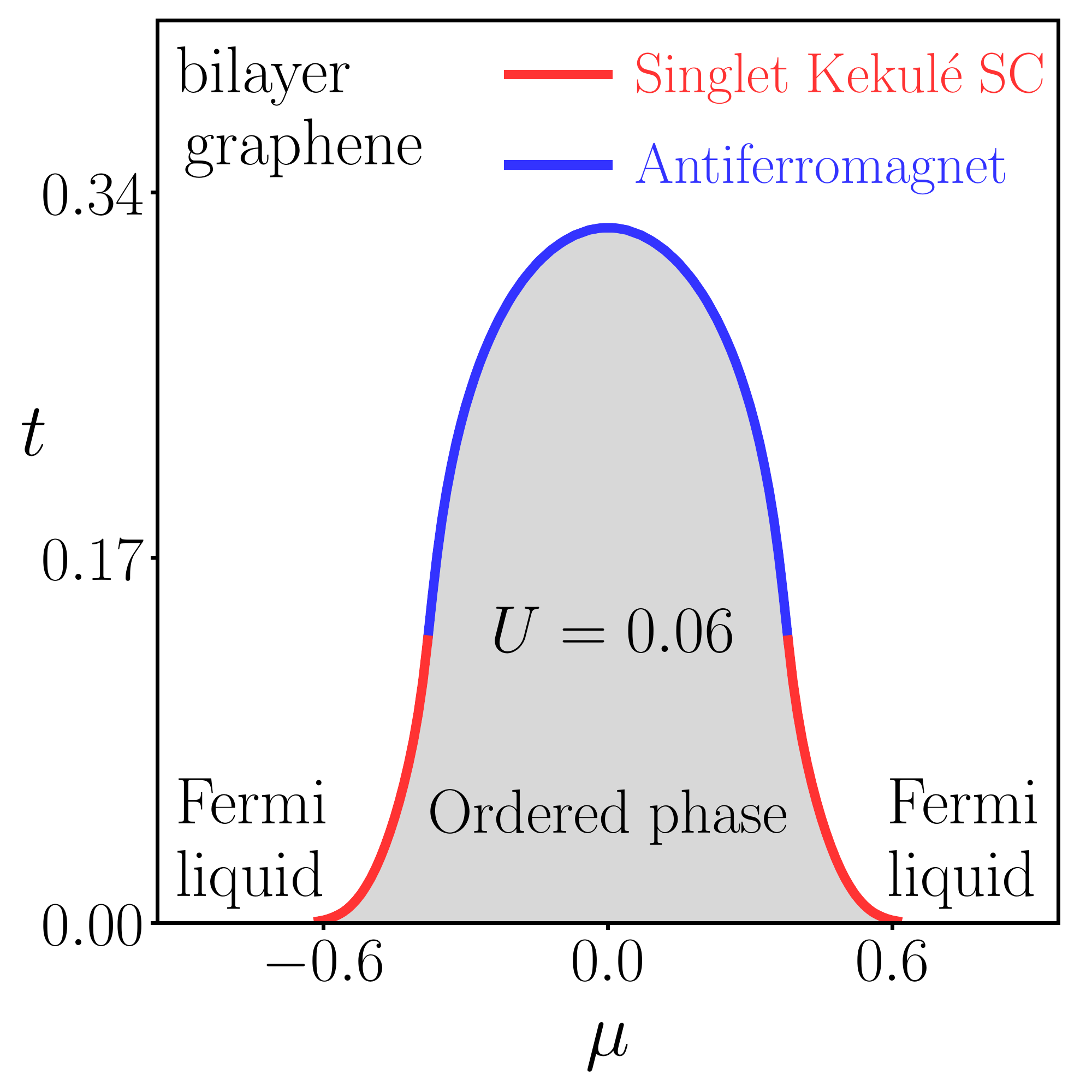} 
\caption{Phase diagrams of the Hubbard model, obtained from a RG analysis in (a) MLG and (b) BLG for a fixed strength of onsite Hubbard repulsion ($U$) as a function of temperature ($t$) and chemical doping ($\mu$), respectively measured from the linear and biquadratic band touching points. All the quantities $U, t$ and $\mu$ are dimensionless (see text for definition). The white region represents disordered chiral Fermi liquid or semimetal at finite or zero doping, respectively, and the gray region depicts an ordered state. Along the phase boundary between them, the system supports an antiferromagnet at and near, and a superconductor (SC) away from the half-filling ($\mu=0$). Nature of the SC is mentioned in each panel. Realizations of distinct paired states in MLG and BLG solely stem from the distinct normal state band structures in these two systems. Inside the ordered state, one expects a regime of coexistence between antiferromagnet and the adjacent SC, due to their internal O(5) symmetry, with pure phases on either side of it. Our RG analysis, however, cannot capture such coexistence, which nevertheless can be demonstrated from a standard mean-field analysis~\cite{PhysRevB.99.121407}. Here, $\mu>0$ ($\mu<0$) correspond to electron (hole) doped systems.}
~\label{fig:2D_HubbardU}
\end{figure}

To relate our findings to a more experimentally feasible scenario, we construct the phase diagram of (un)doped MLG and BLG, when electrons therein interact with each other only through on-site Hubbard repulsion. Results are shown in Figs.~\ref{fig:2D_HubbardU}(a) and \ref{fig:2D_HubbardU}(b), respectively. At zero doping, both systems support antiferromagnetic ordering. However, at finite doping while a spin-singlet \emph{nematic} pairing develops in MLG at the lowest temperature, in doped BLG, the paired state represents a translational symmetry breaking spin-singlet Kekul\'e superconductor. It is an example of a commensurate Fulde-Ferrell-Larkin-Ovchinikov pairing~\cite{PhysRev.135.A550, larkin:1964zz}, also known as pair-density-wave. Our predictions can be experimentally tested on cold atomic systems, for example, where both MLG and BLG have been engineered, and both doping and the strength of the Hubbard repulsion can be tuned efficiently~\cite{annurev-conmatphys-070909-104059, 2012Natur.483..302T, PhysRevLett.111.185307, PhysRevLett.118.240403}. In addition, electronic BLG supports various ordered phases near the half filling that include both insulating and gapless (possibly nematic) states~\cite{Weitz812, Mayorov860, PhysRevLett.108.076602, PhysRevB.85.155412, Velasco2012, PhysRevB.87.161402}. Therefore, increasing the carrier density by applying an external gate voltage one can, at least in principle, induce superconductivity in such systems, where our predictions can be tested directly.

%%%%%%%%%%%%%%%%%%%%%%%%%%%%%%%%%%%%%%%%%%%%%%%%%%%%%%%%%%%%%%%%%%%%%%%%%
%%%%%%%%%%%%%%%%%%%%%%%%%%%%%%%%%%%%%%%%%%%%%%%%%%%%%%%%%%%%%%%%%%%%%%%%%
%%%%%%%%%%%%%%%%%%%%%%%%%%%%% LATTICE FIGURES %%%%%%%%%%%%%%%%%%%%%%%%%%%
%%%%%%%%%%%%%%%%%%%%%%%%%%%%%%%%%%%%%%%%%%%%%%%%%%%%%%%%%%%%%%%%%%%%%%%%%
%%%%%%%%%%%%%%%%%%%%%%%%%%%%%%%%%%%%%%%%%%%%%%%%%%%%%%%%%%%%%%%%%%%%%%%%%
%%%%%%%%%%%%%%%%%%%%%%%%%%%%%%%%%%%%%%%%%%%%%%%%%%%%%%%%%%%%%%%%%%%%%%%%%
%%%%%%%%%%%%%%%%%%%%%%%%%%%%%%%%%%%%%%%%%%%%%%%%%%%%%%%%%%%%%%%%%%%%%%%%%
%%%%%%%%%%%%%%%%%%%%%%%%%%%%%%%%%%%%%%%%%%%%%%%%%%%%%%%%%%%%%%%%%%%%%%%%%
%%%%%%%%%%%%%%%%%%%%%%%%%%%%%%%%%%%%%%%%%%%%%%%%%%%%%%%%%%%%%%%%%%%%%%%%%
%%%%%%%%%%%%%%%%%%%%%%%%%%%%%%%%%%%%%%%%%%%%%%%%%%%%%%%%%%%%%%%%%%%%%%%%%

\section{Extended summary}

In this paper, we examine the effects of short range electronic interactions in MLG and BLG at zero, as well as finite temperature and chemical potential. Our aim in this undertaking is two-fold. On the one hand, we present a standalone perturbative analysis within an unbiased leading-order Wilsonian momentum-shell RG framework. On the other hand, we apply a set of \emph{selection rules}, proposed by the authors in Ref.~\cite{PhysRevB.103.165139}, that organize the nucleating phases along various axes of the phase diagram, e.g., the interaction strength, temperature, and chemical doping, and thereby further demonstrate the versatility and applicability of these rules in general correlated multiband electronic systems. The different band structures but identical underlying point group symmetries of MLG and BLG makes the study of these two systems in tandem specifically suited for this purpose, about which more in Sec.~\ref{sec:extendedsum:selection}. In this section we give an outline of our findings, including the quintessential technical details. Interested readers are invited to consult Secs.~\ref{sec:lattice_models}--\ref{sec:HubbardModel} for more detailed discussion.

%%%%%%%%%%%%%%%%%%%%%%%%%%%%%%%%%%%%%%%%%%%%%%%%%%%%%%%%%%%%%%%%%%%%%%%%%
%%%%%%%%%%%%%%%%%%%%%%%%%%%%%%%%%%%%%%%%%%%%%%%%%%%%%%%%%%%%%%%%%%%%%%%%%
%%%%%%%%%%%%%%%%%%%%%%%%%% 2D HUBBARD U FIGURE %%%%%%%%%%%%%%%%%%%%%%%%%%
%%%%%%%%%%%%%%%%%%%%%%%%%%%%%%%%%%%%%%%%%%%%%%%%%%%%%%%%%%%%%%%%%%%%%%%%%
%%%%%%%%%%%%%%%%%%%%%%%%%%%%%%%%%%%%%%%%%%%%%%%%%%%%%%%%%%%%%%%%%%%%%%%%%
%%%%%%%%%%%%%%%%%%%%%%%%%%%%%%%%%%%%%%%%%%%%%%%%%%%%%%%%%%%%%%%%%%%%%%%%%
%%%%%%%%%%%%%%%%%%%%%%%%%%%%%%%%%%%%%%%%%%%%%%%%%%%%%%%%%%%%%%%%%%%%%%%%%

Our analysis is built on simple tight binding models of MLG and BLG, consisting of a single and double layer of carbon-based honeycomb lattice, respectively. In MLG [see Fig.~\ref{fig:lattices}(a)], we account for the nearest neighbor (NN) hopping only. The BZ of MLG is itself a honeycomb structure, rotated by 90 degrees compared to the real-space lattice, see Fig.~\ref{fig:lattices}(b). The NN hopping results in six Dirac cones located at the six corners of the BZ, only two of which are inequivalent, while the rest are related to them by reciprocal lattice vectors. In Bernal stacked BLG [Fig.~\ref{fig:lattices}(c)], besides the intralayer NN hopping in each layer, we couple the two honeycomb sheets via direct interlayer hopping terms. The associated BZ is analogous to that of MLG, but the band structure now displays biquadratic touchings at the six corners of the BZ, again only two of which are inequivalent. Furthermore, the additional layer degree of freedom gives rise to two gapped bands that are localized to the overlapping dimer sites. These so called split-off bands can be integrated out to obtain an effective description of the low lying biquadratic band touching. The resulting effective lattice is another honeycomb lattice, see Fig.~\ref{fig:lattices}(d). Consequently, MLG and BLG, possessing identical $D_{3d}$ point group symmetry, are described by equal number of low-energy degrees of freedom, while their distinct normal state band structures allow us to investigate its role in the nature of the ordered phases.

Our main object of interest is the low-energy continuum description of interacting spin-1/2 electrons in MLG and BLG, captured by their respective action $S=S_0+S_{\rm int}$, assuming the schematic form
\begin{align}
S_0=&\int \D \tau \int \D^d\vec{r} \Psi^\dag [ \partial_\tau+\hat{h}(\vec{k} \to -i {\boldsymbol \nabla})-\hat{\mu} ] \Psi, \\
S_{\rm int}=& g_{\mu \nu \lambda \rho} \int \D \tau \int \D^d\vec{r} (\Psi^\dag \Gamma_{\mu \nu } \Psi) (\Psi^\dag \Gamma_{\lambda \rho}\Psi),
\end{align}
with $\vec{r}$ being position, $\tau$ imaginary time, $d$ the spatial dimensionality, and $\Psi$ and $\Psi^\dag$ are independent conjugate Grassmann variables. The chemical potential term $\hat{\mu}$ is set to zero unless otherwise mentioned. The operator $\hat{h}(\vec{k})$ in the momentum space describes the band structure, which in the continuum formalism consists of two copies of a linear (quadratic) band touching points in MLG (BLG) (see Sec.~\ref{sec:lowenergy_noninteract}), after accounting for the two inequivalent valleys. Local or momentum-independent, electronic interactions are described by the quartic terms in $S_{\rm int}$, where $\Gamma_{\mu \nu}$ are Hermitian matrices (with $\mu,\nu=0,1,2,3$), $g_{\mu \nu \lambda \rho}$ are coupling constants and the summation over repeated indices is assumed.

Power counting provides the scaling dimension of the coupling constants to be $[g_{\mu \nu \rho \lambda}]=z-d$, where the dynamical critical exponent $z=1 \; (2)$ for MLG (BLG). Therefore, sufficiently weak contact interactions are irrelevant in MLG in the RG sense. However, the corresponding linearly dispersing fermions undergo a continuous quantum phase transition, when the interactions become sufficiently strong. In comparison, local interactions are marginal in BLG and the quadratically dispersing chiral quasiparticles display ``BCS-like'' instabilities in the presence of infinitesimal interactions.

We construct the interacting Lagrangian containing all symmetry-allowed local or intra-unit cell four-fermion terms. We show that momentum-independent electronic interactions are described by only 9 linearly independent quartic terms. See Appendix~\ref{app:Fierz}. On the other hand, spontaneous symmetry breaking is described via 27 local order parameters. Of them, 18 are excitonic, which are grouped into 9 spin singlet and 9 spin triplet orders, see Table~\ref{tab:bilinears_exc}. The pairing sector accommodates in total 9 local pairings, out of which 5 are spin singlets and the remaining 4 are spin triplets, see Table~\ref{tab:bilinears_pair}.

Following the spirit of the $\epsilon$ expansion around the lower critical dimension $d_c=z$, with $\epsilon=d-z$, we carry out a one-loop Wilsonian RG analysis in the individual interaction channels in MLG and BLG. We identify the dominant instabilities, thereby constructing various cuts of the global phase diagram, that are displayed in Figs.~\ref{fig:mass_PD}, \ref{fig:nematic_PD}, \ref{fig:smectic_PD} and \ref{fig:Kekule_PD}. We claim that the selection of ordered phases for a given dominant interaction channel is not arbitrary, rather in general obeys certain algebraic principles, which we discuss in Sec.~\ref{sec:extendedsum:selection}. See also Fig.~\ref{fig:flowchart}.

\begin{figure*}[t]
\includegraphics[width=0.41\linewidth]{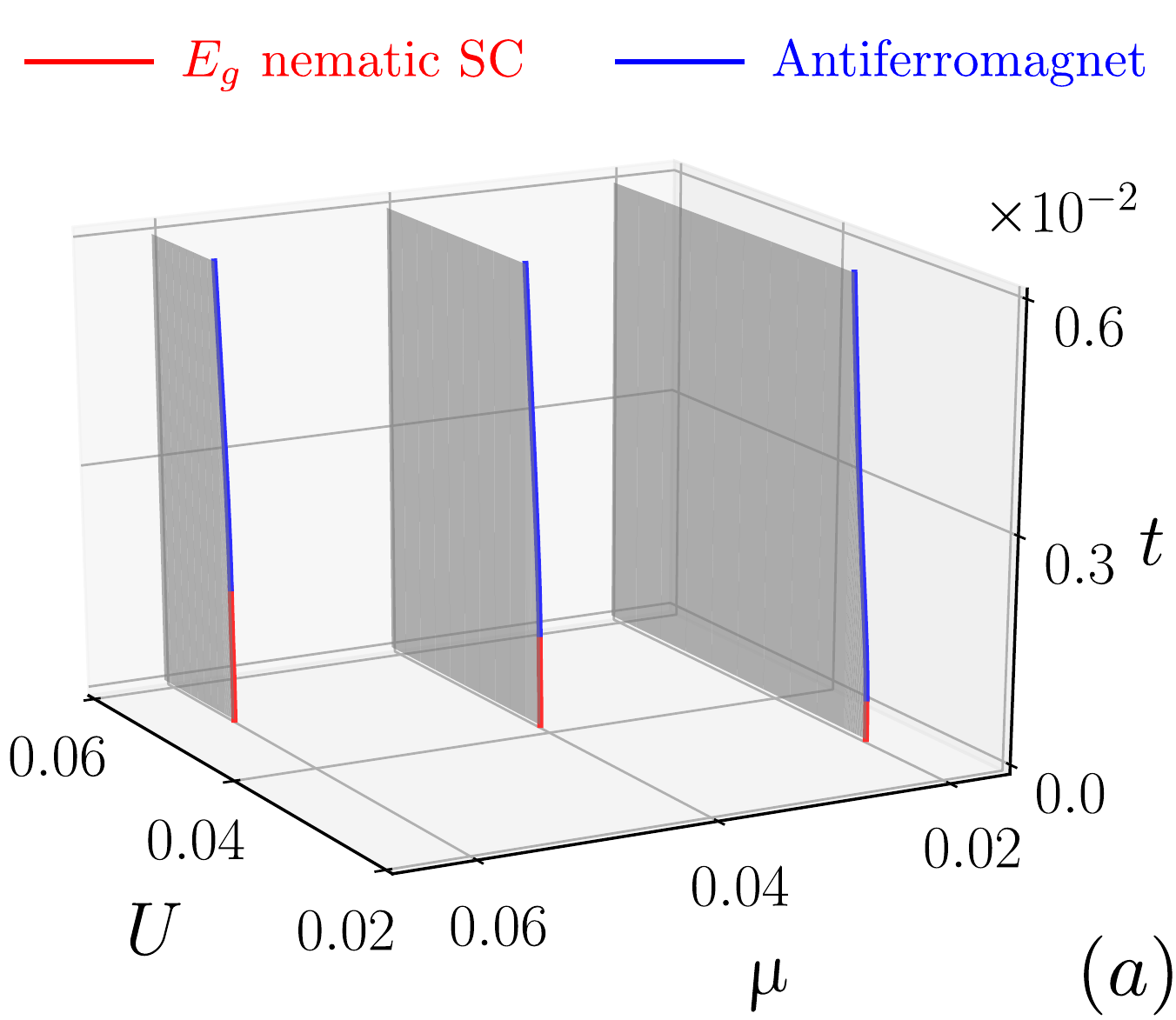}\hspace{1.25cm}
\includegraphics[width=0.425\linewidth]{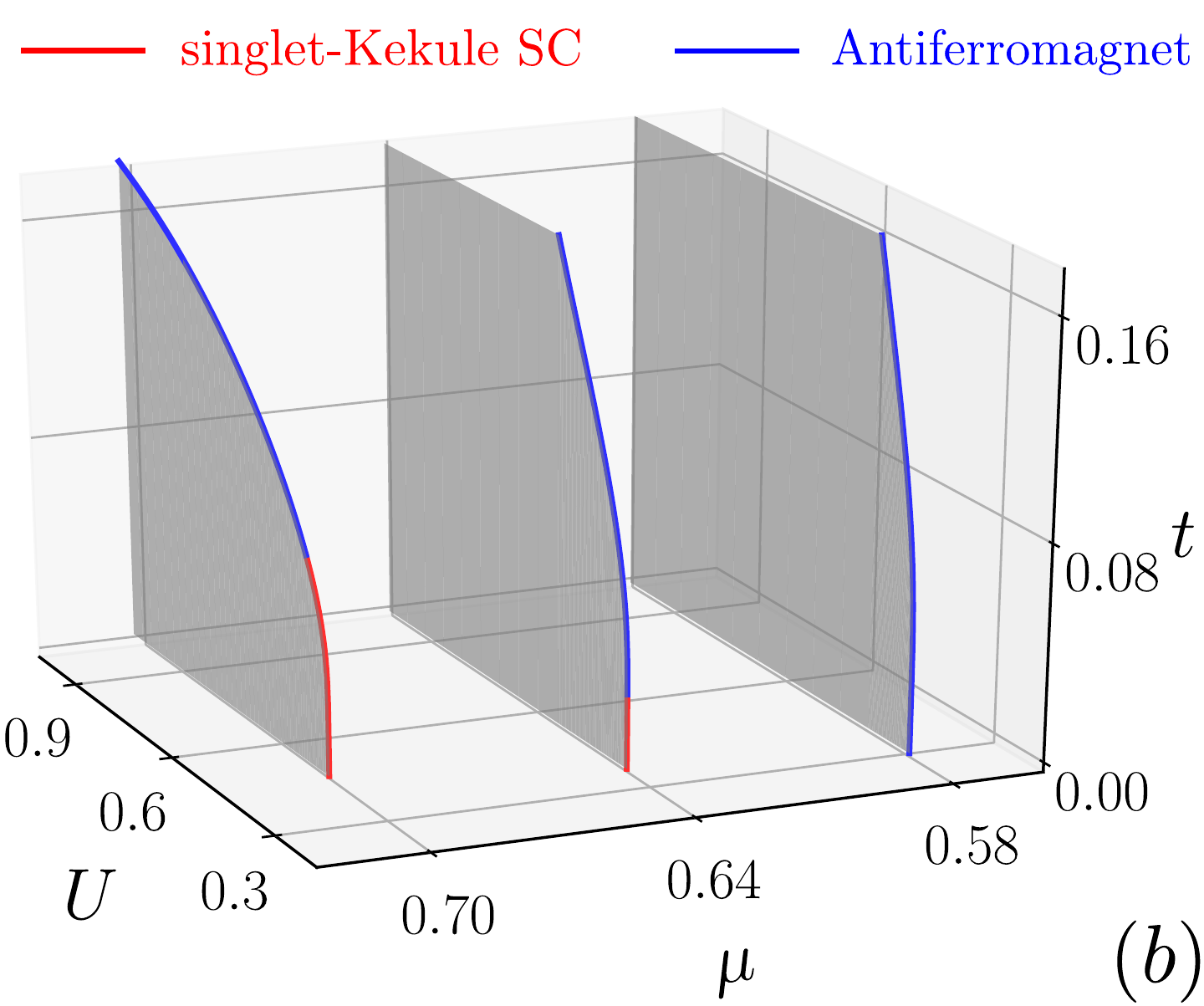}\\
\caption{Representative cuts of the phase diagram of the extended honeycomb Hubbard model described by Eqs.~(\ref{eq:ExtHub1_summary}) and (\ref{eq:ExtHub2_summary}), in the presence of only the on-site repulsion ($U>0$). Each cut is shown in the $(U,t)$ plane, where $t$ is dimensionless temperature and $U$ is the strength of on-site repulsive interaction, for various fixed values of dimensionless chemical potential $\mu$ [see Eq.~(\ref{eq:dimless_pars})]. The shaded region corresponds to the ordered phase, while colored lines indicate phase boundaries of second order phase transitions from the disordered phase (white region), which at finite doping represents chiral Fermi liquids. Along such phase boundaries the blue (red) lines represent excitonic (superconducting) orders, whose nature we explicitly highlight in each panel.  The effects of only $U>0$ are displayed in (a) and (b) for MLG and BLG, respectively. Here $U$ parameterizes the bare dimensionless onsite interaction strength. In both cases, the adjacent superconductor (SC) and excitonic phases form composite O(5) order parameters, shown in Eq.~(\ref{eq:SOP_HubbardU}).}
\label{fig:Hubbard_U}
\end{figure*}

\begin{figure*}[t]
\includegraphics[width=0.45\linewidth]{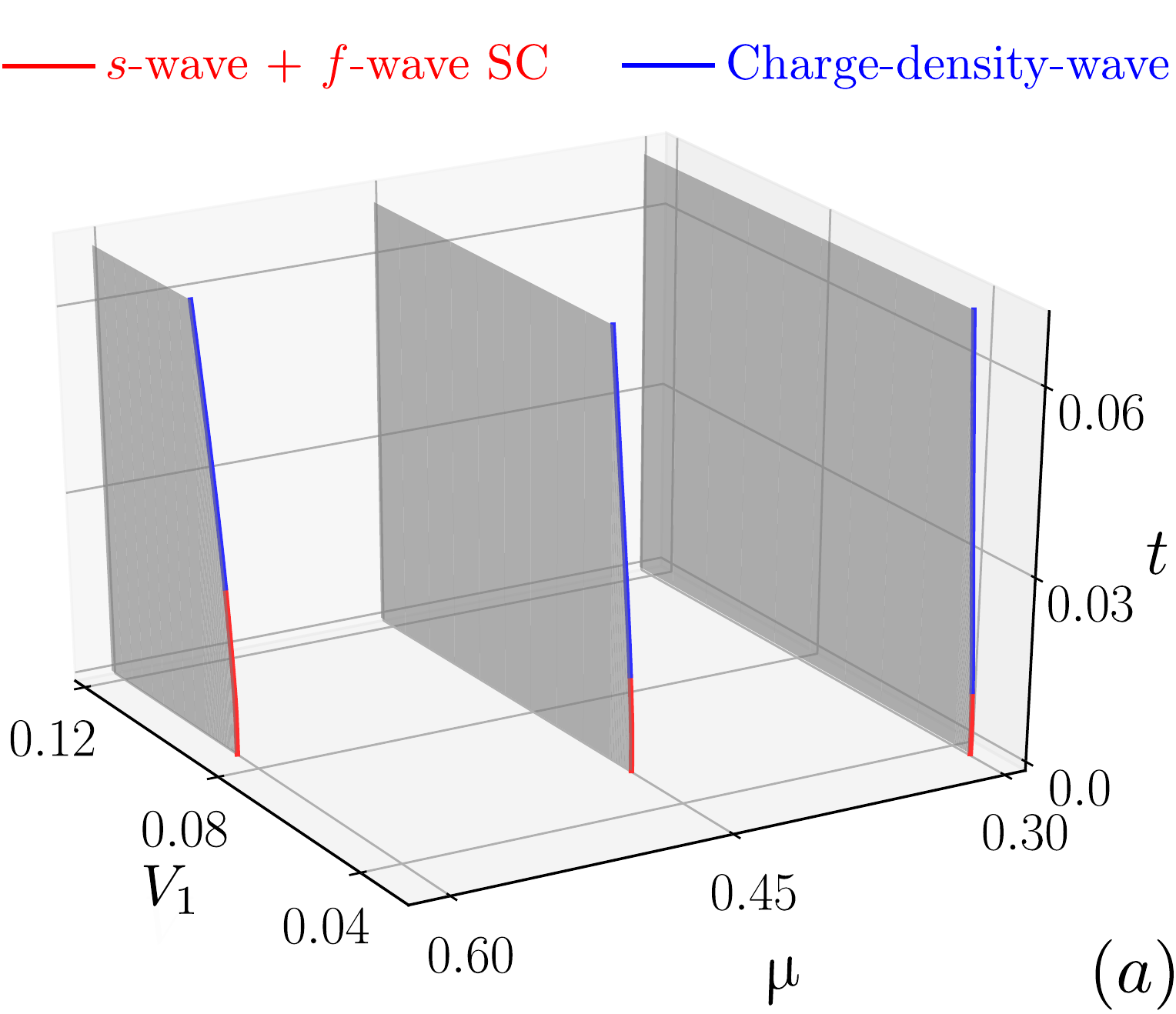}\hspace{1.25cm}
\includegraphics[width=0.41\linewidth]{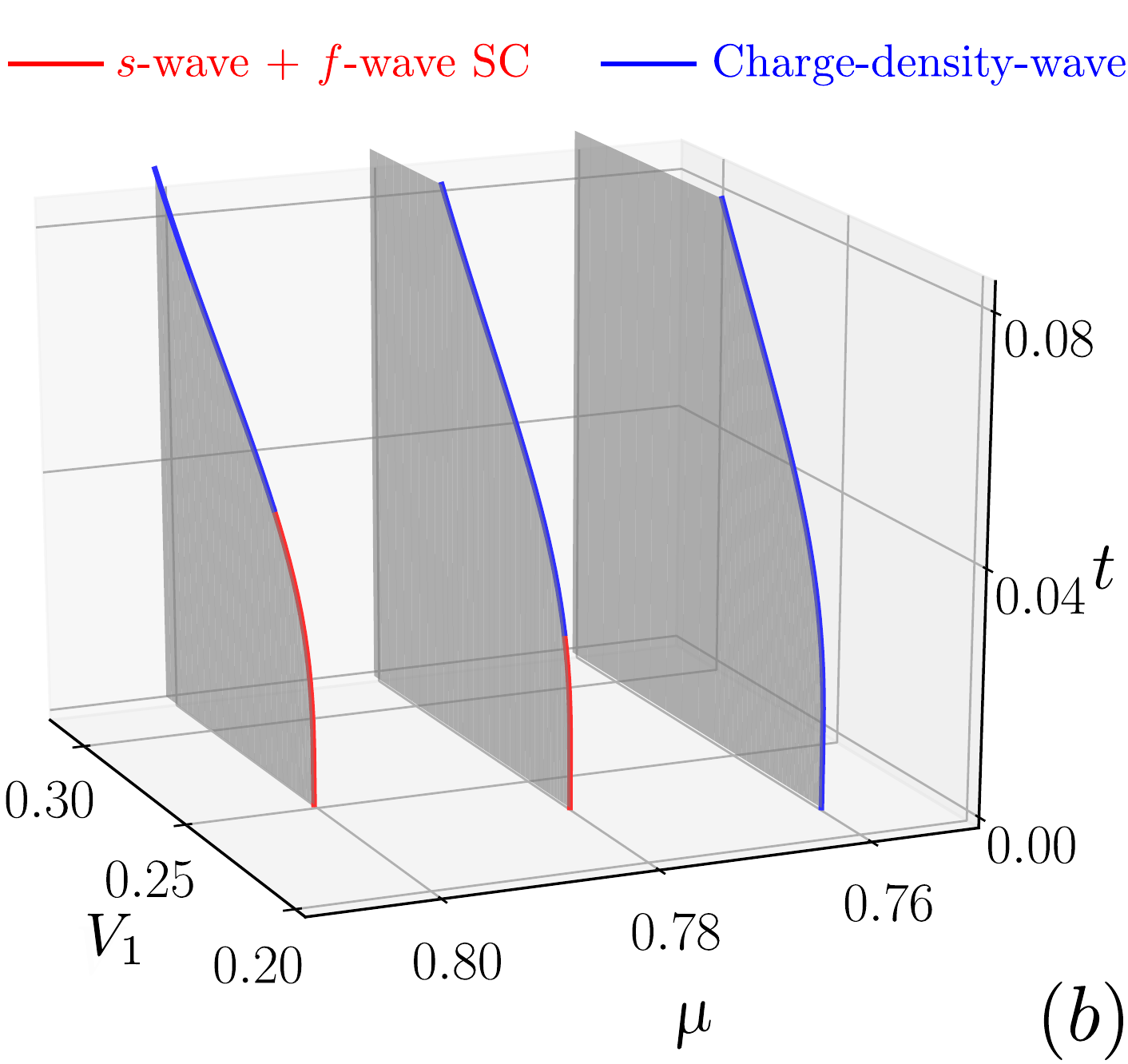}
\caption{Representative cuts of the phase diagram of the extended honeycomb Hubbard model [see Eqs.~(\ref{eq:ExtHub1_summary}) and (\ref{eq:ExtHub2_summary})] in the presence of only the nearest-neighbor (NN) repulsion ($V_1>0$). Each cut is shown in the $(V_1,t)$ plane, for various fixed values of chemical potential $\mu$. The effects of $V_1>0$ are shown in (a) and (b) for MLG and BLG, respectively, where $V_1$ parametrizes the bare dimensionless NN interaction strength. The adjacent $s$-wave and $f$-wave superconductors (SCs) and excitonic phases respectively form O(3) and O(4) supervector order parameters, shown in Eq.~(\ref{eq:SOP_HubbardV}). The rest of the details are identical to the ones in Fig.~\ref{fig:Hubbard_U}.}
\label{fig:Hubbard_V1}
\end{figure*}

\begin{figure*}[t]
\includegraphics[width=0.43\linewidth]{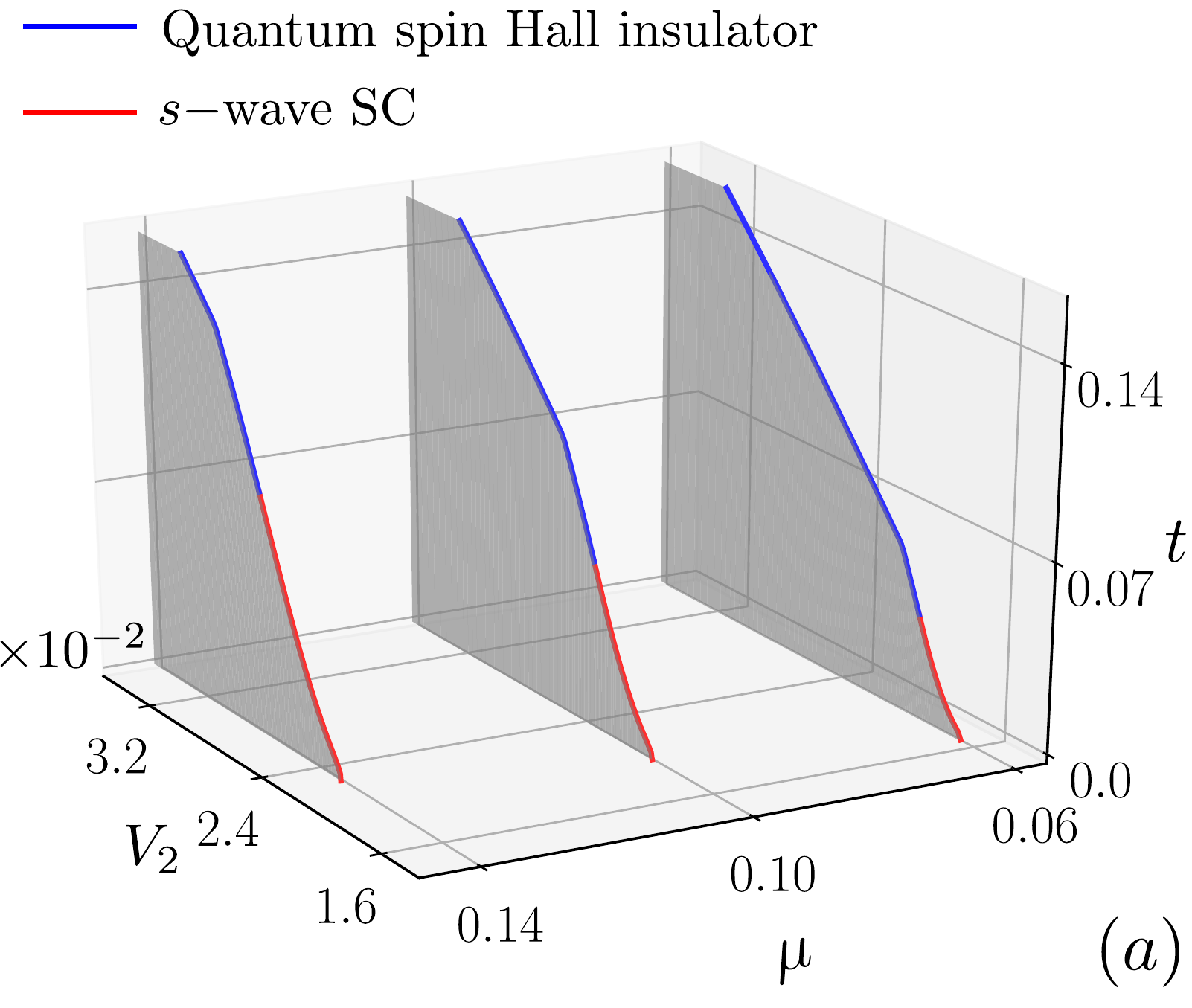}\hspace{1.25cm}
\includegraphics[width=0.43\linewidth]{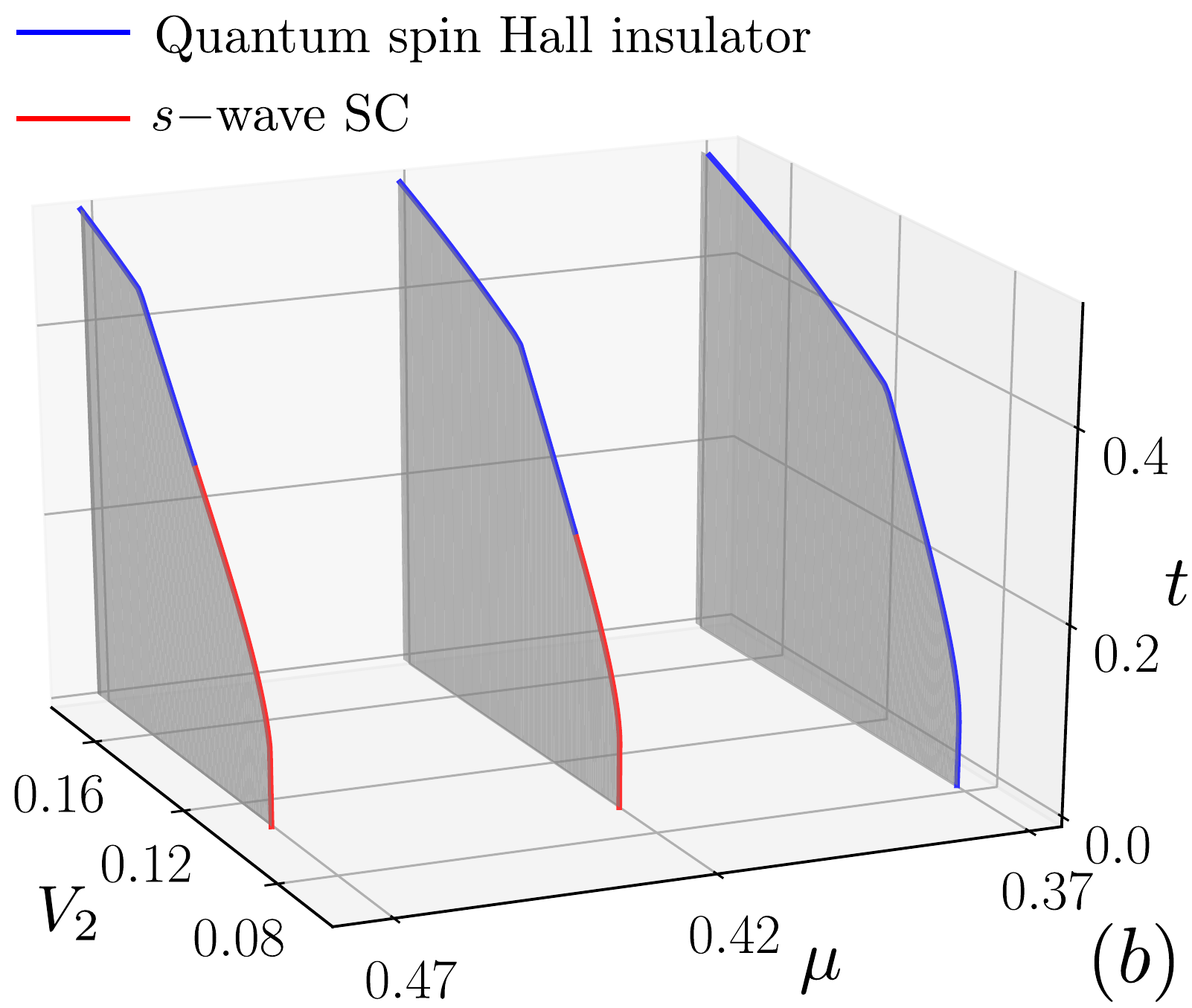}
\caption{Representative cuts of the phase diagram of the extended honeycomb Hubbard model [see Eqs.~(\ref{eq:ExtHub1_summary}) and (\ref{eq:ExtHub2_summary})] in the presence of only the next nearest-neighbor (NNN) repulsion ($V_2>0$). Each cut is shown in the $(V_2,t)$ plane, for various fixed values of chemical potential $\mu$. The effects of $V_2>0$ are shown in (a) and (b) for MLG and BLG, respectively, where $V_2$ parametrizes the bare dimensionless NNN interaction strength. The adjacent superconductor (SC) and excitonic phases form an O(5) supervector order parameter, shown in Eq.~(\ref{eq:SOP_HubbardV2}). The rest of the details are identical to the ones in Figs.~\ref{fig:Hubbard_U} and \ref{fig:Hubbard_V1}.}
\label{fig:Hubbard_V2}
\end{figure*}

Finally, in Sec.~\ref{sec:extendedsum:Hubbard}, we exemplify our findings by focusing on the microscopic extended honeycomb Hubbard model on MLG and BLG, and carry out analogous analyses in three distinct interaction channels, the on-site ($U$), NN ($V_1$), and next-nearest-neighbor ($V_2$) components of the Coulomb repulsion. Instructive cuts of the phase diagram of this model are shown in Figs.~\ref{fig:Hubbard_U}--\ref{fig:Hubbard_V2}.

%%%%%%%%%%%%%%%%%%%%%%%%%%%%%%%%%%%%%%%%%%%%%%%%%%%%%%%%%%%%%%%%%%%%%%%%%%%%%%%%%%%%%%%
%%%%%%%%%%%%%%%%%%%%%%%%%%%%%%%%%%%%%%%%%%%%%%%%%%%%%%%%%%%%%%%%%%%%%%%%%%%%%%%%%%%%%%%
%%%%%%%%%%%%%%%%%%%%%%%%% Onsite HUBBARD PHASE DIAGRAMS %%%%%%%%%%%%%%%%%%%%%%%%%%%%%%%
%%%%%%%%%%%%%%%%%%%%%%%%%%%%%%%%%%%%%%%%%%%%%%%%%%%%%%%%%%%%%%%%%%%%%%%%%%%%%%%%%%%%%%%
%%%%%%%%%%%%%%%%%%%%%%%%%%%%%%%%%%%%%%%%%%%%%%%%%%%%%%%%%%%%%%%%%%%%%%%%%%%%%%%%%%%%%%%
%%%%%%%%%%%%%%%%%%%%%%%%%%%%%%%%%%%%%%%%%%%%%%%%%%%%%%%%%%%%%%%%%%%%%%%%%%%%%%%%%%%%%%%
%%%%%%%%%%%%%%%%%%%%%%%%%%%%%%%%%%%%%%%%%%%%%%%%%%%%%%%%%%%%%%%%%%%%%%%%%%%%%%%%%%%%%%%
%%%%%%%%%%%%%%%%%%%%%%%%%%%%%%%%%%%%%%%%%%%%%%%%%%%%%%%%%%%%%%%%%%%%%%%%%%%%%%%%%%%%%%%
%%%%%%%%%%%%%%%%%%%%%%%%%%%%%%%%%%%%%%%%%%%%%%%%%%%%%%%%%%%%%%%%%%%%%%%%%%%%%%%%%%%%%%%
%%%%%%%%%%%%%%%%%%%%%%%%%%%%%%%%%%%%%%%%%%%%%%%%%%%%%%%%%%%%%%%%%%%%%%%%%%%%%%%%%%%%%%%

\subsection{Selection rules}~\label{sec:extendedsum:selection}

In Ref.~\cite{PhysRevB.103.165139}, the authors introduced a set of directives that organize possible broken symmetry phases in the global phase diagram of interacting (effective) spin-3/2 fermions in a three-dimensional (3D) Luttinger (semi)metal~\cite{PhysRev.102.1030}, displaying bi-quadratic band touching in the normal state.\footnote{In the same spirit, the biquadratic fermions in two dimensions are called Luttinger fermions and Bernal stacked BLG is often referred to as the Luttinger system or semimetal.} These consist of (I) a set of selection rules that allow only two types of fermion bilinear order parameters to be boosted by a certain four-fermion interaction channel, and (II) a generalized energy-entropy argument that organizes these ordered phases along the temperature axis. We here demonstrate the general applicability of these rules by verifying them in a vastly different system. While the Luttinger semimetal is realized in 3D strong spin-orbit coupled systems, MLG and BLG are constituted by two-dimensional honeycomb membranes of carbon atoms. In these systems, due to the low atomic number of carbon, the spin-orbit coupling is negligibly small. In fact we completely neglect spin-orbit coupling throughout this paper. Here we quote these two rules for the sake of completeness.

Let the band structure be described by the Hamiltonian $\hat{h}$, containing Hermitian matrices $H_i$. Consider a local interaction channel and an arbitrary but symmetry allowed local order parameter schematically written as 
\begin{align}
g_{_M}\sum_j(\Psi^\dag M_j \Psi)^2 \hspace{0.5cm}\text{and}\hspace{0.5cm} \Delta_O\sum_k(\Psi^\dag O_k \Psi), \nonumber 
\end{align}
respectively. Here $g_{_M}$ is the coupling constant, and $\Delta_O$ is the conjugate field to the fermion bilinears. Moreover, we define the following quantities. Let $A_M$ and $C_M$ respectively be the number of anticommuting and commuting matrix pairs between an order parameter and a four fermion term. Also, let $A_H$ and $C_H$ respectively be the number of anticommuting and commuting matrix pairs between an order parameter and the single-particle Hamiltonian. We can summarize these relations as~\footnote{We write all matrices as Kronecker products of Pauli matrices as $\Gamma_{\mu \nu \dots}=\sigma_{\mu}\otimes \sigma_{\nu}\dots$, therefore any two matrices either commute or anticommute.}
\begin{align}
&A_H:\ \{O_k,H_i\}=0, & &A_M:\ \{O_k,M_j\}=0, \nonumber\\
&C_H:\ \left[O_k,H_i \right]=0, & &C_M:\ \left[O_k,M_j \right]=0.
\end{align}
Then, various cuts of the global phase diagram in the presence of local interactions is organized according to the following two rules.

\vspace{0.25cm}
\noindent (I) Among the available ordered phases, the interaction channel coupled by $g_{_M}$ maximally boosts nucleation of the ones satisfying
\begin{equation}
\text{(a) }O_i=M_i \hspace{0.3cm} \text{or\ \ (b) } A_M={\rm maximal}.\label{eq:rule_I}
\end{equation}
\noindent (II) Among the phases selected by (I), an ordered phase is energetically (entropically) favorable if
\begin{equation}
A_H-C_H={\rm maximal\ (minimal)}.\label{eq:rule_II}
\end{equation}

We name the two ``or'' conditions in (I) selection rules. They are operative at zero as well as at finite temperature and chemical potential, and only depend on $\hat{h}$ indirectly, in the sense that the interaction terms by construction preserve the microscopic symmetries of the noninteracting Hamiltonian. Of crucial importance is the fact that MLG and BLG are endowed with the same microscopic symmetries. Therefore (I) \emph{cannot distinguish between the two systems}. However, (II) can as we show for the extended honeycomb Hubbard model in Sec.~\ref{sec:extendedsum:Hubbard}.

Besides (I) having exactly two ways [(a) and (b)] of boosting an ordered phase by a given four-fermion interaction, it leads to the following consequence regarding the adjacent or competing phases. Namely, two broken symmetry phases that reside next to each other in a generic cut of the global phase diagram as we tune the interaction strength, temperature or chemical doping, involving the following two sets of matrices say
\begin{align*}
&\{ O^{(1)}_k \}, \quad \ k=\{1,\dots, K\}, \\
&\{ O^{(2)}_l \}, \quad \ l=\{1,\dots, L\},
\end{align*} 
that transform as vectors under O(K) and O(L) rotations, respectively, form \emph{composite} or \emph{supervector} order parameters that transform as vectors under the group O(N), where 
$$K,L<N\leq K+L.$$ 
According to (I) it has to be that $O^{(1)}_k=M_k$ and $\{ O^{(2)}_l, M_k \}=0$, or the other way around, so that the two sets of matrices $\{ O^{(1)}_k \}$ and $\{ O^{(2)}_l \}$ contain exactly $N$ mutually anticommuting matrices.

%%%%%%%%%%%%%%%%%%%%%%%%%%%%%%%%%%%%%%%%%%%%%%%%%%%%%%%%%%%%%%%%%%%%%%%%%%%%%%%%%%%%%%%
%%%%%%%%%%%%%%%%%%%%%%%%%%%%%%%%%%%%%%%%%%%%%%%%%%%%%%%%%%%%%%%%%%%%%%%%%%%%%%%%%%%%%%%
%%%%%%%%%%%%%%%%%%%%%%%%%%%%% NN HUBBARD PHASE DIAGRAMS %%%%%%%%%%%%%%%%%%%%%%%%%%%%%%%
%%%%%%%%%%%%%%%%%%%%%%%%%%%%%%%%%%%%%%%%%%%%%%%%%%%%%%%%%%%%%%%%%%%%%%%%%%%%%%%%%%%%%%%
%%%%%%%%%%%%%%%%%%%%%%%%%%%%%%%%%%%%%%%%%%%%%%%%%%%%%%%%%%%%%%%%%%%%%%%%%%%%%%%%%%%%%%%
%%%%%%%%%%%%%%%%%%%%%%%%%%%%%%%%%%%%%%%%%%%%%%%%%%%%%%%%%%%%%%%%%%%%%%%%%%%%%%%%%%%%%%%
%%%%%%%%%%%%%%%%%%%%%%%%%%%%%%%%%%%%%%%%%%%%%%%%%%%%%%%%%%%%%%%%%%%%%%%%%%%%%%%%%%%%%%%
%%%%%%%%%%%%%%%%%%%%%%%%%%%%%%%%%%%%%%%%%%%%%%%%%%%%%%%%%%%%%%%%%%%%%%%%%%%%%%%%%%%%%%%
%%%%%%%%%%%%%%%%%%%%%%%%%%%%%%%%%%%%%%%%%%%%%%%%%%%%%%%%%%%%%%%%%%%%%%%%%%%%%%%%%%%%%%%
%%%%%%%%%%%%%%%%%%%%%%%%%%%%%%%%%%%%%%%%%%%%%%%%%%%%%%%%%%%%%%%%%%%%%%%%%%%%%%%%%%%%%%%

On the other hand, we name the statement in (II) organizing principle. It arranges the phases, selected by (I), along the temperature axis. Note $C_H=0$ implies a fully and isotropically gapped state, which causes maximal gain of the condensation energy and is, therefore, favorable at the lowest temperature. On the other hand, $C_H>0$ results in gapless states at low energies, which, hence is a configuration of higher entropy. Therefore, such states are typically more prominent at higher temperatures. As such (II) is a generalized energy-entropy argument, that goes beyond the binary distinction of whether an order is gapped or gapless, and capable of organizing more than one ordered phases with $C_H>0$, but varying $A_H$, as the temperature is varied in the system. A schematic representation of the selection rules and the organizing principle is shown in Fig.~\ref{fig:flowchart}.

Chemical potential plays an important role in promoting superconductivity from pure repulsive interactions by enhancing carrier density. In the above outlined algebraic framework, the chemical doping term $\hat{\mu}$ anticommutes with all pairing orders. Therefore, for a $K$-component superconducting order parameter $A_H\to A_H+K$ as $\hat{h}\to \hat{h} - \hat{\mu}$ at finite doping. By contrast, all the excitonic orders commute with $\hat{\mu}$ and thus for an $L$-component excitonic order parameter $A_H\to A_H-L$ as $\hat{h}\to \hat{h} - \hat{\mu}$. Therefore, at finite doping pairing states get energetically favored at the lowest temperature, while excitonic orders being more entropic are favored at higher temperature. Still, the competing excitonic and pairing orders for a given interaction channel follow the selection rule (I).

%%%%%%%%%%%%%%%%%%%%%%%%%%%%%%%%%%%%%%%%%%%%%%%%%%%%%%%%%%%%%%%%%%%%%%%%%%%%%%%%%%%%%%%
%%%%%%%%%%%%%%%%%%%%%%%%%%%%%%%%%%%%%%%%%%%%%%%%%%%%%%%%%%%%%%%%%%%%%%%%%%%%%%%%%%%%%%%
%%%%%%%%%%%%%%%%%%%%%%%%%%%% NNN HUBBARD PHASE DIAGRAMS %%%%%%%%%%%%%%%%%%%%%%%%%%%%%%%
%%%%%%%%%%%%%%%%%%%%%%%%%%%%%%%%%%%%%%%%%%%%%%%%%%%%%%%%%%%%%%%%%%%%%%%%%%%%%%%%%%%%%%%
%%%%%%%%%%%%%%%%%%%%%%%%%%%%%%%%%%%%%%%%%%%%%%%%%%%%%%%%%%%%%%%%%%%%%%%%%%%%%%%%%%%%%%%
%%%%%%%%%%%%%%%%%%%%%%%%%%%%%%%%%%%%%%%%%%%%%%%%%%%%%%%%%%%%%%%%%%%%%%%%%%%%%%%%%%%%%%%
%%%%%%%%%%%%%%%%%%%%%%%%%%%%%%%%%%%%%%%%%%%%%%%%%%%%%%%%%%%%%%%%%%%%%%%%%%%%%%%%%%%%%%%
%%%%%%%%%%%%%%%%%%%%%%%%%%%%%%%%%%%%%%%%%%%%%%%%%%%%%%%%%%%%%%%%%%%%%%%%%%%%%%%%%%%%%%%
%%%%%%%%%%%%%%%%%%%%%%%%%%%%%%%%%%%%%%%%%%%%%%%%%%%%%%%%%%%%%%%%%%%%%%%%%%%%%%%%%%%%%%%
%%%%%%%%%%%%%%%%%%%%%%%%%%%%%%%%%%%%%%%%%%%%%%%%%%%%%%%%%%%%%%%%%%%%%%%%%%%%%%%%%%%%%%%

Note that (I) and (II) still leave room for \emph{degeneracy} among the selected order parameters. For example, when two bilinears gap the quasiparticle spectra and fully anticommute with the matrices in a given interaction channel, then temperature cannot distinguish between them. If, however, one of them is a pairing phase, it will be favored at low temperatures when $\mu>0$, as it optimally gaps the underlying Fermi surface [selection rule (II)]. We demonstrate the applicability of these rules and display the composite order parameters for the various interaction channels for MLG and BLG in Sec.~\ref{sec:phasediagrams}. The adjacent orders are tabulated in Tables~\ref{tab:ptransitions_sgl} and \ref{tab:ptransitions_tr}.

%%%%%%%%%%%%%%%%%%%%%%%%%%%%%%%%%%%%%%%%%%%%%%%%%%%%%%%%%%%%%%%%%%%%%%%%%%%%
%%%%%%%%%%%%%%%%%%%%%%%%%%%%%%%%%%%%%%%%%%%%%%%%%%%%%%%%%%%%%%%%%%%%%%%%%%%%
%%%%%%%%%%%%%%%%%%%%%%%%%%%%%%%%%%%%%%%%%%%%%%%%%%%%%%%%%%%%%%%%%%%%%%%%%%%%
%%%%%%%%%%%%%%%%%%%%%%%%%%%%%%%%%%%%%%%%%%%%%%%%%%%%%%%%%%%%%%%%%%%%%%%%%%%%
%%%%%%%%%%%%%%%%%%%%%%%%%%%%%%%%%%%%%%%%%%%%%%%%%%%%%%%%%%%%%%%%%%%%%%%%%%%%
\subsection{Extended honeycomb Hubbard model}~\label{sec:extendedsum:Hubbard}

We exemplify the selection rules and organizing principle outlined in the previous section via one of the simplest but instructive microscopic models for correlated electrons on the honeycomb lattice, the extended Hubbard model, captured by the Hamiltonian
\begin{align}~\label{eq:ExtHub1_summary}
H&=H_0 + H_U+H_{V_1}+H_{V_2}.
\end{align}
Here $H_0$ describes a collection of noninteracting itinerant electrons on MLG and BLG, where it respectively produces linear and quadratic band touchings, and
\allowdisplaybreaks[4]
\begin{align}
H_U&=\frac{U}{2}\sum_{\vec{R}}n_\uparrow (\vec{R}) n_\downarrow (\vec{R}),\nonumber \\
H_{V_1}&=\frac{V_1}{2} \sum_{\langle\vec{A},\vec{B}\rangle} \sum_{\sigma, \sigma'=\uparrow, \downarrow} n_\sigma(\vec{A}) n_{\sigma'}(\vec{B}), \nonumber \\
H_{V_2}&=\frac{V_2}{2} \sum_{\langle\langle\vec{R},\vec{R}'\rangle\rangle} \sum_{\sigma, \sigma'=\uparrow, \downarrow} n_\sigma(\vec{R}) n_{\sigma'}(\vec{R}'). \label{eq:ExtHub2_summary}
\end{align}
Here, $n_\sigma (\vec{R})$ is the number operator on the site at $\vec{R}$ with spin projection $\sigma=\uparrow, \downarrow$. The sites located at $\vec{A}$ ($\vec{B}$) span the A (B) sublattice of the bipartite honeycomb lattice, while $\langle \cdots \rangle$ and $\langle\langle \cdots \rangle\rangle$ denote all pairs of NN and next-nearest-neighbor (NNN) sites respectively. Note in the case of BLG these sites reside on the emergent honeycomb lattice, resulting after integrating out the dimer sites. The terms $H_U$, $H_{V_1}$, and $H_{V_2}$ then respectively describe the on-site, NN, and NNN interactions with the interaction strengths $U$, $V_1$, and $V_2$.

We address the effects of repulsive $U$, $V_1$, and $V_2$ within a RG framework separately in MLG and BLG. At zero temperature and chemical potential, and in the presence of only repulsive on-site interaction ($U>0$, but $V_1=V_2=0$), we find the nucleation of the antiferromagnetic (AFM) ordering in MLG~\cite{PhysRevLett.97.146401, PhysRevX.3.031010, PhysRevB.91.165108}, as well as in BLG~\cite{PhysRevLett.109.126402, PhysRevB.82.205106}. See Figs.~\ref{fig:Hubbard_U}(a) and \ref{fig:Hubbard_U}(b), respectively. The induced carrier density upon elevating the chemical potential from the band touching points promotes condensation of electrons into Cooper pairs, hence giving rise to superconductivity. Sufficiently low temperatures but $\mu>0$ accommodates $E_g$ nematic pairing in MLG. However, with $\mu>0$ the dominant superconductor in the presence of repulsive Hubbard-$U$ in BLG is the singlet Kekul\'e pairing phase.

Setting $U=V_2=0$, but increasing the strength of repulsive NN interaction ($V_1>0$) at $t=\mu=0$ results in a charge density wave (CDW) ordering in both MLG and BLG, see Figs.~\ref{fig:Hubbard_V1}(a) and \ref{fig:Hubbard_V1}(b). Often in the context of BLG, the CDW phase is also called the layer polarized state. Keeping the temperature sufficiently low and extending the Fermi surface by setting $\mu>0$ gives rise to $s$-wave and $f$-wave pairing phases (appearing in a degenerate fashion) in both systems. Therefore, deep inside the paired state these two systems are expected to sustain $s+if$ or $f+is$ superconductor that optimally gaps the underlying Fermi surface~\cite{PhysRevB.90.041413}.

Finally, setting $U=V_1=0$, but increasing the strength of NNN repulsion ($V_2>0$) at $t=\mu=0$ we find the nucleation of a quantum spin Hall insulator (QSHI) state in both systems, see Figs.~\ref{fig:Hubbard_V2}(a) and \ref{fig:Hubbard_V2}(b). At the same time, finite chemical doping and sufficiently low temperatures result in an $s$-wave pairing phase both in MLG and in BLG. Presently, there is an ongoing debate regarding the fate of topological insulators in MLG resulting from NNN repulsion at the half-filling~\cite{PhysRevB.92.085147, PhysRevB.92.085146, PhysRevB.89.165123, PhysRevB.88.045425, PhysRevB.101.155121}. All the exact numerical diagonalizations have been performed for spinless fermions~\cite{PhysRevB.92.085147, PhysRevB.89.165123, PhysRevB.92.085146}, and the realization or absence of quantum anomalous Hall insulator still remains unsettled. Here, from an unbiased RG analysis we show that at least for spin-1/2 fermions NNN repulsion supports QSHI in MLG. Similar RG analysis for spinless fermions clearly suggest the presence of quantum anomalous Hall insulator in half-filled MLG for strong enough NNN repulsion, which will be discussed in a forthcoming publication.

The resulting broken symmetry phases in the phase diagram of the extended Hubbard model, as well as the similarities or distinctions among the orders realized in MLG and BLG, are in accordance with our proposed selection rules. Namely, the $s$-wave and $f$-wave pairing phases, adjacent to the CDW phase in both systems with a finite NN repulsion, fully anticommute with the CDW order parameter, and form with it O(3) and O(4) composite vectors, respectively, while also opening an isotropic spectral gap in both MLG and BLG. Similarly, the $s$-wave pairing and QSHI orders, appearing adjacent to each other in the presence of NNN repulsion, constitute an O(5) composite order parameter. Like the $s$-wave pairing, QSHI ordering also gaps the spectrum in both MLG and BLG. On the other hand, the AFM order parameter, which is the dominant excitonic phase in both systems with finite on-site repulsion, fully anticommutes with the $E_g$ nematic superconductor (SC) and the singlet Kekul\'e SC, and forms O(5) composite vectors with them. The singlet Kekul\'e pairing fully commutes with the Dirac Hamiltonian in MLG. Therefore in the presence of finite chemical doping a \emph{partially} anticommuting (with the Dirac Hamiltonian) $E_g$ nematic pairing is favored in this system. In contrast, the singlet Kekul\'e SC represents a fully gapped phase in BLG~\cite{PhysRevB.88.075415}, and is thus favored over the partially anticommuting $E_g$ nematic SC. These outcomes stand as paradigmatic examples, where the normal state band structures play the decisive role in determining the ultimate pattern of the symmetry breaking from a soup of incipient competing orders [selection rule (II)], primarily chosen by the dominant local four-fermion interaction [selection rule (I)].

The phase diagrams of the extended Hubbard model share some common features. For example, with increasing chemical doping (1) any ordering sets in at stronger coupling, (2) the requisite strength for the repulsive interaction for the onset of any excitonic order gets pushed toward stronger coupling, and (3) range of interaction over which pairing phase is realized increases.

More detailed and quantitative analysis of the extended honeycomb Hubbard model is presented in Sec.~\ref{sec:HubbardModel}. On a technical note, even though the terms $H_U$, $H_{V_1}$, and $H_{V_2}$ translate to linear combinations of linearly independent interaction terms in the continuum formalism, the effect of these quartic terms can be captured qualitatively by increasing the strength of the local interactions in the AFM, CDW, and QSHI channels, respectively, see Sec.~\ref{sec:phasediagrams}. This is because these coupling constants are the ones that diverge dominantly under coarse grain while increasing the bare strengths of $U$, $V_1$, and $V_2$ of the extended Hubbard model.

Finally, we note that our conclusions for the Hubbardlike model is consistent with the findings from nonperturbative numerical analysis on honeycomb lattice~\cite{PhysRevB.98.045142, wang2020dopinginduced}. Specifically, a functional RG analysis with only NN Coulomb repulsion among spinless fermions (thus accommodating only spin-triplet pairings) reported charge density wave at and near half-filling, and $f$-wave pairing away from it~\cite{PhysRevB.98.045142}, see Fig.~\ref{fig:Hubbard_V1}. A quantum Monte Carlo simulation found nucleation of $s$-wave pairing by doping honeycomb monolayer, residing in the vicinity of a quantum spin Hall insulator~\cite{wang2020dopinginduced}, see Fig.~\ref{fig:Hubbard_V2}. Thus the proposed selection rule and organizing principle, resting on (anti)commutation relations among matrices in the interaction channel, order parameter and noninteracting Hamiltonian, should be decisive in determining the nature of ordering in correlated multiband systems.

\subsection{Organization}

The rest of this paper is organized as follows. In Sec.~\ref{sec:lattice_models} we review the lattice descriptions and the corresponding tightly bound electron models of MLG and BLG. In Sec.~\ref{sec:lowenergy_noninteract} we establish a low-energy continuum theory of the above models by means of Fourier expansion around the band touching points. In doing so, we carry over the microscopic symmetries of the honeycomb lattice so that the two descriptions are symmetry equivalent. We then leave the realm of single-particle physics and introduce short range electron-electron interactions in Sec.~\ref{sec:interactions}, which we address in a perturbative fashion via Wilsonian momentum-shell RG and the $\epsilon$ expansion schemes. A review of the possible broken symmetry phases is then provided in Sec.~\ref{sec:ordered_phases}. Section~\ref{sec:phasediagrams} is dedicated to addressing the various interaction channels one-by-one and demonstrating the realization of ordered states and their compatibility with the selection rules. Finally we revisit the problem of the extended honeycomb Hubbard model in greater details in Sec.~\ref{sec:HubbardModel}. A summary of our findings is presented in Sec.~\ref{sec:summary}. Additional technical details are relegated to the appendices.

%%%%%%%%%%%%%%%%%%%%%%%%%%%%%%%%%%%%%%%%%%%%%%%%%%%%%%%%%%%%%%%%%%%%%%%%%
%%%%%%%%%%%%%%%%%%%%%%%%%%%%%%%%%%%%%%%%%%%%%%%%%%%%%%%%%%%%%%%%%%%%%%%%%
%%%%%%%%%%%%%%%%%%%%%%% LATTICE MODELS %%%%%%%%%%%%%%%%%%%%%%%%%%%%%%%%%%
%%%%%%%%%%%%%%%%%%%%%%%%%%%%%%%%%%%%%%%%%%%%%%%%%%%%%%%%%%%%%%%%%%%%%%%%%
%%%%%%%%%%%%%%%%%%%%%%%%%%%%%%%%%%%%%%%%%%%%%%%%%%%%%%%%%%%%%%%%%%%%%%%%%
\section{Lattice models}~\label{sec:lattice_models}

In this section we introduce single-particle tight binding descriptions of electrons on the honeycomb monolayer and bilayer. Our model includes only NN hopping for MLG. But, in order to capture the coupling between two layers, we incorporate interlayer hopping terms in BLG, besides the intralayer one. In what follows we restrict ourselves to carbon-based honeycomb lattices, and as such will completely neglect spin-orbit coupling, which is extremely small for carbon atoms~\cite{RevModPhys.81.109}.

\subsection{Monolayer graphene}\label{sec:lattice_models:MLG}

The simplest tight-binding Hamiltonian for an isolated sheet of graphene describes a bipartite lattice consisting of two interpenetrating triangular sublattices A and B with the NN hopping amplitude ($t$) as~\cite{PhysRev.71.622, PhysRevLett.53.2449}
\begin{align}
H_{\mathrm{MLG}}=t\sum_{\vec{A},i}a^\dag(\vec{A}) b(\vec{A}+\boldsymbol{\delta}_i) + \mathrm{H.c.}, \label{eq:MLG_tb}
\end{align}
where $a(\vec{A})$ are fermion annihilation operators on the A sublattice, generated by linear combinations of the primitive lattice vectors $\vec{v}_1=\sqrt{3}a (1,0)$ and $\vec{v}_2=a/2 (\sqrt{3},3)$, with $a$ being the distance between neighboring atoms. On the other hand, $b(\vec{B})$ are fermion annihilation operators on the sites belonging to the B sublattice that are obtained as $\vec{B}=\vec{A}+\boldsymbol{\delta}_i$, with $\boldsymbol{\delta}_1=a/2(-\sqrt{3},-1)$, $\boldsymbol{\delta}_2=a/2(\sqrt{3},-1)$ and $\boldsymbol{\delta}_3=a(0,1)$, see Fig.~\ref{fig:lattices}(a). The reciprocal vectors are given by $\vec{q}_1=2\pi/a (1/\sqrt{3},-1/3)$ and $\vec{q}_2=4 \pi/a (0,1/3)$. The BZ has the same hexagonal structure as the real space lattice, only rotated by 90 degrees. For convenience we shift the origin of the BZ to the middle of the hexagon (the $\Gamma$ point), see Fig.~\ref{fig:lattices}(b).

The lack of the inversion symmetry of the honeycomb lattice about any given site (see Fig.~\ref{fig:lattices}(a)) results in Fermi points in the band structure of $H_{\rm MLG}$ instead of an extended Fermi surface, as obtained on a square lattice, for example. The spectra of $H_{\rm MLG}$ read
\begin{align}
\epsilon^{\rm MLG}_{\vec{k}}&=\pm t |f(\vec{k})|,
\end{align}
where
\begin{align}
f(\vec{k})&=\sum_{i=1,2,3} \; \exp \left( \vec{k} \cdot \boldsymbol{\delta}_i \right),
\end{align}
and $f(\vec{k})$ vanishes linearly at the six corners of the BZ~\cite{PhysRev.71.622}. Thus, $\epsilon^{\rm MLG}_{\vec{k}}$ supports six Dirac points, around which the low energy dispersion is linear and isotropic~\cite{PhysRevB.79.193405}. However, only two of these band touching points are inequivalent, which we take to be at $\pm\vec{K}$, with $\vec{K}=\pm\frac{4 \pi}{3\sqrt{3}a}(1,0)$. The rest are connected to these two by primitive reciprocal vectors~\cite{PhysRevLett.53.2449}.

\subsection{Bilayer graphene}

One generates the Bernal stacked BLG by adding another graphene sheet on top of Fig.~\ref{fig:lattices}(a) and shifting it by e.g. $-\boldsymbol{\delta_3}$, see Fig.~\ref{fig:lattices}(c). Due to the two layers the minimal model has twice as many degrees of freedom when compared to MLG. We denote the layer index ($i$) with a subscript on the annihilation ($a_i$ and $b_i$) and creation ($a^\dagger_i$ and $b^\dagger_i$) operators, where $i=1,2$. The tight-binding Hamiltonian including intralayer and interlayer hopping terms is of the form~\cite{PhysRevB.82.205106, PhysRevB.88.075415}
\begin{align}
H_{\mathrm{BLG}}= H^\parallel+ H^\perp_{AB} + H^\perp_{AA} + H^\perp_{BB},
\end{align}
where
\begin{align}~\label{eq:BLG_tb}
&H^\parallel  = t_\parallel \sum_{\vec{A},i} \left[ a_1^\dag (\vec{A}) b_1(\vec{B}_i^+) + b_2^\dag (\vec{A}) a_2(\vec{B}_i^-) \right] + \mathrm{H.c.}, \nonumber \\
&H^\perp_{AB} = t_\perp \sum_{\vec{A}} a_1^\dag (\vec{A}) b_2 (\vec{A}) + \mathrm{H.c.}, \nonumber \\
&H^\perp_{AA} = t_{AA} \sum_{\vec{A},i} a_1^\dag  (\vec{A}) a_2 (\vec{A}-\boldsymbol{\delta}_i) + \mathrm{H.c.}, \nonumber \\
&H^\perp_{BA} = t_{BA} \sum_{\vec{B},i} b_1^\dag  (\vec{B}) a_2 (\vec{B}+\boldsymbol{\delta}_i) + \mathrm{H.c.}.
\end{align}
Here $H^\parallel$ describes the intralayer NN hopping, where $\vec{B}_i^{\pm}=\vec{A} \pm \boldsymbol{\delta}_i$, while the remaining terms encompass the interlayer ones. The terms $H^\perp_{AA}$ and $H^\perp_{BA}$ represent remote interlayer hopping processes, which we will drop from now on. The direct hopping between the dimer sites ($a_1$ and $b_2$) is captured by $H^\perp_{AB}$. After setting $t_{AA}=t_{BA}=0$, the spectra of $H_\mathrm{BLG}$ describe four bands with the dispersions
\begin{align}
\epsilon^{\rm BLG}_{1, \bf k}&=\pm \frac{t_\parallel^2}{2 t_\perp} |f(\vec{k})|^2, \\
\epsilon^{\rm BLG}_{2, \bf k}&=\pm \frac{1}{\sqrt{2}}\sqrt{4 t_\parallel^2 |f(\vec{k})|^2+t_\perp^2+\mathcal{O}(|f(\vec{k})|^4)}. \label{eq:split-off_bands}
\end{align}
Since $f(\vec{k})$ vanishes linearly at $\pm\vec{K}$, $\epsilon^{\rm BLG}_{1, \bf k}$ describes two \emph{quadratic} band touchings, whereas the two $\epsilon^{\rm BLG}_{2, \bf k}$ bands are gapped everywhere, also called the split-off bands. Note that $H^\perp_{BA}$ produces trigonal warping that splits each quadratic band touching into four Dirac points~\cite{PhysRevB.82.201408}. In Sec.~\ref{sec:summary}, we qualitatively discuss the role of trigonal warping on the phase diagrams of interacting electrons in BLG.

In this paper we study these two systems in terms of their low-energy effective theories. As the continuum models are well established within the scientific literature~\cite{RevModPhys.81.109}, in presenting the respective single particle descriptions we will only quote the cornerstones of the derivations.

%%%%%%%%%%%%%%%%%%%%%%%%%%%%%%%%%%%%%%%%%%%%%%%%%%%%%%%%%%%%%%%%%%%%%%%%%
%%%%%%%%%%%%%%%%%%%%%%%%%%%%%%%%%%%%%%%%%%%%%%%%%%%%%%%%%%%%%%%%%%%%%%%%%
%%%%%%%%%%%%%%%%%%%% DERIVE CONTINUUM MODELS %%%%%%%%%%%%%%%%%%%%%%%%%%%%
%%%%%%%%%%%%%%%%%%%%%%%%%%%%%%%%%%%%%%%%%%%%%%%%%%%%%%%%%%%%%%%%%%%%%%%%%
%%%%%%%%%%%%%%%%%%%%%%%%%%%%%%%%%%%%%%%%%%%%%%%%%%%%%%%%%%%%%%%%%%%%%%%%%
\section{Low-energy effective theory}~\label{sec:lowenergy}

We formulate the effective continuum theories by retaining the low energy degrees of freedom of both lattice models, Eqs.~(\ref{eq:MLG_tb}) and (\ref{eq:BLG_tb}). As the split-off bands in BLG are at higher energies, and the eigenfunctions of $\epsilon^{\rm BLG}_{2, \bf k}$ are dominantly localized on the overlapping $a_1$ and $b_2$ sites, we can project out these bands with relative ease~\cite{RevModPhys.81.109, PhysRevB.82.205106, PhysRevB.88.075415}. After integrating out the $a_1$ and $b_2$ sites, the resulting lattice assumes another honeycomb structure, see Fig.~\ref{fig:lattices}(d). As such, the effective BZ has the same structure as that of MLG, with the important distinction that the band touching at the six corners are now \emph{quadratic}, in comparison to the linear band touching in MLG. Moreover, the microscopic symmetries of these two models (reflections, translation, and time reversal) are identical, about which more in a moment. Consequently, the part of the action describing local electron-electron interactions takes the same form in these two systems, see Sec.~\ref{sec:interactions}. This allows us to investigate the role of the normal state band structure in emergent phases within the same symmetry group.

Due to the small atomic number of carbon atoms we neglect the spin-orbit coupling in both MLG and BLG, and introduce the spin degree of freedom as a mere doubling of the Hamiltonian. Besides spin, we will also introduce a particle-hole index using the Nambu doubling, which facilitates the inclusion of both excitonic and superconducting phases within a unified framework. Note that the split-off bands now being projected out, the two effective models have the equal number of low energy degrees of freedom.

\subsection{Non-interacting models}~\label{sec:lowenergy_noninteract}

Let us add a spin index to the fermionic degrees of freedom and expand the fields in terms of their Fourier components as
\begin{align}
r_s(\tau,\vec{r})=\int \frac{\D \omega}{2\pi}\int \frac{\D^2 \vec{k}}{(2 \pi)^2}e^{i(\omega \tau+ \vec{k}\cdot \vec{r})}r_s (\omega,\vec{k}), \label{eq:Fourier}
\end{align}
where $\tau$ and $\omega$ are respectively imaginary time and frequency, $r=a$ and $b$ for electrons on the A and B sublattices, respectively, with the spin orientation $s=\uparrow, \downarrow$. We retain the modes near the two inequivalent valleys at $\pm\vec{K}$, and write $r_s(\omega, \pm \vec{K}+\vec{k})=r^{\pm K}_s(\omega, \vec{k})$. The eight-component spinor in Fourier space is then structured according to
\begin{align}
\Psi_{\omega,\vec{k}}=&\left[ c_{\uparrow}^{+K},  c_{\uparrow}^{-K},  c_{\downarrow}^{+K},  c_{\downarrow}^{-K} \right]^\top, \hspace{0.1cm} c_{s}^{v}=\left[ a_{s}^{v}, b_{s}^{v} \right],\label{eq:spinor}
\end{align}
where $r_{s}^{v}$ are annihilation operators for fermions on the A and B sublattices with valley index $v=\pm \vec{K}$, and spin projection $s=\uparrow, \downarrow$, and $\top$ denotes transposition. Finally, we introduce the particle-hole degree of freedom as the ``outermost'' index and write a sixteen-component Nambu doubled spinors as
\begin{align}
\Psi_{\omega, \vec{k}} \to \Psi_{\rm Nam}=
\begin{pmatrix}
\Psi_{\omega,\vec{k}} \\
\Gamma_{210} \Psi^\ast_{-\omega,-\vec{k}}
\end{pmatrix},~\label{eq:Nambudefinition}
\end{align}
where in the lower block we absorbed the unitary part of the time reversal operator $U=\Gamma_{210}$ [see Sec.~\ref{sec:Symmetries}], so that $\Psi_{\omega,\vec{k}}$ and $\Psi_{\rm Nam}$ transform identically under all symmetries. For the sake of brevity we suppress the subscript ``Nam" from now on, and unless otherwise mentioned, thus $\Psi\equiv \Psi_{\rm Nam}$ and $\Psi^\dag\equiv \Psi^\dag_{\rm Nam}$.

To capture the symmetry properties of MLG and BLG, and subsequently develop the interacting models describing these two systems, we write the matrices acting on the sixteen-dimensional spinors as $\Gamma_{\mu \nu \lambda \rho}=\eta_\mu \otimes \sigma_\nu \otimes \tau_\lambda \otimes \alpha_\rho$ where $\{ \alpha \}$, $\{ \tau \}$, $\{ \sigma \}$ and $\{ \eta \}$ are four sets of Pauli matrices that operate on the sublattice or layer, valley, spin and particle-hole degrees of freedom, respectively. Here, $\mu$, $\nu$, $\lambda$, $\rho=0,1,2,3$, and the index $0$ always corresponds to the unit matrix. In Eq.~(\ref{eq:Nambudefinition}), $\Gamma_{210}=\sigma_2 \otimes \tau_1 \otimes \alpha_0$. Note that having integrated out the overlapping $a_1$ and $b_2$ sites in BLG, the sublattice and layer description is now redundant, as all the remaining $a$ and $b$ sites reside on layers 2 and 1, respectively. Therefore, from now on we refer to this degree of freedom simply as sublattice. Also note that by virtue of writing all sixteen-dimensional matrices in terms of Kronecker products of Pauli matrices, any two such matrices either commute or anticommute.

The non-interacting effective low-energy theory of MLG is composed of linearly dispersing chiral relativistic Dirac fermions, that persist all the way up to a non-universal ultraviolet momentum cutoff $\Lambda \sim 1/a$. In comparison, the corresponding effective theory of BLG is constituted by quadratically dispersing chiral quasiparticles (also up to a ultraviolet momentum cutoff $\Lambda$). We write the Hamiltonians for MLG and BLG respectively as
\begin{align}
\hat{h}^{\rm D}(\vec{k}) &= v \Big[ \Gamma_{3031} p_1(\vec{k}) - \Gamma_{3002} p_2(\vec{k}) \Big]-\mu \Gamma_{3000}, \label{eq:H0_MLG}\\
\hat{h}^{\rm L}(\vec{k}) &= \frac{1}{2m} \Big[ \Gamma_{3001} d_1(\vec{k}) - \Gamma_{3032} d_2(\vec{k})\Big]-\mu \Gamma_{3000},\label{eq:H0_BLG}
\end{align}
where $v=ta$ is the Fermi velocity and $m=t_\perp/(2 t_\parallel^2 a^2)$ is the effective mass. The momentum-dependence is described by the $p$-wave and $d$-wave cubic harmonics, that are related to the spherical harmonics $Y_l^m$ with $l=1$ and $l=2$, respectively, such that 
\begin{align}
p_1(\vec{k})&=k_x, \:\:\: p_2(\vec{k})=k_y, \nonumber \\
d_1(\vec{k})&=  k_x^2 - k_y^2, \:\:\: d_2(\vec{k}) = 2 k_x k_y.
\end{align}

We also introduced a chemical potential ($\mu$) term, which facilitates tuning the Fermi energy to and away from the band touching points. Therefore, $\mu$ plays the analogous role as the gate voltage. In this paper we only consider electron like doping ($\mu > 0$). All the conclusions are equally germane for hole-doped ($\mu<0$) systems. The chemical doping develops an extended (one-dimensional) Fermi surface from the (zero-dimensional) Fermi points, and increases the carrier density, which is conducive for the condensation of electrons into Cooper pairs.

The above models then constitutes the following non-interacting Euclidean (imaginary time) action
\begin{equation}
S_0^j=\int \D \tau \D^d \vec{r} \Psi^\dag(\tau, \vec{r})\Big[ \partial_\tau + \hat{h}^j(\vec{k}\to -i\boldsymbol{\nabla}) \Big]\Psi(\tau, \vec{r}),\label{eq:S0}
\end{equation}
with $j=$D and L respectively for MLG and BLG. Here $d$ is the spatial dimensionality, $\Psi^{\dag}(\tau, \bf r)$ and $\Psi(\tau, \bf r)$ are the inverse Fourier transform of the spinors $\Psi^{\dag}$ and $\Psi$, and $\tau$ is the imaginary time.

%%%%%%%%%%%%%%%%%%%%%%%%%%%%%%%%%%%%%%%%%%%%%%%%%%%%%%%%%%%%%%%%%%%%%%%%%
%%%%%%%%%%%%%%%%%%%%%%%%%%%%%%%%%%%%%%%%%%%%%%%%%%%%%%%%%%%%%%%%%%%%%%%%%
%%%%%%%%%%%%%%%%%%%%%%%%%%%%%%%%%%%%%%%%%%%%%%%%%%%%%%%%%%%%%%%%%%%%%%%%%
\subsection{Symmetries and action}~\label{sec:Symmetries}

The microscopic symmetries of the honeycomb monolayer and bilayer are identical and correspond to the $D_{3d}$ point group. Here we cast these symmetries in terms of reflections and rotations, and show how they manifest in the low-energy theory. We augment the point group symmetries with translation invariance, discrete time reversal symmetry, as well as continuous spin SU(2) and pseudospin SU$_{\rm p}$(2) rotational invariance.

%%%%%%%%%%%%%%%%%%%%%%%%%%%%%%%%%%%%%%%%%%%%%%%%%%%%%%%%%%%%%%%%%%%%%%%%%%%%%%%%%%%%%%%
%%%%%%%%%%%%%%%%%%%%%%%%%%%%%%%%%%%%%%%%%%%%%%%%%%%%%%%%%%%%%%%%%%%%%%%%%%%%%%%%%%%%%%%
%%%%%%%%%%%%%%%%%%%%%%%%%%%%% BILINEARS TABLE Excitonic:  %%%%%%%%%%%%%%%%%%%%%%%%%%%%%
%%%%%%%%%%%%%%%%%%%%%%%%%%%%%%%%%%%%%%%%%%%%%%%%%%%%%%%%%%%%%%%%%%%%%%%%%%%%%%%%%%%%%%%
%%%%%%%%%%%%%%%%%%%%%%%%%%%%%%%%%%%%%%%%%%%%%%%%%%%%%%%%%%%%%%%%%%%%%%%%%%%%%%%%%%%%%%%
\begin{table*}[t!]
\renewcommand{\arraystretch}{1.4}
\begin{tabular}{|c c c c c c c c c c c c c>{\centering}m{0.7cm} m{0.7cm}<{\centering} c|}
\hline

\begin{tabular}{@{}c@{}} IREP \\ ($D_{3d}$) \end{tabular} & Matrix ($N$) & Phase & CC & CF & SB & $S$ & $T$ & tl & $R(\frac{\pi}{2})$ & TR & SU(2) & SU$_{\rm p}$(2) &\multicolumn{2}{c}{ \begin{tabular}{@{}c@{}} Mass \\ \hline MLG\hspace{0.2cm}BLG \end{tabular} } & Fig \\
\hline
$A_{1g}$ & $\Gamma_{3000}$ & fermionic density & $g^s_{_1}$ & $\Delta_{1}^s$ & $A_{1g}^s$ & 
$+$ & $+$ & $+$ & 0 & $+$ & 0 & & \ding{55} & \ding{55} & - \\
\rowcolor{RowColor}
$A_{2g}$ & $\Gamma_{0033}$ & quantum anomalous Hall ins.  & $g^s_{_2}$ & $\Delta_{2}^s$ & $A_{2g}^s$ & 
$-$ & $-$ & $+$ & 0 & $-$ & 0 & & \checkmark & \checkmark & \ref{fig:QAHI} \\

$E_{g}$  & $\Gamma_{3001}, \Gamma_{3032}$ & nematic$_1$  & $g^s_{_3}$ & $\Delta_{3}^s$ & $E_{g}^s$ & 
$+,-$ & $+,-$ & $+,+$ & 1 & $+,+$ & 0 & & \ding{55} & \ding{55} & \ref{fig:Eg_nem} \\
\rowcolor{RowColor}
$A_{1u}$ & $\Gamma_{0030}$ & chiral density & $g^s_{_4}$ & $\Delta_{4}^s$ & $A_{1u}^s$ & 
$+$ & $-$ & $+$ & 0 & $-$ & 0 & & \ding{55} & \ding{55} & - \\

$A_{2u}$ & $\Gamma_{3003}$ & charge density wave  & $g^s_{_5}$ & $\Delta_{5}^s$ & $A_{2u}^s$ & 
$-$ & $+$ & $+$ & 0 & $+$ & 0 & I & \checkmark & \checkmark & \ref{fig:CDW}\\
\rowcolor{RowColor}
$E_{u}$  & $\Gamma_{0031}, \Gamma_{0002}$ & nematic$_2$ & $g^s_{_6}$ & $\Delta_{6}^s$ & $E_{u}^s$ & 
$+,-$ & $-,+$ & $+,+$ & 1 & $-,-$ & 0 & III & \ding{55} &  \ding{55} & \ref{fig:Eu_nem}\\

$A_{1k}$ & $\Gamma_{3011}, \Gamma_{3021}$ & Kekul\' e valence bond solid  & $g^s_{_7}$ & $\Delta_{7}^s$ & $A_{1k}^s$ & 
$+,+$ & $+,-$ & $-,-$ & 0 & $+,+$ & 0 & & \checkmark & \ding{55} & \ref{fig:kek_VBS}\\
\rowcolor{RowColor}
$A_{2k}$ & $\Gamma_{0012}, \Gamma_{0022}$ &  Kekul\' e current & $g^s_{_8}$ & $\Delta_{8}^s$ & $A_{2k}^s$ & 
$-,-$ & $+,-$ & $-,-$ & 0 & $-,-$ & 0 & VI & \ding{55}  & \checkmark & \ref{fig:kek_C}\\

$E_{k}$  &  
\begin{tabular}{@{}c@{}}$\Gamma_{3010}, \Gamma_{3023}$ \\ $\Gamma_{3013}, \Gamma_{3020}$\end{tabular} &
smectic charge density wave  & $g^s_{_9}$ & $\Delta_{9}^s$ & $E_{k}^s$ & 
\begin{tabular}{@{}c@{}} $+,-$ \\ $-,+$ \end{tabular} & 
\begin{tabular}{@{}c@{}} $+,-$ \\ $+,-$ \end{tabular} & 
\begin{tabular}{@{}c@{}} $-,-$ \\ $-,-$ \end{tabular} & 
\begin{tabular}{@{}c@{}} 1 \\ 1 \end{tabular} &
 \begin{tabular}{@{}c@{}} $+,+$ \\ $+,+$ \end{tabular} & 0 & V & \ding{55} & \ding{55} & \ref{fig:smectic}\\
\hline

$A_{1g}$ & $\Gamma_{0s00}$ & ferromagnet  & $g_{_1}^t$ & $\Delta_{1}^t$ & $A_{1g}^t$ & 
$+$ & $+$  & $+$ & 0 & $-$ & 1 & & \ding{55} & \ding{55} & - \\
\rowcolor{RowColor}
$A_{2g}$ & $\Gamma_{3s33}$ & quantum spin Hall ins.  & $g_{_2}^t$ & $\Delta_{2}^t$ & $A_{2g}^t$ & 
$-$ & $-$ & $+$ & 0 & $+$ & 1 & II & \checkmark & \checkmark & \ref{fig:QSHI}\\

$E_{g}$  & $\Gamma_{0s01}, \Gamma_{0s32}$ &  spin-nematic$_1$ & $g_{_3}^t$ & $\Delta_{3}^t$ & $E_{g}^t$ & 
$+,-$ & $+,-$ & $+,+$ & 1 & $-,-$ & 1 & IV & \ding{55} & \ding{55} & \ref{fig:Eg_spin-nem}\\
\rowcolor{RowColor}
$A_{1u}$ & $\Gamma_{3s30}$ & chiral ferromagnet & $g_{_4}^t$ & $\Delta_{4}^t$ & $A_{1u}^t$ & 
$+$ & $-$ & $+$ & 0 & $+$ & 1 & & \ding{55} & \ding{55} & - \\

$A_{2u}$ & $\Gamma_{0s03}$ & antiferromagnet & $g_{_5}^t$ & $\Delta_{5}^t$ & $A_{2u}^t$ & 
$-$ & $+$ & $+$ & 0 & $-$ & 1 & & \checkmark & \checkmark & \ref{fig:AFM}\\
\rowcolor{RowColor}
$E_{u}$  & $\Gamma_{3s31}, \Gamma_{3s02}$ & spin-nematic$_2$  & $g_{_6}^t$ & $\Delta_{6}^t$ & $E_{u}^t$ & 
$+,-$ & $-,+$ & $+,+$ & 1 & $+,+$ & 1 & & \ding{55} & \ding{55} & \ref{fig:Eu_spin-nem}\\

$A_{1k}$ & $\Gamma_{0s11}, \Gamma_{0s21}$ & spin-Kekul\' e solid  & $g_{_7}^t$ & $\Delta_{7}^t$ & $A_{1k}^t$ & 
$+,+$ & $+,-$ & $-,-$ & 0 & $-,-$ & 1 & VII & \checkmark & \ding{55} & \ref{fig:s-kek_S}\\
\rowcolor{RowColor}
$A_{2k}$ & $\Gamma_{3s12}, \Gamma_{3s22}$ & spin-Kekul\' e current & $g_{_8}^t$ & $\Delta_{8}^t$ & $A_{2k}^t$ & 
$-,-$ & $+,-$ & $-,-$ & 0 & $+,+$ & 1 & & \ding{55} & \checkmark & \ref{fig:s-kek_C}\\

$E_{k}$  &  \begin{tabular}{@{}c@{}}
                   $\Gamma_{0s10}, \Gamma_{0s23}$ \\
                   $\Gamma_{0s13}, \Gamma_{0s20}$
            \end{tabular} &
smectic spin-density-wave  & $g_{_9}^t$ & $\Delta_{9}^t$ & $E_{k}^t$ & 
\begin{tabular}{@{}c@{}} $+,-$ \\ $-,+$ \end{tabular} & 
\begin{tabular}{@{}c@{}} $+,-$ \\ $+,-$ \end{tabular}& 
\begin{tabular}{@{}c@{}} $-,-$ \\ $-,-$ \end{tabular} & 
\begin{tabular}{@{}c@{}} 1 \\ 1 \end{tabular} &
\begin{tabular}{@{}c@{}} $-,-$ \\ $-,-$ \end{tabular} & 1 & & \ding{55} & \ding{55} & \ref{fig:spin-smectic}\\
\hline
\end{tabular}
\caption{Properties of the local excitonic order parameters of the schematic form $\Delta_N (\Psi^\dag N \Psi)$ on the honeycomb lattice, where $N$ is a sixteen-dimensional Hermitian matrix with $s=\{1,2,3\}$ and $\alpha=\{1,2\}$. The first (second) nine rows correspond to spin singlet (triplet) orders. The first three columns respectively display the corresponding irreducible representation (IREP) under the $D_{3d}$ group, the order parameter matrices $N$, and the physical nature of the ordered phases. The second three columns respectively show the coupling constant (CC) of quartic interaction formed as $g_{_N}(\Psi^{\dag} N \Psi)^2$, the conjugate field (CF) of the order parameter and the symbol (SB) of the ordered phase in Figs.~\ref{fig:mass_PD}-\ref{fig:Kekule_PD}. Columns 7-13 display the transformation properties of the fermion bilinears under sublattice ($S$) and valley ($T$) reflections, translation (tl), orbital rotation [$R(\pi/2)$], time reversal (TR), and spin SU(2) and pseudospin SU$_{\rm p}$(2) rotations, respectively. Here $+$ ($-$) means even (odd), whereas 0 (1) means the bilinears  transform as scalar (vector) under the corresponding rotation. The order parameters marked by the same roman numeral in the SU$_{\rm p}$(2) column are related to each other by pseudospin SU$_{\rm p}$(2) rotations, which always relates an excitonic order to a pairing one, see Table~\ref{tab:bilinears_pair}. The ``Mass'' columns display whether the order parameter gaps the fermions in MLG and BLG, where \checkmark (\ding{55}) means gapped (gapless) spectrum. The last column ``Fig'' references the figure of the corresponding phase diagram.
 }~\label{tab:bilinears_exc}
\end{table*}
%%%%%%%%%%%%%%%%%%%%%%%%%%%%%%%%%%%%%%%%%%%%%%%%%%%%%%%%%%%%%%%%%%%%%%%%%%%%%%%%%%%%%%%
%%%%%%%%%%%%%%%%%%%%%%%%%%%%%%%%%%%%%%%%%%%%%%%%%%%%%%%%%%%%%%%%%%%%%%%%%%%%%%%%%%%%%%%
%%%%%%%%%%%%%%%%%%%%%%%%%%%%%%%%%%%%%%%%%%%%%%%%%%%%%%%%%%%%%%%%%%%%%%%%%%%%%%%%%%%%%%%
%%%%%%%%%%%%%%%%%%%%%%%%%%%%%%%%%%%%%%%%%%%%%%%%%%%%%%%%%%%%%%%%%%%%%%%%%%%%%%%%%%%%%%%
%%%%%%%%%%%%%%%%%%%%%%%%%%%%%%%%%%%%%%%%%%%%%%%%%%%%%%%%%%%%%%%%%%%%%%%%%%%%%%%%%%%%%%%

The two reflection symmetries of the honeycomb lattice are sublattice ($S$) and valley ($T$) reflections~\cite{PhysRevB.79.085116}. They respectively take the form
\begin{align}
S = \Gamma_{0001} \oplus
\Bigg(
\begin{array}{c}
k_x \to k_x \\
k_y \to -k_y
\end{array}
\Bigg),
\end{align}
and
\begin{align}
T = \Gamma_{0010} \oplus
\Bigg(
\begin{array}{c}
k_x \to -k_x \\
k_y \to k_y
\end{array}
\Bigg).
\end{align}
Reflection $S$ exchanges the sublattices A$\leftrightarrow$B, but leaves the valleys invariant. Whereas under $T$, $\vec{K} \leftrightarrow -\vec{K}$, and hence it exchanges the Dirac points, but does not affect the sublattice degree of freedom.

Both the linear and quadratic band touching points are rotationally invariant at low energies. Let us denote the orbitals as $v_i=p_i$ or $d_i$ for $i=1,2$. Then, rotations by an angle $\theta$ in orbital space take the form $R(\theta)=\exp(i\theta \Gamma_{0033}/2)$, and a rotation of the Hamiltonian corresponds to an ordinary vector rotation of the two-component vector $(v_1,v_2)^\top$
\begin{align}
R(\theta) \hat{h}^j R^{-1}(\theta) \Leftrightarrow
\begin{pmatrix}
\cos \theta & -\sin \theta \\
\sin \theta & \cos \theta 
\end{pmatrix}
\begin{pmatrix}
v_1 \\
v_2
\end{pmatrix}.
\end{align}
For example, under rotation by $\theta=\pi/2$
\begin{align}
R(\pi/2) \hat{h}^j R^{-1}(\pi/2):\ 
\begin{pmatrix}
v_1 \\
v_2
\end{pmatrix}
\rightarrow
\begin{pmatrix}
-v_2\\
v_1
\end{pmatrix}.
\end{align}
The corresponding rotation of momentum axes by an angle $\theta=-\pi/(2l)$ takes the form
\begin{align}
\begin{pmatrix}
k_x\\
k_y
\end{pmatrix} \rightarrow
\begin{pmatrix}
\cos \theta & -\sin \theta \\
\sin \theta & \cos \theta 
\end{pmatrix}
\begin{pmatrix}
k_x \\
k_y
\end{pmatrix},
\end{align}
where the angular momentum $l=1$ and 2 for $p$ and $d$ orbitals, respectively. The requisite rotation by the angle $\theta=-\pi/(2l)$ for the invariance of the Hamiltonian indicates that respectively the linear and biquadratic band touching represent momentum space vortices of vorticity $l$, and the wave functions of the valence and conduction bands are eigenstates of orbital angular momentum $l$.

Translation by a vector $\vec{R}$ acts on the electron fields as
\begin{align}
r_s(\omega, \vec{k})\to e^{i\vec{k\cdot R}}r_s(\omega, \vec{k}).
\end{align}
Therefore the matrix transforming the 16-component Nambu spinors under translation has to pick up opposite signs at the valleys at $\pm \vec{K}$. A translation of the spinor $\Psi$ is then written as
\begin{align}
\Psi(\omega, \vec{k})\to e^{i\vec{K\cdot R}\ \Gamma_{0030}} e^{i\vec{k\cdot R}}\Psi(\omega, \vec{k}),
\end{align}
with $\Gamma_{0030}$ being the generator of translations. More intuitively, the above transformation in real space reads
\begin{align}
\Psi(\tau, \vec{r})\to e^{i\vec{K\cdot R}\ \Gamma_{0030}} \Psi(\tau, \vec{r}+\vec{R}).
\end{align}

The reversal of time is implemented by the antiunitary operator $\mathcal{T}=U \mathcal{K}$. Here $U=\Gamma_{0210}$ is a unitary operator and $\mathcal{K}$ is complex conjugation. For spin-1/2 electrons, we indeed find $\mathcal{T}^2=-1$.

The Hamiltonian operators in Eqs.~(\ref{eq:H0_MLG}) and (\ref{eq:H0_BLG}) are manifestly invariant under SU(2) spin rotations, generated by $\Gamma_{0s00}$, with $s=\{1,2,3\}$. Finally, the pseudospin SU$_{\rm p}$(2) transformations are generated by $\{ \Gamma_{3000},\Gamma_{1003}, \Gamma_{2003} \}$. Here $\Gamma_{1003}$ and $\Gamma_{2003}$ rotate between an excitonic (such as, charge density wave) and two (real and imaginary) components of a paired state (such as, $s$-wave pairing). While the number operator $\Gamma_{3000}$ rotates the U(1) superconducting phase~\cite{PhysRevX.8.011049}. See Tables~\ref{tab:bilinears_exc} and~\ref{tab:bilinears_pair}. Having established the non-interacting continuum model and the operative symmetries, in the next section we build on this foundation and examine the effects of electron-electron interactions.

%%%%%%%%%%%%%%%%%%%%%%%%%%%%%%%%%%%%%%%%%%%%%%%%%%%%%%%%%%%%%%%%%%%%%%%%%
%%%%%%%%%%%%%%%%%%%%%%%%%%%%%%%%%%%%%%%%%%%%%%%%%%%%%%%%%%%%%%%%%%%%%%%%%
%%%%%%%%%%%%%%%%%%%%%%%%%%%%%%%%%%%%%%%%%%%%%%%%%%%%%%%%%%%%%%%%%%%%%%%%%
\subsection{Electron-electron interactions}~\label{sec:interactions}

We incorporate \emph{short range} or \emph{local} electron-electron interactions by adding all symmetry-allowed momentum-independent four fermion terms to the action $S_0^j$. The schematic form of the interacting part of the action in the presence of such a local quartic term is
\begin{equation}
S_{\rm int}= \int \D \tau \D^d \vec{r} L_{\rm int} \equiv \int \D \tau \D^d \vec{r} g_{_{MN}} (\Psi^\dag M \Psi) (\Psi^\dag N \Psi),
\end{equation}
where $\Psi\equiv \Psi_{\tau,\vec{r}}$ and $\Psi^\dag\equiv \Psi^\dag_{\tau,\vec{r}}$, $M$ and $N$ are 16 dimensional (in the Nambu doubled basis) Hermitian matrices, and $g_{_{MN}}$ is the corresponding coupling constant. Note from Eq.~(\ref{eq:S0}) the scaling dimension of the spinors is $[ \Psi^{\dag}_{\tau, \bf r}]=[ \Psi_{\tau, \bf r}]=d/2$, such that $S^j_0$ is scale invariant for $j=$D and L. Consequently, a coupling constant of local quartic terms scales as $\left[ g_{_{MN}} \right]= z-d$, where $z$ is the dynamical scaling exponent, determining the relative scaling between energy and momentum according to $\epsilon_{\vec{k}} \sim |\vec{k}|^z$. Therefore $z=1$ ($2$) in case of Dirac (Luttinger) fermions. Consequently, $S_{\rm int}$ remains scale invariant. As $d=2$, sufficiently weak short range electron-electron interactions among two-dimensional Dirac fermions are \emph{irrelevant}, and the Dirac cones remain stable up to a critical strength of interactions, where, however, the system undergoes a quantum phase transition into an ordered phase. On the other hand, in a two-dimensional system of Luttinger fermions local quartic interactions are \emph{marginal}, and the quadratic band touching point is unstable in the presence of infinitesimal interactions. In terms of the density of states (DOS), the Dirac system exhibits linearly vanishing DOS, namely $\rho(E)\sim|E|$, while in the case of Luttinger fermions the low-energy DOS is constant i.e., $\rho(E)\sim\text{constant}$. The scaling of the DOS determines the (ir)relevance of local interactions.

Throughout we neglect the long range tail of the Coulomb interaction. In an undoped Dirac liquid, the long range Coulomb interaction is a \emph{marginally irrelevant} coupling (due to the vanishing DOS), irrespective of its bare strength, that otherwise causes only a logarithmic enhancement of the Fermi velocity~\cite{PhysRevB.59.R2474, PhysRevB.75.235423, PhysRevLett.102.026802, PhysRevLett.113.105502, PhysRevLett.118.026403, PhysRevB.87.045425, 2016JHEP...04..018R, PhysRevLett.97.146401, zhao2021coulomb, 2011NatPh...7..701E}. On the other hand, in undoped Luttinger liquid long-range Coulomb interaction gets screened due to the finite DOS, resulting in local density-density interaction, thus only renormalizing the bare strength of the coupling $g^s_{_1}$~\cite{PhysRevB.85.245451} [see Eq.~(\ref{eq:L_int})]. At finite temperature this approximation is further justified due to the thermal screening~\cite{kapusta-gale-2006}, while at finite doping conventional Thomas-Fermi screening sets in. Nevertheless, long-range Coulomb interaction can shift the phase boundaries between the ordered and disordered phases, without altering the nature of the former one, as has been shown for both two- and three-dimensional Dirac liquids~\cite{PhysRevB.80.081405, PhysRevB.94.115137}.

%%%%%%%%%%%%%%%%%%%%%%%%%%%%%%%%%%%%%%%%%%%%%%%%%%%%%%%%%%%%%%%%%%%%%%%%%
%%%%%%%%%%%%%%%%%%%%%%%%%%%%%%%%%%%%%%%%%%%%%%%%%%%%%%%%%%%%%%%%%%%%%%%%%
%%%%%%%%%%%%%%%%%%%%%%% PHASE DIAGRAMS - MASS CHANNELS %%%%%%%%%%%%%%%%%%
%%%%%%%%%%%%%%%%%%%%%%%%%%%%%%%%%%%%%%%%%%%%%%%%%%%%%%%%%%%%%%%%%%%%%%%%%
%%%%%%%%%%%%%%%%%%%%%%%%%%%%%%%%%%%%%%%%%%%%%%%%%%%%%%%%%%%%%%%%%%%%%%%%%
\begin{figure*}[t!]
\subfloat[Phase diagrams for the quartic interaction in the quantum anomalous Hall insulator or $A_{2g}$ singlet channel.]{
\includegraphics[width=0.21\linewidth]{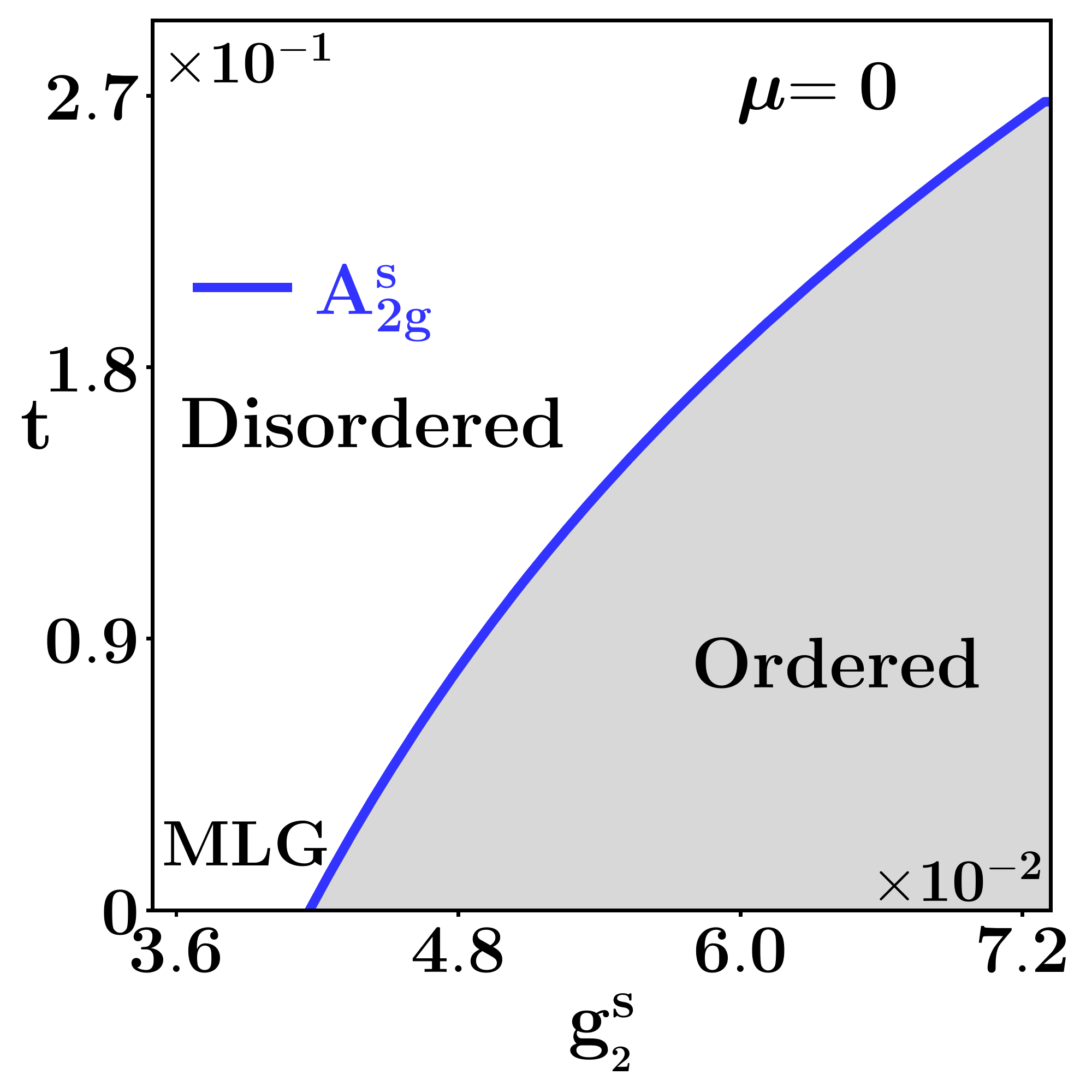}\hspace{0.5cm}
\includegraphics[width=0.21\linewidth]{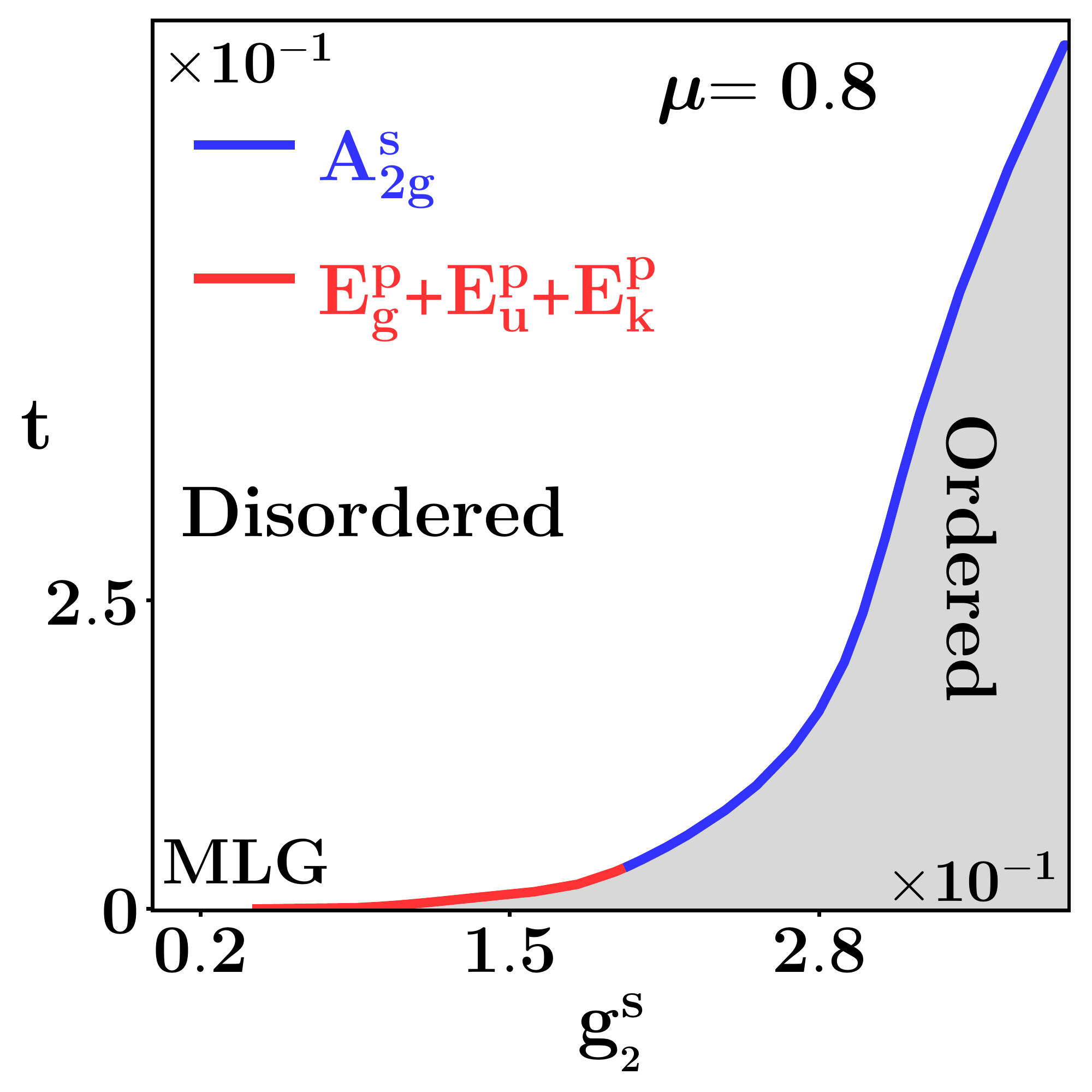}\hspace{0.5cm}
\includegraphics[width=0.21\linewidth]{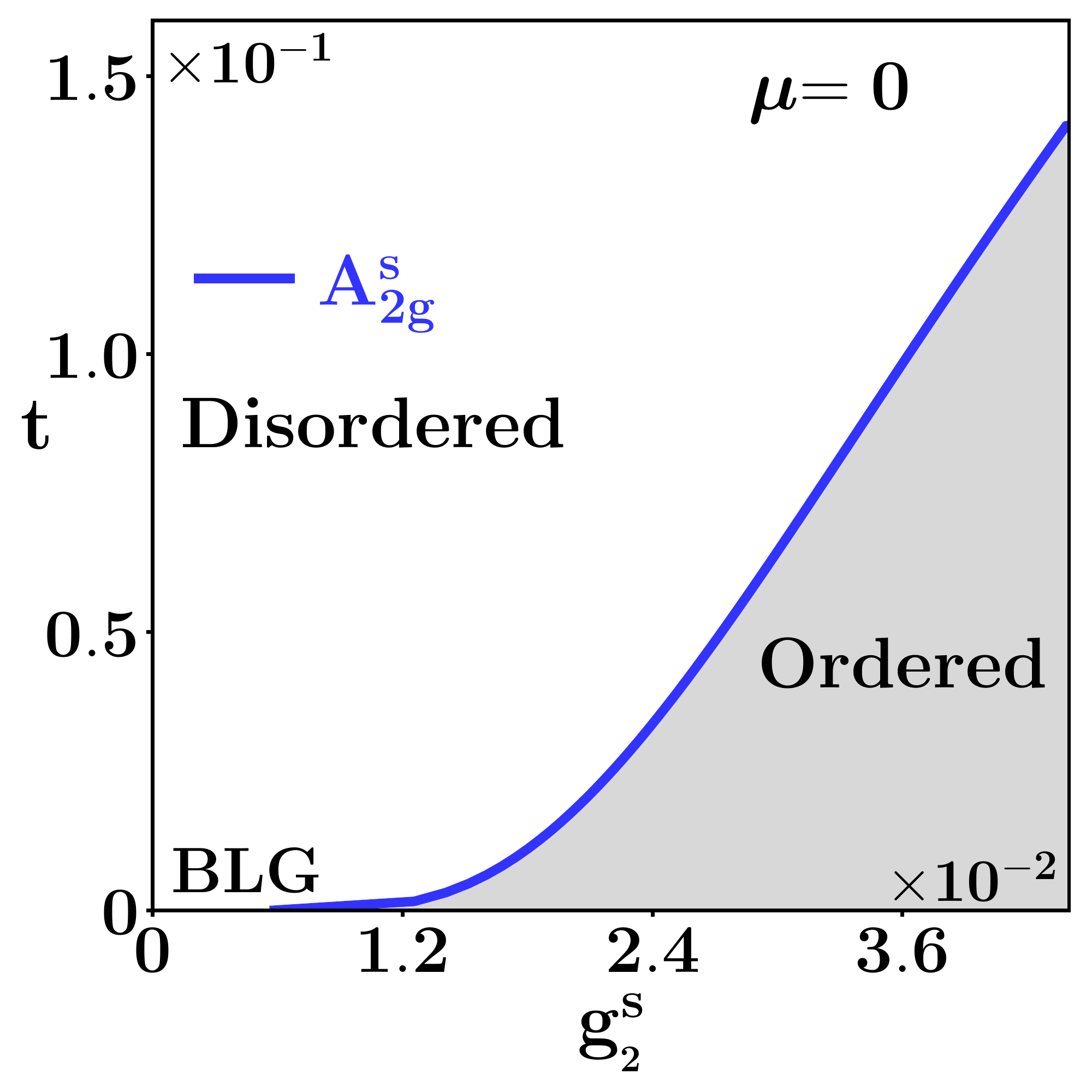}\hspace{0.5cm}
\includegraphics[width=0.21\linewidth]{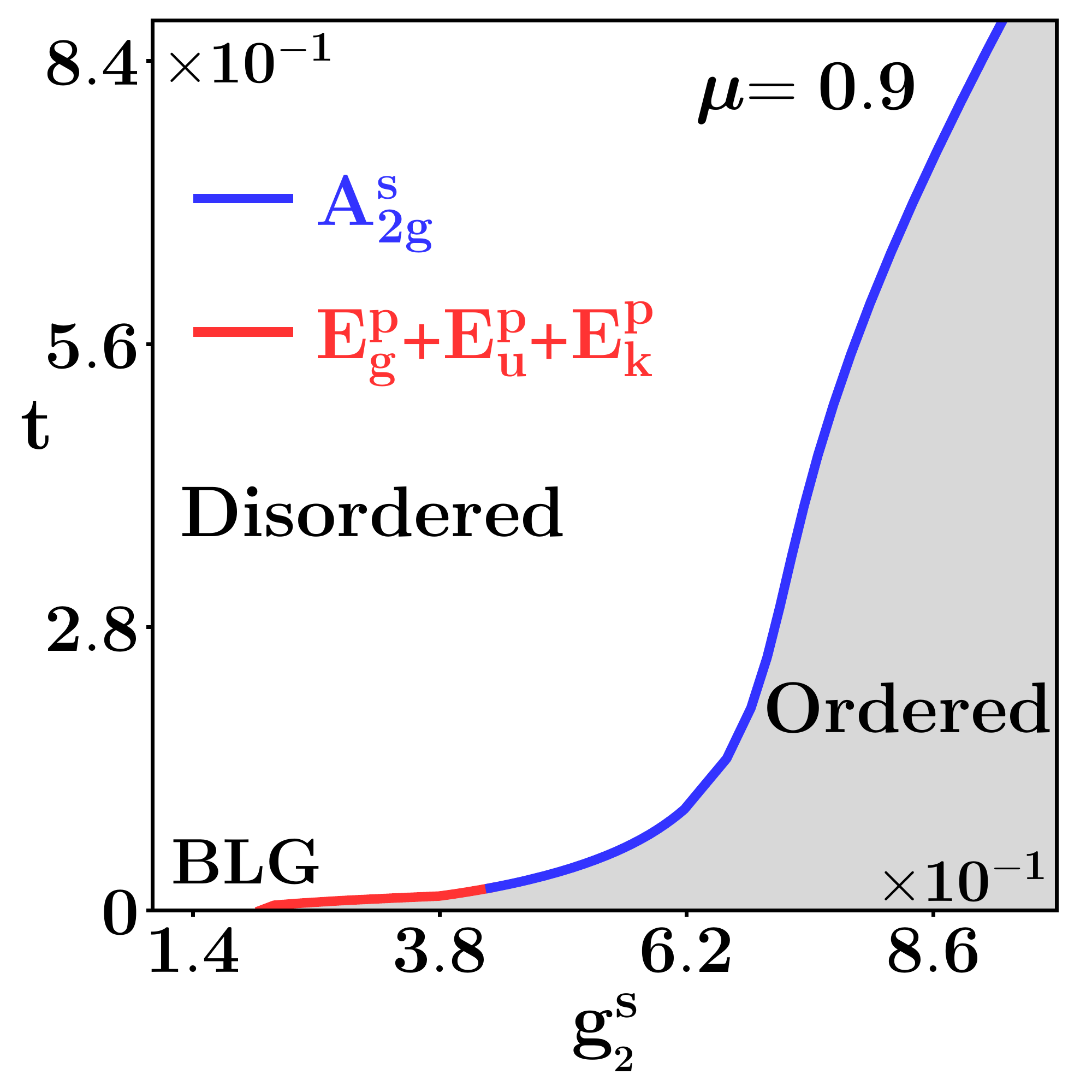}
\label{fig:QAHI}}\\
\subfloat[Phase diagrams for the quartic interactions in the charge density wave or $A_{2u}$ singlet channel.]{
\includegraphics[width=0.21\linewidth]{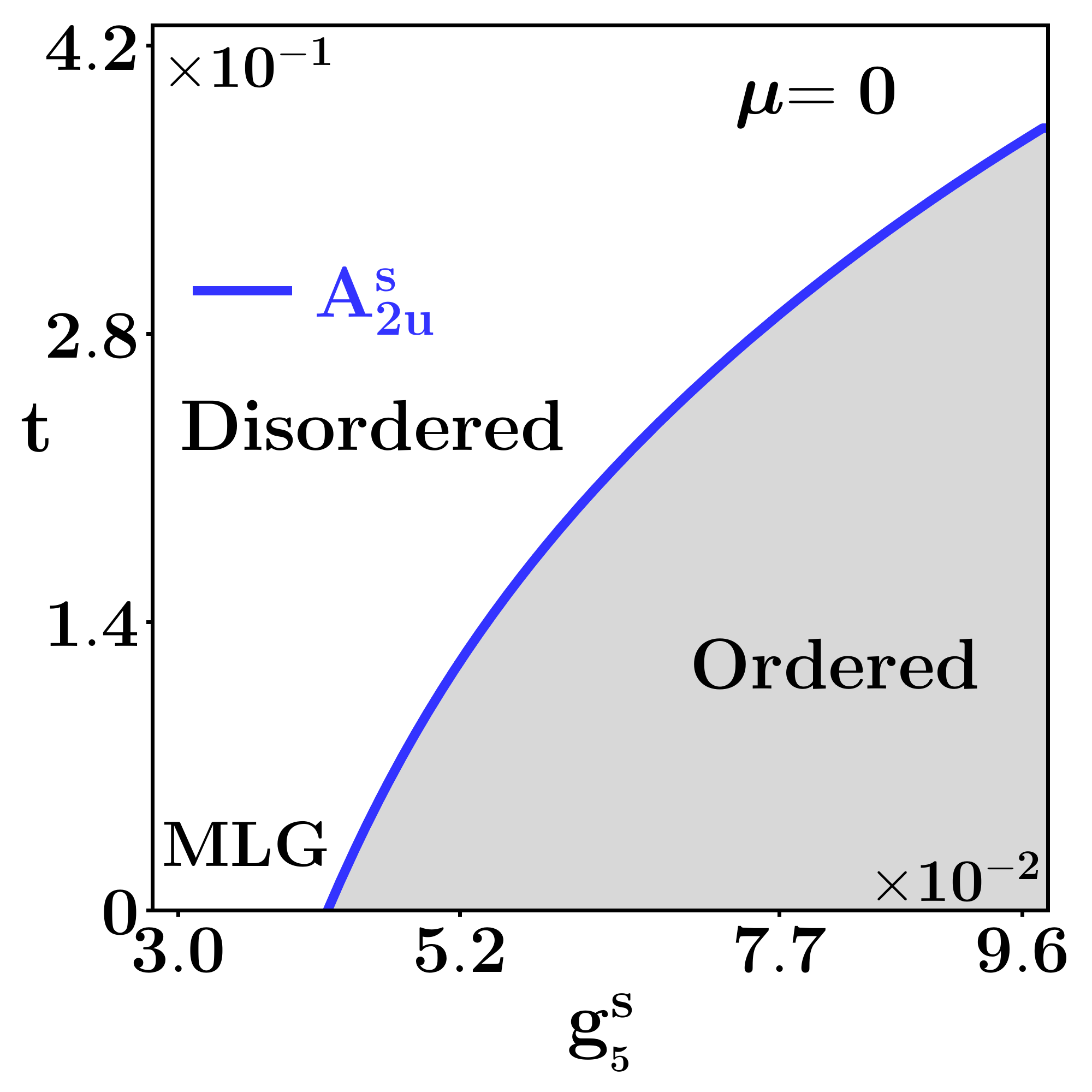}\hspace{0.5cm}
\includegraphics[width=0.21\linewidth]{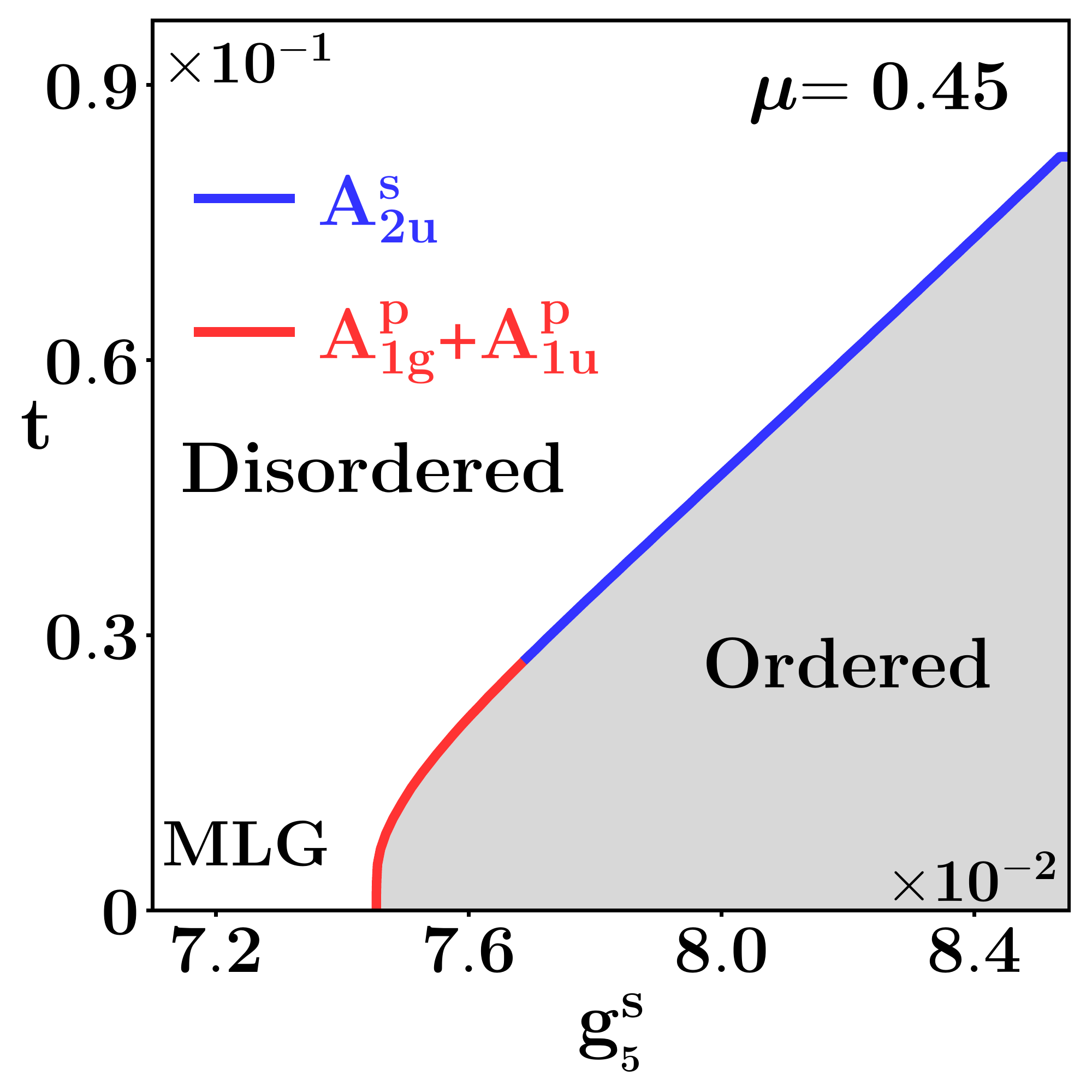}\hspace{0.5cm}
\includegraphics[width=0.21\linewidth]{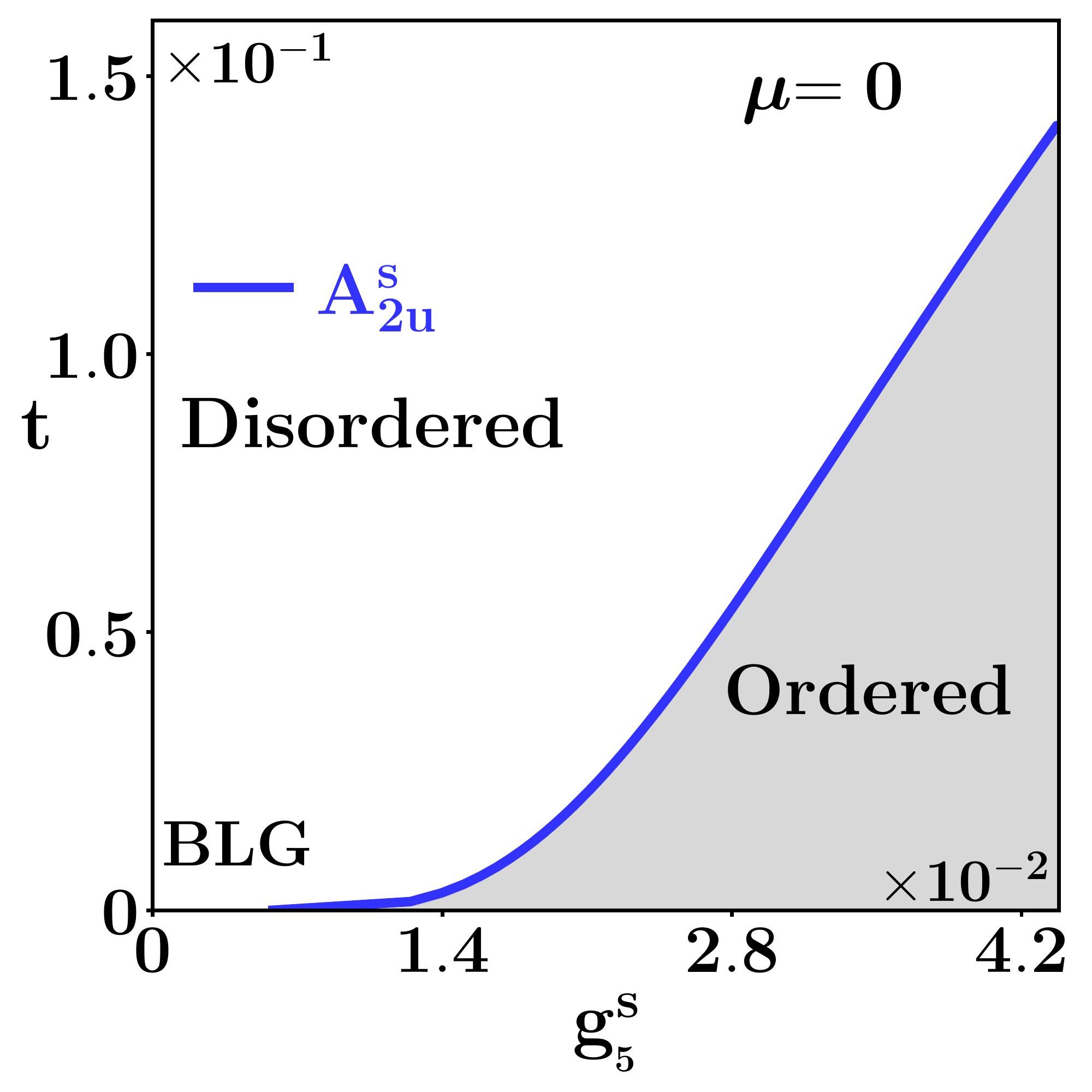}\hspace{0.5cm}
\includegraphics[width=0.21\linewidth]{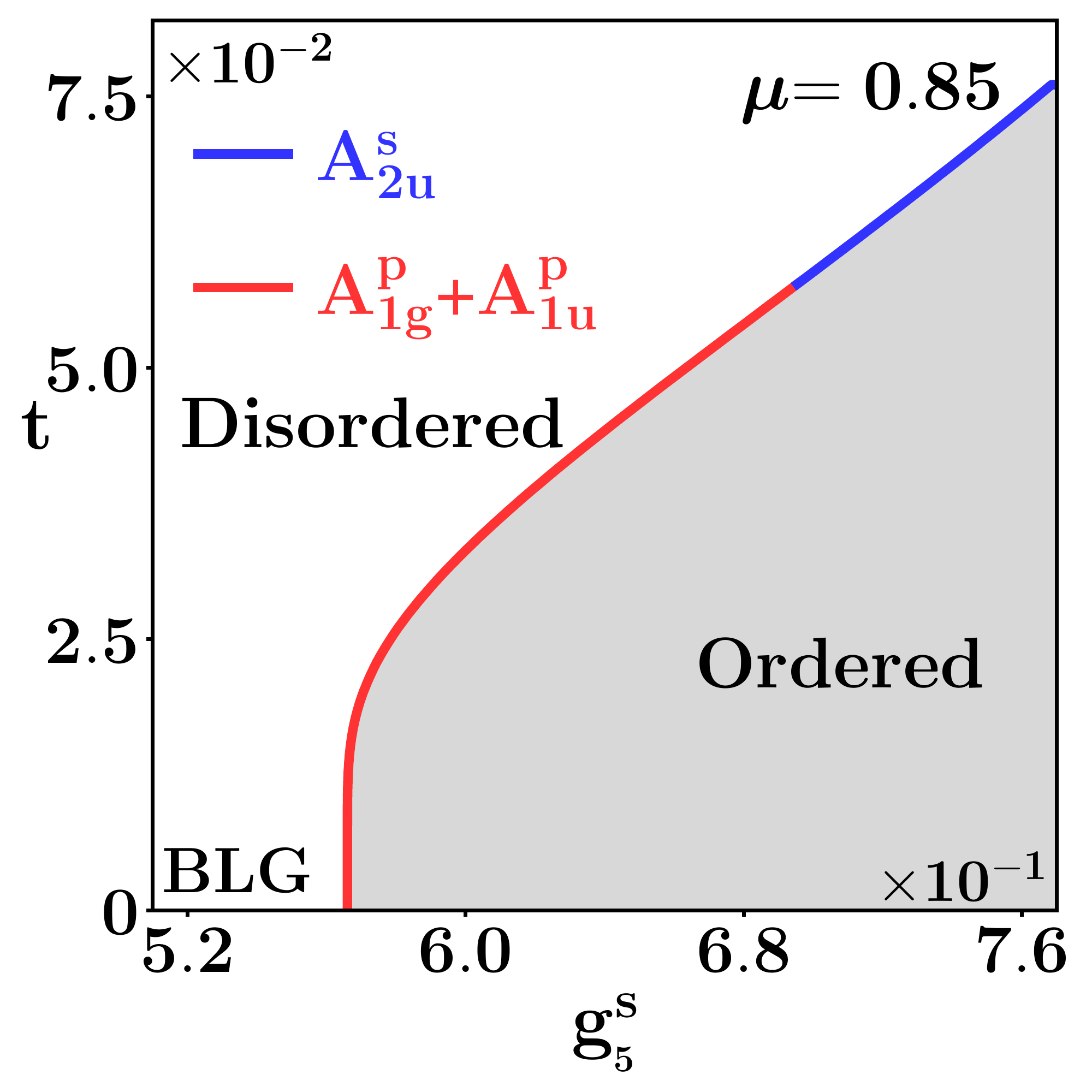}
\label{fig:CDW}}\\
\subfloat[Phase diagrams for the quartic interaction in the quantum spin Hall insulator or $A_{2g}$ triplet channel.]{
\includegraphics[width=0.21\linewidth]{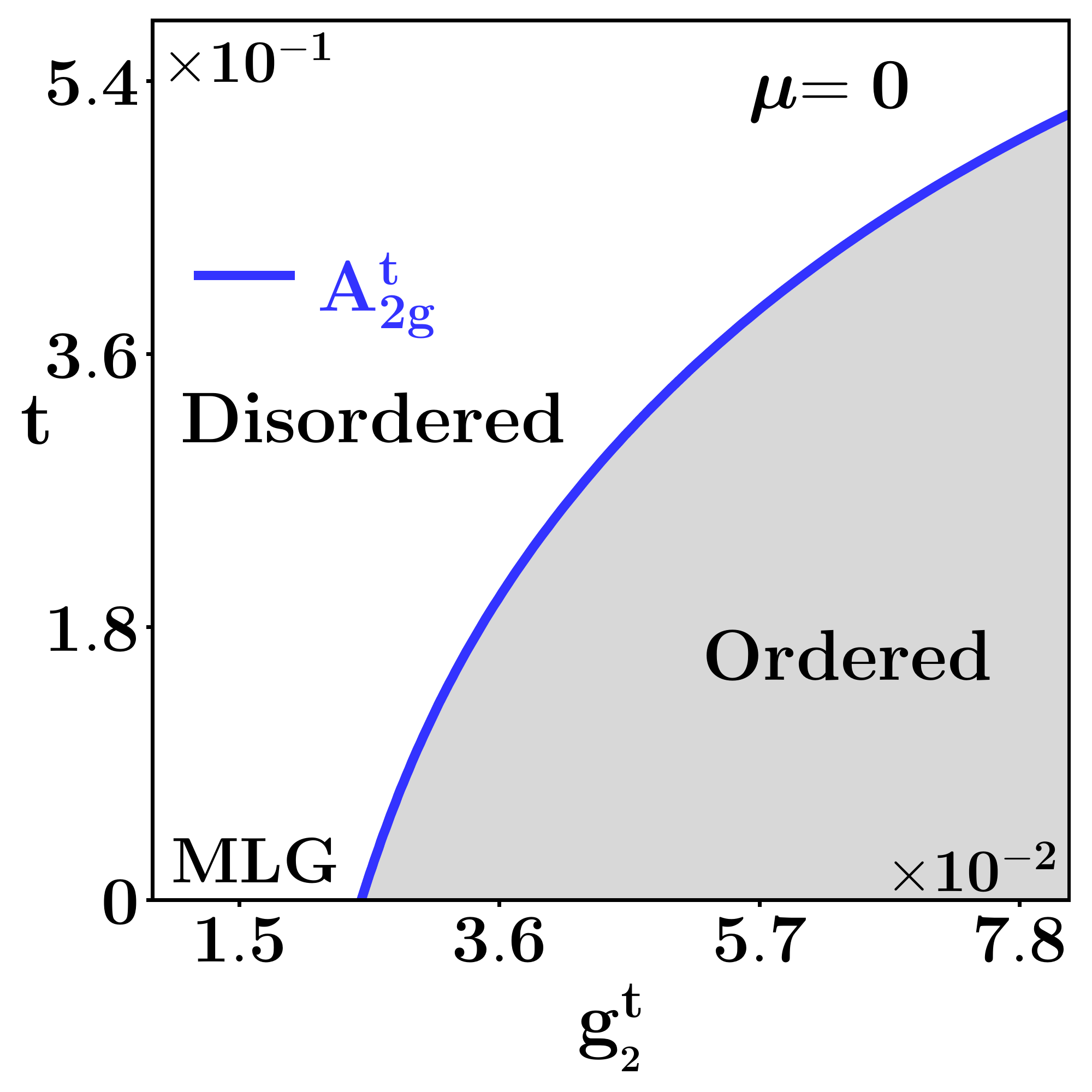}\hspace{0.5cm}
\includegraphics[width=0.21\linewidth]{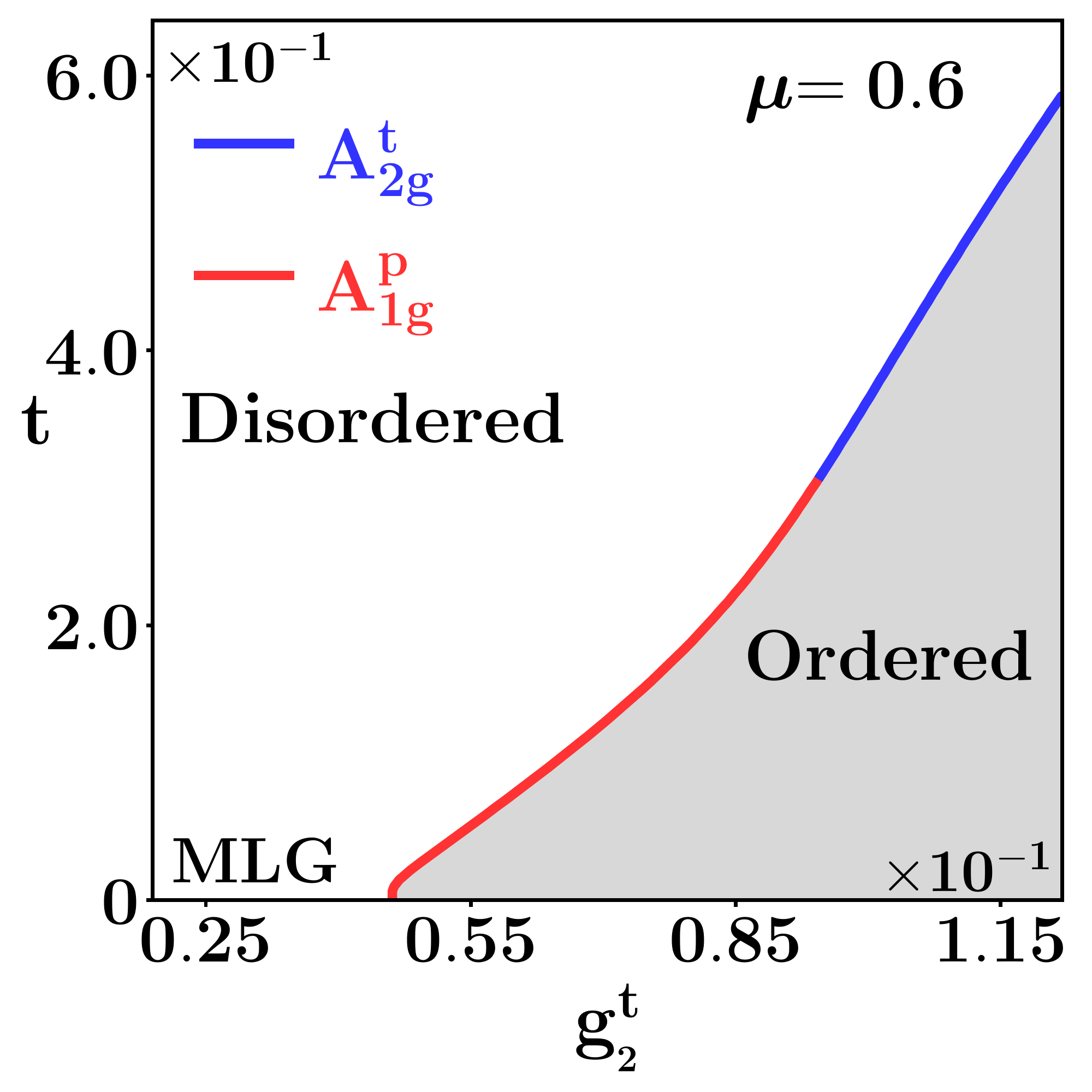}\hspace{0.5cm}
\includegraphics[width=0.21\linewidth]{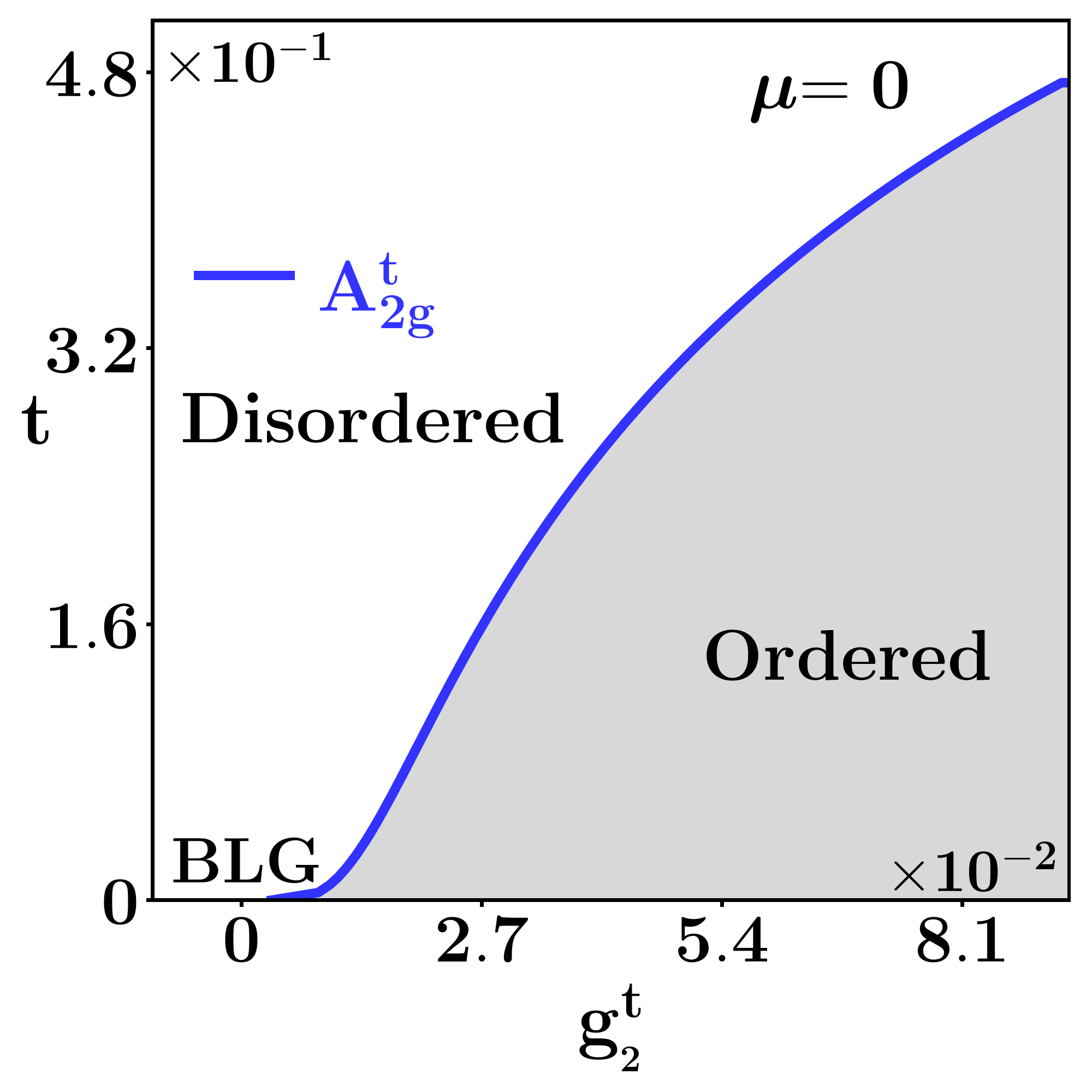}\hspace{0.5cm}
\includegraphics[width=0.21\linewidth]{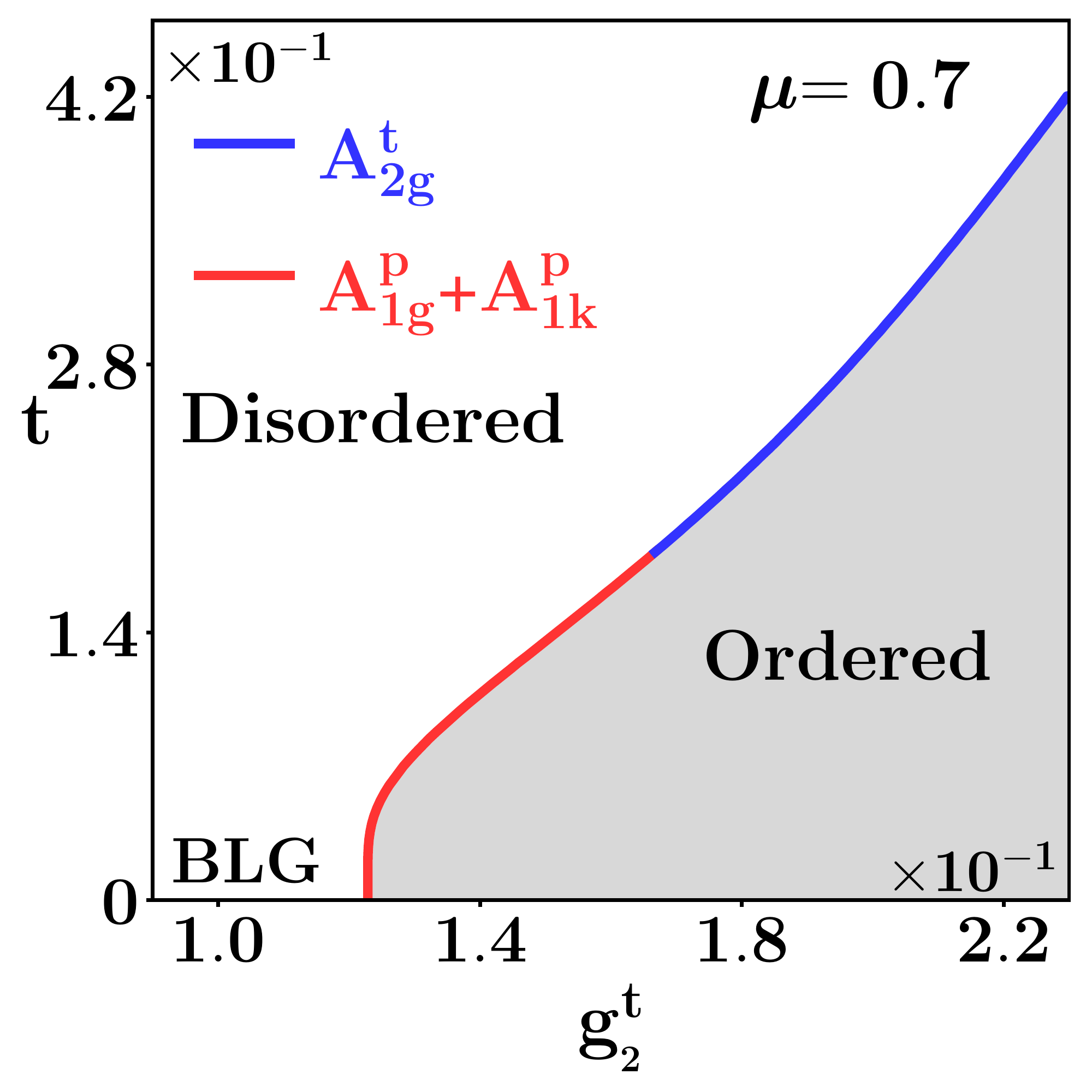}
\label{fig:QSHI}}\\
\subfloat[Phase diagrams for the quartic interaction in the antiferromagnet or $A_{2u}$ triplet channel.]{
\includegraphics[width=0.21\linewidth]{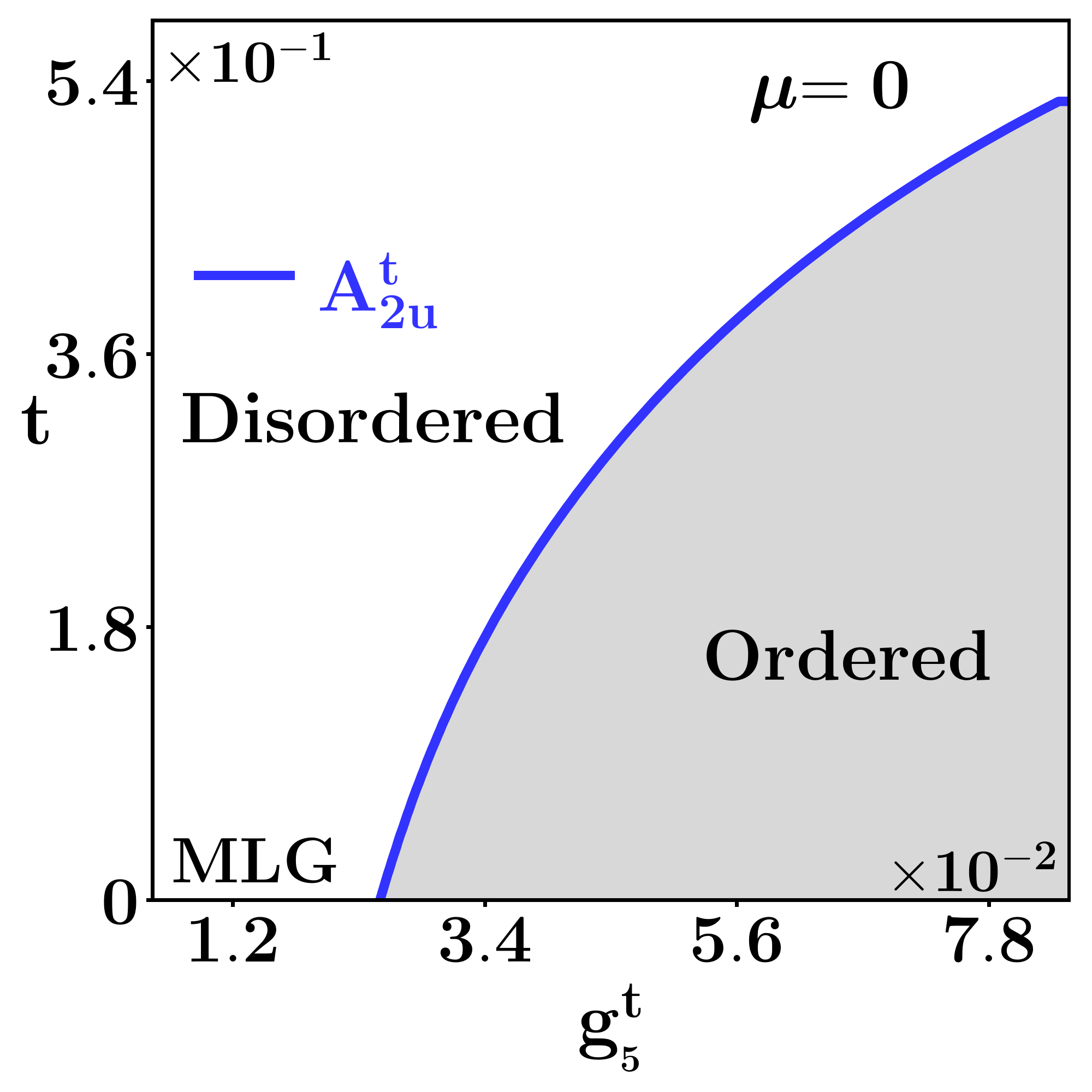}\hspace{0.5cm}
\includegraphics[width=0.21\linewidth]{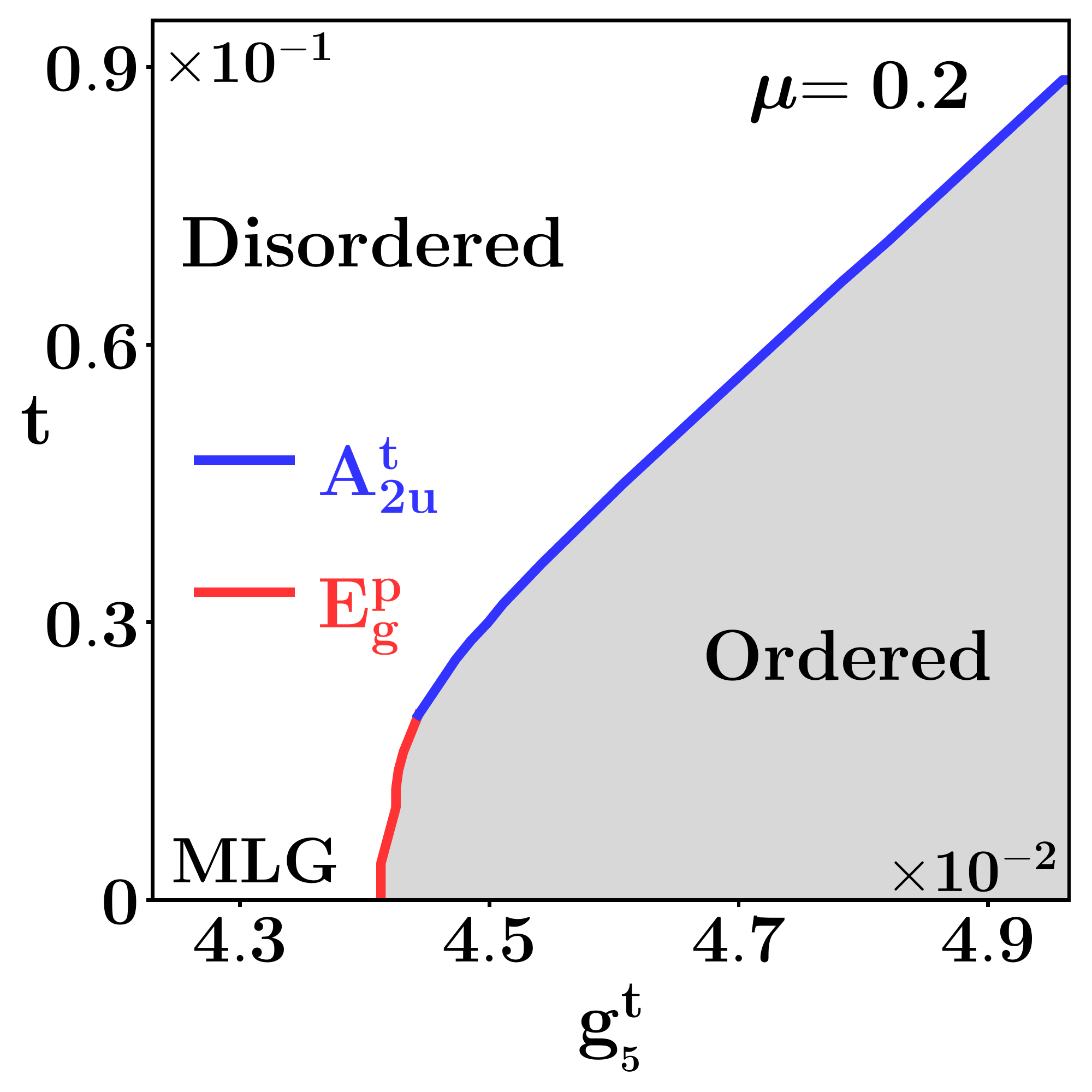}\hspace{0.5cm}
\includegraphics[width=0.21\linewidth]{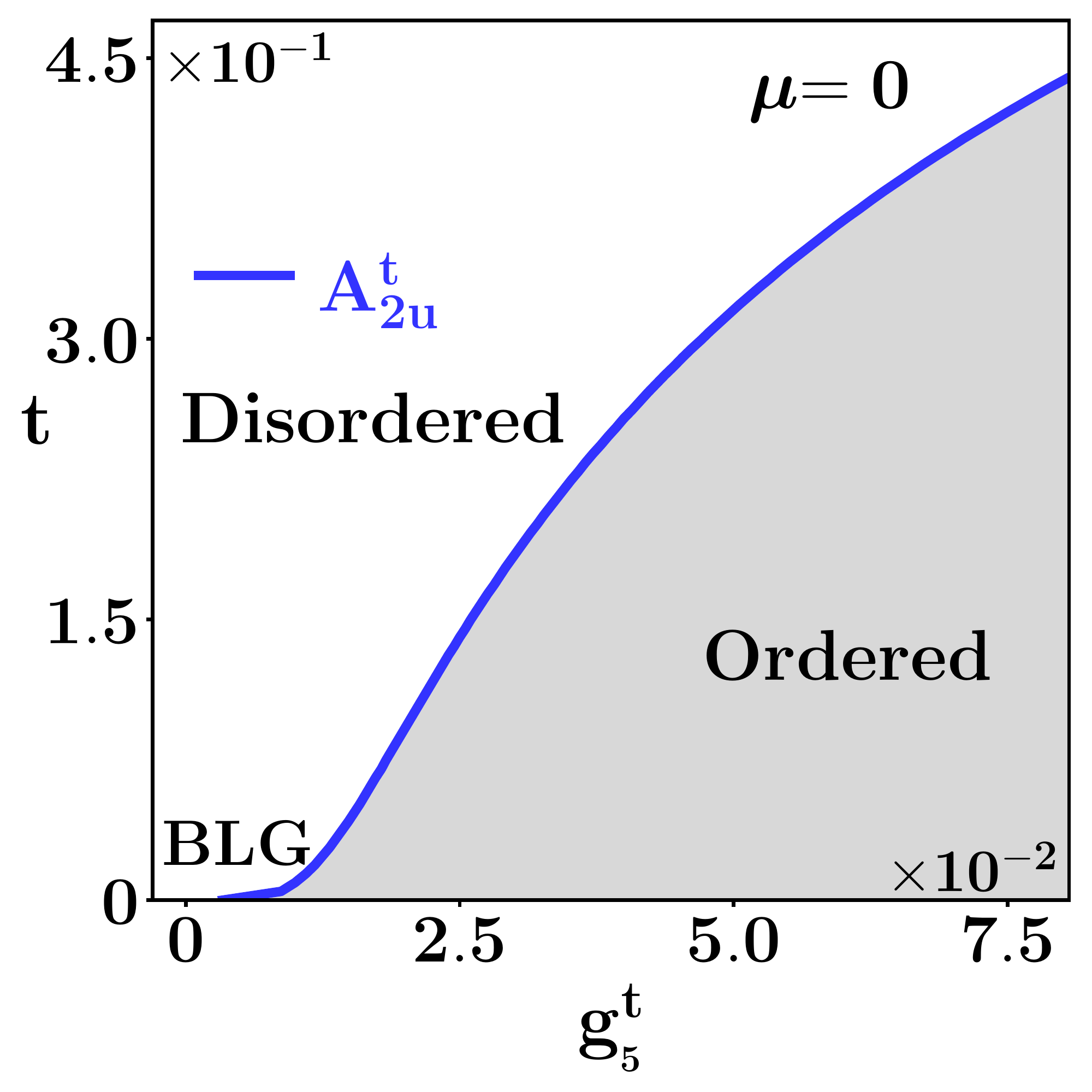}\hspace{0.5cm}
\includegraphics[width=0.21\linewidth]{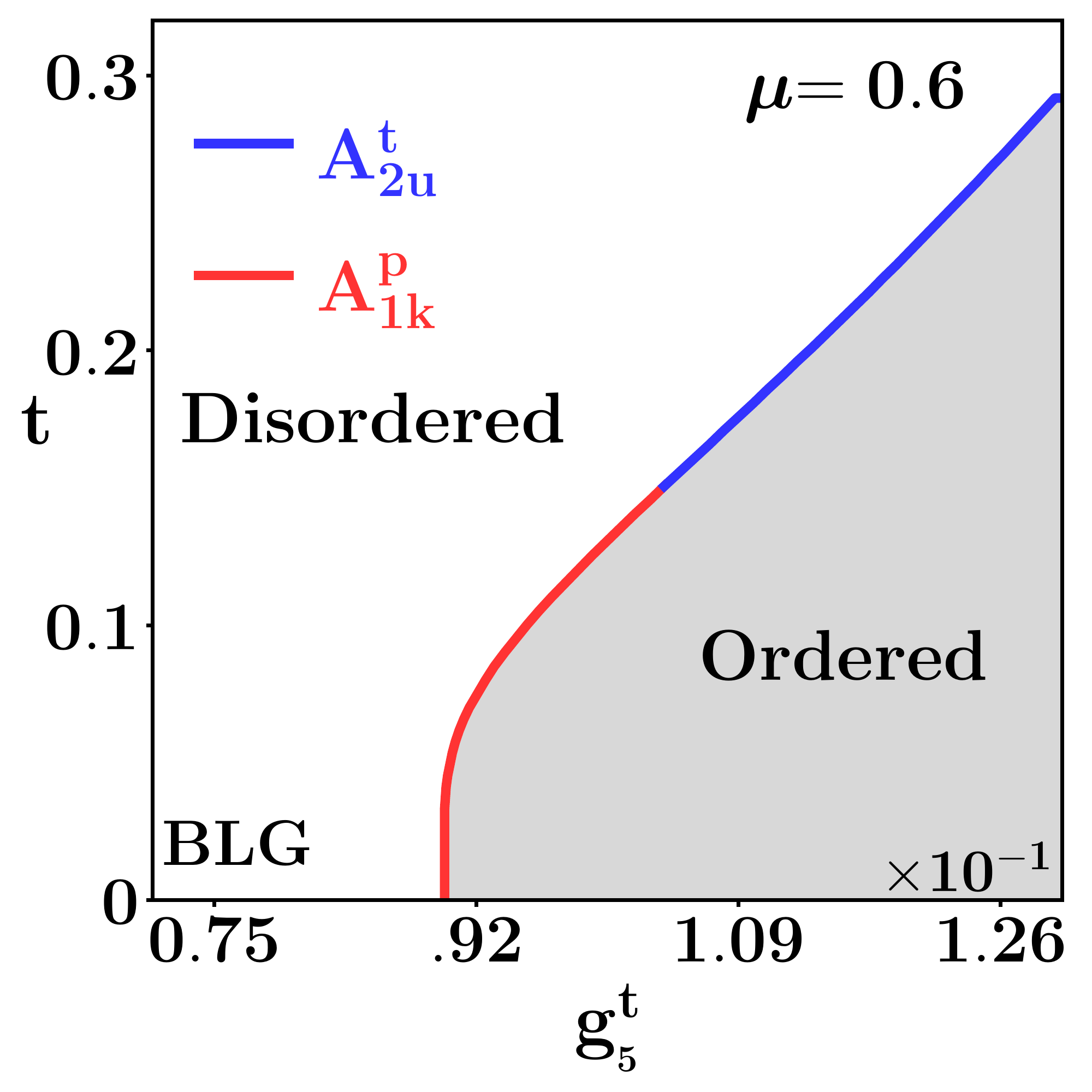}
\label{fig:AFM}}
\caption{Cuts of the global phase diagram with quartic interactions in the mass channels for both MLG and BLG at zero ($\mu=0$) and finite ($\mu>0$) chemical doping. The horizontal axis is bare interaction strength, while the vertical axis is dimensionless temperature [see Eq.~(\ref{eq:dimless_pars})]. The value of the chemical potential is indicated in each panel. Multiple orders diverging in a degenerate manner is indicated by a ``+'' sign. For the labeling of various ordered phases consult Tables~\ref{tab:bilinears_exc} and \ref{tab:bilinears_pair}. The bilinears of adjacent phases form composite order parameters, as shown in Tables~\ref{tab:ptransitions_sgl} and \ref{tab:ptransitions_tr}. See also Sec.~\ref{sec:PD_mass}.}  
\label{fig:mass_PD}
\end{figure*}
%%%%%%%%%%%%%%%%%%%%%%%%%%%%%%%%%%%%%%%%%%%%%%%%%%%%%%%%%%%%%%%%%%%%%%%%%
%%%%%%%%%%%%%%%%%%%%%%%%%%%%%%%%%%%%%%%%%%%%%%%%%%%%%%%%%%%%%%%%%%%%%%%%%
%%%%%%%%%%%%%%%%%%%%%%%%%%%%%%%%%%%%%%%%%%%%%%%%%%%%%%%%%%%%%%%%%%%%%%%%%
%%%%%%%%%%%%%%%%%%%%%%%%%%%%%%%%%%%%%%%%%%%%%%%%%%%%%%%%%%%%%%%%%%%%%%%%%
%%%%%%%%%%%%%%%%%%%%%%%%%%%%%%%%%%%%%%%%%%%%%%%%%%%%%%%%%%%%%%%%%%%%%%%%%

All the local quartic terms appearing in the interacting Lagrangian must transform as scalars under all operative symmetries introduced in Sec.~\ref{sec:Symmetries}, which severely restricts the number of couplings constants to 18. The interacting Lagrangian ($L_{\rm int}$) containing all symmetry-allowed local quartic terms reads~\cite{PhysRevB.79.085116}
\begin{equation}
L_{\rm int}=L_{\rm int}^{\rm sing}+L_{\rm int}^{\rm trip},
\end{equation}
where
\widetext
\allowdisplaybreaks[4]
\begin{align}~\label{eq:L_int}
&L_{\rm int}^{\rm sing}=
g^s_{_1} (\Psi^\dag \Gamma_{3000} \Psi)^2+
 g^s_{_2} (\Psi^\dag \Gamma_{0033} \Psi)^2+
 g^s_{_3} \Big[ (\Psi^\dag \Gamma_{3001} \Psi)^2+ (\Psi^\dag \Gamma_{3032} \Psi)^2 \Big] +
 g^s_{_4} (\Psi^\dag \Gamma_{0030} \Psi)^2+
 g^s_{_5} (\Psi^\dag \Gamma_{3003} \Psi)^2 \nonumber \\
&+g^s_{_6} \Big[ (\Psi^\dag \Gamma_{0031} \Psi)^2+(\Psi^\dag \Gamma_{0002} \Psi)^2 \Big] 
+\sum_{j=1}^2\bigg\{ 
g^s_{_7}(\Psi^\dag \Gamma_{30j1} \Psi)^2 + g^s_{_8}(\Psi^\dag \Gamma_{00j2} \Psi)^2 + 
g^s_{_9}\Big[ (\Psi^\dag \Gamma_{30j0} \Psi)^2 + (\Psi^\dag \Gamma_{30j3} \Psi)^2 \Big] 
\bigg\}, \nonumber \\
&L_{\rm int}^{\rm trip}=\sum_{s=1}^3 \Bigg[ g_{_1}^t (\Psi^\dag \Gamma_{0s00} \Psi)^2+ 
 g_{_{2}}^t (\Psi^\dag \Gamma_{3s33} \Psi)^2+
 g_{_3}^t \Big[ (\Psi^\dag \Gamma_{0s01} \Psi)^2+ (\Psi^\dag \Gamma_{0s32} \Psi)^2 \Big] +
 g_{_4}^t (\Psi^\dag \Gamma_{3s30} \Psi)^2+
 g_{_5}^t (\Psi^\dag \Gamma_{0s03} \Psi)^2 \nonumber \\
&+g_{_6}^t \Big[ (\Psi^\dag \Gamma_{3s31} \Psi)^2+ (\Psi^\dag \Gamma_{3s02} \Psi)^2 \Big] +
\sum_{j=1}^2 \bigg\{
 g_{_7}^t (\Psi^\dag \Gamma_{0sj1} \Psi)^2+ 
 g_{_8}^t (\Psi^\dag \Gamma_{3sj2} \Psi)^2
+g_{_9}^t \Big[ (\Psi^\dag \Gamma_{0sj0} \Psi)^2+ (\Psi^\dag \Gamma_{0sj3} \Psi)^2 \Big] \bigg\} \Bigg].
\end{align}

\twocolumngrid
\noindent Visibly, $L_{\rm int}$ separates into spin singlet ($L_{\rm int}^{\rm sing}$) and spin triplet ($L_{\rm int}^{\rm trip}$) parts, containing the matrices $\sigma_0$ and $\sigma_s$ (with $s=1,2,3$), acting on the spin index, respectively. The sublattice and valley degrees of freedom (accompanied by $\alpha$ and $\tau$ matrices, respectively) repeat for the two irreducible representations of the SU(2) spin algebra, while the particle-hole index ($\eta$ matrix) is exactly the opposite between singlet and triplet channels. Since there is only one operative SU(2) algebra in the spin sector, this suggests a redundancy in the above expression of the interacting Lagrangian. This expectation can be verified using the Fierz identity for eight-dimensional Hermitian matrices (note the particle-hole degree of freedom in this sense is an artificial doubling), which reveals that there are altogether only nine linearly independent four fermion terms. Without the loss of generality, we choose these to be the spin singlet interactions ($g^s_{_i}$). For a detailed presentation of the Fierz reduction see Appendix~\ref{app:Fierz}. Throughout, here we focus on repulsive electron-electron interactions, which in our notation corresponds to $g_\mu^j>0$ for $\mu=1,\dots,9$ and $j=s,t$.

To shed light on the structure of the global phase diagram of interacting two-dimensional Dirac and Luttinger fermions, we perform Wilsonian momentum shell RG analysis at zero, as well as finite temperature ($T$) and chemical potential ($\mu$). We already established the scaling of the coupling constants to be $\left[ g_\mu^j \right] =z-d$, which pins the lower critical dimension of the corresponding theories at $d=z$ and facilitates an $\epsilon$ expansion around the marginal dimensionality with $\epsilon=d-z$~\cite{book-QFT-critical-phenomena,PRep.385.69,2021JHEP...01..004S}. Notice that the physical values of $\epsilon$ are 1 and 0 for two-dimensional Dirac and Luttinger fermions, respectively. However, our conclusions are insensitive to the exact value of $\epsilon$, except the nonuniversal locations of the phase boundaries between the ordered and disordered phases.

To proceed further, we introduce an ultraviolet momentum cutoff $\Lambda$, that replaces the (hexagonal) lattice Brillouin zone with a spherical one around two inequivalent valleys in the continuum theory. Moreover, we introduce the dimensionless temperature and chemical potential. These two quantities respectively for Dirac (D) and Luttinger (L) systems are defined as 
\begin{align}~\label{eq:dimless_pars}
t^{\rm D}&=T/(\Lambda v ), & t^{\rm L}&=2mT/\Lambda^2, \nonumber \\
\tilde{\mu}^{\rm D}&=\mu/(\Lambda v ), & \tilde{\mu}^{\rm L}&=2m\mu/\Lambda^2.
\end{align}
From now on we omit the tilde and denote the dimensionless chemical potential by $\mu$. Furthermore, for brevity we suppress the D and L indices. The scaling dimensions of these two quantities are $\left[ t\right]=\left[ \mu \right]=z$, and hence they are \emph{relevant} parameters. The RG flow equation or the $\beta$ function of a quantity describes its behavior under coarse grain as we integrate out the fast Fourier modes within a Wilsonian momentum shell $\Lambda e^{-\ell}<|\vec{k}|<\Lambda$, where $\ell$ is the logarithm of the RG scale. The flow equations for dimensionless temperature, chemical potential and coupling constants can be summarized respectively as
\begin{align}
\frac{\D t}{\D \ell}&=z t, \hspace{1cm} \frac{\D \mu}{\D \ell}= z \mu, \label{eq:beta_tmu} \\
\frac{\D g^s_{_i}}{\D \ell}&= -\epsilon g^s_{_i} + \sum_{j,k} g^s_{_j} g^s_{_k} H^i_{jk}(t,\mu). \label{eq:beta_g}
\end{align}
The coupling constants appearing in the RG flow equations are also dimensionless, obtained by taking $g^s_\mu \Lambda^\epsilon/(4 \pi) \to g_\mu^s$. For the exact form of the functions $H^i_{jk}(t,\mu)$ see Appendix~\ref{app:beta_func}. Eq.~(\ref{eq:beta_g}) implies that all orderings take place at critical couplings $g^s_{_i} \sim \epsilon$, which is consistent with our observation of local interactions being irrelevant and marginal in MLG and BLG, respectively. Therefore, in a Dirac system all orderings set in beyond a finite strength of the interactions, while a Luttinger system is conducive for ordering even for sufficiently weak interactions.

Our methodology of detecting phase transitions is as follows. As temperature and chemical potential ``grow'' under the RG transformation, they provide an infrared cutoff when either one of them becomes of order of \emph{unity}. The renormalized or scale dependent temperature and chemical doping take the closed analytic forms $t(\ell)=t(0)\exp[z\ell]$ and $\mu(\ell)=\mu(0)\exp[z\ell]$, respectively. Thus, with finite bare values (at $\ell=0$) of $t$ and $\mu$, denoted by $t(0)$ and $\mu(0)$, respectively, we must terminate the RG flows of the coupling constants at a scale $\ell^\ast=\min.(\ell^\ast_t, \ell^\ast_\mu)$, where
~\cite{PhysRevB.99.121407,PhysRevB.103.165139,PhysRevLett.112.147002,PhysRevB.39.2344}
\begin{align}
\ell^\ast_t = \frac{1}{z}\ln \left( \frac{1}{t(0)}\right), \hspace{1cm} \ell^\ast_\mu = \frac{1}{z}\ln \left( \frac{1}{\mu(0)}\right).
\end{align}
For a given interaction channel ($g_{_i}$), we integrate the differential equations in Eq.~(\ref{eq:beta_g}) between $0\leq\ell \leq \ell^\ast$ for increasing $g_{_i}(0)$. Its smallest bare value for which at least one coupling constant diverges (not necessarily $g_{_i}$) indicates a phase transition to an ordered phase and contributes a data point $(t,\mu,g^\ast_{_i})$ to the phase boundary, where $g^\ast_{_i}$ is the critical interaction strength in the given interaction channel. However, to reveal the type of instability and the nature of the ordered phase we have to add order parameter fields to the Lagrangian and investigate their fate under the coarse grain, which we do next.
It is worth pointing out that here we determine the infrared cutoff ($\ell^\ast$) for the RG flow of coupling constants by setting the renormalized temperature $t(\ell)=1$ and chemical potential $\mu(\ell)=1$. However, the nature of the ordered states is insensitive to such choice, as long as $t(\ell) \sim 1$ and $\mu(\ell) \sim 1$. Only the nonuniversal details, such as the transition temperature ($t_c$), interaction coupling at which ordering switches from pairing to excitonic, depend on it.

%%%%%%%%%%%%%%%%%%%%%%%%%%%%%%%%%%%%%%%%%%%%%%%%%%%%%%%%%%%%%%%%%%%%%%%%%%%%%%%%%%%%%%%
%%%%%%%%%%%%%%%%%%%%%%%%%%%%%%%%%%%%%%%%%%%%%%%%%%%%%%%%%%%%%%%%%%%%%%%%%%%%%%%%%%%%%%%
%%%%%%%%%%%%%%%%%%%%%%%%%%%%%% BILINEARS TABLE Pairing:  %%%%%%%%%%%%%%%%%%%%%%%%%%%%%%
%%%%%%%%%%%%%%%%%%%%%%%%%%%%%%%%%%%%%%%%%%%%%%%%%%%%%%%%%%%%%%%%%%%%%%%%%%%%%%%%%%%%%%%
%%%%%%%%%%%%%%%%%%%%%%%%%%%%%%%%%%%%%%%%%%%%%%%%%%%%%%%%%%%%%%%%%%%%%%%%%%%%%%%%%%%%%%%
\begin{table*}[t!]
\renewcommand{\arraystretch}{1.4}
\begin{tabular}{|c c c c c c c c c c c c>{\centering}m{0.7cm}m{0.7cm}<{\centering}|}
\hline
\begin{tabular}{@{}c@{}} IREP \\ ($D_{3d}$) \end{tabular} & Matrix ($N$) & Phase &  CF & SB & $S$ & $T$ & tl & $R(\frac{\pi}{2})$ & TR & SU(2) & SU$_{\rm p}$(2) & \multicolumn{2}{c|}{ \begin{tabular}{@{}c@{}} Mass \\ \hline MLG\hspace{0.2cm}BLG \end{tabular} } \\
\hline 
$A_{1g}$ & $\Gamma_{\alpha 000}$ & $s$-wave SC  &  $\Delta_{1}^p$ & $A_{1g}^p$ & 
$+$ & $+$ & $+$ & 0 & $+\ (-)$ & 0 & I & \checkmark & \checkmark \\
\rowcolor{RowColor}
$A_{2g}$ & $\Gamma_{\alpha s33}$ & fully gapless SC$_1$ &  $\Delta_{2}^p$ & $A_{2g}^p$ & 
$-$ & $-$ & $+$ & 0 & $+\ (-)$ & 1 & & \ding{55} & \ding{55} \\

$E_{g}$  & $\Gamma_{\alpha 001}, \Gamma_{\alpha 032}$ & nematic SC$_1$ &  $\Delta_{3}^p$ & $E_{g}^p$ & 
$+,-$ & $+,-$ & $+,+$ & 1 & $+,+\ (-,-)$ & 0 & III & \ding{55} & \ding{55} \\
\rowcolor{RowColor}
$A_{1u}$ & $\Gamma_{\alpha s30}$ & $f$-wave SC &  $\Delta_{4}^p$ & $A_{1u}^p$ & 
$+$ & $-$ & $+$ & 0 & $+\ (-)$ & 1 & II & \checkmark & \checkmark \\

$A_{2u}$ & $\Gamma_{\alpha 003}$ & fully gapless SC$_2$ &  $\Delta_{5}^p$ & $A_{2u}^p$ & 
$-$ & $+$ & $+$ & 0 & $+\ (-)$ & 0 & & \ding{55} & \ding{55} \\
\rowcolor{RowColor}
$E_{u}$  & $\Gamma_{\alpha s31}, \Gamma_{\alpha s02}$ & nematic SC$_2$ &  $\Delta_{6}^p$ & $E_{u}^p$ & 
$+,-$ & $-,+$ & $+,+$ & 1 & $+,+\ (-,-)$ & 1 & IV & \ding{55} & \ding{55} \\

$A_{1k}$ & $\Gamma_{\alpha 011}, \Gamma_{\alpha 021}$ & singlet Kekul\' e SC & $\Delta_{7}^p$ & $A_{1k}^p$ & 
$+,+$ & $+,-$ & $-,-$ & 0 & $+,+\ (-,-)$ & 0 & VI & \ding{55} & \checkmark \\
\rowcolor{RowColor}
$A_{2k}$ & $\Gamma_{\alpha s12}, \Gamma_{\alpha s22}$ &  triplet Kekul\' e SC & $\Delta_{8}^p$ & $A_{2k}^p$ & 
$-,-$ & $+,-$ & $-,-$ & 0 & $+,+\ (-,-)$ & 1 & VII & \checkmark & \ding{55} \\

$E_{k}$  & 
\begin{tabular}{@{}c@{}}$\Gamma_{\alpha 010}, \Gamma_{\alpha 023}$ \\ $\Gamma_{\alpha 013}, \Gamma_{\alpha 020}$\end{tabular} &
smectic SC & $\Delta_{9}^p$ & $E_{k}^p$ & 
\begin{tabular}{@{}c@{}} $+,-$ \\ $-,+$ \end{tabular} & 
\begin{tabular}{@{}c@{}} $+,-$ \\ $+,-$ \end{tabular} & 
\begin{tabular}{@{}c@{}} $-,-$ \\ $-,-$ \end{tabular} & 
\begin{tabular}{@{}c@{}} 1 \\ 1 \end{tabular} &
\begin{tabular}{@{}c@{}} $+,+\ (-,-)$ \\ $+,+\ (-,-)$ \end{tabular}
 & 0 & V & \ding{55} & \ding{55} \\
\hline
\end{tabular}
\caption{Properties of the local superconducting order parameters of the schematic form $\Delta_N (\Psi^\dag N \Psi)$ on the honeycomb lattice, where $N$ is a sixteen-dimensional Hermitian matrix with $s=\{1,2,3\}$ and $\alpha=\{1,2\}$. The first three columns respectively display the corresponding irreducible representation (IREP) under the $D_{3d}$ group, the order parameter matrices $N$, and the physical nature of the paired states. The fourth and fifth columns respectively show the conjugate field (CF) of the order parameter and the symbol (SB) of the ordered phase in Figs.~\ref{fig:mass_PD}-\ref{fig:Kekule_PD}. Columns 6-12 display the transformation properties of the fermion bilinears under sublattice ($S$) and valley ($T$) reflections, translation (tl), orbital rotation [$R(\pi/2)$], time reversal (TR), and spin SU(2) and pseudospin SU$_{\rm p}$(2) rotations, respectively. Here $+$ ($-$) means even (odd), whereas 0 (1) means the bilinears transform as scalar (vector) under the corresponding rotation. In the TR column a symbol outside (within) parentheses corresponds to $\alpha=1$ (2). The order parameters marked by the same roman numeral in the SU$_{\rm p}$(2) column are related to each other by pseudospin SU$_{\rm p}$(2) rotations, which always relates a pairing order to an excitonic one, see Table~\ref{tab:bilinears_exc}. The last two columns display whether the order parameter fully gaps fermions in MLG and BLG, where \checkmark (\ding{55}) means gapped (gapless) spectrum.
 }~\label{tab:bilinears_pair}
\end{table*}
%%%%%%%%%%%%%%%%%%%%%%%%%%%%%%%%%%%%%%%%%%%%%%%%%%%%%%%%%%%%%%%%%%%%%%%%%%%%%%%%%%%%%%%
%%%%%%%%%%%%%%%%%%%%%%%%%%%%%%%%%%%%%%%%%%%%%%%%%%%%%%%%%%%%%%%%%%%%%%%%%%%%%%%%%%%%%%%
%%%%%%%%%%%%%%%%%%%%%%%%%%%%%%%%%%%%%%%%%%%%%%%%%%%%%%%%%%%%%%%%%%%%%%%%%%%%%%%%%%%%%%%
%%%%%%%%%%%%%%%%%%%%%%%%%%%%%%%%%%%%%%%%%%%%%%%%%%%%%%%%%%%%%%%%%%%%%%%%%%%%%%%%%%%%%%%
%%%%%%%%%%%%%%%%%%%%%%%%%%%%%%%%%%%%%%%%%%%%%%%%%%%%%%%%%%%%%%%%%%%%%%%%%%%%%%%%%%%%%%%

%%%%%%%%%%%%%%%%%%%%%%%%%%%%%%%%%%%%%%%%%%%%%%%%%%%%%%%%%%%%%%%%%%%%%%%%%
%%%%%%%%%%%%%%%%%%%%%%%%%%%%%%%%%%%%%%%%%%%%%%%%%%%%%%%%%%%%%%%%%%%%%%%%%
%%%%%%%%%%%%%%%%%%%%%%%%%%%%%%%%%%%%%%%%%%%%%%%%%%%%%%%%%%%%%%%%%%%%%%%%%
%%%%%%%%%%%%%%%%%%%%%%%%%%%%%%%%%%%%%%%%%%%%%%%%%%%%%%%%%%%%%%%%%%%%%%%%%
%%%%%%%%%%%%%%%%%%%%%%%%%%%%%%%%%%%%%%%%%%%%%%%%%%%%%%%%%%%%%%%%%%%%%%%%%
\subsection{Broken symmetry phases}~\label{sec:ordered_phases}

To identify the nature of the broken symmetry phase in an unbiased fashion we introduce all symmetry-allowed local fermion bilinears to the action, which assume the schematic form $\Delta_M (\Psi^\dag M \Psi)$, where $\Delta_M$ is the corresponding conjugate field and $M$ is a \emph{sixteen}-dimensional Hermitian matrix. The ordered state is characterized by the expectation value of the fermion bilinear $\langle \Psi^\dag M \Psi \rangle\neq 0$. Note while the RG analysis of the coupling constants $g_{_i}$ can be performed using eight-dimensional matrices (without invoking the Nambu doubling), one has to utilize the Nambu basis in the renormalization of order parameter fields to account for the superconducting channels. The effective single-particle Hamiltonian containing all symmetry allowed local orders reads
\begin{align}
H_{\rm local}= \int \D^d \vec{r}  ( h_{\rm exc} + h_{\rm pair} ),
\end{align}
where $h_{\rm exc}=h_{\rm exc}^{\rm sing}+h_{\rm exc}^{\rm trip}$, with
\widetext
\allowdisplaybreaks[4]
\begin{align}
h_{\rm exc}^{\rm sing}
 =&\Delta^s_1 \Psi^\dag \Gamma_{3000} \Psi
 +\Delta^s_2 \Psi^\dag \Gamma_{0033} \Psi
 +\Delta^s_3 \Big( \Psi^\dag \Gamma_{3001} \Psi+ \Psi^\dag \Gamma_{3032} \Psi \Big)
 +\Delta^s_4 \Psi^\dag \Gamma_{0030} \Psi
 +\Delta^s_5 \Psi^\dag \Gamma_{3003} \Psi \nonumber \\
 +&\Delta^s_6 \Big( \Psi^\dag \Gamma_{0031} \Psi + \Psi^\dag \Gamma_{0002} \Psi \Big)
 +\sum_{j=1}^2\bigg[ \Delta^s_7  \Psi^\dag \Gamma_{30j1} \Psi 
 +\Delta^s_8 \Psi^\dag \Gamma_{00j2} \Psi 
 +\Delta^s_9 \Big( \Psi^\dag \Gamma_{30j0} \Psi + \Psi^\dag \Gamma_{30j3} \Psi \Big)\bigg], 
\end{align}
and
\begin{align}
h_{\rm exc}^{\rm trip}
=&\sum_{s=1}^3 \bigg\{ \Delta_{1}^t \Psi^\dag \Gamma_{0s00} \Psi 
 +\Delta_{2}^t \Psi^\dag \Gamma_{3s33} \Psi
 +\Delta_{3}^t \Big( \Psi^\dag \Gamma_{0s01} \Psi + \Psi^\dag \Gamma_{0s32} \Psi \Big)
 +\Delta_{4}^t \Psi^\dag \Gamma_{3s30} \Psi
 +\Delta_{5}^t \Psi^\dag \Gamma_{0s03} \Psi \nonumber \\
 +&\Delta_{6}^t \Big( \Psi^\dag \Gamma_{3s31} \Psi+ \Psi^\dag \Gamma_{3s02} \Psi \Big)
 +\sum_{j=1}^2 \bigg[ \Delta_{7}^t  \Psi^\dag \Gamma_{0sj1} \Psi
 +\Delta_{8}^t  \Psi^\dag \Gamma_{3sj2} \Psi +\Delta_{9}^t \Big( \Psi^\dag \Gamma_{0sj0} \Psi+  \Psi^\dag \Gamma_{0sj3} \Psi \Big) \bigg] \bigg\},
\end{align}
whereas
\begin{align}
h_{\rm pair}
=&\sum_{\alpha=1}^2 \bigg\{ 
 \Delta_{1,\alpha}^p \Psi^\dag \Gamma_{\alpha 000} \Psi
 +\Delta_{3,\alpha}^p \Big( \Psi^\dag \Gamma_{\alpha 001} \Psi + \Psi^\dag \Gamma_{\alpha 032} \Psi \Big) 
 +\Delta_{5,\alpha}^p \Psi^\dag \Gamma_{\alpha 003} \Psi \nonumber \\
+&\sum_{j=1}^2 \bigg[ \Delta_{7,\alpha}^p  \Psi^\dag \Gamma_{\alpha 0j1} \Psi 
 +\Delta_{9,\alpha}^p \Big( \Psi^\dag \Gamma_{\alpha 0j0} \Psi +\Psi^\dag \Gamma_{\alpha 0j3} \Psi \Big) \bigg] \nonumber \\
+&\sum_{s=1}^3 \bigg[\Delta_{2,\alpha}^p \Psi^\dag \Gamma_{\alpha s33} \Psi
 +\Delta_{4,\alpha}^p \Psi^\dag \Gamma_{\alpha s30} \Psi
 +\Delta_{6,\alpha}^p \Big( \Psi^\dag \Gamma_{\alpha s31} \Psi+ \Psi^\dag \Gamma_{\alpha s02} \Psi \Big) 
 +\sum_{j=1}^2 \Delta_{8,\alpha}^p \Psi^\dag \Gamma_{\alpha sj2} \Psi \bigg]  \bigg\}.
\end{align}
\twocolumngrid

\noindent Here $\alpha=1,2$, $\Delta^p_{i,1}=\Delta^p_i \cos\phi$, $\Delta^p_{i,2}=\Delta^p_i \sin \phi$, and $\phi$ is the U(1) superconducting phase. According to $H_{\rm local}$, 27 order parameters are organized into three groups, each containing nine entries. Namely $h_{\rm exc}^{\rm sing}$ and $h_{\rm exc}^{\rm trip}$ contain nine spin singlet and nine spin triplet excitonic bilinears, respectively, while $h_{\rm pair}$ accommodates five singlet and four triplet pairing orders. The properties of fermion bilinears e.g. their irreducible point group representations, physical meanings, and transformations under various discrete and continuous symmetries are displayed in Tables~\ref{tab:bilinears_exc} and \ref{tab:bilinears_pair}, where we also collect the notation that is used throughout this paper.

\begin{figure*}[t]
\subfloat[Phase diagrams for the quartic interaction in the nematic$_1$ or $E_g$ singlet channel.]{
\includegraphics[width=0.21\linewidth]{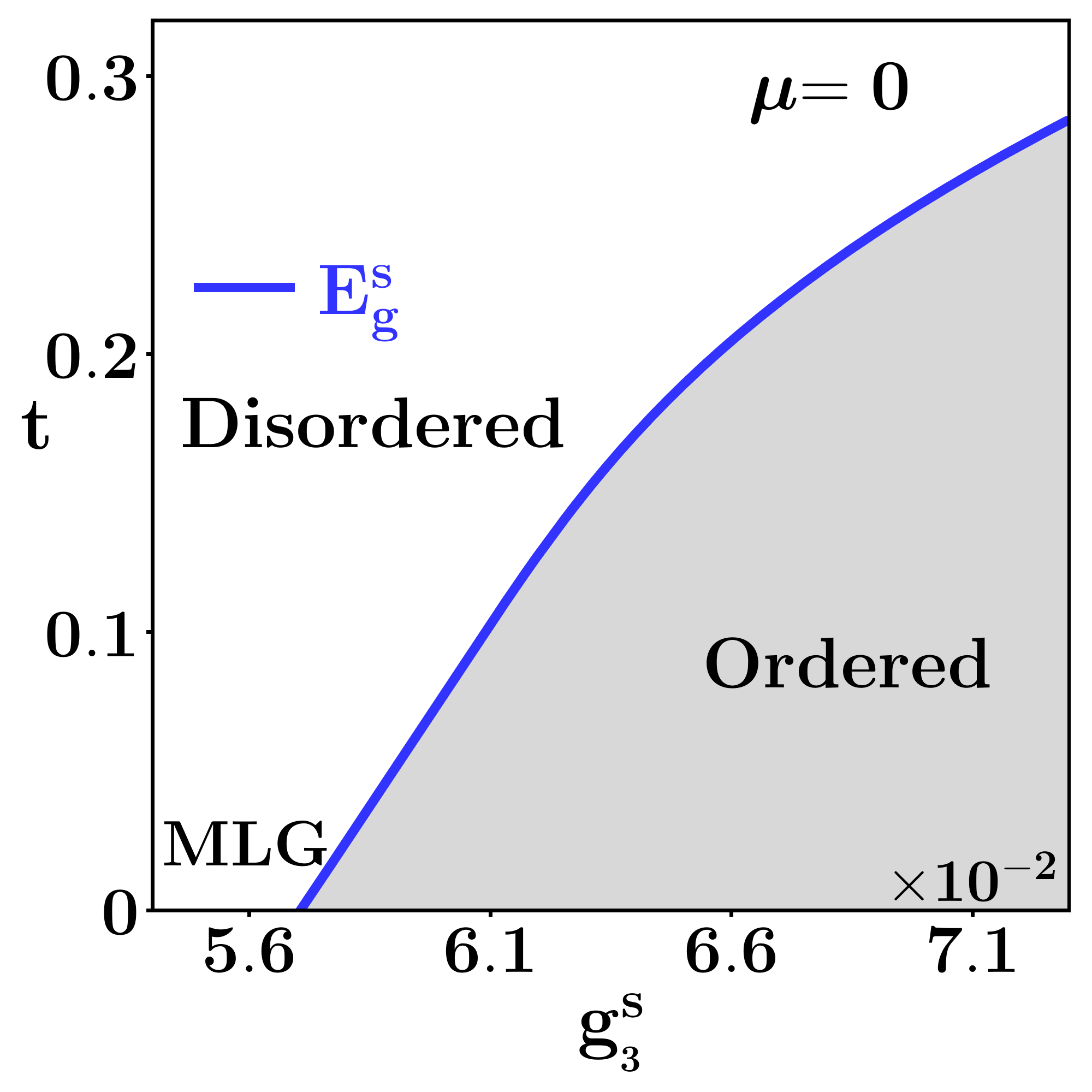}\hspace{0.5cm}
\includegraphics[width=0.21\linewidth]{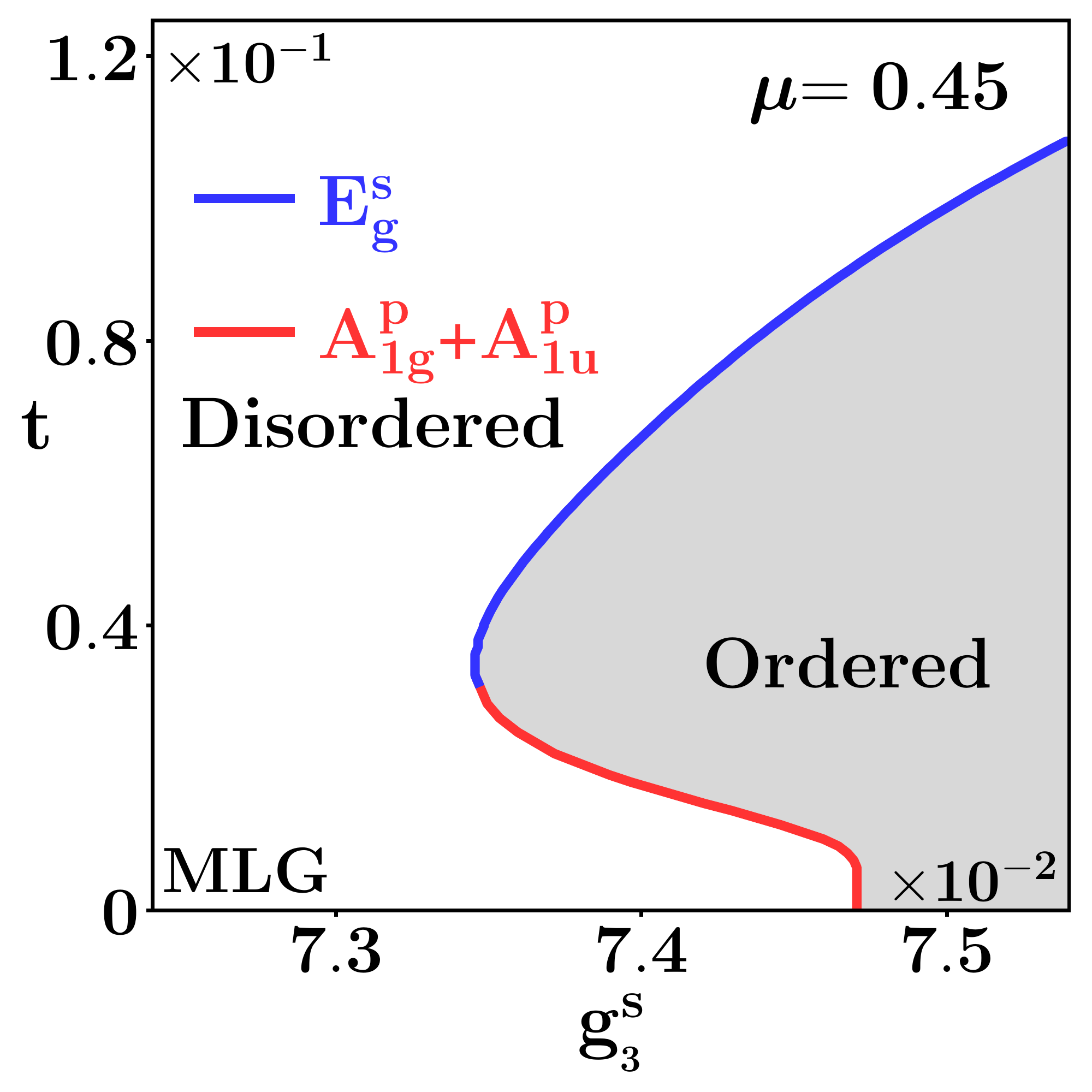}\hspace{0.5cm}
\includegraphics[width=0.21\linewidth]{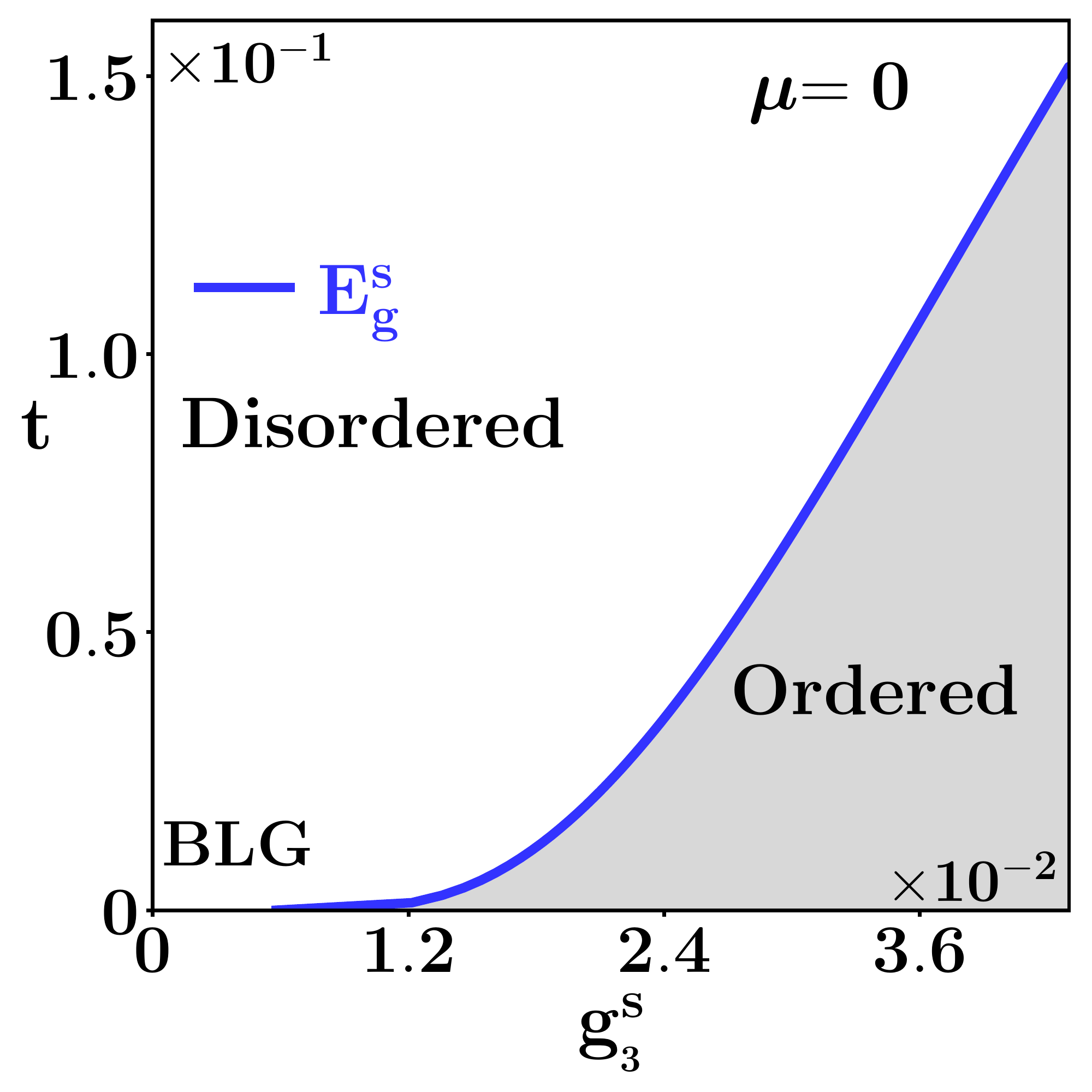}\hspace{0.5cm}
\includegraphics[width=0.21\linewidth]{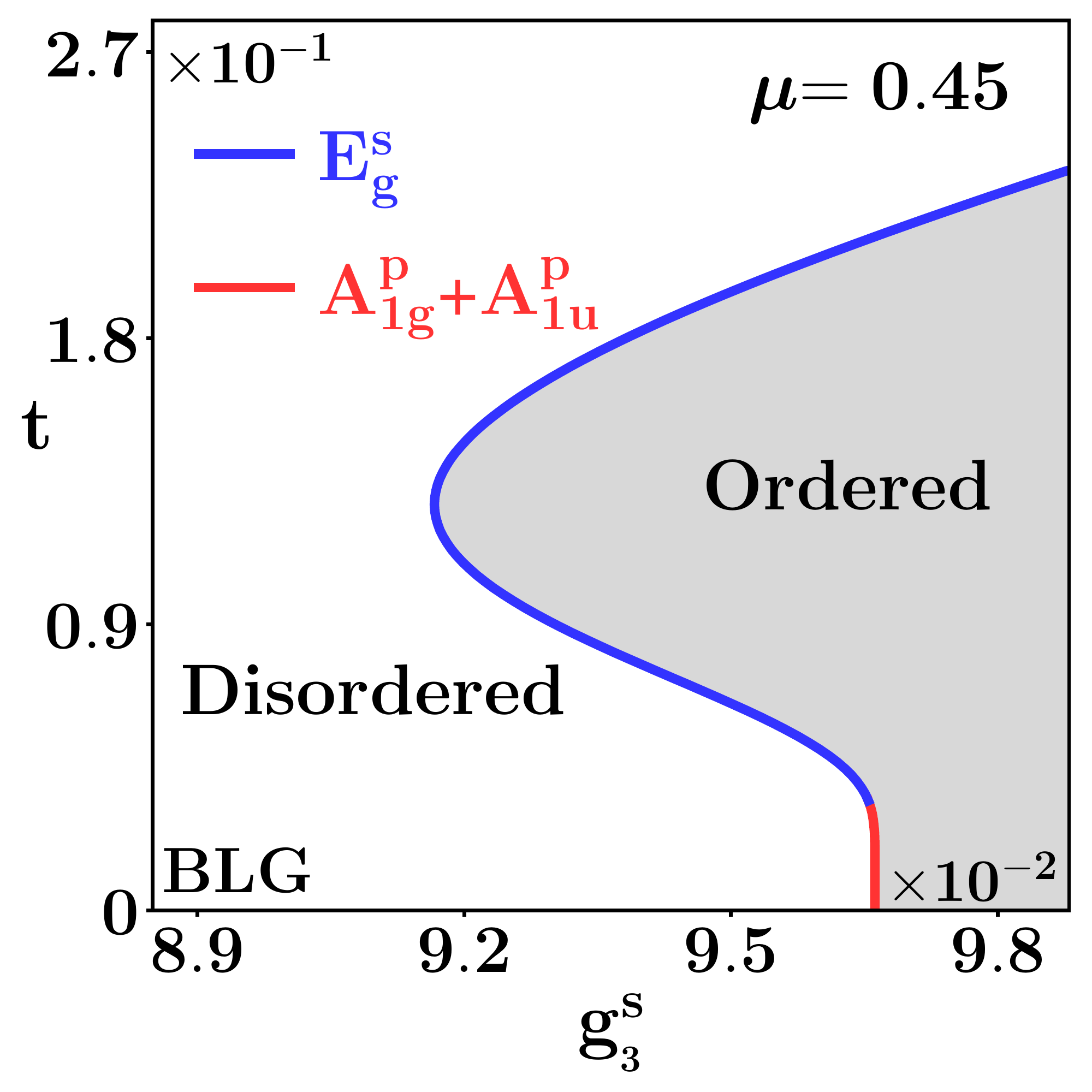}
\label{fig:Eg_nem}}\\
\subfloat[Phase diagrams for the quartic interaction in the nematic$_2$ or $E_u$ singlet channel.]{
\includegraphics[width=0.21\linewidth]{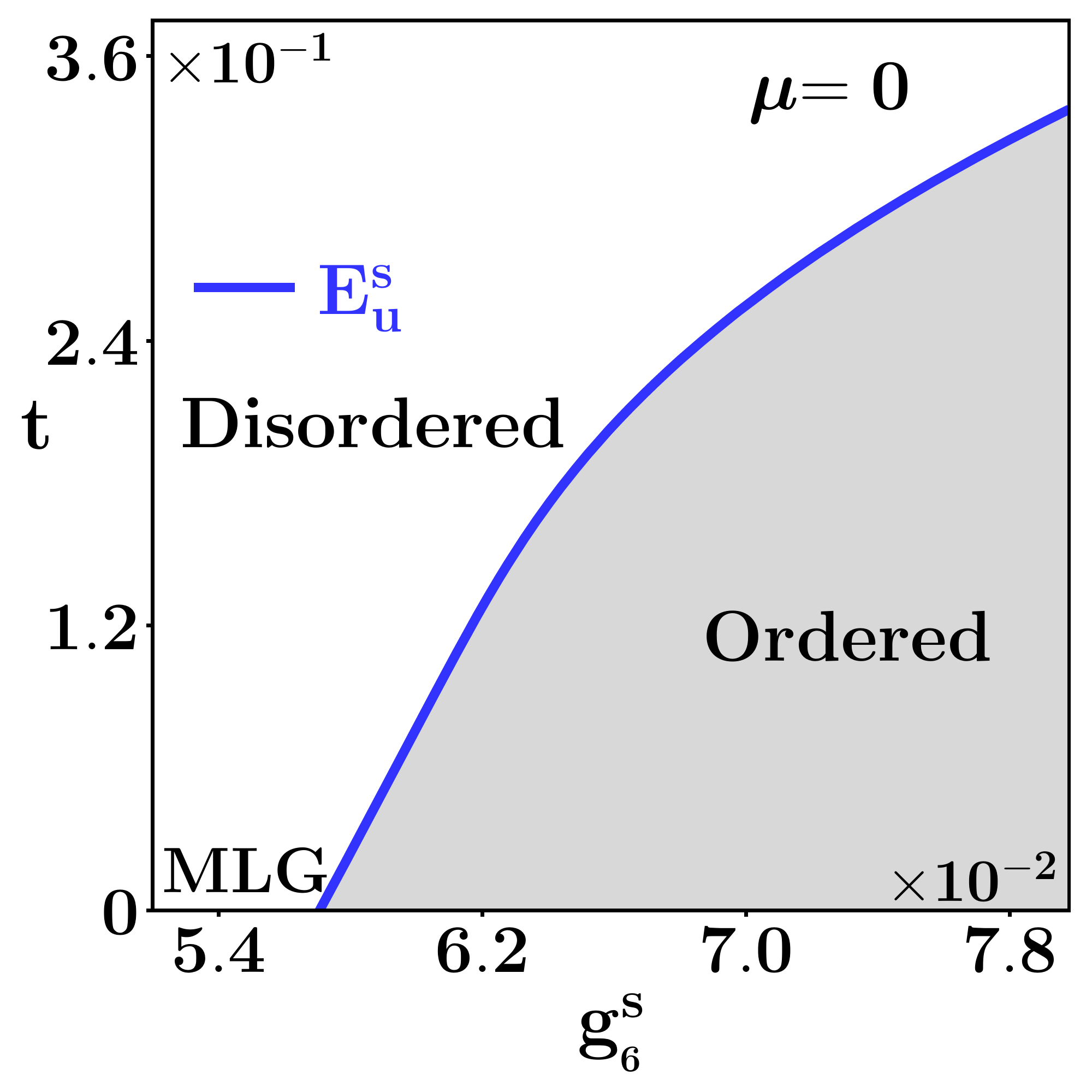}\hspace{0.5cm}
\includegraphics[width=0.21\linewidth]{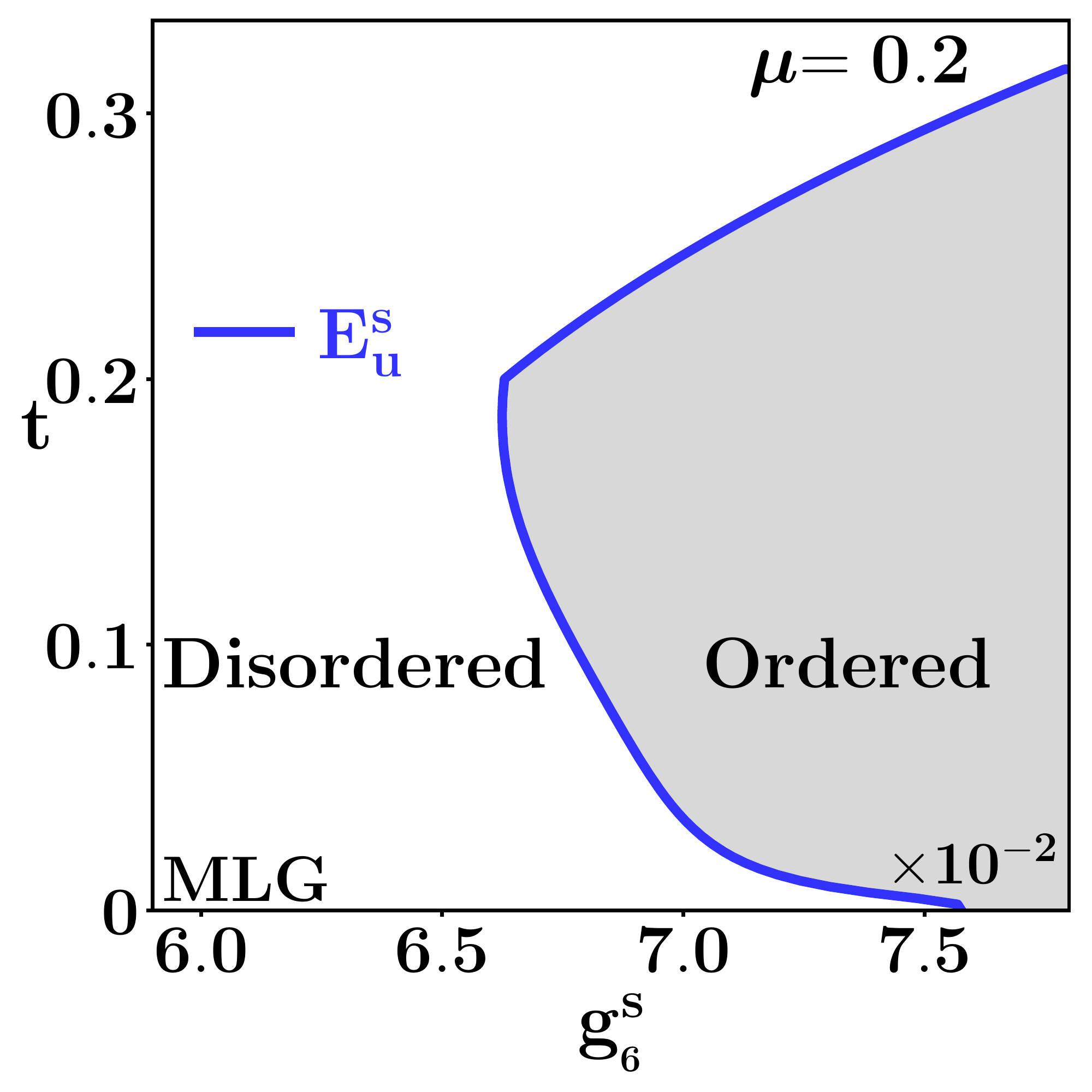}\hspace{0.5cm}
\includegraphics[width=0.21\linewidth]{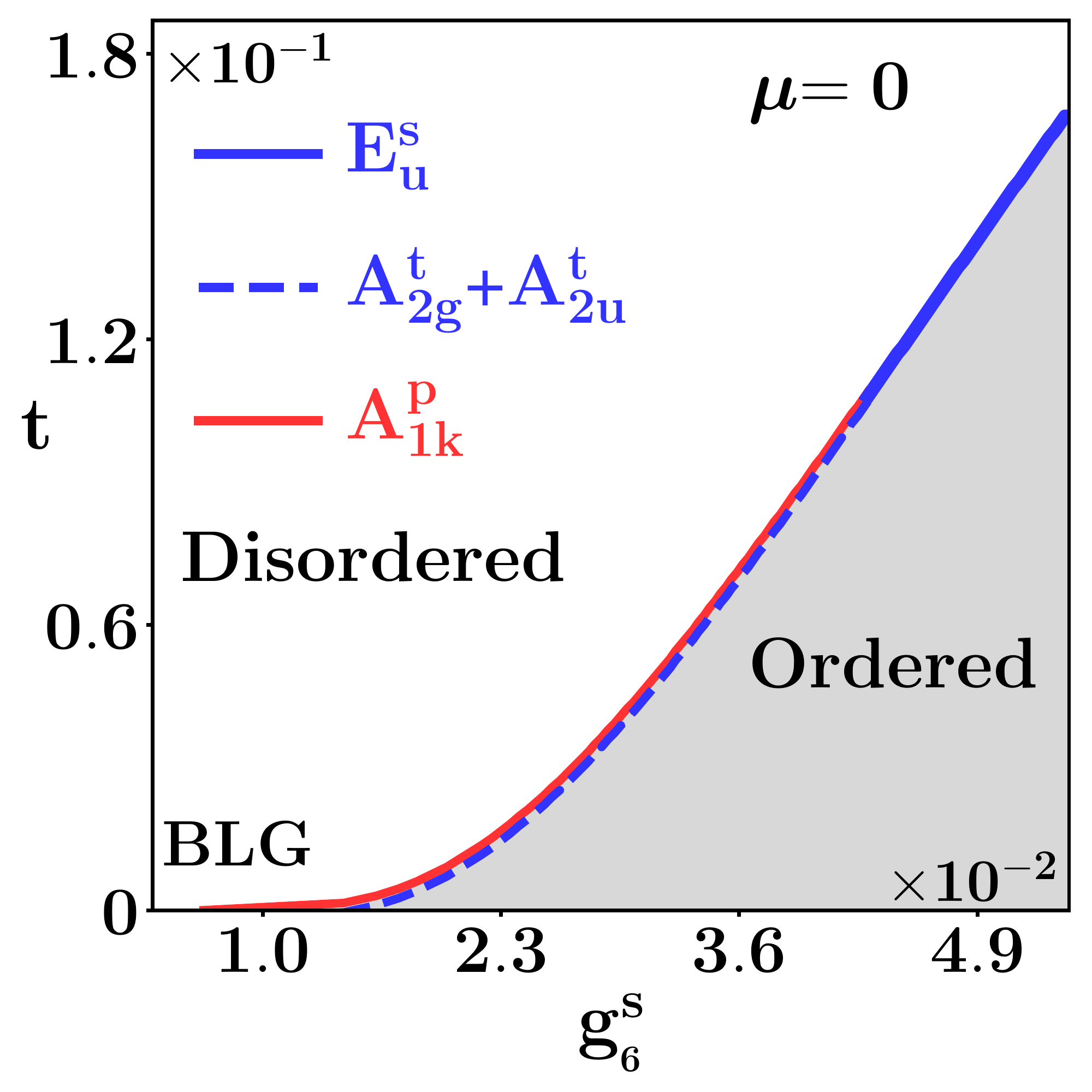}\hspace{0.5cm}
\includegraphics[width=0.21\linewidth]{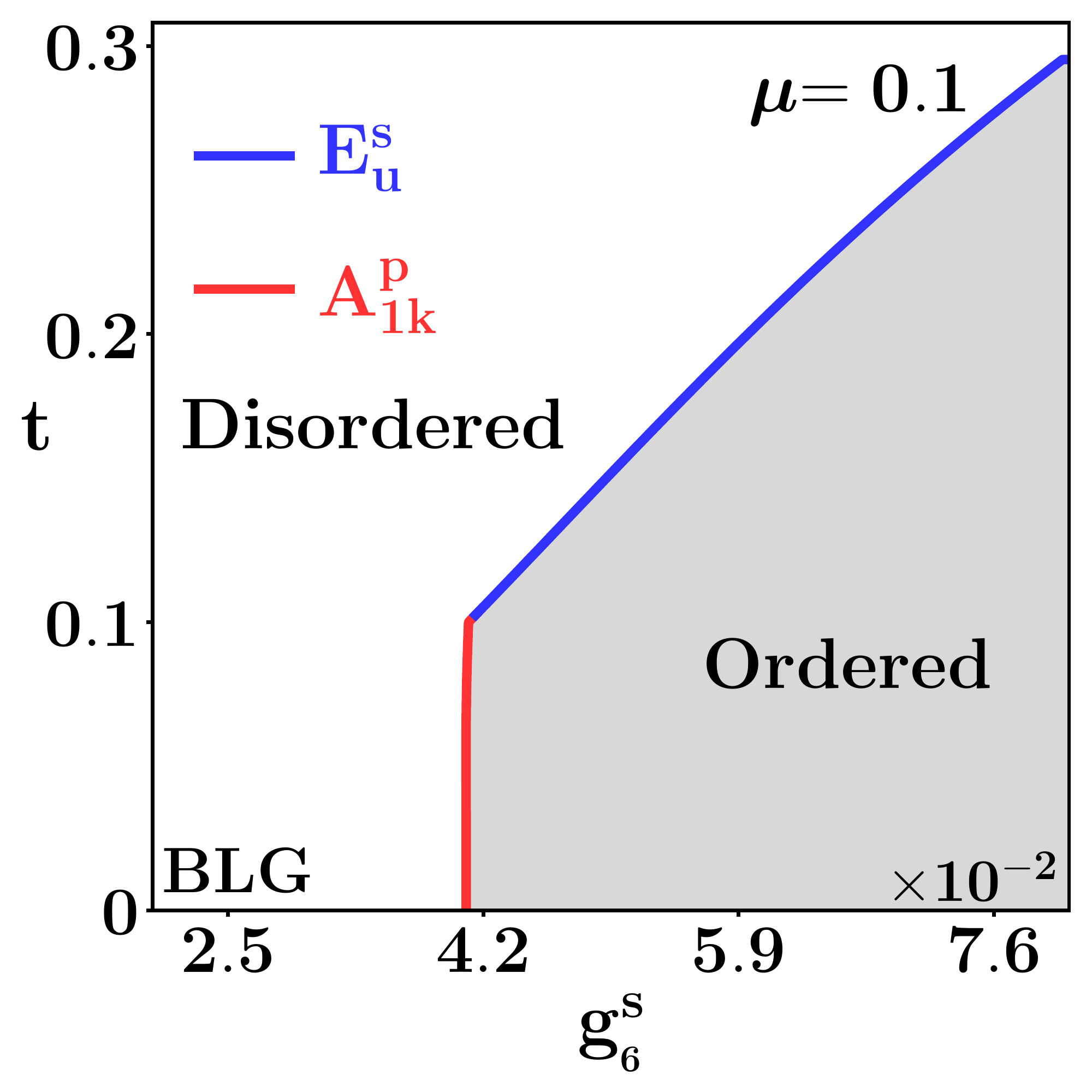}
\label{fig:Eu_nem}}\\
\subfloat[Phase diagrams for the quartic interaction in the spin-nematic$_1$ or $E_g$ triplet channel.]{
\includegraphics[width=0.21\linewidth]{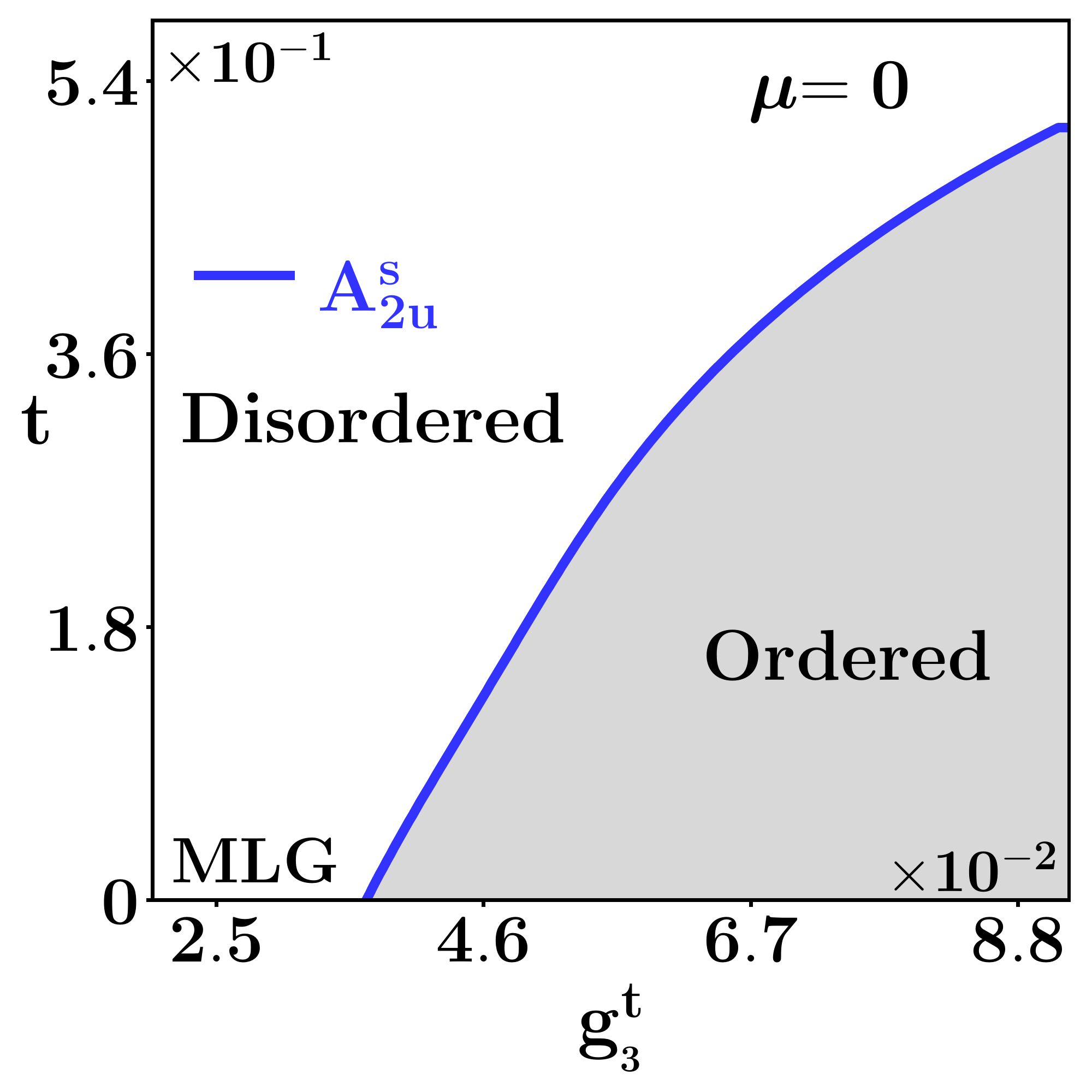}\hspace{0.5cm}
\includegraphics[width=0.21\linewidth]{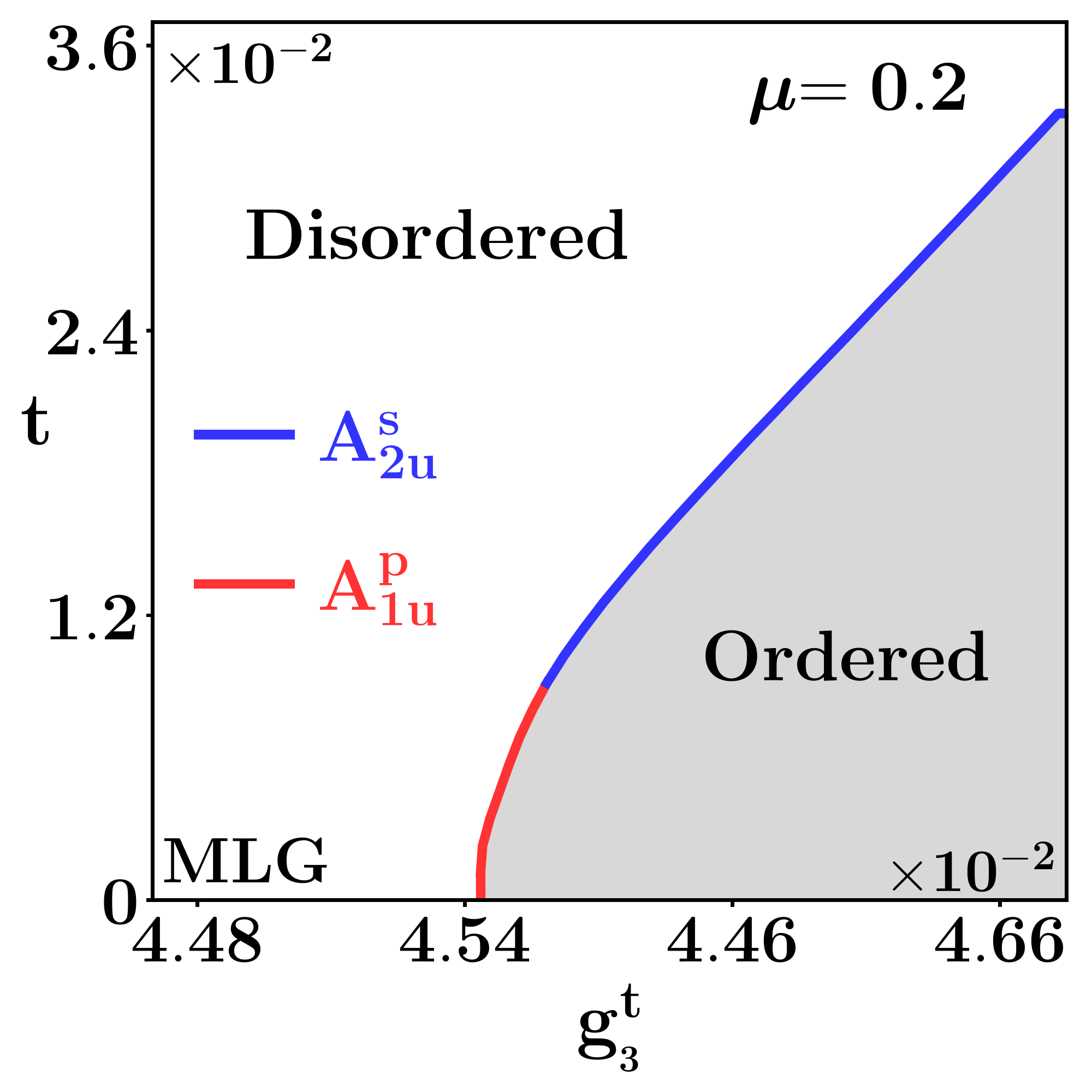}\hspace{0.5cm}
\includegraphics[width=0.21\linewidth]{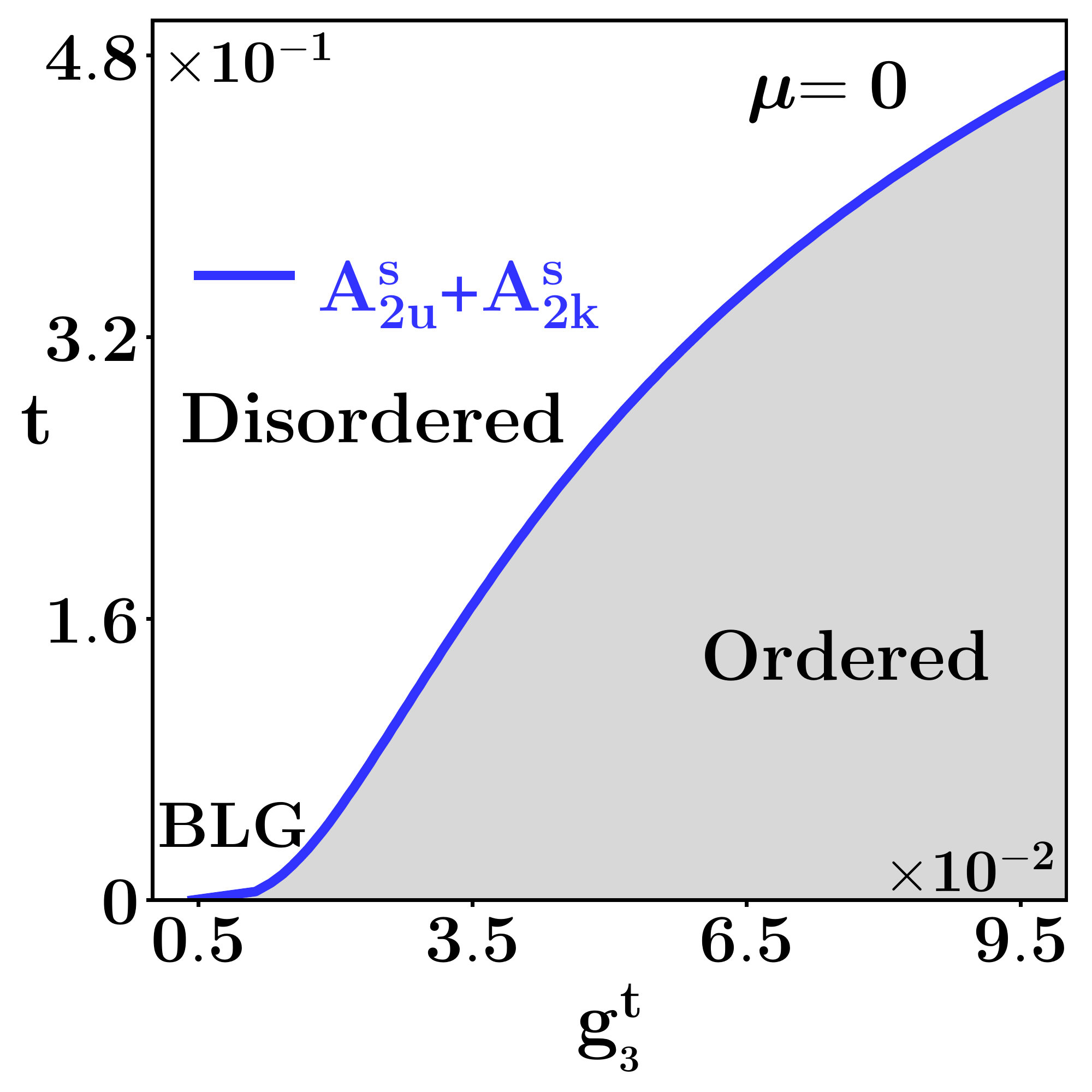}\hspace{0.5cm}
\includegraphics[width=0.21\linewidth]{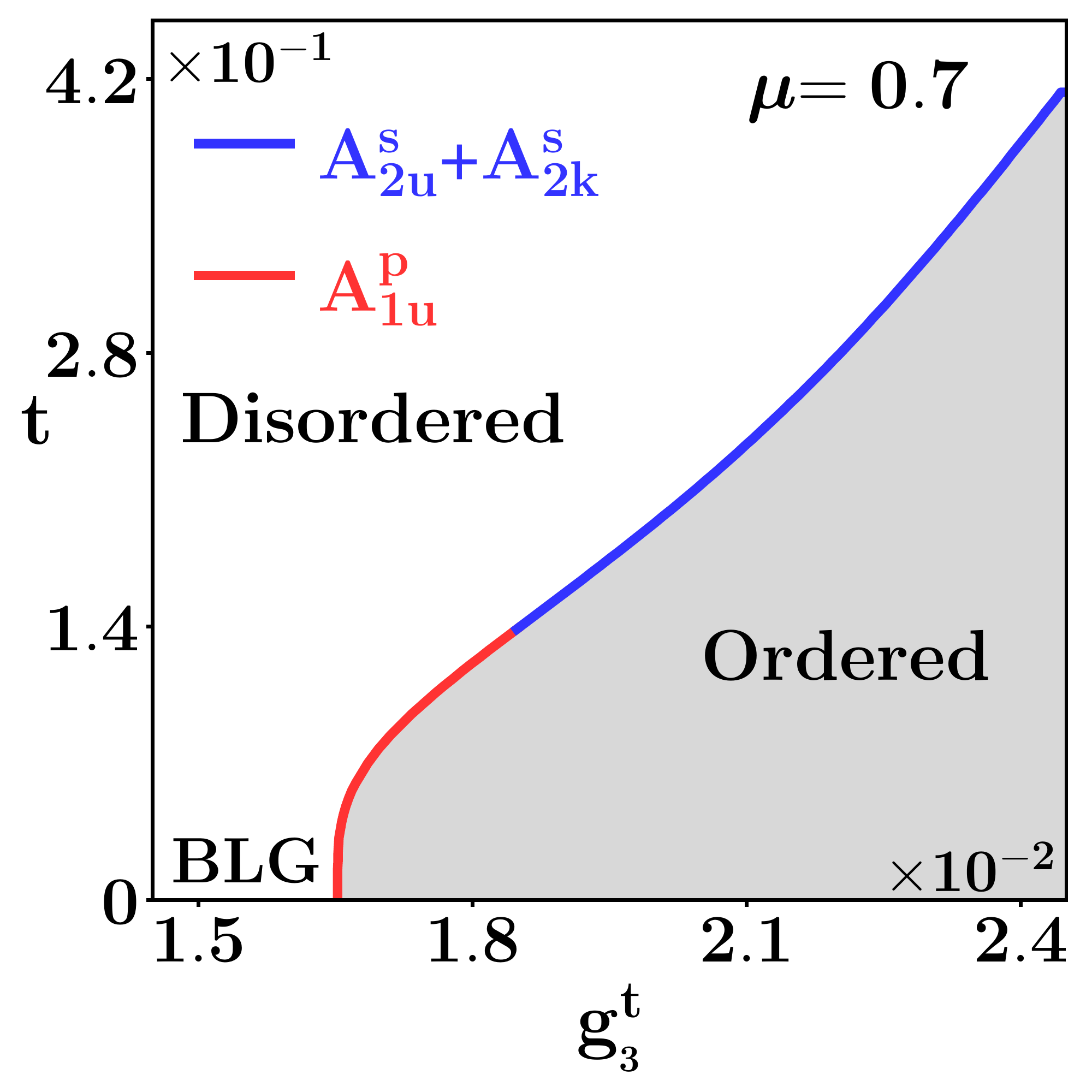}
\label{fig:Eg_spin-nem}}\\
\subfloat[Phase diagrams for the quartic interaction in the spin-nematic$_2$ or $E_u$ triplet channel.]{
\includegraphics[width=0.21\linewidth]{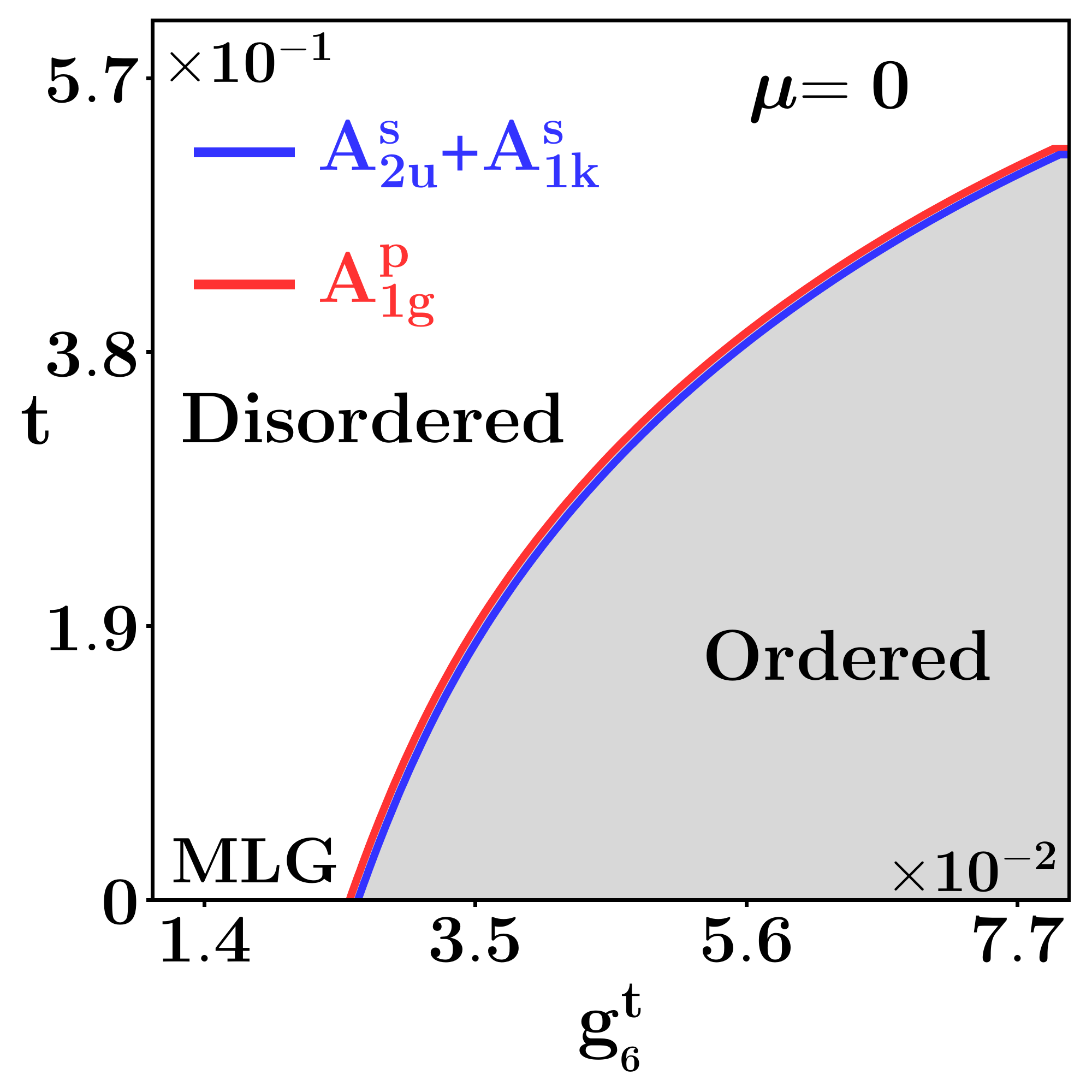}\hspace{0.5cm}
\includegraphics[width=0.21\linewidth]{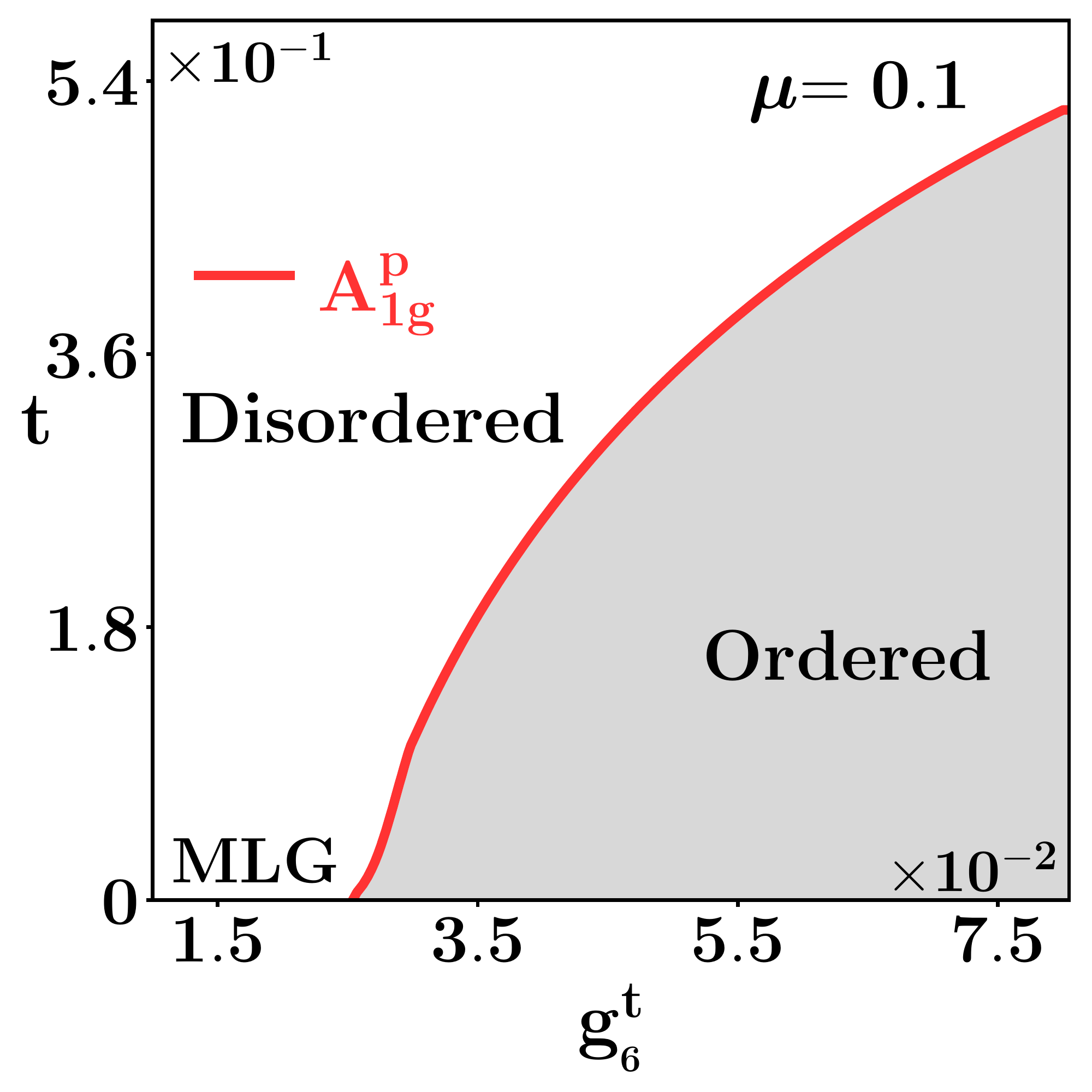}\hspace{0.5cm}
\includegraphics[width=0.21\linewidth]{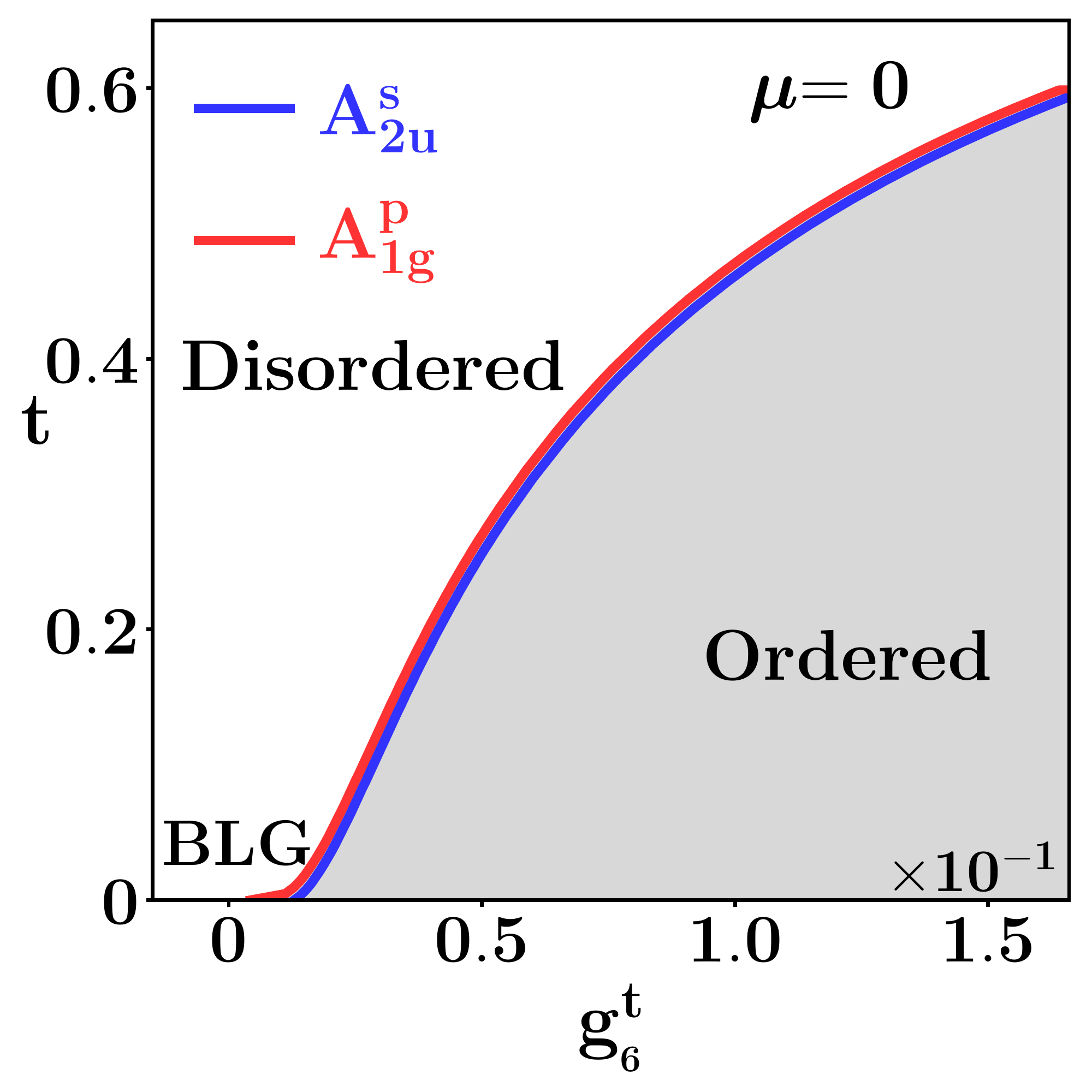}\hspace{0.5cm}
\includegraphics[width=0.21\linewidth]{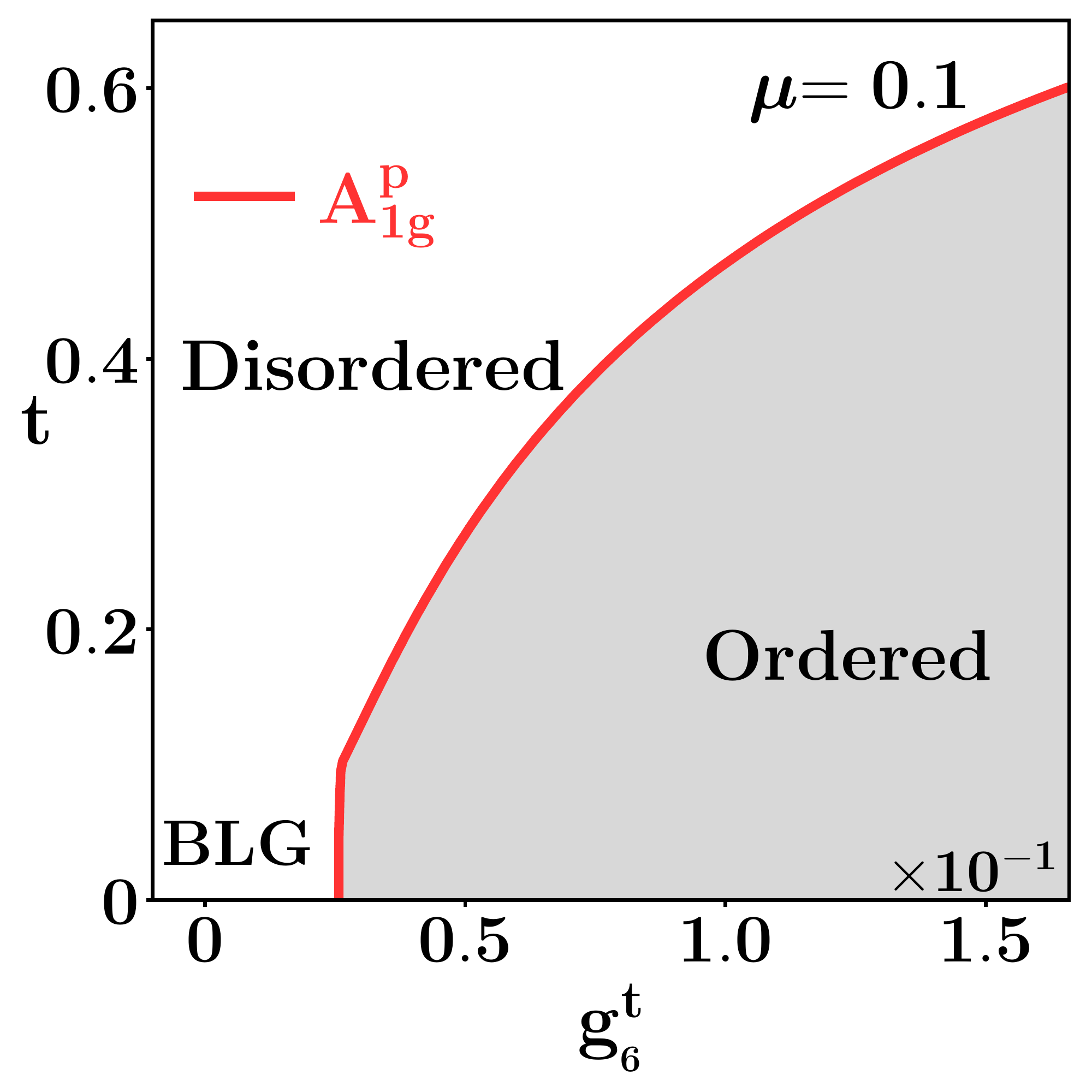}
\label{fig:Eu_spin-nem}}
\caption{Cuts of the global phase diagram with quartic interactions in the (spin-)nematic channels. The rest of the details are identical to those in Fig.~\ref{fig:mass_PD}. For the constructions of composite order parameters from adjacent phases, see Sec.~\ref{sec:PD_nematic}.}~\label{fig:nematic_PD}
\end{figure*}

Besides the 9 independent coupling constants $g^s_{_i}$ we also renormalize the 27 order parameter fields from $H_{\rm local}$ and arrive at their $\beta$-functions taking the schematic form
\begin{align}
\bar{\beta}^j_{\Delta_k}\equiv\frac{\D \Delta_k}{\D \ell}-z=\sum_{l} F_{k,l}^j (t,\mu) g^s_{_l},\label{eq:beta_D}
\end{align}
where $j=$D and L respectively for Dirac and Luttinger fermions in MLG and BLG. Here we absorb the contribution from the scaling dimension ($z$) of the conjugate fields $\Delta$ into its $\beta$-function. For the exact form of $\bar{\beta}^j_{\Delta_k}$ consult Appendix~\ref{app:beta_func}. To identify the nature of the ordered phases, we integrate 27 $\beta$-functions from Eq.~(\ref{eq:beta_D}) simultaneously with the 9 flow equations from Eq.~(\ref{eq:beta_g}). Upon reaching a phase boundary, where $g_{_j}(\ell_d)\sim \mathcal{O}(1)$ for $0<\ell_d \leq \ell^\ast$ for some bare value $g_{_i}(0)$, the largest positive $\Delta_k (\ell_d)$ indicates the dominant instability. The phase diagrams constructed this way are displayed in Figs.~\ref{fig:mass_PD}, \ref{fig:nematic_PD}, \ref{fig:smectic_PD} and \ref{fig:Kekule_PD}, and the phase transitions occurring in various interaction channels at zero and finite temperature and chemical potential are tabulated in Tables.~\ref{tab:ptransitions_sgl} and \ref{tab:ptransitions_tr}.

Notice that along some phase boundaries multiple order parameter fields diverge in a degenerate fashion. However, as we cut the multi-dimensional space of couplings in a very specific way (e.g. along the axes $g^\mu_{_i}$), this does not necessarily indicate an enlargement of symmetry among these orders at the governing quantum critical point, which will be discussed in a forthcoming paper. Rather, this suggests that the respective order parameters simultaneously diverge with a specific phase locking of the internal degrees of freedom. For example, when the divergences of the conjugate fields for both $s$-wave and $f$-wave pairings are degenerate, the ordered state is expected to support $s+if$ or $f+is$ pairing~\cite{PhysRevB.90.041413}, which produces a maximal gap in the quasiparticle spectra.

To facilitate further discussion on order selection in two-dimensional Dirac and Luttinger fermions, let us briefly outline some properties of the available broken symmetry phases. In this section we quote the symbol of the respective orderings, with which they are referred to in the phase diagrams (Figs.~\ref{fig:mass_PD}--\ref{fig:Kekule_PD}). The symbols, together with nomenclature and various symmetry properties of fermion bilinears are collected in Tables~\ref{tab:bilinears_exc} and \ref{tab:bilinears_pair}.

%%%%%%%%%%%%%%%%%%%%%%%%%%%%%%%%%%%%%%%%%%%%%%%%%%%%%%%%%%%%%%%%%%%%%
%%%%%%%%%%%%%%%%%%%%%%%%%%%%%%%%%%%%%%%%%%%%%%%%%%%%%%%%%%%%%%%%%%%%%
%%%%%%%%%%%%%%%%%%%%%%%%%%%%%%%%%%%%%%%%%%%%%%%%%%%%%%%%%%%%%%%%%%%%%
%%%%%%%%%%%%%%%%%%%%%%%%%%%%%%%%%%%%%%%%%%%%%%%%%%%%%%%%%%%%%%%%%%%%%
%%%%%%%%%%%%%%%%%%%%%%%%%%%%%%%%%%%%%%%%%%%%%%%%%%%%%%%%%%%%%%%%%%%%%
\subsubsection{Particle-hole or excitonic orders}

Due to the neglected spin-orbit coupling the excitonic order parameters (just like the interaction terms) fall into two distinct categories. Namely, for each spin singlet order, where all matrices come with the $\sigma_0$ matrix, there exists a spin triplet analog appearing with the $\sigma_s$ matrices, where $s=\{1,2,3\}$, operating on the spin index. We indicate this distinction by $s$ (singlet) and $t$ (triplet) superscripts on the corresponding conjugate fields. Since the Hamiltonians in Eqs.~(\ref{eq:H0_MLG}) and (\ref{eq:H0_BLG}) are oblivious to spin or invariant under the rotation of the spin quantization axis, this distinction does not affect the emergent fermionic quasiparticle spectra inside the ordered state. Note, however, that any triplet ordering breaks SU(2) spin rotational symmetry and is hence accompanied by two massless Goldstone modes.

The fermionic density and spin density (respectively denoted by $A_{1g}^s$ and $A_{1g}^t$) do not break any discrete lattice symmetries, while the chiral and spin-chiral chemical potential ( denoted by $A_{1u}^s$ and $A_{1u}^t$, respectively) break valley reflection symmetry ($T$). All four orders commute with both the Dirac and Luttinger Hamiltonians, resulting in a trivial renormalization of their conjugate fields at zero temperature, see Appendix~\ref{sec:App_diagrams}. Consequently, none of them is realized in the ordered phase at $t=0$.

%%%%%%%%%%%%%%%%%%%%%%%%%%%%%%%%%%%%%%%%%%%%%%%%%%%%%%%%%%%%%%%%%%%%%%%%%
%%%%%%%%%%%%%%%%%%%%%%%%%%%%%%%%%%%%%%%%%%%%%%%%%%%%%%%%%%%%%%%%%%%%%%%%%
%%%%%%%%%%%%%%%%%%%% PHASE DIAGRAMS - NEMATIC CHANNELS %%%%%%%%%%%%%%%%%%
%%%%%%%%%%%%%%%%%%%%%%%%%%%%%%%%%%%%%%%%%%%%%%%%%%%%%%%%%%%%%%%%%%%%%%%%%
%%%%%%%%%%%%%%%%%%%%%%%%%%%%%%%%%%%%%%%%%%%%%%%%%%%%%%%%%%%%%%%%%%%%%%%%%
%%%%%%%%%%%%%%%%%%%%%%%%%%%%%%%%%%%%%%%%%%%%%%%%%%%%%%%%%%%%%%%%%%%%%%%%%
%%%%%%%%%%%%%%%%%%%%%%%%%%%%%%%%%%%%%%%%%%%%%%%%%%%%%%%%%%%%%%%%%%%%%%%%%
%%%%%%%%%%%%%%%%%%%%%%%%%%%%%%%%%%%%%%%%%%%%%%%%%%%%%%%%%%%%%%%%%%%%%%%%%
%%%%%%%%%%%%%%%%%%%%%%%%%%%%%%%%%%%%%%%%%%%%%%%%%%%%%%%%%%%%%%%%%%%%%%%%%
%%%%%%%%%%%%%%%%%%%%%%%%%%%%%%%%%%%%%%%%%%%%%%%%%%%%%%%%%%%%%%%%%%%%%%%%%

Both the Dirac and Luttinger systems altogether accommodate \emph{six} mass orders in the particle-hole channel~\cite{PhysRevB.80.205319, PhysRevB.88.075415}. These phases introduce an isotropic gap in the quasiparticle spectrum and are thus favorable at zero and low temperature. All of them fully anticommute with the noninteracting Hamiltonian. The quantum Hall states break both sublattice ($S$) and valley ($T$) reflection symmetries. The quantum anomalous Hall insulator (QAHI), denoted by $A_{2g}^s$, additionally breaks time reversal symmetry, while it is restored in its spin triplet counterpart, the quantum spin Hall insulator (QSHI), denoted by $A_{2g}^t$~\cite{PhysRevLett.61.2015, PhysRevLett.100.156401, PhysRevLett.95.226801}. The charge density wave (CDW) and antiferromagnet (AFM) order parameters break sublattice reflection, but preserve valley reflection symmetry~\cite{PhysRevLett.97.146401, PhysRevB.79.085116}. They are respectively denoted by $A_{2u}^s$ and $A_{2u}^t$. The QAHI, QSHI, CDW and AFM orderings represent massive phases in both Dirac and Luttinger liquids. Both models accommodate two additional masses in the family of Kekul\'e orders.

The Kekul\'e orderings break translational symmetry of the honeycomb lattice and result in the enlargement of the unit cell, while preserving the rotational invariance~\cite{PhysRevLett.98.186809}. In MLG, the Kekul\'e valence bond solid (KVBS) and spin-Kekul\'e solid (sKS) are fully gapped phases, while the Kekul\'e current (KC) and spin-Kekul\' e current (sKC) order parameters fully commute with $\hat{h}^D$. Exactly the opposite is true in BLG, namely KC and sKC represent spin singlet and spin triplet masses for Luttinger fermions, respectively, while $\hat{h}^L$ fully commutes with the KVBS and sKS order parameters~\cite{PhysRevB.88.075415}. The KVBS, sKS, KC and sKC phases are respectively denoted by $A_{1k}^s$, $A_{1k}^t$, $A_{2k}^s$, and $A_{2k}^t$.

All rotational symmetry breaking phases are gapless, and due to the higher density of states at low energies they are more entropically favorable than their gapped counterparts. Such orderings in MLG merely shift the location of the Dirac points, while in BLG it splits each quadratic band touching into two linear Dirac cones~\cite{PhysRevB.88.075415}. Consequently, all the nematic and spin nematic orders cause power-law suppression of the DOS for Luttinger fermions, yielding $\rho(E) \sim |E|$ at low energies. Four nematic phases, transforming under the $E_g$ and $E_u$ representations of the $D_{3d}$ point group, preserve translational invariance. Two of them are spin singlets (denoted by $E_g^s$ and $E_u^s$), while two spin-nematic order parameters (denoted by $E_g^t$ and $E_u^t$) additionally breaks the spin-rotational invariance. In addition to rotational invariance, translational symmetry is also broken in the two smectic phases, the smectic charge density wave and smectic spin density wave orderings, respectively denoted by $E_k^s$ and $E_k^t$. The smectic phases also produce gapless, but anisotropic Dirac points in the ordered phases.

%%%%%%%%%%%%%%%%%%%%%%%%%%%%%%%%%%%%%%%%%%%%%%%%%%%%%%%%%%%%%%%%%%%%%
%%%%%%%%%%%%%%%%%%%%%%%%%%%%%%%%%%%%%%%%%%%%%%%%%%%%%%%%%%%%%%%%%%%%%
%%%%%%%%%%%%%%%%%%%%%%%%%%%%%%%%%%%%%%%%%%%%%%%%%%%%%%%%%%%%%%%%%%%%%
%%%%%%%%%%%%%%%%%%%%%%%%%%%%%%%%%%%%%%%%%%%%%%%%%%%%%%%%%%%%%%%%%%%%%
%%%%%%%%%%%%%%%%%%%%%%%%%%%%%%%%%%%%%%%%%%%%%%%%%%%%%%%%%%%%%%%%%%%%%
\subsubsection{Particle-particle or superconducting orders}

When we add chemical doping to these nodal systems, instead of Fermi points, they develop extended Fermi surfaces around the two independent band touching points. The chemical potential term is of the form $\mu \Gamma_{3000}$, which commutes with all excitonic orders, therefore they cannot gap such metallic systems. On the other hand, all pairing orders include $\eta_\alpha=\{\eta_1,\eta_2\}$ in the particle-hole sector, and hence they can conceivably anticommute with $h^j(\vec{k})$ when $\mu$ is finite. Therefore, low temperatures (when a gapped spectrum is favorable due to the gain of condensation energy) and finite $\mu$ are conducive for superconducting orders, as the increased carrier density is conducive for the condensation of electrons into Cooper pairs.

The Nambu basis altogether accommodates nine local or momentum-independent pairing orders, which we mark with a superscript $p$ in Table~\ref{tab:bilinears_pair}. The singlet $s$-wave (denoted by $A_{1g}^p$) and triplet $f$-wave (denoted by $A_{1u}^p$)~\cite{PhysRevLett.100.146404} superconductors (SCs) represent fully gapped phases in MLG and BLG. The singlet and triplet Kekul\'e SCs break translational symmetry. The singlet Kekul\' e SC (denoted by $A^p_{1k}$) represents a mass order in the Luttinger system (BLG), but fully commutes with the Dirac Hamiltonian~\cite{PhysRevB.88.075415}. In the exact opposite way, the triplet Kekul\'e SC (denoted by $A^p_{2k}$) gaps a Dirac liquid, but fully commutes with $\hat{h}^{\rm L}$~\cite{PhysRevB.82.035429}. Two nematic superconductors, transforming under the $E_g$ and $E_u$ representations of the $D_{3d}$ group (denoted by $E_g^p$ and $E_u^p$, respectively), break rotational symmetry and produce gapless quasiparticle spectra. Furthermore, the smectic pairing (denoted by $E_k^p$) breaks rotational and translational invariance, and also leads to gapless bands. While the nematic and smectic pairings produce point nodes in the spectra of emergent Bogoliubov-de Gennes quasiparticles, the remaining two pairings transforming under the $A_{2g}$ and $A_{2u}$ representations (denoted by $A^p_{2g}$ and $A^p_{2u}$, respectively) produces Fermi surfaces inside the ordered phases.

\section{Phase diagrams}~\label{sec:phasediagrams}

In this section, we scrutinize multiple cuts of the global phase diagram of interacting fermions in MLG and BLG by systematically increasing the strength of interactions in one channel at a time, both at zero and finite chemical potential. The fact that the form of contact interactions in MLG and BLG is identical and they only differ in their noninteracting Hamiltonians gives us an opportunity to further dissect the selection rules outlined in Sec.~\ref{sec:extendedsum:selection} and assess the imprint of normal state band structure on the global phase diagram. This is because selection rule (I) only depends on $S_{\rm int}$ while (II) only on $S_0$. Consequently, (I) cannot distinguish between MLG and BLG, and therefore any deviation must be rooted in (II).

Our methodology, involving Wilsonian momentum shell RG analysis and $\epsilon$ expansion around the lower critical dimension, is as outlined in Secs.~\ref{sec:interactions} and \ref{sec:ordered_phases}. Each phase diagram is displayed in the $(g_\mu^i,t)$ plane, where $g_\mu^i$ is a dimensionless (bare) coupling constant with $\mu=1,\dots,9$ and $i=$ singlet ($s$) and triplet ($t$), and $t$ is dimensionless temperature. We scan for the phase boundaries by keeping temperature constant and increasing the bare interaction strength, and detect the phase boundary and the nature of symmetry breaking \emph{between disordered and ordered phases}, marked in the phase diagrams respectively as white and gray regions. Respectively, at zero and finite doping the disordered phase represents chiral nodal Fermi liquids (with point nodes) and regular Fermi liquids (with extended Fermi surface). Phase boundaries from the disordered state into an excitonic (superconducting) order are indicated with blue (red) lines. The RG procedure is only equipped to detect the divergence of some conjugate field that directly couples with a fermion bilinear, indicating onset of the corresponding ordered state and hence a phase transition. But, it does not tell us about potential regions of coexistence of adjacent phases deep inside the ordered phase. Also note that the critical interaction strength and the transition temperature are non-universal quantities.

%%%%%%%%%%%%%%%%%%%%%%%%%%%%%%%%%%%%%%%%%%%%%%%%%%%%%%%%%%%%%%%%%%%%%%%%%%%%%%%%%%%%%%%%%%%%
%%%%%%%%%%%%%%%%%%%%%%%%%%%%%%%%%%%%%%%%%%%%%%%%%%%%%%%%%%%%%%%%%%%%%%%%%%%%%%%%%%%%%%%%%%%%
%%%%%%%%%%%%%%%%%% PHASE TRANSITIONS TABLE SINGLET: %%%%%%%%%%%%%%%%%%%%%%%%%%%%%%%%%%%%%%%%
%%%%%%%%%%%%%%%%%%%%%%%%%%%%%%%%%%%%%%%%%%%%%%%%%%%%%%%%%%%%%%%%%%%%%%%%%%%%%%%%%%%%%%%%%%%%
%%%%%%%%%%%%%%%%%%%%%%%%%%%%%%%%%%%%%%%%%%%%%%%%%%%%%%%%%%%%%%%%%%%%%%%%%%%%%%%%%%%%%%%%%%%%
\begin{table*}[t!]
\renewcommand{\arraystretch}{1.3}
\begin{tabular}{|c | c c c c c c|c c c c c c|}
\hline
\multirow{3}{*}{CC} & \multicolumn{6}{c|}{ Monolayer graphene (MLG) } & \multicolumn{6}{c|}{ Bilayer graphene (BLG) } \\ 
%\cline{2-13}
 & \multicolumn{3}{c}{$\mu=0$} & \multicolumn{3}{c|}{$\mu > 0$} & \multicolumn{3}{c}{$\mu=0$} & \multicolumn{3}{c|}{$\mu > 0$}  \\
%\cline{2-13}
& Low $t$ & High $t$ & Symmetry & Low $t$ & High $t$ & Symmetry & Low $t$ & High $t$ & Symmetry & Low $t$ & High $t$ & Symmetry  \\
\hline
%$g_{_{1}}^s$ & 
%- & $A_{1g}^s$ & & - & $A_{1g}^s$ & &
%$A_{1g}^p,A_{1u}^p,A_{1k}^p$ & $A_{1g}^s$ & & $A_{1g}^p,A_{1u}^p,A_{1k}^p$ & $A_{1g}^s$ & \\
%\hline
$g_{_2}^s$ & 
$A_{2g}^s$ & $A_{2g}^s$ & - & \begin{tabular}{@{}c@{}} $E_{g}^p$ \\ $E_{u}^p$ \\ $E_{k}^p$ \end{tabular}  &  $A_{2g}^s$ & \begin{tabular}{@{}c@{}} $2\otimes$O(3) \\ $4\otimes$O(4) \\ $4\otimes$O(3) \end{tabular} &
$A_{2g}^s$ &  $A_{2g}^s$ & - & \begin{tabular}{@{}c@{}} $E_{g}^p$ \\ $E_{u}^p$ \\ $E_{k}^p$ \end{tabular} & $A_{2g}^s$ & \begin{tabular}{@{}c@{}} $2\otimes$O(3) \\ $4\otimes$O(4) \\ $4\otimes$O(3) \end{tabular}\\
\rowcolor{RowColor}
$g_{_3}^s$ & 
$E_{g}^s$ & $E_{g}^s$ & - & \begin{tabular}{@{}c@{}} $A_{1g}^p$ \\ $A_{1u}^p$ \end{tabular}
 & $E_{g}^s$ & \begin{tabular}{@{}c@{}} $1\otimes$O(4) \\ $2\otimes$O(5) \end{tabular} &
$E_{g}^s$ & $E_{g}^s$ & - & \begin{tabular}{@{}c@{}} $A_{1g}^p$ \\ $A_{1u}^p$ \end{tabular} & $E_{g}^s$ & \begin{tabular}{@{}c@{}} $1\otimes$O(4) \\ $2\otimes$O(5) \end{tabular}\\
%
%
%$g_{_4}^s$ & 
%- & $A_{1u}^s$ & & - & $A_{1u}^s$ & &
%$A_{2k}^s,A_{2k}^t,A_{1k}^p$  & $A_{1u}^s$ & & $A_{1k}^p$ & $A_{1u}^s$ & \\
%\hline
%
%
$g_{_5}^s$ &
$A_{2u}^s$ & $A_{2u}^s$ & - & \begin{tabular}{@{}c@{}} $A_{1g}^p$ \\ $A_{1u}^p$ \end{tabular} & $A_{2u}^s$ & \begin{tabular}{@{}c@{}} $1\otimes$O(3) \\ $1\otimes$O(3) \end{tabular} &
$A_{2u}^s$ & $A_{2u}^s$ & - & \begin{tabular}{@{}c@{}} $A_{1g}^p$ \\ $A_{1u}^p$ \end{tabular} & $A_{2u}^s$ & \begin{tabular}{@{}c@{}} $1\otimes$O(3) \\ $1\otimes$O(3) \end{tabular} \\
\rowcolor{RowColor}
%
% g6:
$g_{_6}^s$ &
$E_{u}^s$ & $E_{u}^s$ & - & $E_{u}^s$ & $E_{u}^s$ & - &
\begin{tabular}{@{}c@{}} $A_{2g}^t$ \\ $A_{2u}^t$ \\ $A_{1k}^p$ \end{tabular}    & $E_{u}^s$ & \begin{tabular}{@{}c@{}} $1\otimes$O(5) \\ $1\otimes$O(5) \\ $2\otimes$O(4) \end{tabular} & $A_{1k}^p$ & $E_{u}^s$ & $1\otimes$O(4)\\
$g_{_7}^s$ &
$A_{1k}^s$ & $A_{1k}^s$ & - & $A_{1g}^p$ & $A_{1k}^s$ & $1\otimes$O(4) &
\begin{tabular}{@{}c@{}} $A_{2u}^s$ \\$A_{2u}^t$ \\ $A_{1g}^p$ \end{tabular}  & $A_{1k}^s$ & \begin{tabular}{@{}c@{}} $1\otimes$O(3) \\ $1\otimes$O(5) \\ $1\otimes$O(4) \end{tabular} & $A_{1g}^p$ & $A_{1k}^s$ & $1\otimes$O(4)\\
\cellcolor{RowColor}$g_{_8}^s$ & 
\begin{tabular}{@{}c@{}} \cellcolor{RowColor} $A_{1u}^p$ \\ \cellcolor{RowColor}$A_{2u}^s$ \\ \cellcolor{RowColor}$A_{2u}^t$ \\  \cellcolor{RowColor2} $E_g^s$ \end{tabular} & 
\begin{tabular}{@{}c@{}} \cellcolor{RowColor} \\ \cellcolor{RowColor} $E_g^s$ \\  \cellcolor{RowColor} \\ \cellcolor{RowColor2} $A_{2k}^s$ \end{tabular} & 
\begin{tabular}{@{}c@{}} \cellcolor{RowColor} $2\otimes$O(5) \\ \cellcolor{RowColor} $1\otimes$O(3) \\ \cellcolor{RowColor}$1\otimes$O(5) \\ \cellcolor{RowColor2} $1\otimes$O(4) \end{tabular}&
\begin{tabular}{@{}c@{}} \rowcolor{RowColor} $A_{1u}^p$ \\  $E_{g}^s$ \end{tabular} &
\begin{tabular}{@{}c@{}}  $E_{g}^s$ \\  $A_{2k}^s$ \end{tabular} &
\begin{tabular}{@{}c@{}}  $2\otimes$O(5) \\  $1\otimes$O(4) \end{tabular} &
$A_{2k}^s$ & $A_{2k}^s$ & - & $A_{1u}^p$ & $A_{2k}^s$ & $2\otimes$O(5) \\
$g_{_9}^s$ &
\begin{tabular}{@{}c@{}} $A_{2g}^t$ \\ $A_{1g}^p$ \end{tabular} & $E_{k}^s$ & \begin{tabular}{@{}c@{}} $2\otimes$O(5) \\ $2\otimes$O(4) \end{tabular} & $A_{1g}^p$ & $E_{k}^s$ & $2\otimes$O(4) &
\begin{tabular}{@{}c@{}} $A_{2g}^t$ \\ $A_{1g}^p$ \end{tabular} &  $E_{k}^s$ & \begin{tabular}{@{}c@{}} $2\otimes$O(5) \\ $2\otimes$O(4) \end{tabular} & $A_{1g}^p$ & $E_{k}^s$ & $2\otimes$O(4) \\
\hline
\end{tabular}
\caption{Dominant instabilities in the presence of quartic interactions in the spin singlet channels in MLG and BLG, at zero and finite chemical potential ($\mu$). The blue cells correspond to the second set of adjacent phases for interaction in the $A_{2k}$ singlet channel, which hosts three different phase transition between disordered and ordered phases in MLG for $\mu=0$, see Fig.~\ref{fig:kek_C}. The low and the high temperature ($t$) phases are indicated in each scenario, suggesting adjacent phases when the two orderings are different. We then display the nature of the composite order parameters in columns ``Symmetry'', where $k\otimes$O(N) indicates $k$ copies of an O(N) algebra. The first column shows the coupling constant (CC) of the interaction channel. See Tables~\ref{tab:bilinears_exc} and \ref{tab:bilinears_pair} for notations. When the low and high temperature phases are identical the notion of composite order parameter becomes moot. See Sec.~\ref{sec:phasediagrams} for detailed discussion and Figs.~\ref{fig:mass_PD}-\ref{fig:Kekule_PD} for various cuts of the phase diagrams.}~\label{tab:ptransitions_sgl}
\end{table*}
%%%%%%%%%%%%%%%%%%%%%%%%%%%%%%%%%%%%%%%%%%%%%%%%%%%%%%%%%%%%%%%%%%%%%%%%%%%%%%%%%%%%%%%
%%%%%%%%%%%%%%%%%%%%%%%%%%%%%%%%%%%%%%%%%%%%%%%%%%%%%%%%%%%%%%%%%%%%%%%%%%%%%%%%%%%%%%%
%%%%%%%%%%%%%%%%%%%%%%%%%%%%%%%%%%%%%%%%%%%%%%%%%%%%%%%%%%%%%%%%%%%%%%%%%%%%%%%%%%%%%%%
%%%%%%%%%%%%%%%%%%%%%%%%%%%%%%%%%%%%%%%%%%%%%%%%%%%%%%%%%%%%%%%%%%%%%%%%%%%%%%%%%%%%%%%
%%%%%%%%%%%%%%%%%%%%%%%%%%%%%%%%%%%%%%%%%%%%%%%%%%%%%%%%%%%%%%%%%%%%%%%%%%%%%%%%%%%%%%%

\begin{figure*}[t]
\subfloat[Phase diagrams for the quartic interaction in the smectic charge density wave or $E_k$ singlet channel.]{
\includegraphics[width=0.21\linewidth]{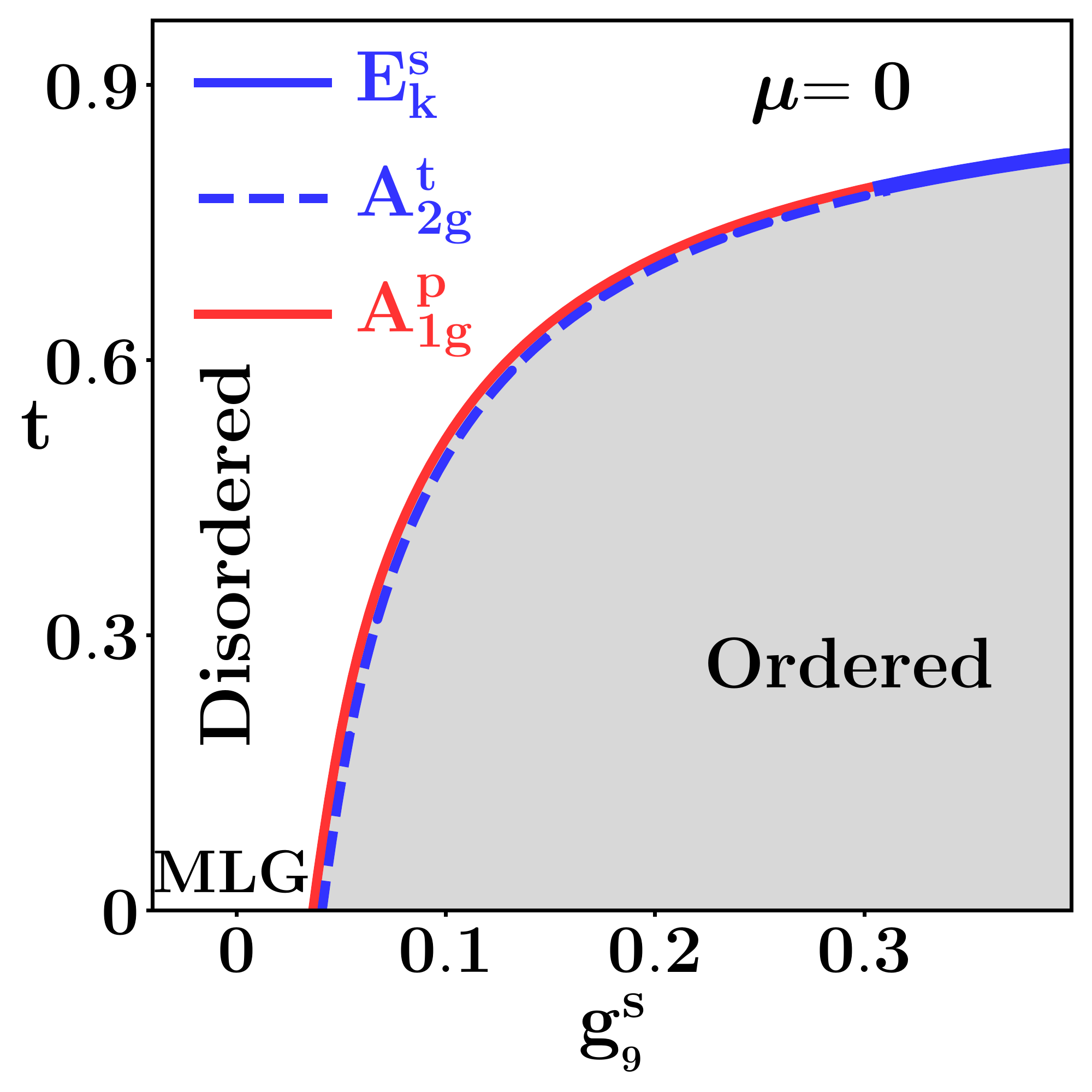}\hspace{0.5cm}
\includegraphics[width=0.21\linewidth]{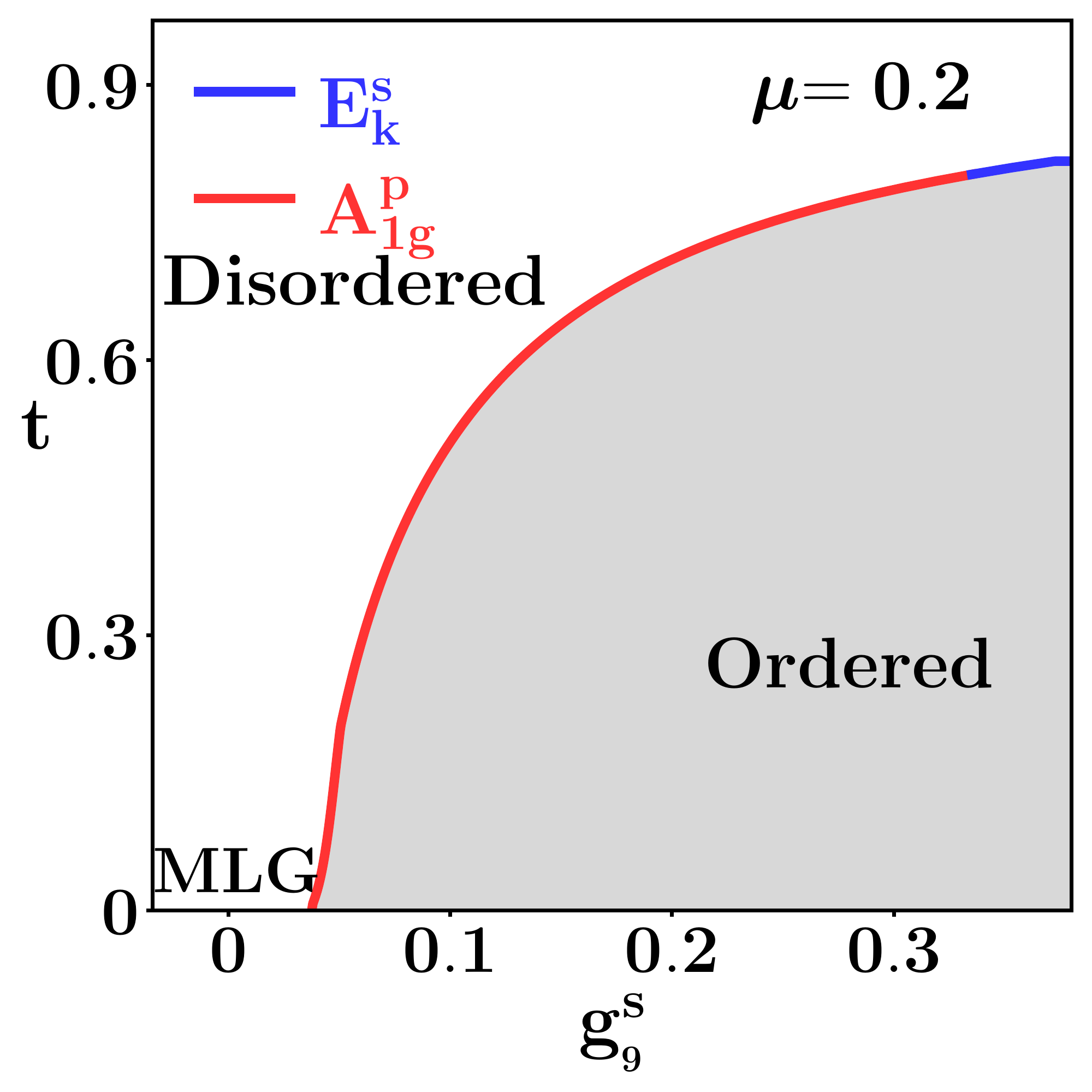}\hspace{0.5cm}
\includegraphics[width=0.21\linewidth]{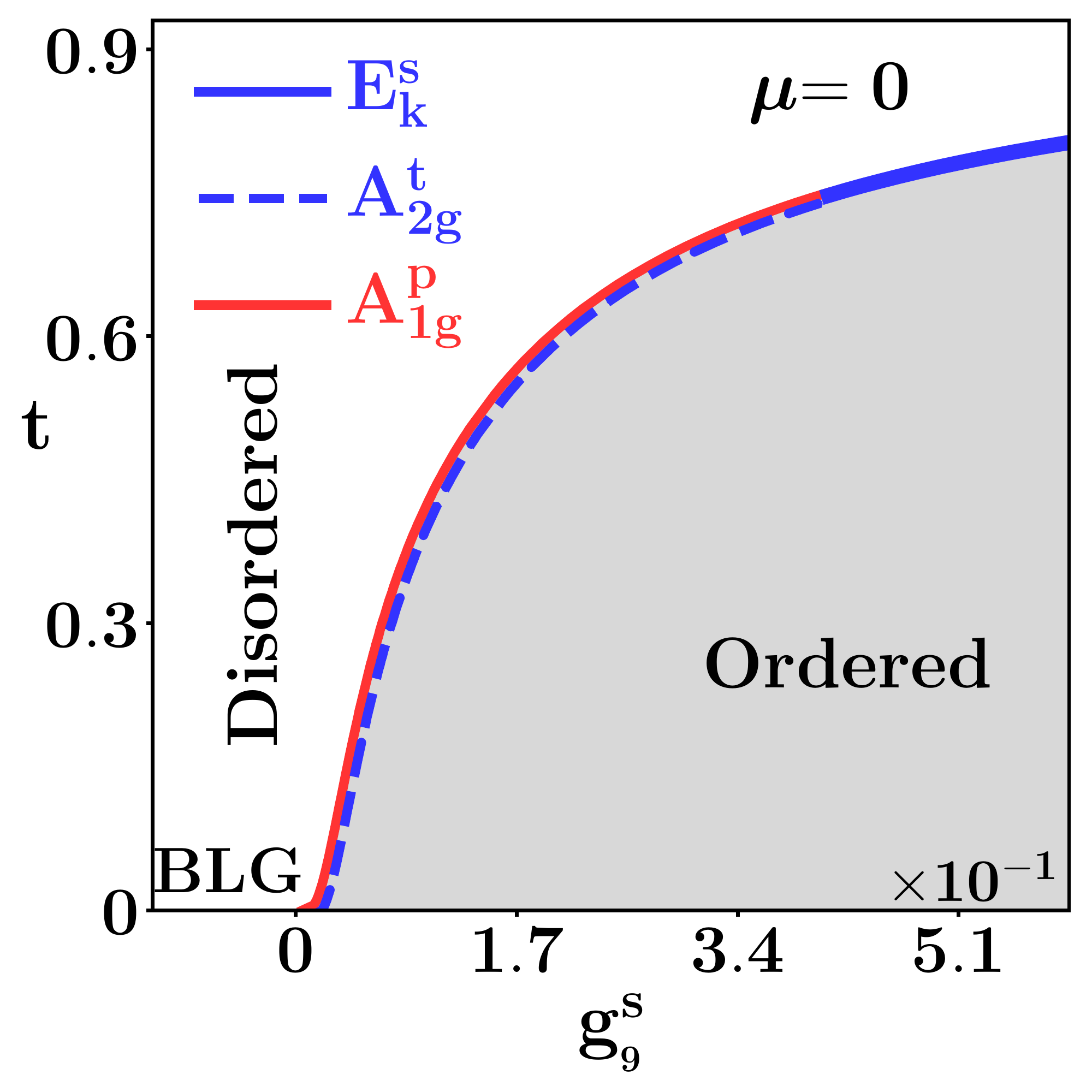}\hspace{0.5cm}
\includegraphics[width=0.21\linewidth]{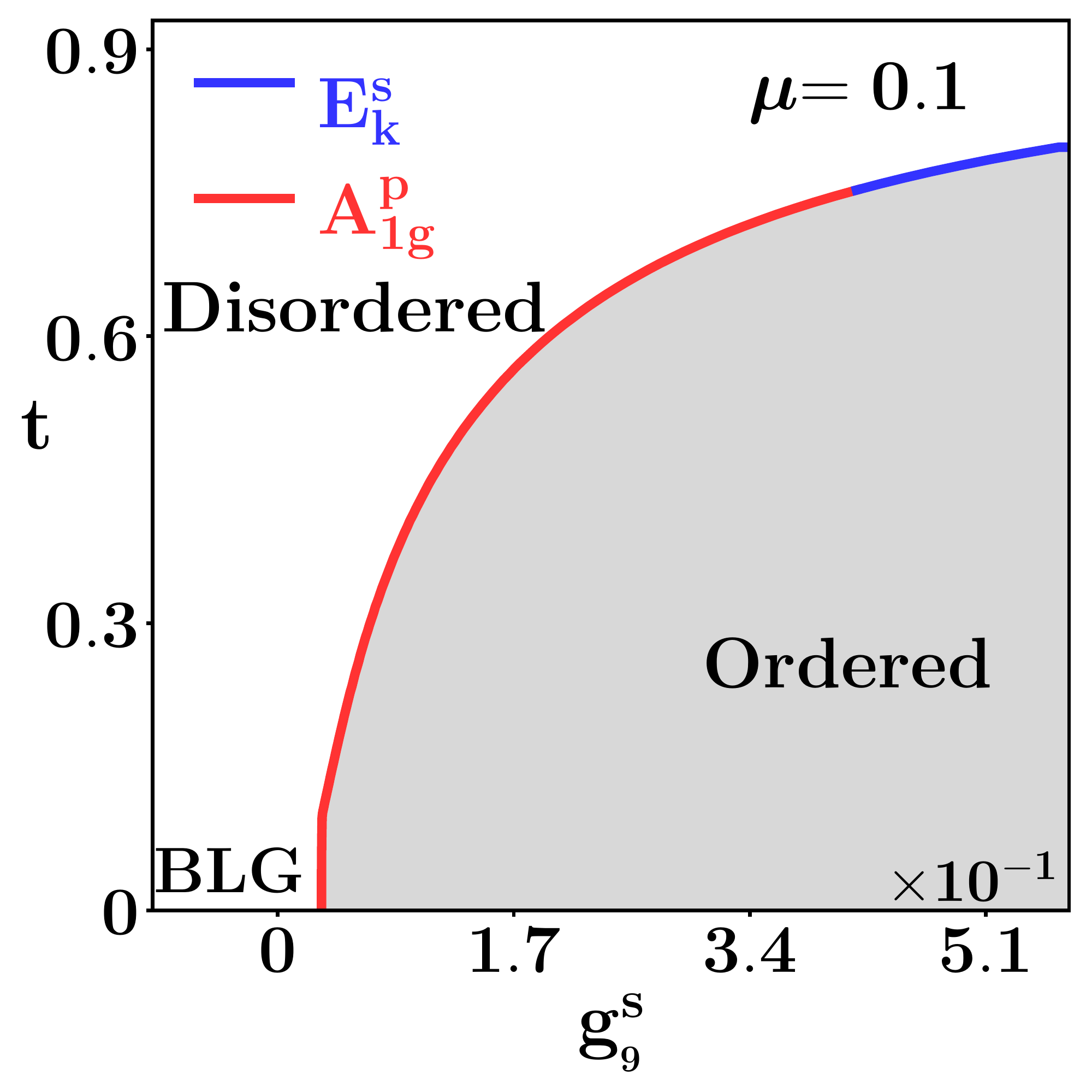}
\label{fig:smectic}}\\
\subfloat[Phase diagrams for the quartic interaction in the smectic spin-density-wave or $E_k$ triplet channel.]{
\includegraphics[width=0.21\linewidth]{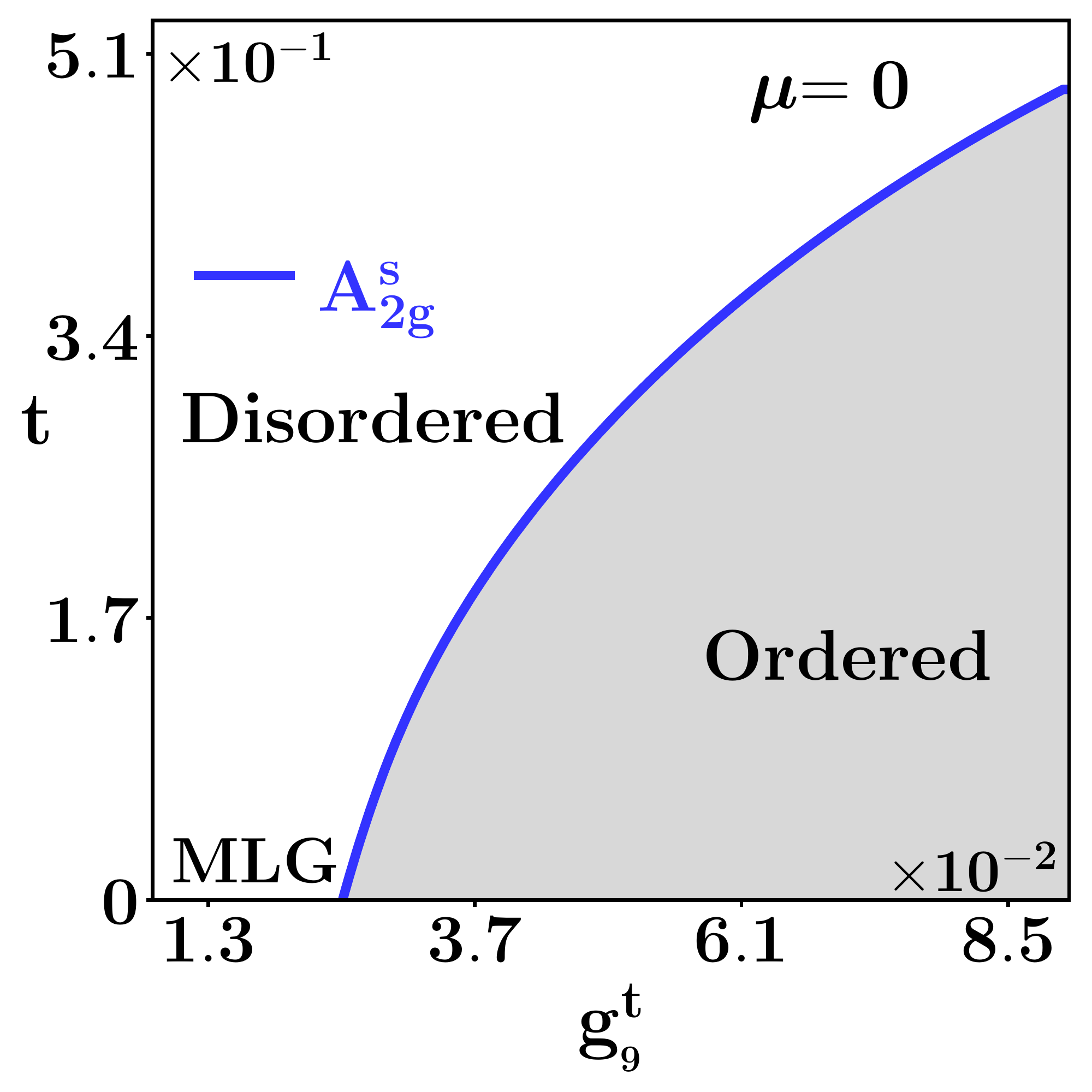}\hspace{0.5cm}
\includegraphics[width=0.21\linewidth]{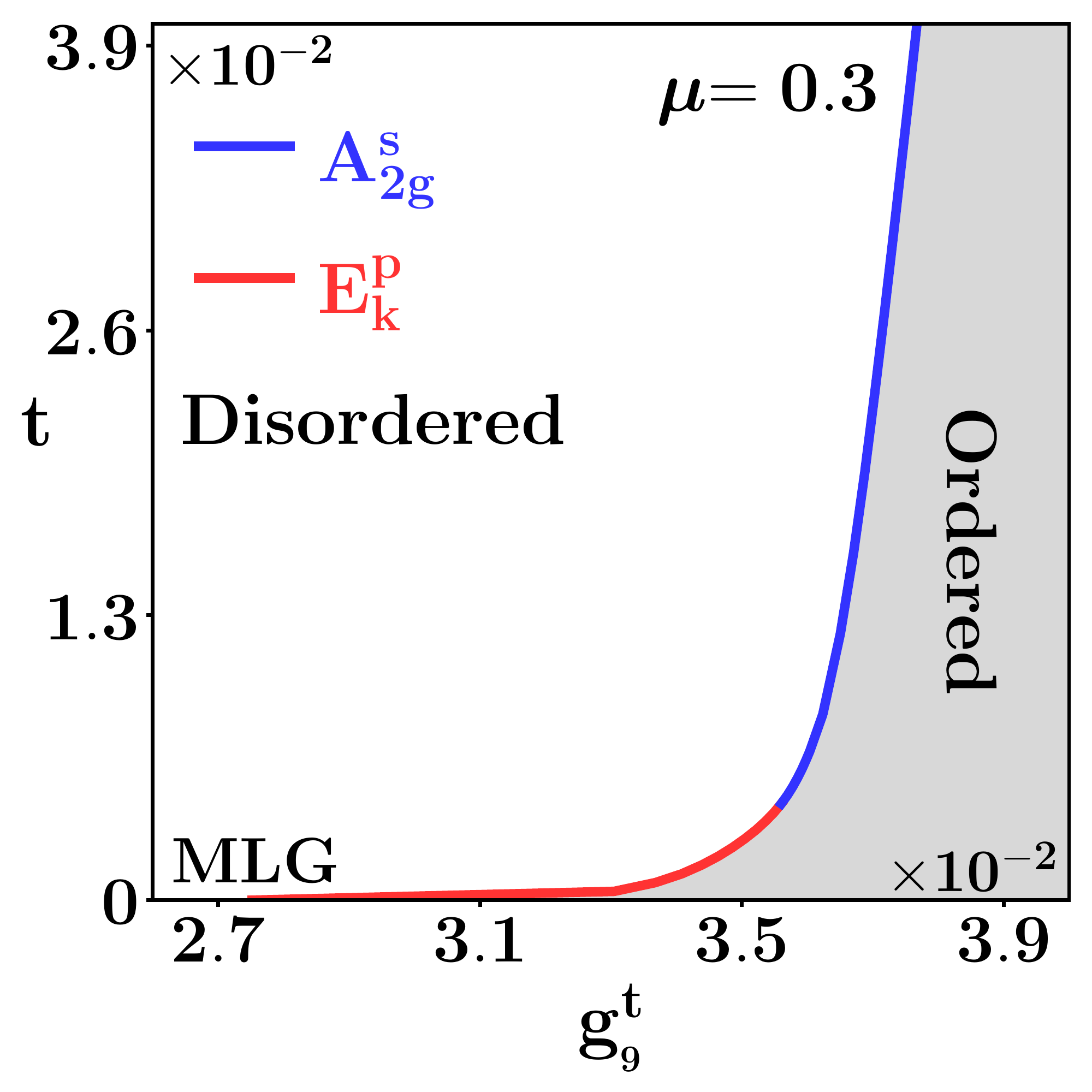}\hspace{0.5cm}
\includegraphics[width=0.21\linewidth]{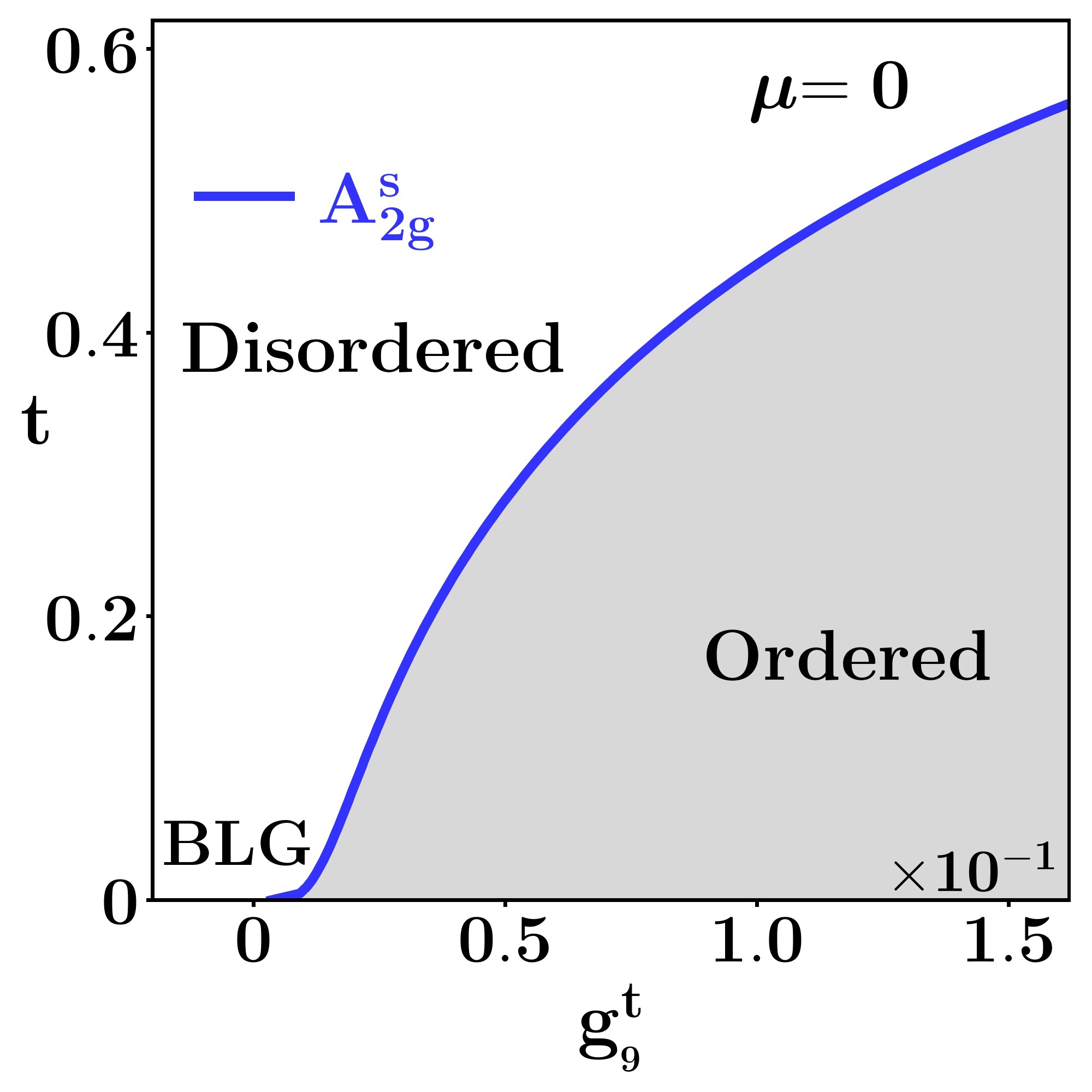}\hspace{0.5cm}
\includegraphics[width=0.21\linewidth]{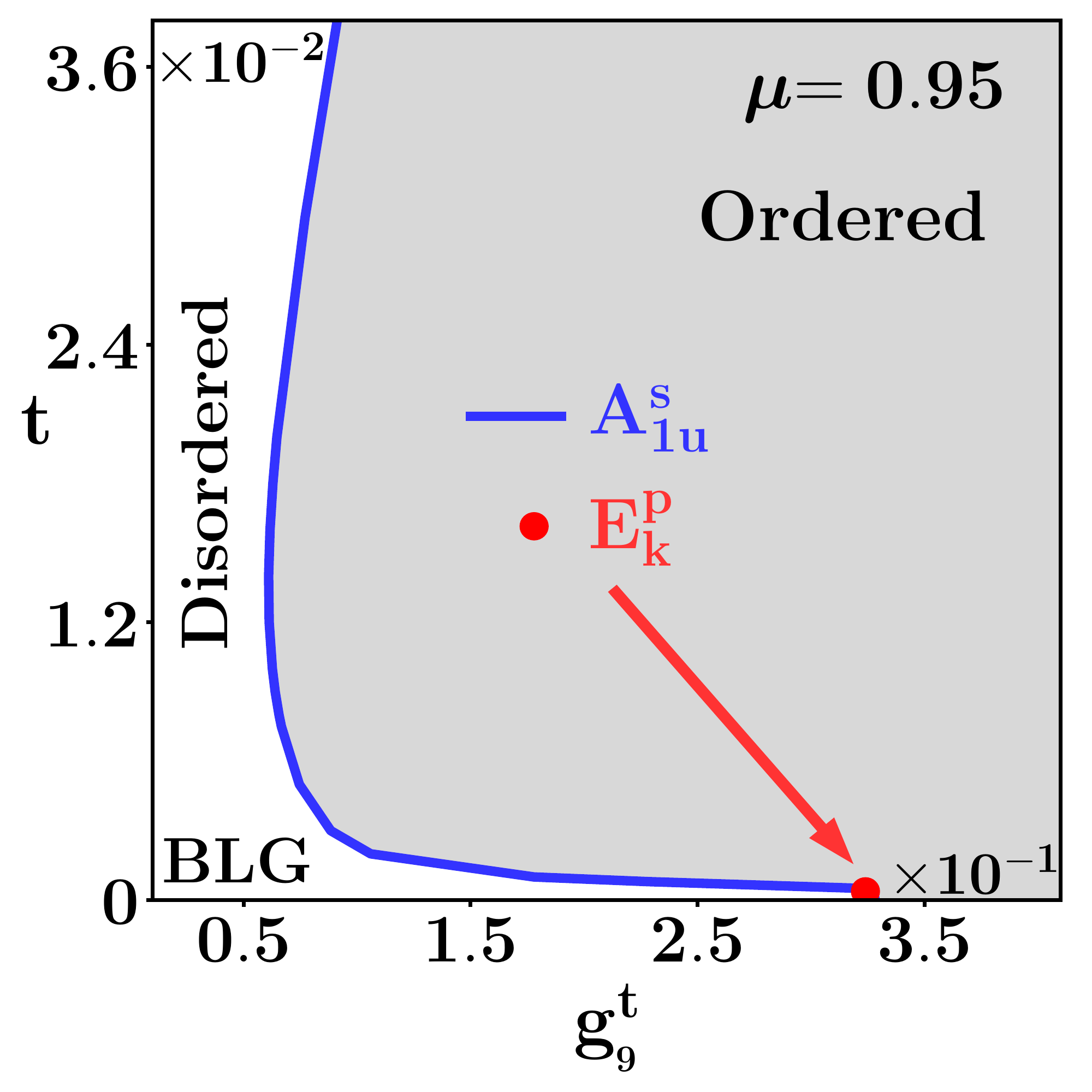}
\label{fig:spin-smectic}}
\caption{Cuts of the global phase diagram with the quartic interactions in the (spin-)smectic channels. The rest of the details are identical to those in Fig.~\ref{fig:mass_PD}. For the constructions of composite order parameters from adjacent phases, see Sec.~\ref{sec:PD_smectic}.} 
\label{fig:smectic_PD}
\end{figure*}

Four cuts of the phase diagram for each interaction channel (for MLG and BLG, at zero and finite chemical potential) are shown in Figs.~\ref{fig:mass_PD}--\ref{fig:Kekule_PD}. All phase transitions and the nature of the supervector order parameters of adjacent phases are summarized in Tables~\ref{tab:ptransitions_sgl} and \ref{tab:ptransitions_tr}. Notice that we always cut the global phase diagram along one certain axis in the space of the coupling constants, which is generally not how a physical system behaves. However, unless fine tuned, among multiple running couplings there will always be one that diverges the fastest. The selection rules are then determined by the strongest diverging coupling constant. However, effective quartic interactions in a certain channel can in principle be tuned in Determinantal quantum Monte Carlo (DQMC) simulations~\cite{PhysRevD.24.2278}, obtained by integrating out bosonic order parameter fluctuations. For brevity, when quoting a set of matrices we will use $\alpha=\{1,2\}$ in the particle-hole or Nambu sector of the Hilbert space, indicating off-diagonal or pairing orders, and $s=\{1,2,3\}$ in the spin sector, indicating a spin triplet order parameter.

\subsection{Quartic interactions: mass channels}~\label{sec:PD_mass}

For quartic interactions in the four excitonic mass channels that are common across the two systems, selection rule (Ia) is the decisive at $\mu=0$, namely $O_i=M_i$, where $O_i$ ($M_i$) are the matrices involved in the order parameter of the broken symmetry phase (four fermion term of the given interaction channel). At $\mu>0$ and low temperatures, we observe nucleation of ``adjacent'' superconducting phase(s). Next we discuss these cases.

%%%%%%%%%%%%%%%%%%%%%%%%%%%%%%%%%%%%%%%%%%%%%%%%%%%%%%%%%%%%%%%%%%%%%%
%%%%%%%%%%%%%%%%%%%%%%%%%%%%%%% QAHI %%%%%%%%%%%%%%%%%%%%%%%%%%%%%%%%%
%%%%%%%%%%%%%%%%%%%%%%%%%%%%%%%%%%%%%%%%%%%%%%%%%%%%%%%%%%%%%%%%%%%%%%
\vspace{0.3cm}
(1) For quartic interaction in the QAHI channel $M=\Gamma_{0033}$, which fully \emph{commutes} with all superconducting masses. These orders are therefore not available to condense into. Instead, in both systems we observe nucleation of the $E_u$ and $E_g$ nematic SCs (denoted by nematic SC$_1$ and nematic SC$_2$, respectively, in Tables~\ref{tab:bilinears_exc} and \ref{tab:bilinears_pair}) and the smectic SC phases. These pairing order parameters indeed fully anticommute with $M$ [selection rule (Ib)]. The high temperature phase is the QAHI, in accordance with selection rule (Ia). The multicomponent order parameters combining QAHI and various nematic and smectic SCs in this case constitute the following supervectors
\begin{align}
\vec{V}^{\rm QAHI}_1&=\left\{
\overbrace{
\underbrace{\Gamma_{\alpha 001}, \Gamma_{\alpha 032} }_{\mbox{$E_{g}^p$}}, 
\underbrace{\Gamma_{0033} }_{\mbox{$A_{2g}^s$}} 
}^{\mbox{2 copies of O(3) vectors}}
\right\},\nonumber\\
\vec{V}^{\rm QAHI}_2&=\left\{
\overbrace{
\underbrace{\Gamma_{\alpha s31}, \Gamma_{\alpha s02} }_{\mbox{$E_{u}^p$}},
\underbrace{\Gamma_{0033} }_{\mbox{$A_{2g}^s$}} 
}^{\mbox{4 copies of O(4) vectors}}
\right\},\nonumber \\
\vec{V}^{\rm QAHI}_3&=\left\{
\overbrace{
\underbrace{\Gamma_{\alpha 010}, \Gamma_{\alpha 020},\Gamma_{\alpha 013}, \Gamma_{\alpha 023} }_{\mbox{$E_{k}^p$}},
\underbrace{\Gamma_{0033} }_{\mbox{$A_{2g}^s$}} 
}^{\mbox{4 copies of O(3) vectors}}
\right\}.
\end{align}
The corresponding cuts of the phase diagram are displayed in Fig.~\ref{fig:QAHI}.

%%%%%%%%%%%%%%%%%%%%%%%%%%%%%%%%%%%%%%%%%%%%%%%%%%%%%%%%%%%%%%%%%%%%%%
%%%%%%%%%%%%%%%%%%%%%%%%%%%%%%%% CDW %%%%%%%%%%%%%%%%%%%%%%%%%%%%%%%%%
%%%%%%%%%%%%%%%%%%%%%%%%%%%%%%%%%%%%%%%%%%%%%%%%%%%%%%%%%%%%%%%%%%%%%%
\vspace{0.3cm}
(2) For quartic interaction in the CDW channel $M=\Gamma_{3003}$, which fully anticommutes with two superconducting masses [selection rule (Ib)], the singlet $s$-wave and triplet $f$-wave pairings, which form following the composite order parameters with CDW:
\begin{align}
\vec{V}^{\rm CDW}_1&=\left\{
\overbrace{
\underbrace{\Gamma_{\alpha 000} }_{\mbox{$A_{1g}^p$}},
\underbrace{\Gamma_{3003} }_{\mbox{$A_{2u}^s$}} 
}^{\mbox{O(3) vector}}
\right\},\nonumber\\
\vec{V}^{\rm CDW}_2&=\left\{
\overbrace{
\underbrace{\Gamma_{\alpha s30} }_{\mbox{$A_{1u}^p$}},
\underbrace{\Gamma_{3003} }_{\mbox{$A_{2u}^s$}} 
}^{\mbox{2 copies of O(4) vectors}}
\right\}.\label{eq:SOP_CDW}
\end{align}
Indeed we observe degenerate nucleation of these two pairing phases at low temperatures in MLG and BLG, when $\mu>0$. The adjacent excitonic ordering is the CDW [selection rule (Ia)], which sets in at higher $t$ where superconductivity is destroyed, see Fig.~\ref{fig:CDW}. The absence of order differentiation between Dirac and Luttinger systems with quartic interaction in the CDW channel is due to the fact that all dominant order parameters (one excitonic and two superconducting) fully anticommute with $\hat{h}^{\rm D}$ as well as with $\hat{h}^{\rm L}$. At zero doping, we only find CDW ordering in both MLG and BLG [selection rule (Ia)]. In Sec.~\ref{sec:HubbardModel} we demonstrate the same result for NN repulsion.

%%%%%%%%%%%%%%%%%%%%%%%%%%%%%%%%%%%%%%%%%%%%%%%%%%%%%%%%%%%%%%%%%%%%%%
%%%%%%%%%%%%%%%%%%%%%%%%%%%%%%% QSHI %%%%%%%%%%%%%%%%%%%%%%%%%%%%%%%%%
%%%%%%%%%%%%%%%%%%%%%%%%%%%%%%%%%%%%%%%%%%%%%%%%%%%%%%%%%%%%%%%%%%%%%%
\vspace{0.3cm}
(3) For quartic interaction in the QSHI channel $M=\Gamma_{3s33}$, which fully anticommutes with two superconducting order parameters. One of them, the $s$-wave pairing, is a mass order in both systems, while the singlet Kekul\'e SC gaps only Luttinger fermions.  Accordingly, in the Dirac system at low temperatures and finite chemical potential only $s$-wave SC can be observed. On the other hand, we find degenerate nucleation of $s$-wave and singlet Kekul\'e pairings in BLG. The nucleating orders and their differentiation through the underlying band structure follows from the selection rules (Ib) and (II), respectively. The high temperature phase in either case is the adjacent QSHI order, which follows from selection rule (Ia). The relevant cuts of the phase diagram are shown in Fig.~\ref{fig:QSHI}. The supervector order parameters formed by adjacent phases read 
\begin{align}
\vec{V}^{\rm QSHI}_1&=\left\{
\overbrace{
\underbrace{\Gamma_{\alpha 000} }_{\mbox{$A_{1g}^p$}},
\underbrace{\Gamma_{3s33} }_{\mbox{$A_{2g}^t$}} 
}^{\mbox{O(5) vector}}
\right\},\nonumber\\
\vec{V}^{\rm QSHI}_2&=\left\{
\overbrace{
\underbrace{\Gamma_{\alpha 011}, \Gamma_{\alpha 021} }_{\mbox{$A_{1k}^p$}},
\underbrace{\Gamma_{3s33} }_{\mbox{$A_{2g}^t$}} 
}^{\mbox{2 copies of O(5) vectors}}
\right\}.\label{eq:SOP_QSHI}
\end{align}
At zero doping we always find QSHI in MLG and BLG, according to selection rule (Ia).

%%%%%%%%%%%%%%%%%%%%%%%%%%%%%%%%%%%%%%%%%%%%%%%%%%%%%%%%%%%%%%%%%%%%%%
%%%%%%%%%%%%%%%%%%%%%%%%%%%%%%%% AFM %%%%%%%%%%%%%%%%%%%%%%%%%%%%%%%%%
%%%%%%%%%%%%%%%%%%%%%%%%%%%%%%%%%%%%%%%%%%%%%%%%%%%%%%%%%%%%%%%%%%%%%%
\vspace{0.3cm}
(4) For quartic interaction in the AFM channel $M=\Gamma_{0s03}$, which fully anticommutes with the $E_g$ nematic and singlet Kekul\'e SCs. Latter one represents a mass order for Luttinger fermions, but fully commutes with the Dirac Hamiltonian. Therefore Kekul\' e SC is not a viable candidate for Dirac fermions at low temperatures. Rather, the maximally anticommuting pairing for the linearly dispersing Dirac electrons is the $E_g$ nematic SC. The composite order parameters in these two cases read
\begin{align}
\vec{V}^{\rm AFM}_1&=\left\{
\overbrace{
\underbrace{\Gamma_{\alpha 011}, \Gamma_{\alpha 021} }_{\mbox{$A_{1k}^p$}},
\underbrace{\Gamma_{0s03} }_{\mbox{$A_{2u}^t$}} 
}^{\mbox{2 copies of O(5) vectors}}
\right\},\nonumber\\
\vec{V}^{\rm AFM}_2&=\left\{
\overbrace{
\underbrace{\Gamma_{\alpha 001}, \Gamma_{\alpha 032} }_{\mbox{$E_{g}^p$}},
\underbrace{\Gamma_{0s03} }_{\mbox{$A_{2u}^t$}} 
}^{\mbox{2 copies of O(5) vectors}}
\right\}.\label{eq:SOP_AFM}
\end{align}
Therefore, at finite $\mu$ the low temperature paring state in MLG (BLG) is singlet $E_g$ nematic (Kekul\'e) SC, which follows from selection rule (Ib). The high temperature phase at finite $\mu$ in both systems is the AFM [selection rule (Ia)], see Fig.~\ref{fig:AFM}. At zero doping we always find AFM in both MLG and BLG. In Sec.~\ref{sec:HubbardModel} we again support this observation in the phase diagram of the honeycomb Hubbard model with only on-site repulsion.

%%%%%%%%%%%%%%%%%%%%%%%%%%%%%%%%%%%%%%%%%%%%%%%%%%%%%%%%%%%%%%%%%%%%%%%%%
%%%%%%%%%%%%%%%%%%%%%%%%%%%%%%%%%%%%%%%%%%%%%%%%%%%%%%%%%%%%%%%%%%%%%%%%%
%%%%%%%%%%%%%%%%%%%% PHASE DIAGRAMS - Smectic CHANNELS %%%%%%%%%%%%%%%%%%
%%%%%%%%%%%%%%%%%%%%%%%%%%%%%%%%%%%%%%%%%%%%%%%%%%%%%%%%%%%%%%%%%%%%%%%%%
%%%%%%%%%%%%%%%%%%%%%%%%%%%%%%%%%%%%%%%%%%%%%%%%%%%%%%%%%%%%%%%%%%%%%%%%%
%%%%%%%%%%%%%%%%%%%%%%%%%%%%%%%%%%%%%%%%%%%%%%%%%%%%%%%%%%%%%%%%%%%%%%%%%
%%%%%%%%%%%%%%%%%%%%%%%%%%%%%%%%%%%%%%%%%%%%%%%%%%%%%%%%%%%%%%%%%%%%%%%%%
%%%%%%%%%%%%%%%%%%%%%%%%%%%%%%%%%%%%%%%%%%%%%%%%%%%%%%%%%%%%%%%%%%%%%%%%%
%%%%%%%%%%%%%%%%%%%%%%%%%%%%%%%%%%%%%%%%%%%%%%%%%%%%%%%%%%%%%%%%%%%%%%%%%
%%%%%%%%%%%%%%%%%%%%%%%%%%%%%%%%%%%%%%%%%%%%%%%%%%%%%%%%%%%%%%%%%%%%%%%%%

%%%%%%%%%%%%%%%%%%%%%%%%%%%%%%%%%%%%%%%%%%%%%%%%%%%%%%%%%%%%%%%%%%%%%%%%%%%%%%%%%%%%%%%
%%%%%%%%%%%%%%%%%%%%%%%%%%%%%%%%%%%%%%%%%%%%%%%%%%%%%%%%%%%%%%%%%%%%%%%%%%%%%%%%%%%%%%%
%%%%%%%%%%%%%%%%%%%%%%%%%%%%%%%%%%%%%%%%%%%%%%%%%%%%%%%%%%%%%%%%%%%%%%%%%%%%%%%%%%%%%%%
%%%%%%%%%%%%%%%%%%%%%%%%%%%%%%%%%%%%%%%%%%%%%%%%%%%%%%%%%%%%%%%%%%%%%%%%%%%%%%%%%%%%%%%
%%%%%%%%%%%%%%%%%%%%%%%%%%%%%%%%%%%%%%%%%%%%%%%%%%%%%%%%%%%%%%%%%%%%%%%%%%%%%%%%%%%%%%%
\subsection{Quartic interactions: nematic channels}~\label{sec:PD_nematic}

Nematic order parameters break rotational symmetry and they do not introduce a gap in the Luttinger or Dirac spectrum. Therefore other \emph{gapped} phases might be more favorable at low temperatures when we tune the quartic interactions in the nematic channels, especially if they satisfy the selection rule (Ib). Also note that in general spin fluctuations may induce charge ordering, while the opposite process is comparatively suppressed. This is because in the spin sector of the Hilbert space $\sigma_i \sigma_j$ (with $i,j=1,2,3$) can conceivably equal to $\sigma_0$ when $i=j$, but $\sigma_0 \sigma_0\neq\sigma_i$. Here two Pauli matrices are appearing from the interaction vertices, and their product results from the corresponding one-loop Feynman diagrams (see Fig.~\ref{fig:Feynmann_int}). As a result, the spin singlet nematic interactions at zero chemical doping or high temperatures follow selection rule (Ia). In contrast, the appearing excitonic phases for interactions in their spin triplet counterparts follow selection rule (Ib). Next we systematically discuss these outcomes.

\vspace{0.3cm}
(1) For quartic interaction in the $E_g$ nematic (nematic$_1$) channel $M=(\Gamma_{3001}, \Gamma_{3032})$, which fully anticommutes with the $s$-wave and $f$-wave pairings, and forms the following supervector order parameters:
\begin{align}
\vec{V}^{E_g^s}_1&=\left\{
\overbrace{
\underbrace{\Gamma_{\alpha 000}}_{\mbox{$A_{1g}^p$}},
\underbrace{\Gamma_{3001},\Gamma_{3032}}_{\mbox{$E_{g}^s$}} 
}^{\mbox{O(4) vector}}
\right\},\nonumber\\
\vec{V}^{E_g^s}_2&=\left\{
\overbrace{
\underbrace{\Gamma_{\alpha s30}}_{\mbox{$A_{1u}^p$}},
\underbrace{\Gamma_{3001},\Gamma_{3032}}_{\mbox{$E_{g}^s$}} 
}^{\mbox{2 copies of O(5) vectors}}
\right\}.
\end{align}
These two pairing phases are fully gapped in both MLG and BLG, and indeed we observe degenerate appearance of them with increased chemical doping in both systems [selection rule (Ib)]. At $\mu=0$ or at high temperatures we find nucleation of the $E_g$ nematic order [selection rule (Ia)] in both systems, see Fig.~\ref{fig:Eg_nem}.

\vspace{0.3cm}
(2) For quartic interaction in the $E_u$ nematic (nematic$_2$) channel $M=(\Gamma_{0031}, \Gamma_{0002})$. These interaction matrices anticommute with the singlet Kekul\' e SC, which gaps the Luttinger fermions, but not the Dirac fermions. Therefore, in BLG at low temperatures we observe the singlet Kekul\'e SC together with the degenerate QSHI and AFM orders [selection rule (Ib)]. Adding carrier density ($\mu>0$) favors the singlet Kekul\'e pairing at sufficiently low temperatures. The high temperature phase is the $E_u$ nematic order, favored by the gain in entropy, see Fig.~\ref{fig:Eu_nem}. At the same time, in MLG we find only the $E_u$ nematic phase to be the dominant broken symmetry phase for zero and finite $\mu$ [selection rule (Ia)]. The adjacent phases then form the following composite order parameters:
\begin{align}
\vec{V}^{E_u^s}_1&=\left\{
\overbrace{
\underbrace{\Gamma_{\alpha 011},\Gamma_{\alpha 021}}_{\mbox{$A_{1k}^p$}},
\underbrace{\Gamma_{0031},\Gamma_{0002}}_{\mbox{$E_{u}^s$}} 
}^{\mbox{2 copies of O(4) vector}}
\right\},\nonumber\\
\vec{V}^{E_u^s}_2&=\left\{
\overbrace{
\underbrace{\Gamma_{3s33}}_{\mbox{$A_{2g}^t$}},
\underbrace{\Gamma_{0031},\Gamma_{0002}}_{\mbox{$E_{u}^s$}} 
}^{\mbox{O(5) vector}}
\right\},\nonumber \\
\vec{V}^{E_u^s}_3&=\left\{
\overbrace{
\underbrace{\Gamma_{0s03}}_{\mbox{$A_{2u}^t$}},
\underbrace{\Gamma_{0031},\Gamma_{0002}}_{\mbox{$E_{u}^s$}} 
}^{\mbox{O(5) vector}}
\right\}.
\end{align}

\vspace{0.3cm}
(3) For quartic interaction in the $E_g$ spin-nematic (spin-nematic$_1$) channel $M=(\Gamma_{0s01}, \Gamma_{0s32})$. The maximally (but not fully) anticommuting pairing mass is the $f$-wave pairing.  Among the excitonic masses, the CDW and KC orders fully anticommute with $M$. Note that KC is a mass order in BLG, but commutes with the Dirac Hamiltonian. Accordingly, at $\mu=0$ or at sufficiently high temperature KC and CDW appear together in BLG, whereas in MLG we only find CDW order in this parameter regime. The $f$-wave pairing can be observed at $\mu>0$ and low temperatures in both MLG and BLG, while higher $t$ destroys the superconductivity and the system instead enters into the respective excitonic phases, see Fig.~\ref{fig:Eg_spin-nem}. Note that the $E_g$ spin-nematic ordering is completely absent in our range of parameters, thus all appearing broken symmetry phases follow selection rule (Ib). The corresponding composite order parameters constructed from the adjacent phases read
\begin{align}
\vec{V}^{E_g^t}_1&=\left\{
\overbrace{
\underbrace{\Gamma_{\alpha s30}}_{\mbox{$A_{1u}^p$}},
\underbrace{\Gamma_{3003}}_{\mbox{$A_{2u}^s$}} 
}^{\mbox{2 copies of O(4) vector}}
\right\},\nonumber\\
\vec{V}^{E_g^t}_2&=\left\{
\overbrace{
\underbrace{\Gamma_{\alpha s30}}_{\mbox{$A_{1u}^p$}},
\underbrace{\Gamma_{0012},\Gamma_{0022}}_{\mbox{$A_{2k}^s$}} 
}^{\mbox{2 copies of O(5) vectors}}
\right\}.
\end{align}

\vspace{0.3cm}
(4) For quartic interaction in the $E_u$ spin-nematic (or spin-nematic$_2$) channel $M=(\Gamma_{3s31}, \Gamma_{3s02})$. These interaction matrices fully anticommute with the $s$-wave pairing and CDW orders, which indeed appear simultaneously on the phase diagram of MLG and BLG at zero doping, see Fig.~\ref{fig:Eu_spin-nem}. Furthermore, $M$ also anticommutes with the KVBS order parameter, which represents a massive phase for the Dirac fermions, but commutes with the Luttinger Hamiltonian. Accordingly, KVBS accompanies CDW and $s$-wave pairing in MLG, but not in BLG. As all these phases represent mass orders and they fully anticommute with $M$, temperature does not lift the degeneracy among them. However, by adding chemical potential we select the pairing phase: $s$-wave SC. As only selection rule (Ib) is operative in these phase diagrams, we do not obtain any composite order parameter.

%%%%%%%%%%%%%%%%%%%%%%%%%%%%%%%%%%%%%%%%%%%%%%%%%%%%%%%%%%%%%%%%%%%%%%%%%%%%%%%%%%%%%%%
%%%%%%%%%%%%%%%%%%%%%%%%%%%%%%%%%%%%%%%%%%%%%%%%%%%%%%%%%%%%%%%%%%%%%%%%%%%%%%%%%%%%%%%
%%%%%%%%%%%%%%%%%%%%%%%%%%%%%%%%%%%%%%%%%%%%%%%%%%%%%%%%%%%%%%%%%%%%%%%%%%%%%%%%%%%%%%%
%%%%%%%%%%%%%%%%%%%%%%%%%%%%%%%%%%%%%%%%%%%%%%%%%%%%%%%%%%%%%%%%%%%%%%%%%%%%%%%%%%%%%%%
%%%%%%%%%%%%%%%%%%%%%%%%%%%%%%%%%%%%%%%%%%%%%%%%%%%%%%%%%%%%%%%%%%%%%%%%%%%%%%%%%%%%%%%
\subsection{Quartic interactions: smectic channels}~\label{sec:PD_smectic}

Smectic orders break rotational and translational symmetries, but they do not support any mass gap. Selection rule (Ia) is, therefore, not conducive for a fully gapped phase, and low temperature phases are typically inhabited by mass orders, obeying selection rule (Ib).

\vspace{0.3cm}
\noindent 1. For quartic interaction in the smectic CDW channel $M=(\Gamma_{3010}, \Gamma_{3020},\Gamma_{3013},\Gamma_{3023})$. The low temperature region of the phase diagrams in Fig.~\ref{fig:smectic} is occupied by the massive phases that fulfill selection rule (Ib): the $s$-wave SC and QSHI in both MLG and BLG. While at zero doping they nucleate in a degenerate fashion, finite doping favors the $s$-wave pairing at low temperatures. At high temperatures, we observe the smectic CDW phase at both zero and finite $\mu$ in MLG and BLG, following selection rule (Ia). The adjacent phases then form the following supervector order parameters:
\begin{align}
\vec{V}^{E_k^s}_1&=\left\{
\overbrace{
\underbrace{\Gamma_{\alpha 000}}_{\mbox{$A_{1g}^p$}},
\underbrace{\Gamma_{3010},\Gamma_{3020},\Gamma_{3013},\Gamma_{3023}}_{\mbox{$E_{k}^s$}} 
}^{\mbox{2 copies of O(4) vector}}
\right\},\nonumber\\
\vec{V}^{E_k^s}_2&=\left\{
\overbrace{
\underbrace{\Gamma_{3 s33}}_{\mbox{$A_{2g}^t$}},
\underbrace{\Gamma_{3010},\Gamma_{3020},\Gamma_{3013},\Gamma_{3023}}_{\mbox{$E_{k}^s$}} 
}^{\mbox{2 copies of O(5) vectors}}.
\right\}.
\end{align}

\vspace{0.3cm}
\noindent 2. For quartic interaction in the smectic spin density wave (SDW) channel $M=(\Gamma_{0s10}, \Gamma_{0s20},\Gamma_{0s13},\Gamma_{0s23})$, which fully anticommutes with the QAHI mass order. As such we observe nucleation of this phase in both systems at zero chemical potential for low, as well as high $t$ [selection rule (Ib)]. On the other hand, there is no fully anticommuting available pairing phase, and by elevating the chemical potential we induce the only partially anticommuting smectic SC. Once superconductivity is destroyed, in MLG the system goes back into the QAHI phase. However, in BLG it is the chiral density order parameter that develops a finite expectation value at high $t$. This is possibly due to the high chemical potential ($\mu=0.95$), required to induce superconductivity, which causes the system to act like a metal, as opposed to a semimetal. The relevant cuts of the global phase diagram are displayed in Fig.~\ref{fig:spin-smectic}. The corresponding composite order parameters are
\begin{align}
\vec{V}^{E_k^t}_1&=\left\{
\overbrace{
\underbrace{\Gamma_{\alpha 010},\Gamma_{\alpha 020},\Gamma_{\alpha 013},\Gamma_{\alpha 023}}_{\mbox{$E_{k}^p$}},
\underbrace{\Gamma_{0033}}_{\mbox{$A_{2g}^s$}} 
}^{\mbox{4 copies of O(3) vector}}
\right\},\nonumber\\
\vec{V}^{E_k^t}_2&=\left\{
\overbrace{
\underbrace{\Gamma_{\alpha 010},\Gamma_{\alpha 020},\Gamma_{\alpha 013},\Gamma_{\alpha 023}}_{\mbox{$E_{k}^p$}},
\underbrace{\Gamma_{0030}}_{\mbox{$A_{1u}^s$}} 
}^{\mbox{4 copies of O(3) vector}}
\right\}.
\end{align}

%%%%%%%%%%%%%%%%%%%%%%%%%%%%%%%%%%%%%%%%%%%%%%%%%%%%%%%%%%%%%%%%%%%%%%%%%%%%%%%%%%%%%%%%%%%%
%%%%%%%%%%%%%%%%%%%%%%%%%%%%%%%%%%%%%%%%%%%%%%%%%%%%%%%%%%%%%%%%%%%%%%%%%%%%%%%%%%%%%%%%%%%%
%%%%%%%%%%%%%%%%%%%% PHASE TRANSITIONS TABLE TRIPLET: %%%%%%%%%%%%%%%%%%%%%%%%%%%%%%%%%%%%%%
%%%%%%%%%%%%%%%%%%%%%%%%%%%%%%%%%%%%%%%%%%%%%%%%%%%%%%%%%%%%%%%%%%%%%%%%%%%%%%%%%%%%%%%%%%%%
%%%%%%%%%%%%%%%%%%%%%%%%%%%%%%%%%%%%%%%%%%%%%%%%%%%%%%%%%%%%%%%%%%%%%%%%%%%%%%%%%%%%%%%%%%%%
\begin{table*}[t!]
\renewcommand{\arraystretch}{1.3}
\begin{tabular}{|c | c c c c c c|c c c c c c|}
\hline
\multirow{3}{*}{CC} & \multicolumn{6}{c|}{ Monolayer graphene (MLG) } & \multicolumn{6}{c|}{ Bilayer graphene (BLG) } \\ 
%\cline{2-13}
 & \multicolumn{3}{c}{$\mu=0$} & \multicolumn{3}{c|}{$\mu > 0$} & \multicolumn{3}{c}{$\mu=0$} & \multicolumn{3}{c|}{$\mu > 0$}  \\
%\cline{2-13}
& Low $t$ & High $t$ & Symmetry & Low $t$ & High $t$ & Symmetry & Low $t$ & High $t$ & Symmetry &  Low $t$ & High $t$ & Symmetry  \\
\hline
%$g_{_1}^t$ &
%$A_{2g}^s$ & $A_{2g}^s$ & & $E_{u}^p$ & $E_{u}^p$ & &
%\begin{tabular}{@{}c@{}} $A_{2g}^t,A_{2u}^t$, \\ $A_{2k}^t,A_{1u}^p$ \end{tabular} & 
%$E_{g}^s$ & & $A_{1u}^p$ & $E_{g}^s$ & \\
%\hline
%
%
$g_{_2}^t$ &
$A_{2g}^t$ & $A_{2g}^t$ & - & $A_{1g}^p$ & $A_{2g}^t$ & $1\otimes$O(5) &
$A_{2g}^t$ & $A_{2g}^t$ & - & \begin{tabular}{@{}c@{}} $A_{1g}^p$ \\ $A_{1k}^p$\end{tabular}  & $A_{2g}^t$ &  \begin{tabular}{@{}c@{}} $1\otimes$O(5) \\ $2\otimes$O(5)\end{tabular}\\
\rowcolor{RowColor}
%
% g12:
$g_{_3}^t$ &
$A_{2u}^s$ & $A_{2u}^s$ & - & $A_{1u}^p$ & $A_{2u}^s$ & $2\otimes$O(4) &
\begin{tabular}{@{}c@{}} $A_{2u}^s$ \\ $A_{2k}^s$\end{tabular} & \begin{tabular}{@{}c@{}}$A_{2u}^s$ \\ $A_{2k}^s$\end{tabular} & - & $A_{1u}^p$ & \begin{tabular}{@{}c@{}} $A_{2u}^s$ \\ $A_{2k}^s$\end{tabular} & \begin{tabular}{@{}c@{}} $2\otimes$O(4) \\ $2\otimes$O(5) \end{tabular}\\
%
%
%$g_{_4}^t$ &
%$A_{1k}^s,A_{1g}^p$ & $A_{1u}^t$ & & $A_{1g}^p$ & $A_{1u}^t$ & & 
%$A_{2k}^s,A_{1g}^p$ & $A_{1u}^t$ & & $A_{1g}^p$ & $A_{1u}^t$ & \\
%\hline
%
%
$g_{_5}^t$ & 
$A_{2u}^t$ & $A_{2u}^t$ & - & $E_{g}^p$ & $A_{2u}^t$ & $2\otimes$O(5) &
$A_{2u}^t$ & $A_{2u}^t$ & - & $A_{1k}^p$ & $A_{2u}^t$ & $2\otimes$O(5)\\
\rowcolor{RowColor}
%
% g15:
$g_{_6}^t$ & 
\begin{tabular}{@{}c@{}} $A_{2u}^s$ \\ $A_{1k}^s$ \\ $A_{1g}^p$ \end{tabular} & 
\begin{tabular}{@{}c@{}} $A_{2u}^s$ \\ $A_{1k}^s$ \\ $A_{1g}^p$ \end{tabular} & - & $A_{1g}^p$ & $A_{1g}^p$ & - &
\begin{tabular}{@{}c@{}} $A_{2u}^s$ \\ $A_{1g}^p$ \end{tabular} & 
\begin{tabular}{@{}c@{}} $A_{2u}^s$ \\ $A_{1g}^p$ \end{tabular} & - & $A_{1g}^p$ & $A_{1g}^p$ & - \\
$g_{_7}^t$ & 
$A_{2u}^s$ & $A_{2u}^s$ & - & 
$E_{k}^p$ & 
$E_{u}^s$ & 
$4\otimes$O(3) &
$A_{2u}^s$ & $A_{2u}^s$ & - & 
\begin{tabular}{@{}c@{}} $E_{k}^p$ \\ $E_{u}^s$ \end{tabular} & 
\begin{tabular}{@{}c@{}} $E_{u}^s$ \\ $A_{1k}^t$ \end{tabular} & 
\begin{tabular}{@{}c@{}} $4\otimes$O(3) \\ $2\otimes$O(5) \end{tabular} \\
\rowcolor{RowColor}
$g_{_8}^t$ & 
\begin{tabular}{@{}c@{}} $A_{2u}^s$ \\ $A_{1g}^p$ \end{tabular} & \begin{tabular}{@{}c@{}} $A_{2u}^s$ \\ $A_{1g}^p$ \end{tabular} & - & $A_{1g}^p$ & $A_{1g}^p$ & - &
\begin{tabular}{@{}c@{}} $A_{2u}^s$ \\ $A_{1g}^p$ \end{tabular} & \begin{tabular}{@{}c@{}} $A_{2u}^s$ \\ $A_{1g}^p$ \end{tabular} & - & $A_{1g}^p$ & $A_{1g}^p$ & - \\
$g_{_9}^t$ &
$A_{2g}^s$ & $A_{2g}^s$ & - & $E_{k}^p$ & $A_{2g}^s$ & $4\otimes$O(3) &
$A_{2g}^s$ & $A_{2g}^s$ & - & $E_{k}^p$ & $A_{1u}^s$ & $4\otimes$O(3) \\
\hline
\end{tabular}
\caption{Dominant instabilities in the presence of quartic interactions in spin triplet channels in MLG and BLG, at zero and finite chemical potential ($\mu$). The low and the high temperature ($t$) phases are indicated in each scenario, suggesting adjacent phases when the two orderings are different. We then display the nature of the composite order parameters in columns ``Symmetry'', where $k\otimes$O(N) indicates $k$ copies of an O(N) algebra. The first column shows the coupling constant (CC) of the interaction channel (see Tables~\ref{tab:bilinears_exc} and \ref{tab:bilinears_pair}). When the low and high temperature phases are identical the notion of composite order parameter becomes moot. See Sec.~\ref{sec:phasediagrams} for detailed discussion and Figs.~\ref{fig:mass_PD}-\ref{fig:Kekule_PD} for various cuts of the phase diagrams.}~\label{tab:ptransitions_tr}
\end{table*}
%%%%%%%%%%%%%%%%%%%%%%%%%%%%%%%%%%%%%%%%%%%%%%%%%%%%%%%%%%%%%%%%%%%%%%%%%%%%%%%%%%%%%%%
%%%%%%%%%%%%%%%%%%%%%%%%%%%%%%%%%%%%%%%%%%%%%%%%%%%%%%%%%%%%%%%%%%%%%%%%%%%%%%%%%%%%%%%
%%%%%%%%%%%%%%%%%%%%%%%%%%%%%%%%%%%%%%%%%%%%%%%%%%%%%%%%%%%%%%%%%%%%%%%%%%%%%%%%%%%%%%%
%%%%%%%%%%%%%%%%%%%%%%%%%%%%%%%%%%%%%%%%%%%%%%%%%%%%%%%%%%%%%%%%%%%%%%%%%%%%%%%%%%%%%%%
%%%%%%%%%%%%%%%%%%%%%%%%%%%%%%%%%%%%%%%%%%%%%%%%%%%%%%%%%%%%%%%%%%%%%%%%%%%%%%%%%%%%%%%

%%%%%%%%%%%%%%%%%%%%%%%%%%%%%%%%%%%%%%%%%%%%%%%%%%%%%%%%%%%%%%%%%%%%%%%%%%%%%%%%%%%%%%%
%%%%%%%%%%%%%%%%%%%%%%%%%%%%%%%%%%%%%%%%%%%%%%%%%%%%%%%%%%%%%%%%%%%%%%%%%%%%%%%%%%%%%%%
%%%%%%%%%%%%%%%%%%%%%%%%%%%%%%%%%%%%%%%%%%%%%%%%%%%%%%%%%%%%%%%%%%%%%%%%%%%%%%%%%%%%%%%
%%%%%%%%%%%%%%%%%%%%%%%%%%%%%%%%%%%%%%%%%%%%%%%%%%%%%%%%%%%%%%%%%%%%%%%%%%%%%%%%%%%%%%%
%%%%%%%%%%%%%%%%%%%%%%%%%%%%%%%%%%%%%%%%%%%%%%%%%%%%%%%%%%%%%%%%%%%%%%%%%%%%%%%%%%%%%%%
\subsection{Quartic interactions: Kekul\'e channels}~\label{sec:PD_kekule}

Finally, we focus on the quartic interactions in the Kekul\' e channels. Note that all Kekul\' e orders (including both bond and current) at least break the translational symmetry, but preserve the rotational invariance.

\begin{figure*}[t]
\subfloat[Phase diagrams for the quartic interaction in the Kekul\' e valence bond solid or $A_{1k}$ singlet channel.]{
\includegraphics[width=0.21\linewidth]{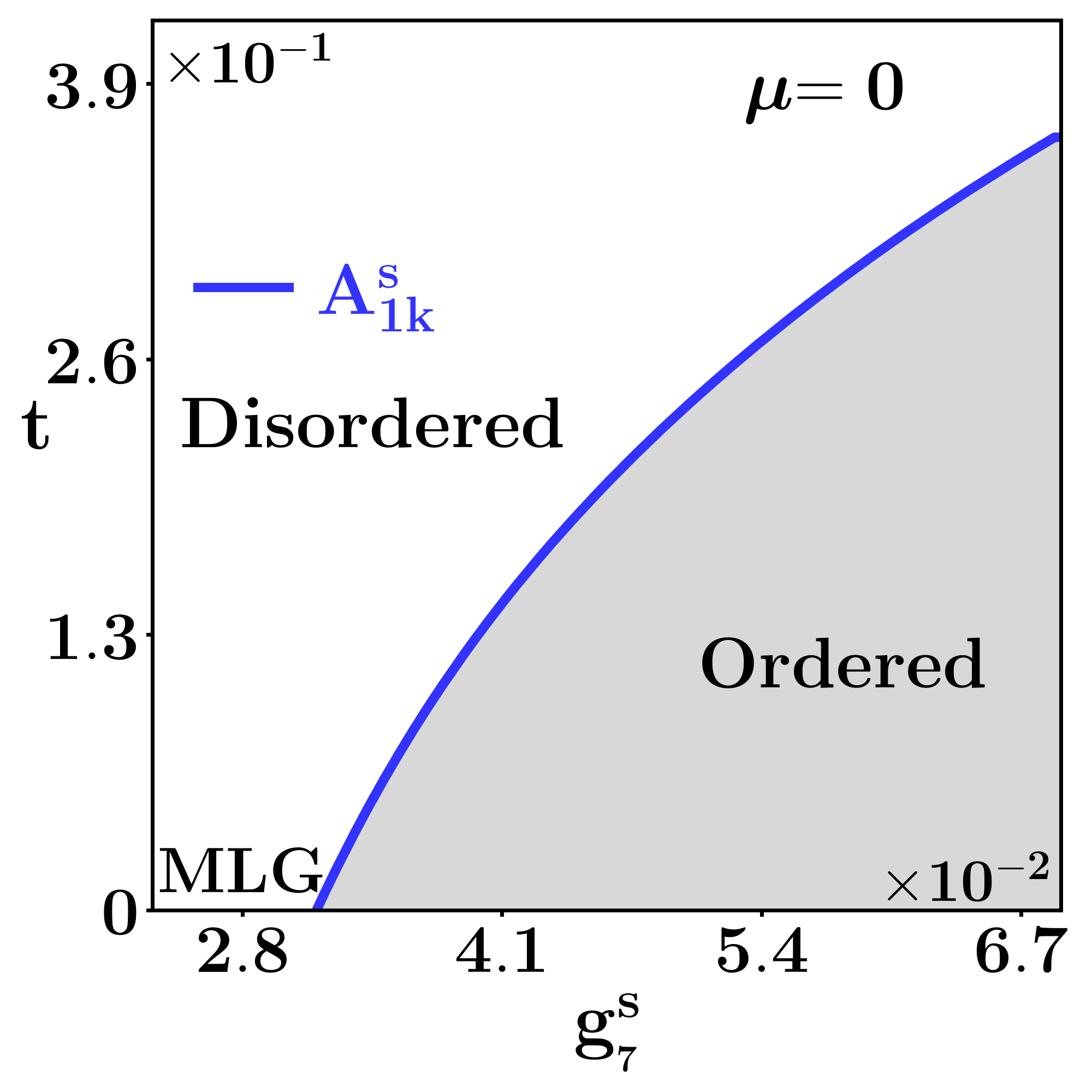}\hspace{0.5cm}
\includegraphics[width=0.21\linewidth]{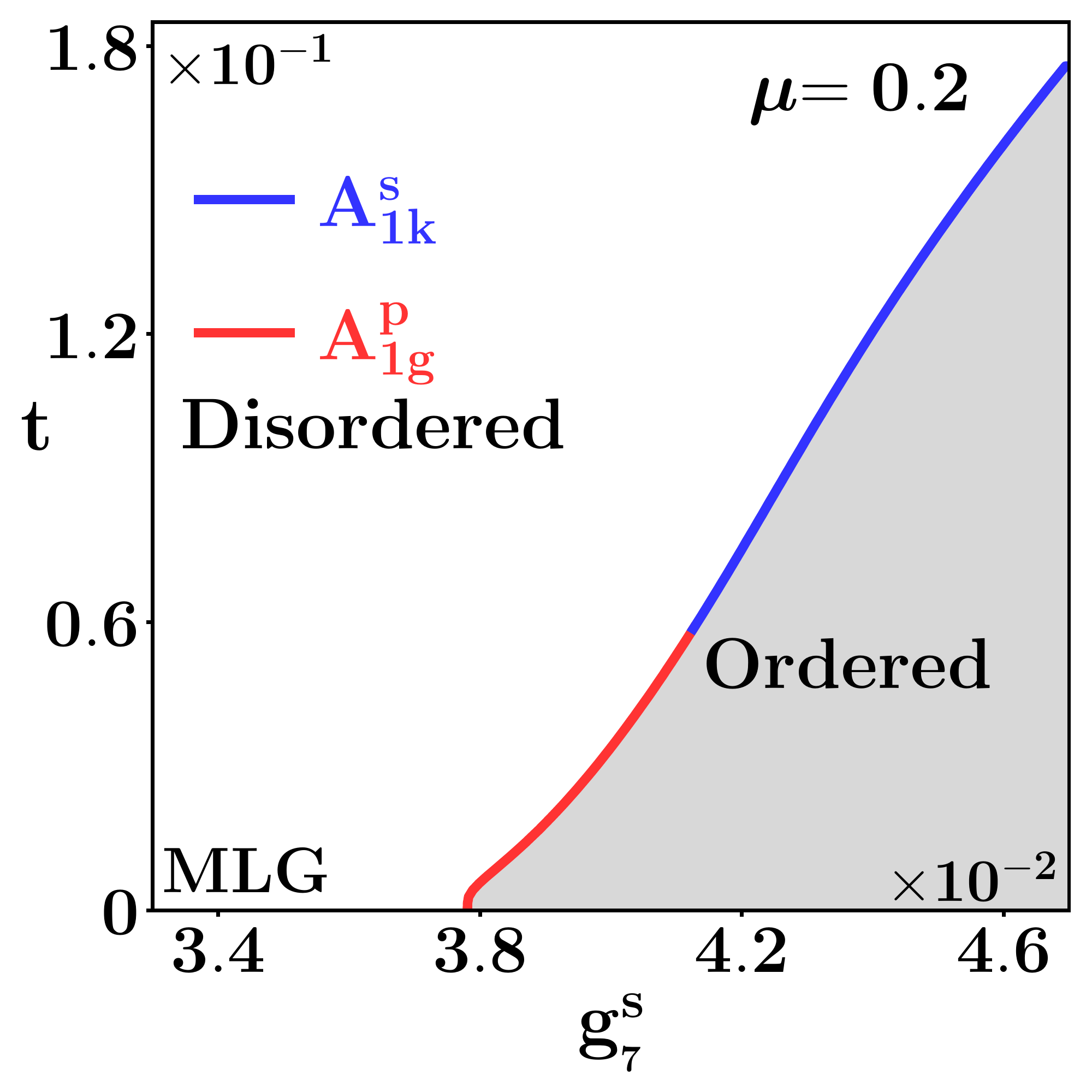}\hspace{0.5cm}
\includegraphics[width=0.21\linewidth]{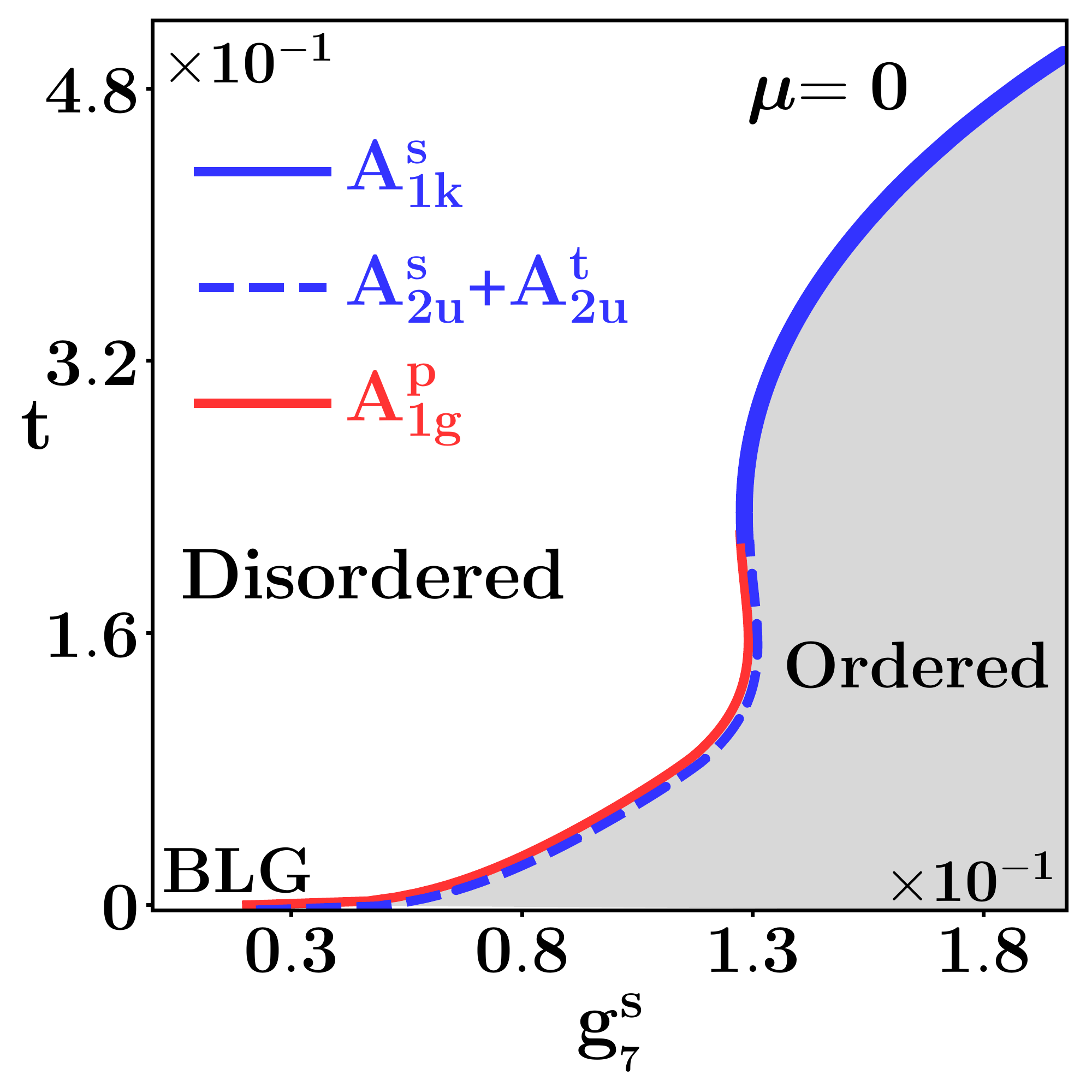}\hspace{0.5cm}
\includegraphics[width=0.21\linewidth]{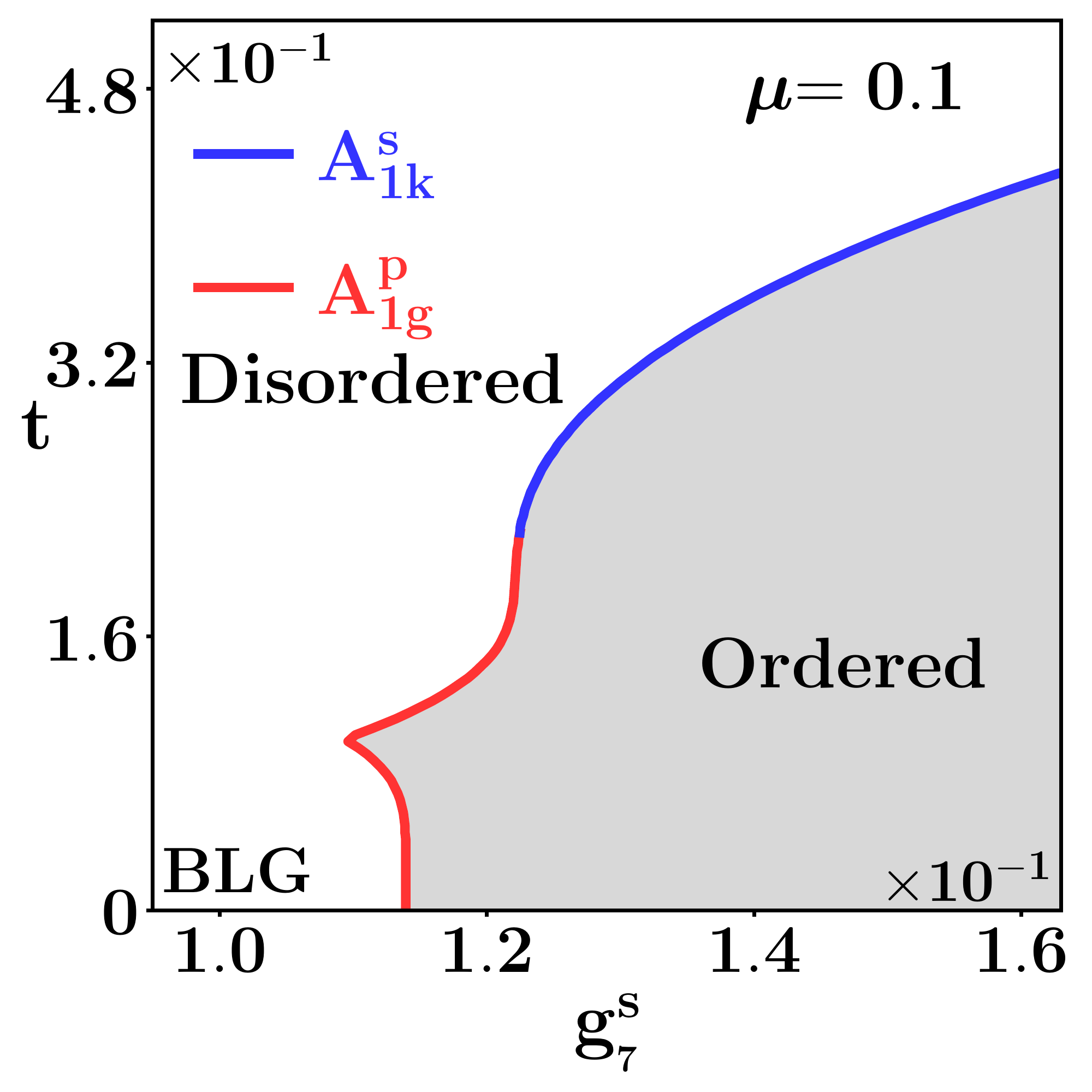}
\label{fig:kek_VBS}}\\
\subfloat[Phase diagrams for the quartic interaction in the Kekul\' e current or $A_{2k}$ singlet channel.]{
\includegraphics[width=0.21\linewidth]{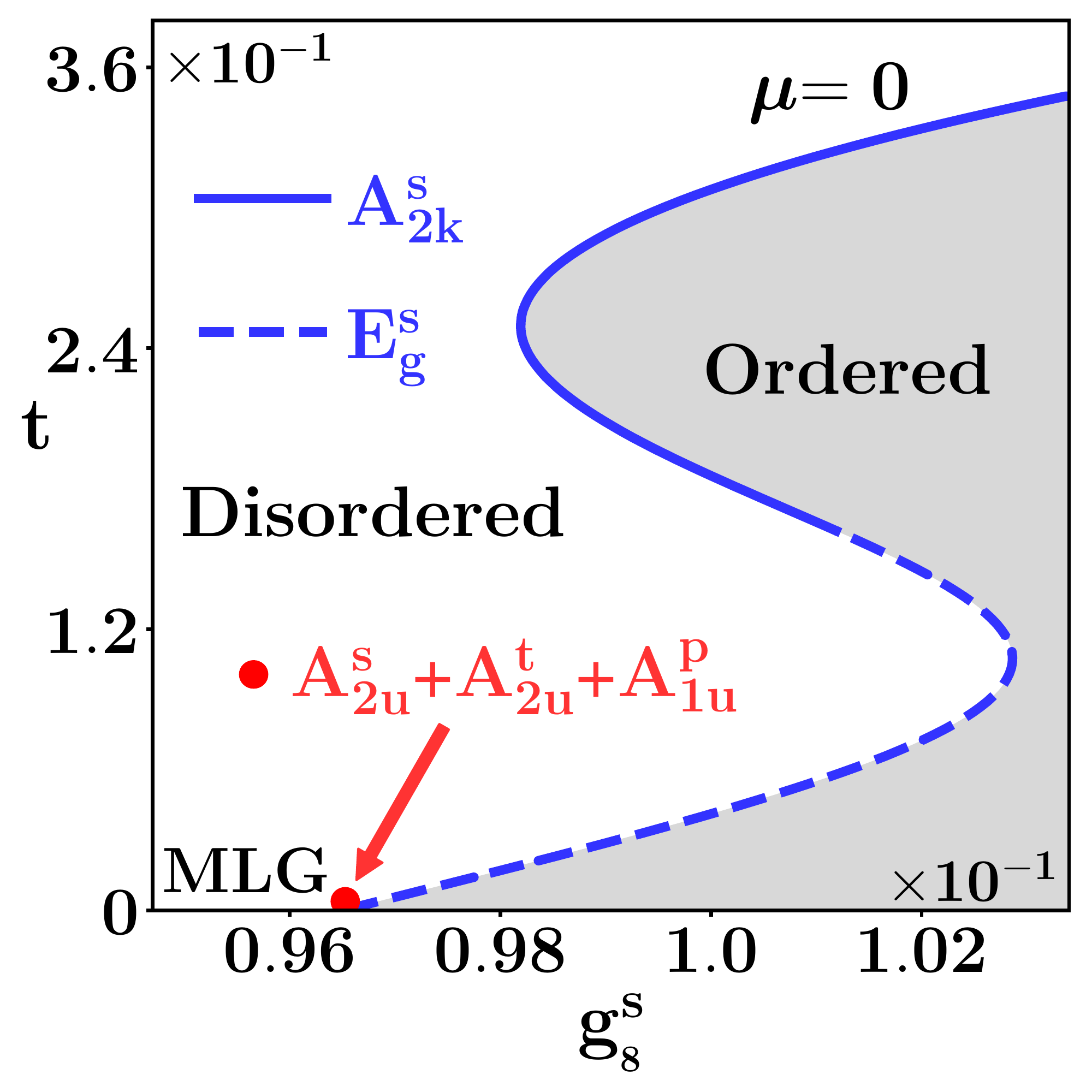}\hspace{0.5cm}
\includegraphics[width=0.21\linewidth]{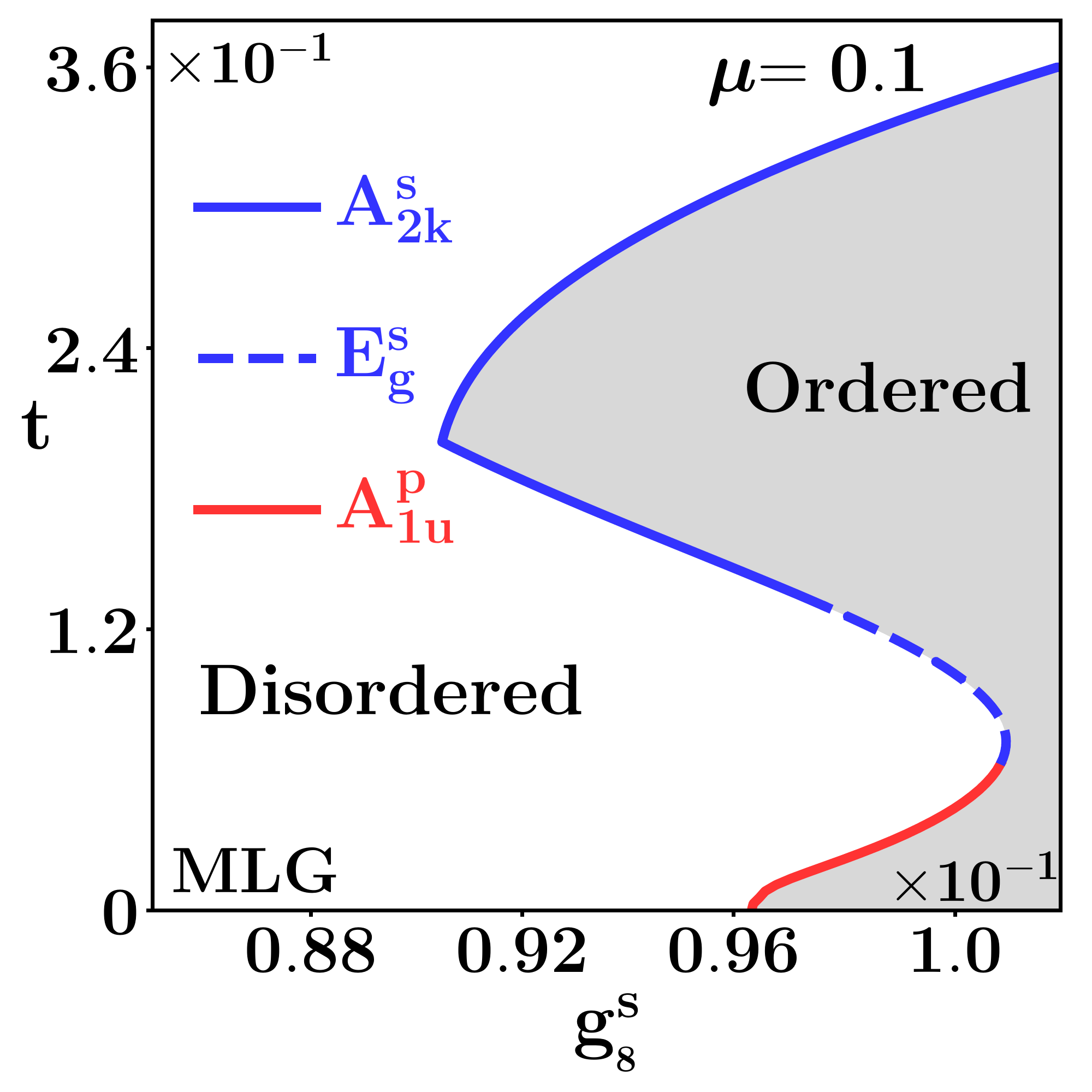}\hspace{0.5cm}
\includegraphics[width=0.21\linewidth]{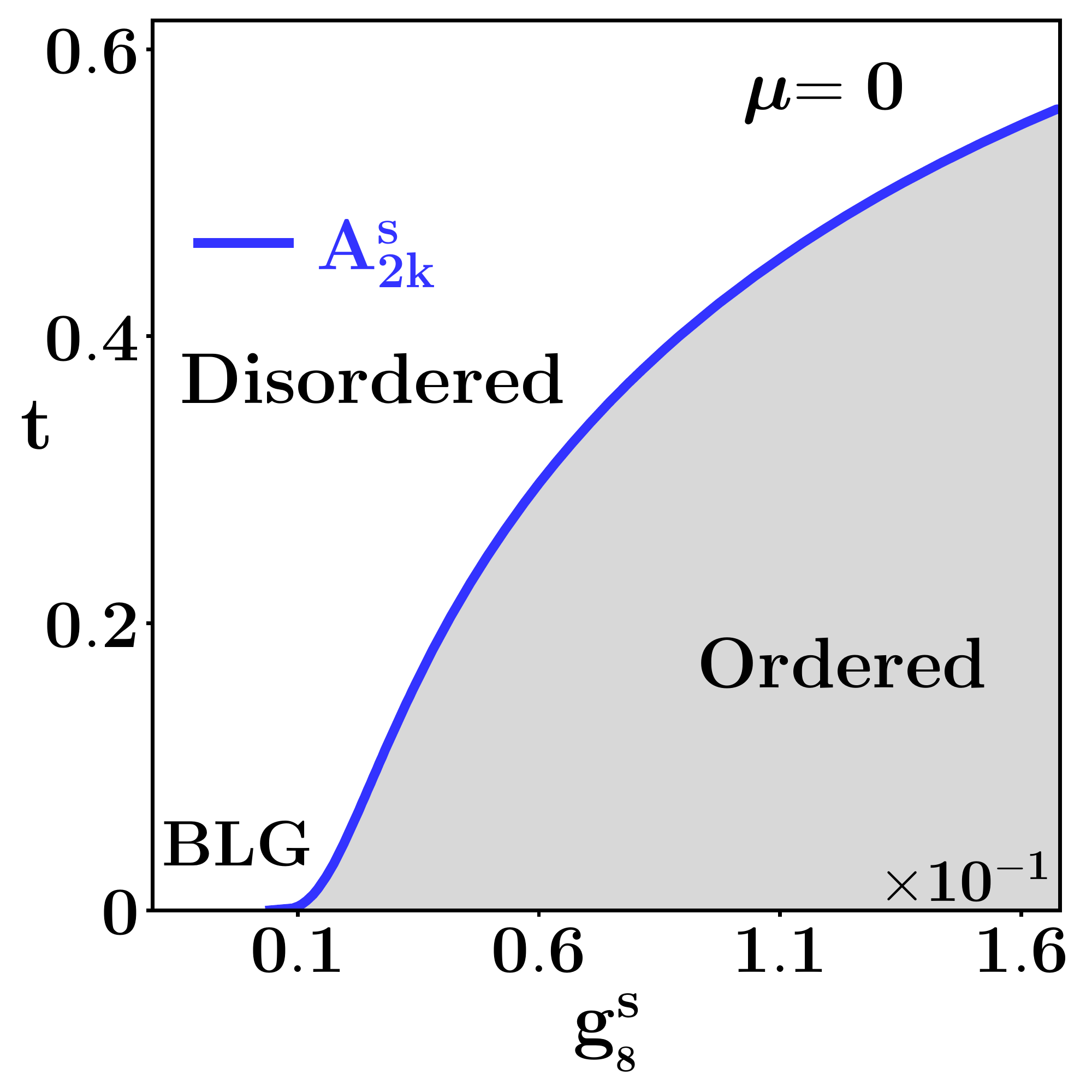}\hspace{0.5cm}
\includegraphics[width=0.21\linewidth]{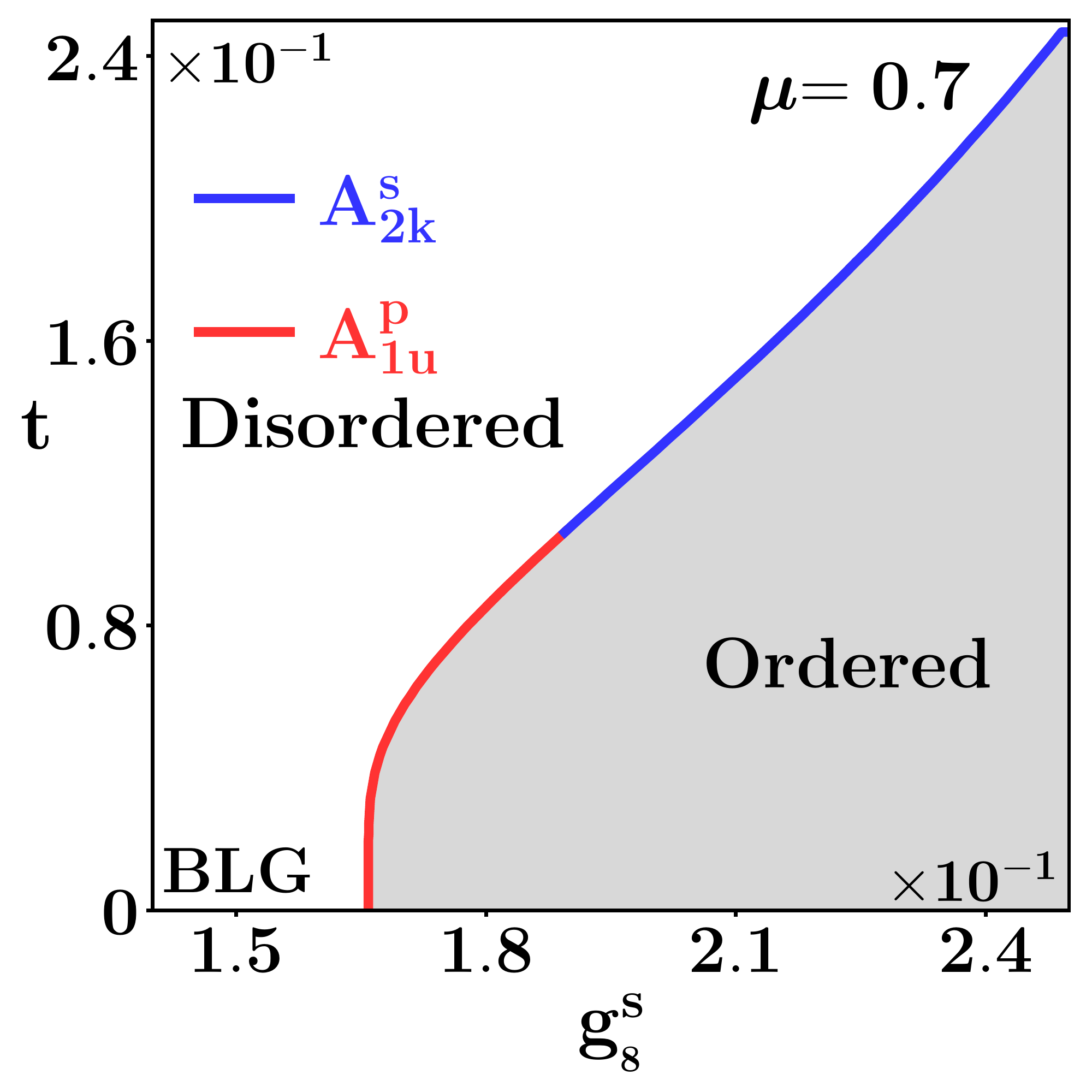}
\label{fig:kek_C}}\\
\subfloat[Phase diagrams for the quartic interaction in the spin Kekul\' e solid or $A_{1k}$ triplet channel.]{ 
\includegraphics[width=0.21\linewidth]{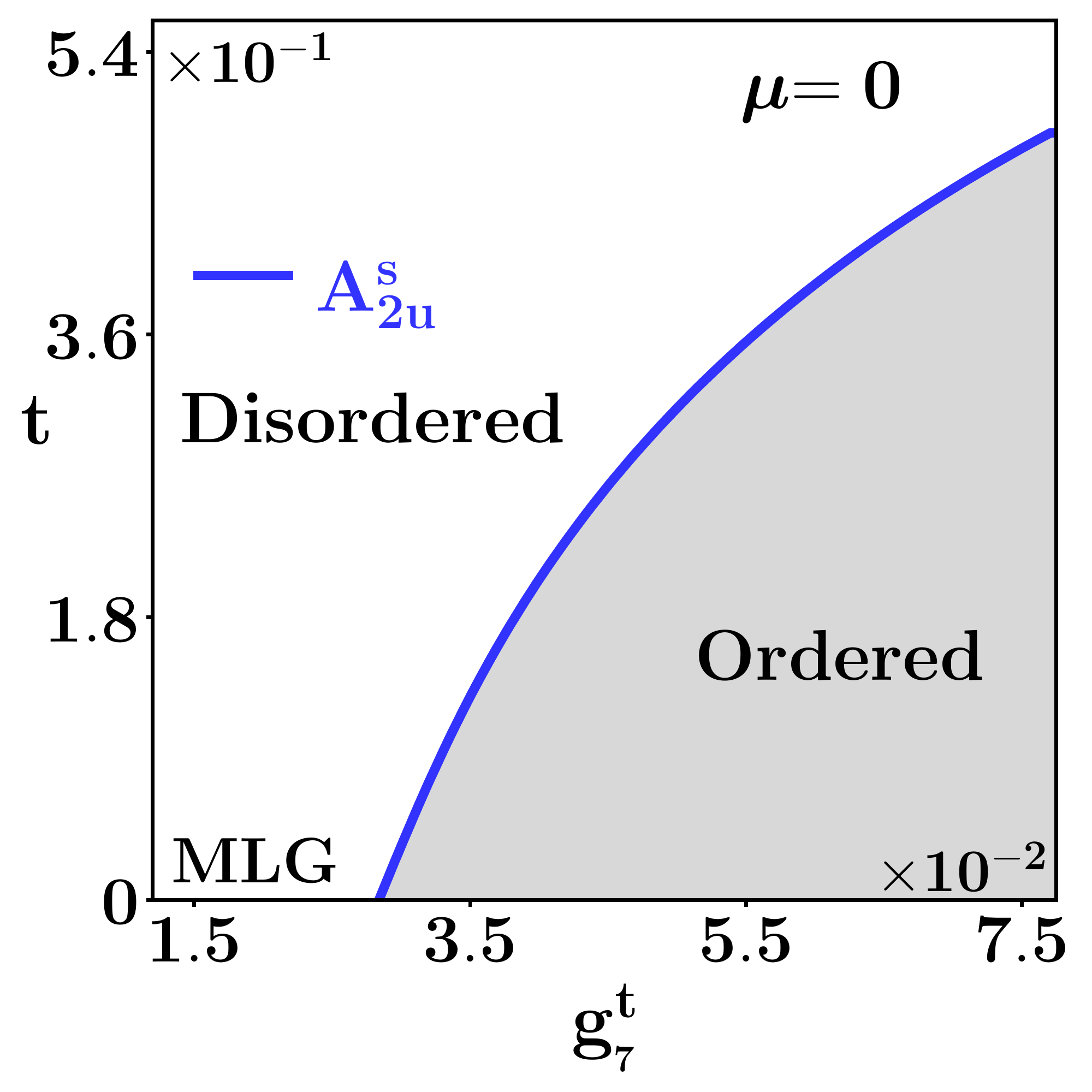}\hspace{0.5cm}
\includegraphics[width=0.21\linewidth]{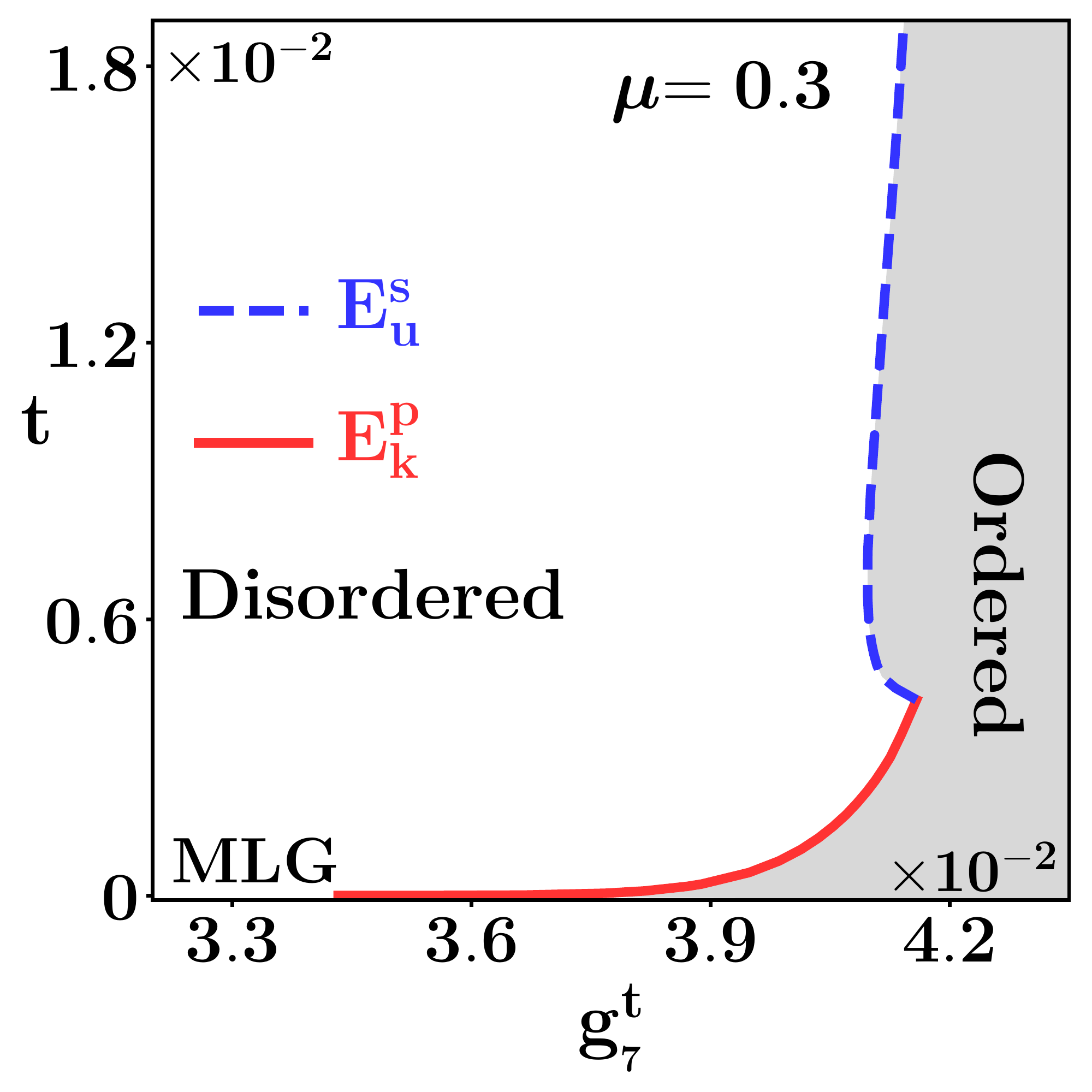}\hspace{0.5cm}
\includegraphics[width=0.21\linewidth]{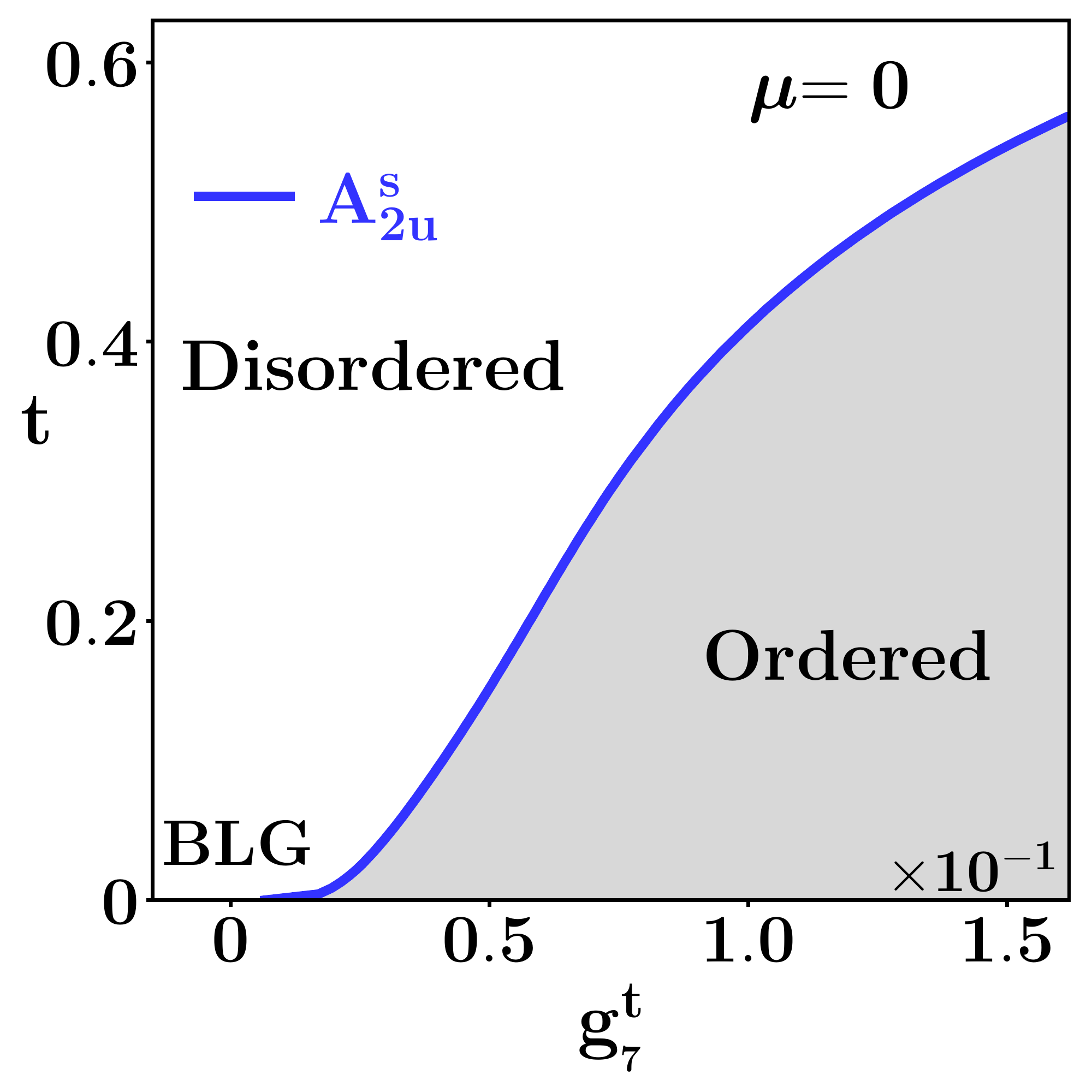}\hspace{0.5cm}
\includegraphics[width=0.21\linewidth]{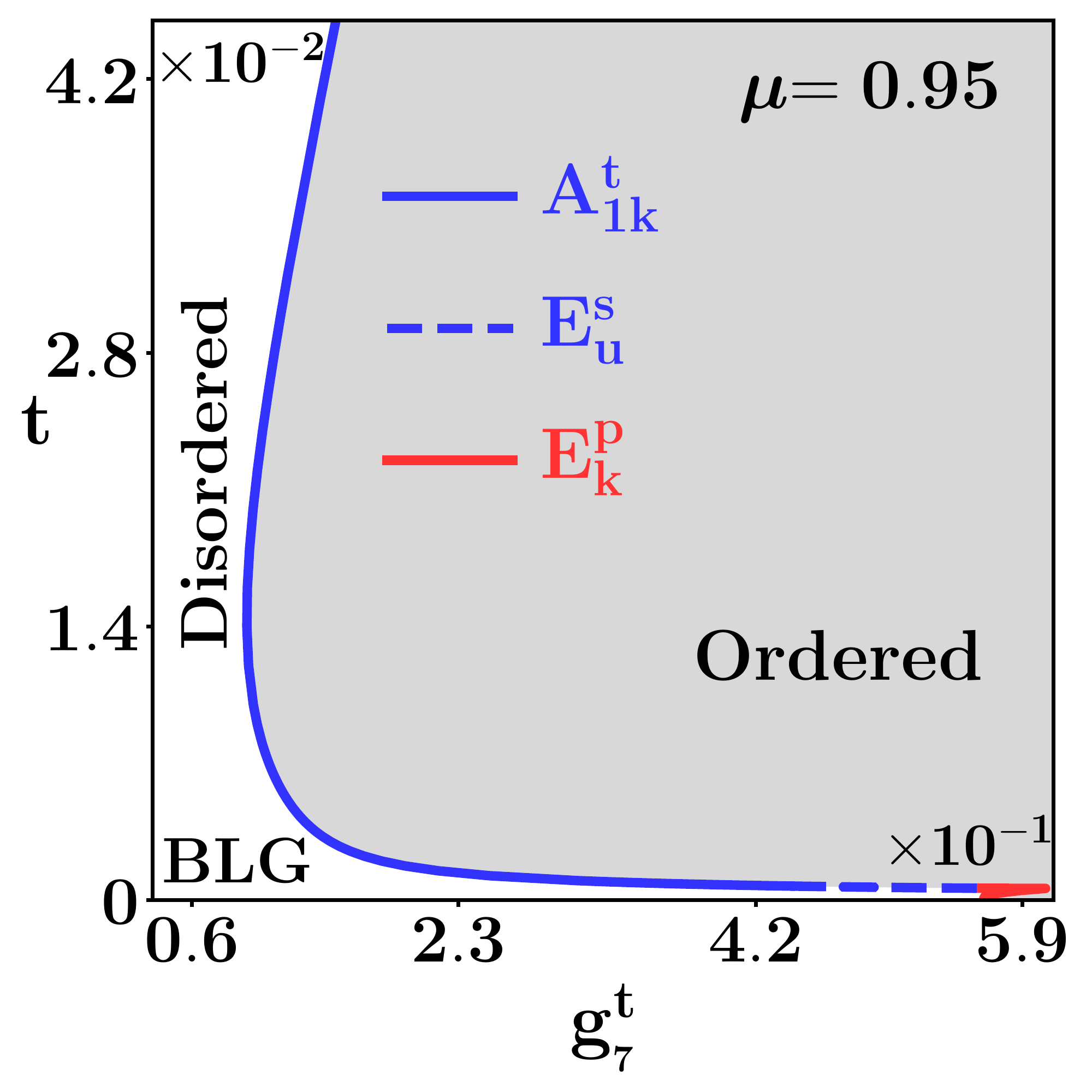}
\label{fig:s-kek_S}}\\
\subfloat[Phase diagrams for the quartic interaction in the spin Kekul\' e current or $A_{2k}$ triplet channel.]{
\includegraphics[width=0.21\linewidth]{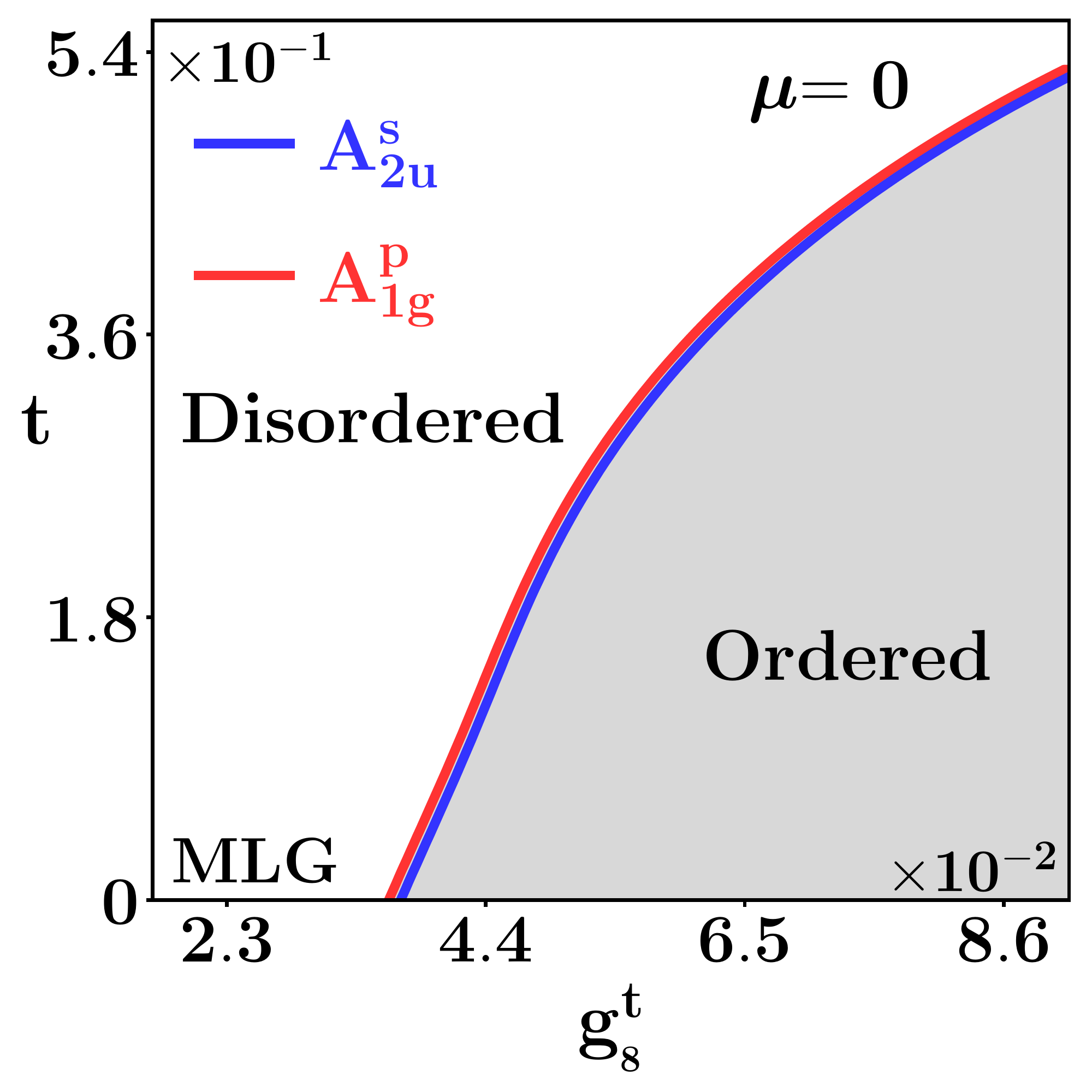}\hspace{0.5cm}
\includegraphics[width=0.21\linewidth]{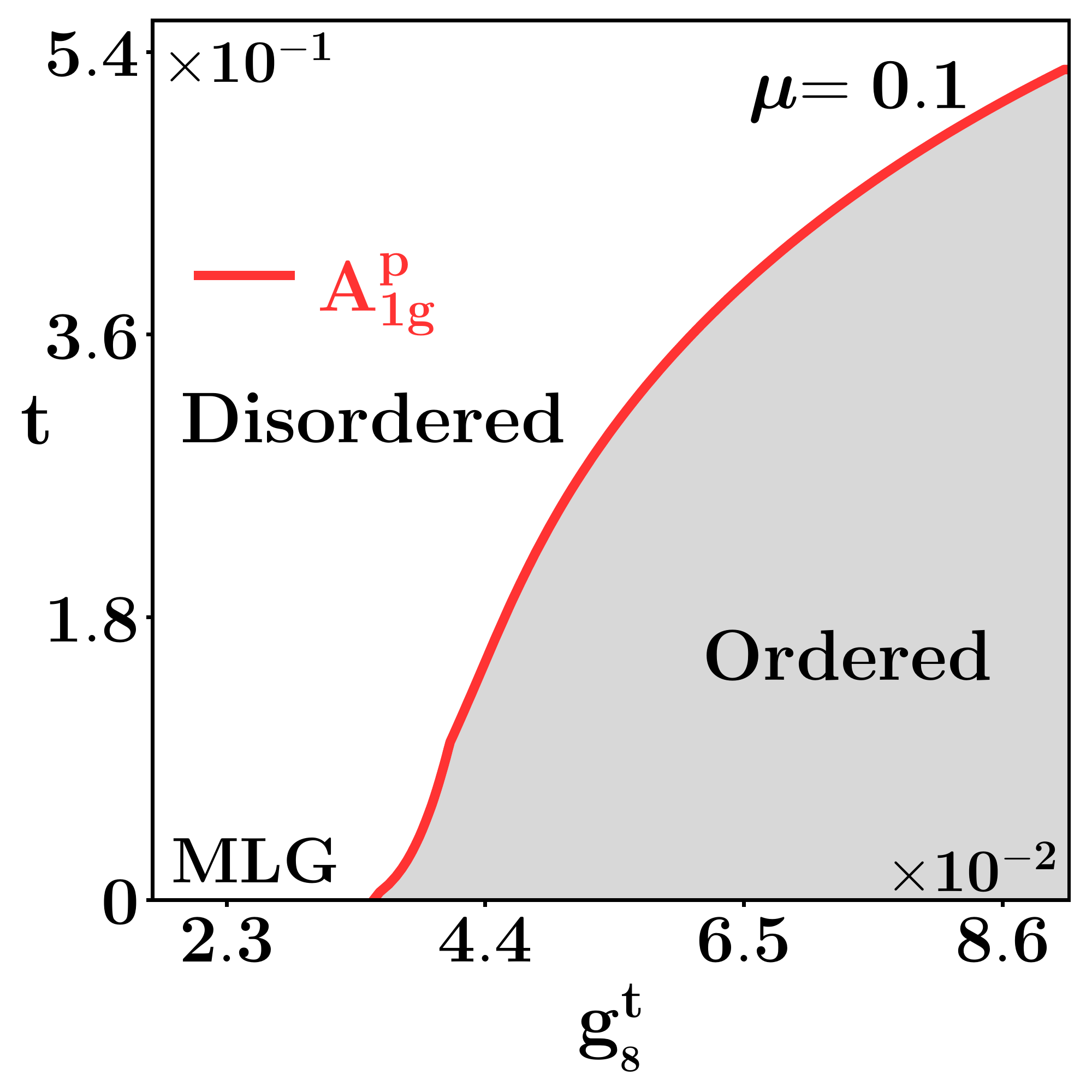}\hspace{0.5cm}
\includegraphics[width=0.21\linewidth]{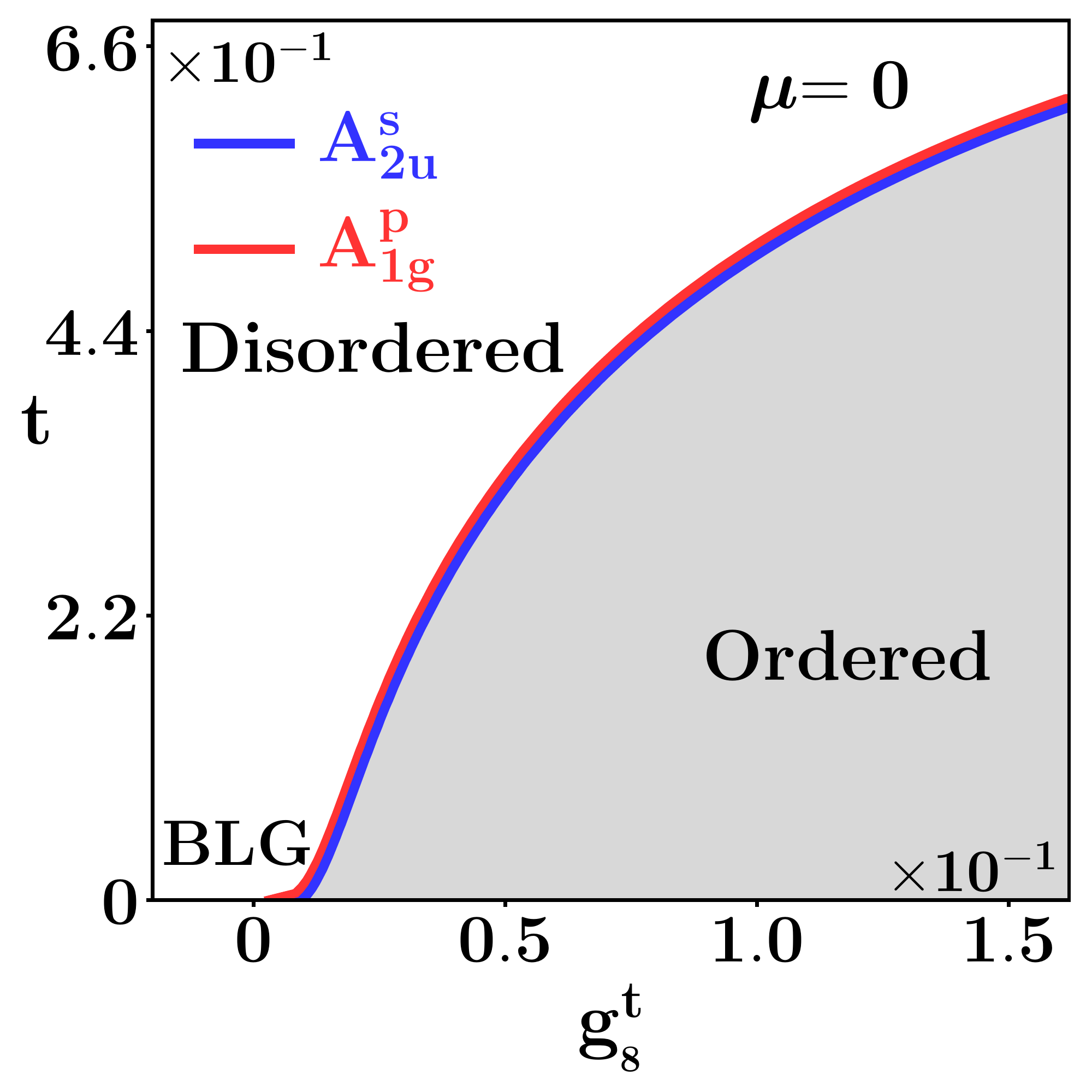}\hspace{0.5cm}
\includegraphics[width=0.21\linewidth]{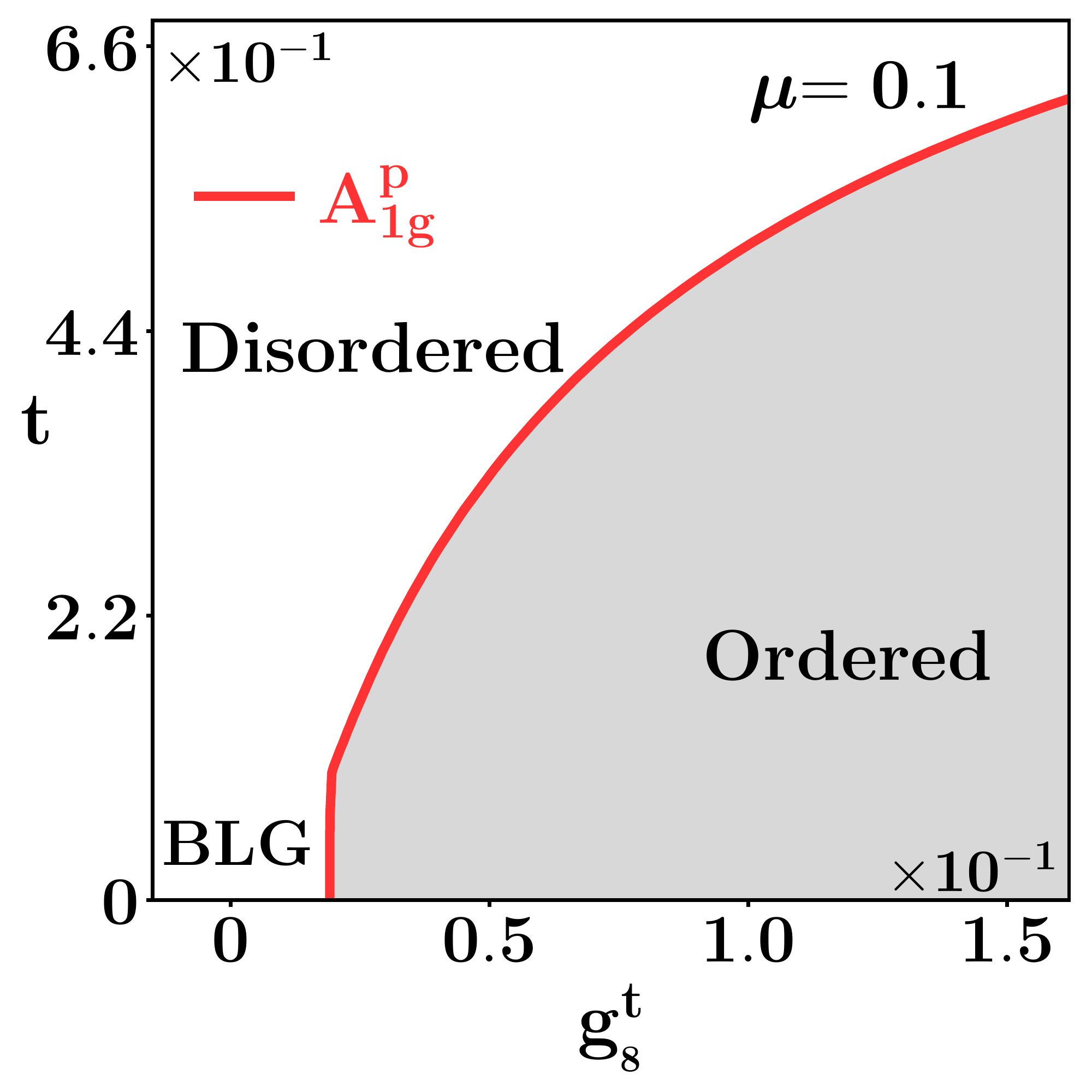}
\label{fig:s-kek_C}}
\caption{Cuts of the global phase diagram with the quartic interactions in the Kekul\'e channels. The rest of the details are identical to those in Fig.~\ref{fig:mass_PD}. For the constructions of composite order parameters from adjacent phases, see Sec.~\ref{sec:PD_kekule}.} 
\label{fig:Kekule_PD}
\end{figure*}

\noindent 1. For quartic interaction in the KVBS channel $M=(\Gamma_{3011}, \Gamma_{3021})$. The high temperature phase in both systems is KVBS itself, fulfilling selection rule (Ia). Note that for the Dirac fermions KVBS is a mass, therefore at $\mu=0$ in MLG it persists all the way down to the lowest temperature. However, the KVBS order parameter fully commutes with $\hat{h}^{\rm L}$ and in BLG this phase is replaced by the competing massive phases, namely the CDW, AFM and $s$-wave SC orderings at $\mu=0$. Since $M$ fully anticommutes with the $s$-wave pairing order parameter, finite $\mu$ selects this phase in MLG as well as BLG, see Fig.~\ref{fig:kek_VBS}. The corresponding supervectors are
\begin{align}
\vec{V}^{\rm KVBS}_1&=\left\{
\overbrace{
\underbrace{\Gamma_{\alpha 000}}_{\mbox{$A_{1g}^p$}},
\underbrace{\Gamma_{3011},\Gamma_{3021}}_{\mbox{$A_{1k}^s$}} 
}^{\mbox{O(4) vector}}
\right\},\nonumber\\
\vec{V}^{\rm KVBS}_2&=\left\{
\overbrace{
\underbrace{\Gamma_{3003}}_{\mbox{$A_{2u}^s$}},
\underbrace{\Gamma_{3011},\Gamma_{3021}}_{\mbox{$A_{1k}^s$}} 
}^{\mbox{O(3) vector}}
\right\}, \nonumber \\
\vec{V}^{\rm KVBS}_3&=\left\{
\overbrace{
\underbrace{\Gamma_{0s03}}_{\mbox{$A_{2u}^t$}},
\underbrace{\Gamma_{3011},\Gamma_{3021}}_{\mbox{$A_{1k}^s$}} 
}^{\mbox{O(5) vector}}
\right\}.
\end{align}

\vspace{0.3cm}
\noindent 2. The quartic interaction in the KC channel displays analogous behavior to that in the KVBS channel, but the mass nature of the phase is flipped between MLG and BLG. Here $M=(\Gamma_{0012}, \Gamma_{0022})$ and at high temperatures the dominant instability is KC in both systems [selection rule (Ia)]. In BLG the KC is a massive phase and at $\mu=0$ it persists down to the lowest temperature. In MLG at zero chemical doping the low temperature regime is occupied by phases that follow selection rule (Ib). At the lowest temperature and $\mu=0$ we observe degenerate nucleation of CDW, AFM and $f$-wave pairing states, all representing mass orders with $C_H=0$. Increasing the temperature the broken symmetry phase becomes the $E_g$ nematic, for which $C_H=A_H$. See section \ref{sec:extendedsum:selection} for the definitions of $C_H$ and $A_H$. As the KC order parameter fully commutes with the Dirac Hamiltonian ($A_H=0$), this ordering only occurs at high temperatures. The onset of these orderings along the temperature axis is consistent with selection rule (II). In both systems, finite chemical potential and low temperatures favor the adjacent $f$-wave pairing, following selection rule (Ib). The corresponding cuts of the phase diagram are displayed in Fig.~\ref{fig:kek_C}, and the composite order parameters are 
\begin{align}
\vec{V}^{\rm KC}_1&=\left\{
\overbrace{
\underbrace{\Gamma_{\alpha s30}}_{\mbox{$A_{1u}^p$}},
\underbrace{\Gamma_{0012},\Gamma_{0022}}_{\mbox{$A_{2k}^s$}} 
}^{\mbox{2 copies of O(5) vector}}
\right\},\nonumber\\
\vec{V}^{\rm KC}_2&=\left\{
\overbrace{
\underbrace{\Gamma_{3003}}_{\mbox{$A_{2u}^s$}},
\underbrace{\Gamma_{3001},\Gamma_{3032}}_{\mbox{$E_{g}^s$}} 
}^{\mbox{O(3) vector}}
\right\},\nonumber\\
\vec{V}^{\rm KC}_3&=\left\{
\overbrace{
\underbrace{\Gamma_{0s03}}_{\mbox{$A_{2u}^t$}},
\underbrace{\Gamma_{3001},\Gamma_{3032}}_{\mbox{$E_{g}^s$}}
}^{\mbox{O(5) vector}}
\right\}, \nonumber \\
\vec{V}^{\rm KC}_4&=\left\{
\overbrace{
\underbrace{\Gamma_{\alpha s30}}_{\mbox{$A_{1u}^p$}},
\underbrace{\Gamma_{3001},\Gamma_{3032}}_{\mbox{$E_{g}^s$}}
}^{\mbox{2 copies of O(5) vector}}
\right\},\nonumber\\
\vec{V}^{\rm KC}_5&=\left\{
\overbrace{
\underbrace{\Gamma_{3001},\Gamma_{3032}}_{\mbox{$E_{g}^s$}},
\underbrace{\Gamma_{0012},\Gamma_{0022}}_{\mbox{$A_{2k}^s$}} 
}^{\mbox{O(4) vector}}
\right\}.
\end{align}

\vspace{0.3cm}
\noindent 3. For quartic interaction in the spin Kekul\'e solid (KS) channel $M=(\Gamma_{0s11}, \Gamma_{0s21})$. At zero chemical potential, following selection rule (Ib), we observe nucleation of the excitonic CDW order in both MLG and BLG. For $\mu>0$ and low temperatures the Dirac, as well as the Luttinger fermions condense into the smectic superconductor, whereas intermediate $t$ destroys the pairing phase and gives rise to the $E_u$ nematic ordering, for which $C_H=A_H$. Furthermore, at even higher temperature BLG displays the spin KS phase, according to selection rule (Ia), that maximally commutes with the Luttinger Hamiltonian, $A_H=0$. Once again, onset of these phases by increasing temperature is in accordance with selection rule (II). For the phase diagrams in this interaction channel see Fig.~\ref{fig:s-kek_S}. The composite order parameters in these cases are
\begin{align}
\vec{V}^{\rm sKS}_1&=\left\{
\overbrace{
\underbrace{\Gamma_{\alpha 010},\Gamma_{\alpha 020},\Gamma_{\alpha 013},\Gamma_{\alpha 023}}_{\mbox{$E_{k}^p$}},
\underbrace{\Gamma_{0031},\Gamma_{0002}}_{\mbox{$E_{u}^s$}} 
}^{\mbox{4 copies of O(3) vector}}
\right\},\nonumber\\
\vec{V}^{\rm sKS}_2&=\left\{
\overbrace{
\underbrace{\Gamma_{0031},\Gamma_{0002}}_{\mbox{$E_{u}^s$}},
\underbrace{\Gamma_{0s11},\Gamma_{0s21}}_{\mbox{$A_{1k}^t$}} 
}^{\mbox{2 copies of O(5) vector}}
\right\}.
\end{align}

\vspace{0.3cm}
\noindent 4. For quartic interaction in the spin KC channel $M=(\Gamma_{3s12}, \Gamma_{3s22})$, which fully anticommutes with the $s$-wave pairing and CDW order parameters. At $\mu=0$ we observe degenerate nucleation of these two phases for the entire range of temperature. Note that for both order parameters $\{O_i,M_j\}=0$ and $C_H=0$ (i.e., they represent fully gapped phases), and thus they are not distinguished by temperature. Only by setting $\mu>0$ we select the $s$-wave SC in both MLG and BLG, see Fig.~\ref{fig:s-kek_C}. In the absence of an adjacent high temperature phase we do not obtain composite order parameters in this case. Note that this is due to the fact that only selection rule (Ib) is operative in this case for both MLG and BLG.

%%%%%%%%%%%%%%%%%%%%%%%%%%%%%%%%%%%%%%%%%%%%%%%%%%%%%%%%%%%%%%%%%%%%%%%%%
%%%%%%%%%%%%%%%%%%%%%%%%%%%%%%%%%%%%%%%%%%%%%%%%%%%%%%%%%%%%%%%%%%%%%%%%%
%%%%%%%%%%%%%%%%%%%% PHASE DIAGRAMS - KEKULE CHANNELS %%%%%%%%%%%%%%%%%%%
%%%%%%%%%%%%%%%%%%%%%%%%%%%%%%%%%%%%%%%%%%%%%%%%%%%%%%%%%%%%%%%%%%%%%%%%%
%%%%%%%%%%%%%%%%%%%%%%%%%%%%%%%%%%%%%%%%%%%%%%%%%%%%%%%%%%%%%%%%%%%%%%%%%
%%%%%%%%%%%%%%%%%%%%%%%%%%%%%%%%%%%%%%%%%%%%%%%%%%%%%%%%%%%%%%%%%%%%%%%%%
%%%%%%%%%%%%%%%%%%%%%%%%%%%%%%%%%%%%%%%%%%%%%%%%%%%%%%%%%%%%%%%%%%%%%%%%%
%%%%%%%%%%%%%%%%%%%%%%%%%%%%%%%%%%%%%%%%%%%%%%%%%%%%%%%%%%%%%%%%%%%%%%%%%
%%%%%%%%%%%%%%%%%%%%%%%%%%%%%%%%%%%%%%%%%%%%%%%%%%%%%%%%%%%%%%%%%%%%%%%%%
%%%%%%%%%%%%%%%%%%%%%%%%%%%%%%%%%%%%%%%%%%%%%%%%%%%%%%%%%%%%%%%%%%%%%%%%%

%%%%%%%%%%%%%%%%%%%%%%%%%%%%%%%%%%%%%%%%%%%%%%%%%%%%%%%%%%%%%%%%%%%%%%%%%%%%%%%%%%%%%%%
%%%%%%%%%%%%%%%%%%%%%%%%%%%%%%%%%%%%%%%%%%%%%%%%%%%%%%%%%%%%%%%%%%%%%%%%%%%%%%%%%%%%%%%
%%%%%%%%%%%%%%%%%%%%%%%%%%%%%%%%%%%%%%%%%%%%%%%%%%%%%%%%%%%%%%%%%%%%%%%%%%%%%%%%%%%%%%%
%%%%%%%%%%%%%%%%%%%%%%%%%%%%%%%%%%%%%%%%%%%%%%%%%%%%%%%%%%%%%%%%%%%%%%%%%%%%%%%%%%%%%%%
%%%%%%%%%%%%%%%%%%%%%%%%%%%%%%%%%%%%%%%%%%%%%%%%%%%%%%%%%%%%%%%%%%%%%%%%%%%%%%%%%%%%%%%
\section{Honeycomb Hubbard model}~\label{sec:HubbardModel}

In the preceding sections we scrutinized the honeycomb monolayer and bilayer in the presence of generic short range repulsive electron-electron interactions. We demonstrated through numerous examples the validity of the selection rules and organizing principle (see Sec.~\ref{sec:extendedsum:selection}), by tuning the strength of interactions in one specific channel at a time at zero and finite temperature and chemical doping. However, the bare values of the interaction couplings in a physically realistic microscopic model would in principle be non-zero simultaneously in multiple interaction channels. To substantiate our findings, we here consider one of the simplest yet realistic microscopic models for correlated electrons on the honeycomb lattice, the extended Hubbard model.

In Sec.~\ref{sec:extendedsum:Hubbard}, we present the Hamiltonian containing on-site, NN and NNN repulsion, see Eqs.~(\ref{eq:ExtHub1_summary}) and (\ref{eq:ExtHub2_summary}), while the resulting phase diagrams for MLG and BLG are shown in Figs.~\ref{fig:Hubbard_U}, \ref{fig:Hubbard_V1} and \ref{fig:Hubbard_V2}. Here we give a detailed walkthrough on how to translate between the Hubbard Hamiltonian, written in terms of a lattice model, and our effective low-energy description. The three distinct interaction terms (on-site, NN and NNN repulsion) are addressed separately in three subsequent sections. In Bernal BLG as the split-off bands are dominantly localized on the $a_1$ and $b_2$ sites and here we consider only density-density interactions, one can integrate out fermionic density on these two sites to arrive at an effective or renormalized extended honeycomb Hubbard model defined on the low-energy sites, at the cost of renormalized strengths of the corresponding coupling constants~\cite{PhysRevB.82.205106}. Such trivial renormalization is, however, omitted here.

\subsection{On-site repulsion ($U$)}

To make a connection with Eq.~(\ref{eq:ExtHub2_summary}), we first make use of the Hamman decomposition to rewrite $H_U$ as~\cite{PhysRevLett.97.146401, PhysRevB.2.1373}
\widetext
\begin{align}
H_U=\frac{U}{16}\sum_{\vec{A}} \Big\{ \left[ n(\vec{A})+n(\vec{A}+\boldsymbol{\delta}_i) \right]^2+\left[ n(\vec{A})-n(\vec{A}+\boldsymbol{\delta}_i) \right]^2
-\left[ \vec{m}(\vec{A})+\vec{m}(\vec{A}+\boldsymbol{\delta}_i) \right]^2-\left[ \vec{m}(\vec{A})-\vec{m}(\vec{A}+\boldsymbol{\delta}_i) \right]^2 \Big\},
\end{align}
where $n(\vec{R})=r^\dag_\sigma (\vec{R}) r_\sigma (\vec{R})$ is the number operator, $\vec{m}=r^\dag_\sigma (\vec{R}) \boldsymbol{\sigma}_{\sigma \sigma'}r_{\sigma'} (\vec{R})$ is the magnetization, with $r^\dag_\sigma(\vec{R})$ and $r_\sigma(\vec{R})$ being fermionic creation and annihilation operators with spin projection $\sigma=\uparrow, \downarrow$ on the site at $\vec{R}$, and summation over repeated indices is assumed. The first (second) term in the Hamman decomposition corresponds to the total (staggered) density, while the third (fourth) term to the total (staggered) magnetization. Recalling the Fourier expansion of the fermionic fields in Eq.~(\ref{eq:Fourier}) and the spinor structure in Eq.~(\ref{eq:spinor}), they are written as
\begin{align}
n(\vec{A})+n(\vec{B})&=
 \Psi^\dag \Gamma_{3000} \Psi
+\cos(2 \vec{K}\cdot \vec{r})\Psi^\dag \Gamma_{3010} \Psi
+\sin(2 \vec{K}\cdot \vec{r})\Psi^\dag \Gamma_{3020} \Psi,\label{eq:density} \\
n(\vec{A})-n(\vec{B})&=
 \Psi^\dag \Gamma_{3003} \Psi
+\cos(2 \vec{K}\cdot \vec{r}) \Psi^\dag \Gamma_{3013} \Psi 
+\sin(2 \vec{K}\cdot \vec{r}) \Psi^\dag \Gamma_{3023} \Psi,\label{eq:staggered_density} \\
m_s(\vec{A})+m_s(\vec{B})&=
\Psi^\dag \Gamma_{3s00} \Psi
+\cos(2 \vec{K}\cdot \vec{r}) \Psi^\dag \Gamma_{3s10} \Psi
+\sin(2 \vec{K}\cdot \vec{r}) \Psi^\dag \Gamma_{3s20} \Psi,\label{eq:magnetization} \\
m_s(\vec{A})-m_s(\vec{B})&=
\Psi^\dag \Gamma_{3s03} \Psi
+\cos(2 \vec{K}\cdot \vec{r}) \Psi^\dag \Gamma_{3s13} \Psi 
+\sin(2 \vec{K}\cdot \vec{r}) \Psi^\dag \Gamma_{3s23} \Psi,\label{eq:staggered_magnetization} 
\end{align}
where $m_s(\vec{R})$ is the $s$ component of the magnetization ($s=1,2,3$) on the site at $\vec{R}$.

Upon squaring the above quantities we neglect the oscillatory contributions, as any position (and thus wave-number) dependent term will be less relevant in the RG sense (as can be shown via power counting). Then the on-site repulsion in the continuum theory takes the form
\begin{align}
H_U=\frac{Ua^2}{16}\Bigg\{ &(\Psi^\dag \Gamma_{3000}\Psi)^2 + (\Psi^\dag \Gamma_{3003} \Psi)^2 + \frac{1}{2}\Big[(\Psi^\dag \Gamma_{3010}\Psi)^2
+(\Psi^\dag \Gamma_{3020}\Psi)^2+(\Psi^\dag \Gamma_{3013}\Psi)^2+(\Psi^\dag \Gamma_{3023}\Psi)^2 \Big] \nonumber \\
-&(\Psi^\dag \Gamma_{0s00}\Psi)^2 - (\Psi^\dag \Gamma_{0s03} \Psi)^2 - \frac{1}{2}\Big[(\Psi^\dag \Gamma_{0s10}\Psi)^2 
+(\Psi^\dag \Gamma_{0s20}\Psi)^2+(\Psi^\dag \Gamma_{0s13}\Psi)^2+(\Psi^\dag \Gamma_{0s23}\Psi)^2 \Big] \Bigg\}.
\end{align}
Hence the initial condition (or bare value) of the coupling constants as a function of $U$ is
\begin{align}
g^s_1(\ell=0)=g^s_5(\ell=0)=-\frac{Ua^2}{16},\hspace{0.3cm} g^s_9(\ell=0)=-\frac{1}{2}\frac{Ua^2}{16},\hspace{0.3cm}
g^t_1(\ell=0)=g^t_5(\ell=0)=\frac{Ua^2}{16},\hspace{0.3cm}  g^t_9(\ell=0)=\frac{1}{2}\frac{Ua^2}{16}.
\end{align}
We then use the Fierz constraints to rewrite the spin triplet four fermion terms as a linear combinations of spin singlet terms, eliminating $g_i^t$, and subsequently rescale according to $Ua^2/16 \rightarrow U$. Doing so we arrive at
\begin{align}~\label{eq:HubbardU_init}
g^s_1(\ell=0)=-U, \hspace{0.6cm} g^s_5(\ell=0)=-U, \hspace{0.6cm} g^s_9(\ell=0)=-\frac{1}{2}U, 
\end{align}
\twocolumngrid

\noindent and all other linearly independent coupling constants $g_i^s(\ell=0)=0$ for $i= 2,3,4,6,7,8$.

To study the phase diagram of the honeycomb Hubbard model in the presence of on-site repulsion we apply the same methodology as in Sec.~\ref{sec:phasediagrams}. But instead of scanning along one certain axis in the space of coupling constants, we use Eq.~(\ref{eq:HubbardU_init}) and tune $U$ until we detect a phase transition. The $(U,t)$ cuts of the phase diagram for various values of $\mu$ are displayed in Figs.~\ref{fig:Hubbard_U}(a) and \ref{fig:Hubbard_U}(b) for MLG and BLG, respectively. At $\mu=t=0$ both the Dirac and Luttinger fermions display antiferromagnetic ($A_{2u}^t$) ordering, which is to be expected on a half-filled bipartite lattice in the presence of repulsive Hubbard interactions. For finite chemical doping the induced pairing phase for linearly dispersing Dirac fermions is the $E_g$ nematic superconductor [Fig.~\ref{fig:Hubbard_U}(a)], while for the quadratic Luttinger fermions it is the $A_{1k}$ singlet Kekul\' e pairing [Fig.~\ref{fig:Hubbard_U}(b)]. Note, the adjacent excitonic and pairing phases in both cases fully anticommute with each other and form O(5) supervector order parameters given by
\begin{align}~\label{eq:SOP_HubbardU}
\vec{V}^{\rm U}_{\rm MLG}&=\left\{
\overbrace{
\underbrace{\Gamma_{\alpha 001}, \Gamma_{\alpha 032} }_{\mbox{$E_{g}^p$}},
\underbrace{\Gamma_{0s03} }_{\mbox{$A_{2u}^t$}} 
}^{\mbox{2 copies of O(5) vectors}}
\right\}, \nonumber\\
\vec{V}^{\rm U}_{\rm BLG}&=\left\{
\overbrace{
\underbrace{\Gamma_{\alpha 011}, \Gamma_{\alpha 021} }_{\mbox{$A_{1k}^p$}},
\underbrace{\Gamma_{0s03} }_{\mbox{$A_{2u}^t$}} 
}^{\mbox{2 copies of O(5) vectors}}
\right\}. 
\end{align}

One can immediately recognize the pattern where two adjacent and fully (or partially) anticommuting phases fulfill the selection rules (Ia) and (Ib). Selection rule (Ia) in Eq.~(\ref{eq:rule_I}) requires that the kernel of the interaction term and the bilinear order parameter are the same. Since the quartic interactions are written in the particle-hole basis, the four-fermion term in the AFM channel is the dominant interaction with $U>0$. Indeed the phase diagrams with finite repulsive Hubbard interactions accommodate the same ordered phases as those corresponding to the purely antiferromagnetic interaction channel [Fig.~\ref{fig:AFM}]. Consequently, the resulting composite order parameters $\vec{V}^{\rm U}_{\rm MLG}$ and $\vec{V}^{\rm U}_{\rm BLG}$ are the same as $\vec{V}^{\rm AFM}_1$ and $\vec{V}^{\rm AFM}_2$ in Eq.~(\ref{eq:SOP_HubbardV}), respectively. As mentioned in Sec.~\ref{sec:extendedsum:Hubbard}, the different pairing orders arising in MLG and BLG are due to the fact that the singlet Kekul\' e pairing is a gapped phase only in BLG, while in MLG the maximally anticommuting pairing order parameter is the $E_g$ nematic superconductor. Such a distinction in the paired state at low temperature \emph{solely} stems from the differences in the normal state band structures in these two systems, in agreement with selection rule (II).

%%%%%%%%%%%%%%%%%%%%%%%%%%%%%%%%%%%%%%%%%%%%%%%%%%%%%%%%%%%%%%%%%%%%%%%%%%%%%%%%%%%%%%%
%%%%%%%%%%%%%%%%%%%%%%%%%%%%%%%%%%%%%%%%%%%%%%%%%%%%%%%%%%%%%%%%%%%%%%%%%%%%%%%%%%%%%%%
%%%%%%%%%%%%%%%%%%%%%%%%%%%%%%%%%%%%%%%%%%%%%%%%%%%%%%%%%%%%%%%%%%%%%%%%%%%%%%%%%%%%%%%
%%%%%%%%%%%%%%%%%%%%%%%%%%%%%%%%%%%%%%%%%%%%%%%%%%%%%%%%%%%%%%%%%%%%%%%%%%%%%%%%%%%%%%%
%%%%%%%%%%%%%%%%%%%%%%%%%%%%%%%%%%%%%%%%%%%%%%%%%%%%%%%%%%%%%%%%%%%%%%%%%%%%%%%%%%%%%%%

\subsection{Nearest neighbor repulsion ($V_1$)}

Next we turn our focus to the $H_{V_1}$ term in Eq.~(\ref{eq:ExtHub2_summary}), which models (with $V_1>0$) a finite ranged Coulomb repulsion between electrons on the NN sites. We derive the initial conditions similarly to those for the on-site repulsion. The Hamman decomposition in this case reads
\begin{align}~\label{eq:HubbardV_init}
H_{V_1}=\frac{V_1}{4} \sum_{\vec{A}} \sum_{\tau=\pm} \tau \big[ n(\vec{A})+\tau \sum_{i=1}^3 n(\vec{A}+\boldsymbol{\delta}_i)\big]^2.
\end{align}
In our basis the second term takes the form
\begin{align}
\sum_{i=1}^3 n(\vec{A}+\boldsymbol{\delta}_i)=\frac{3}{2} \Psi^\dag \Gamma_{3000} \Psi - \frac{3}{2} \Psi^\dag \Gamma_{3003}\Psi,
\end{align}
while the number operator on the $\vec{A}$ sites can be written as
\widetext
\begin{align}
&n(\vec{A})=\frac{1}{2} \Big( \Psi^\dag \Gamma_{3000} \Psi + \Psi^\dag \Gamma_{3003} \Psi \Big) + 
\frac{\cos(2 \vec{K}\cdot \vec{x})}{2}\Big( \Psi^\dag \Gamma_{3010} \Psi + \Psi^\dag \Gamma_{3013} \Psi \Big) + 
\frac{\sin(2 \vec{K}\cdot \vec{x})}{2}\Big( \Psi^\dag \Gamma_{3020} \Psi + \Psi^\dag \Gamma_{3023} \Psi \Big)
\end{align}
\twocolumngrid
\noindent We again neglect the oscillatory terms when taking the square of the density and the staggered density, and arrive at the expression
\begin{align}
H_{V_1}=3 \frac{V_1 a^2}{4} \Big[ (\Psi^\dag \Gamma_{3000} \Psi)^2 - (\Psi^\dag \Gamma_{3003} \Psi)^2 \Big].
\end{align}
After rescaling as $3 V_1 a^2/4 \to V_1$ we obtain for the bare values of the coupling constants to be
\begin{align}
g^s_1(\ell=0)&=-V_1, & g^s_5(\ell=0)&=V_1, 
\end{align}
and $g_i^s(\ell=0)=0$ for $i= 2,3,4,6,7,8,9$.

At $\mu=t=0$ we find the charge density wave ($A_{2u}^s$) ordering in both MLG [Fig.~\ref{fig:Hubbard_V1}(a)] and BLG [Fig.~\ref{fig:Hubbard_V1}(b)], while setting $\mu>0$ results in a \emph{degenerate} nucleation of singlet $s$-wave and triplet $f$-wave pairing phases. The corresponding composite order parameters are
\begin{align}~\label{eq:SOP_HubbardV}
\vec{V}^{\rm V_1}_1&=\left\{
\overbrace{
\underbrace{\Gamma_{1000}, \Gamma_{2000} }_{\mbox{$A_{1g}^p$}},
\underbrace{\Gamma_{3003} }_{\mbox{$A_{2u}^s$}} 
}^{\mbox{O(3) vector}}
\right\},\nonumber\\
\vec{V}^{\rm V_1}_2&=\left\{
\overbrace{
\underbrace{\Gamma_{1s30}, \Gamma_{2s30} }_{\mbox{$A_{1u}^p$}},
\underbrace{\Gamma_{3003} }_{\mbox{$A_{2u}^s$}} 
}^{\mbox{2 copies of O(4) vectors}}
\right\}.
\end{align}
Following analogous logic to the on-site repulsion case, we find that the quartic term in the CDW channel is the dominant interaction that fulfills the selection rule (Ia), while the adjacent pairing orders are the $s$-wave and $f$-wave superconductors, satisfying the selection rule (Ib). This can be anchored by comparing  $\vec{V}^{\rm V_1}_1$ and $\vec{V}^{\rm V_1}_2$ with Eq.~(\ref{eq:SOP_CDW}), as well as Figs.~\ref{fig:Hubbard_V1}(a) and \ref{fig:Hubbard_V1}(b) with Fig.~\ref{fig:CDW}. Note the appearing pairing phases fully gap both systems, and therefore we do not see an order differentiation between MLG and BLG, unlike the situation for the on-site repulsion. Below the transition temperature, we expect the pure superconducting state to display either $s+if$ or $f+is$ symmetry that maximally gaps the underlying Fermi surface~\cite{PhysRevB.90.041413}.

\subsection{Next-nearest neighbor repulsion ($V_2$)}

Finally, we examine the effect of the $H_{V_2}$ term in Eq.~(\ref{eq:ExtHub2_summary}), that describes NNN repulsion. Each site on the honeycomb lattice has six NNNs, and the repulsion in this case acts between sites belonging to the same sublattice. The six vectors separating NNNs are $\pm \vec{v}_i$, $i=1,2,3$, where $\vec{v}_1$ and $\vec{v}_2$ are the primitive lattice vectors defined in Sec.~\ref{sec:lattice_models:MLG}, and $\vec{v}_3=\vec{v}_2-\vec{v}_1$ [Fig.~\ref{fig:lattices}]. Once again we decompose the quartic terms and write
\begin{align}
H_{V_2}&= \frac{V_2}{4}\sum_{\vec{A},\vec{B}} \sum^3_{i=1} \sum_{\tau,\lambda=\pm} \Big[ n(\vec{A}) + \tau n(\vec{B}) \Big] \nonumber \\
&\times \Big[ n(\vec{A}+\lambda \vec{v}_i) + \tau n(\vec{B}+\lambda \vec{v}_i) \Big].
\end{align}
The first factor of the product contains the density and staggered density, the form of which we already derived in Eqs.~(\ref{eq:density}) and (\ref{eq:staggered_density}), respectively, while the two terms in the second factor can be written as
\widetext
\begin{align}
\sum^3_{i=1}\sum_{\lambda=\pm} \Big[ n(\vec{A}+\lambda \vec{v}_i) + n(\vec{B}+\lambda \vec{v}_i) \Big] = 
6 \Psi^\dag \Gamma_{3000} \Psi - 3 \cos(2 \vec{K}\cdot \vec{r}) \Psi^\dag \Gamma_{3010} \Psi - 3 \sin(2 \vec{K}\cdot \vec{r}) \Psi^\dag \Gamma_{3020} \Psi, \\
\sum^3_{i=1}\sum_{\lambda=\pm} \Big[ n(\vec{A}+\lambda \vec{v}_i) - n(\vec{B}+\lambda \vec{v}_i) \Big] = 
6 \Psi^\dag \Gamma_{3003} \Psi - 3 \cos(2 \vec{K}\cdot \vec{r}) \Psi^\dag \Gamma_{3013} \Psi - 3 \sin(2 \vec{K}\cdot \vec{r}) \Psi^\dag \Gamma_{3023} \Psi.
\end{align}
When performing the multiplication we again neglect any oscillatory terms, and write the continuum Hamiltonian as
\begin{align}
H_{V_2}=3 \frac{V_2 a^2}{4} \Big\{ 2 (\Psi^\dag \Gamma_{3000} \Psi)^2 + 2 (\Psi^\dag \Gamma_{3003} \Psi)^2 -
\frac{1}{2} \big[ (\Psi^\dag \Gamma_{3010} \Psi)^2 + (\Psi^\dag \Gamma_{3013} \Psi)^2 +(\Psi^\dag \Gamma_{3020} \Psi)^2 +(\Psi^\dag \Gamma_{3023} \Psi)^2 \big] \Big\}.
\end{align}
\twocolumngrid
\noindent Rescaling $3 V_2 a^2/4 \to V_2$, the initial conditions of NNN repulsion in the continuum formalism read
\begin{align}
g^s_1(\ell=0)=g^s_5(\ell=0)=-2 V_2,\hspace{0.3cm} g^s_9(\ell=0)=\frac{1}{2} V_2,
\end{align}
and $g^s_i(\ell=0)=0$ for $i=2,3,4,6,7,8$.

At zero temperature and chemical potential we find quantum spin Hall insulator ($A_{2g}^t$) phase in both MLG and BLG, see Figs.~\ref{fig:Hubbard_V2}(a) and \ref{fig:Hubbard_V2}(b) respectively. At finite chemical doping and sufficiently low temperatures, the adjacent pairing phase in both systems is the $s$-wave superconductor. These phases constitute the composite order parameter
\begin{align}
\vec{V}^{\rm V_2}&=\left\{
\overbrace{
\underbrace{\Gamma_{\alpha 000} }_{\mbox{$A_{1g}^p$}},
\underbrace{\Gamma_{3s33} }_{\mbox{$A_{2g}^t$}} 
}^{\mbox{O(5) vector}}
\right\}.\label{eq:SOP_HubbardV2}
\end{align}
Note that both nucleating orders gap the Dirac, as well as the Luttinger fermions. The dominant interaction channel in the presence of NNN repulsion is the QSHI, and indeed the phase diagrams are analogous to Fig.~\ref{fig:QSHI}, while the composite order parameter $\vec{V}^{\rm V_2}$ is the same as $\vec{V}^{\rm QSHI}_1$ in Eq.~(\ref{eq:SOP_QSHI}). Note, however, in contrast to the pure QSHI interaction channel, repulsive NNN interactions do not support singlet Kekul\' e pairing in BLG. This is likely due to the smectic charge density wave component in the corresponding initial conditions ($g^s_{_9}$).

%%%%%%%%%%%%%%%%%%%%%%%%%%%%%%%%%%%%%%%%%%%%%%%%%%%%%%%%%%%%%%%%%%%%%%%%%%%%%%%%%%%%%%%
%%%%%%%%%%%%%%%%%%%%%%%%%%%%%%%%%%%%%%%%%%%%%%%%%%%%%%%%%%%%%%%%%%%%%%%%%%%%%%%%%%%%%%%
%%%%%%%%%%%%%%%%%%%%%%%%%%%%%%%%%%%%%%%%%%%%%%%%%%%%%%%%%%%%%%%%%%%%%%%%%%%%%%%%%%%%%%%
%%%%%%%%%%%%%%%%%%%%%%%%%%%%%%%%%%%%%%%%%%%%%%%%%%%%%%%%%%%%%%%%%%%%%%%%%%%%%%%%%%%%%%%
%%%%%%%%%%%%%%%%%%%%%%%%%%%%%%%%%%%%%%%%%%%%%%%%%%%%%%%%%%%%%%%%%%%%%%%%%%%%%%%%%%%%%%%
\section{Summary and discussion}~\label{sec:summary}

Here we analyze a system of spinful fermions on monolayer and bilayer graphene, interacting via short range (or momentum-independent) Coulomb repulsion. We start by writing down the respective lattice models featuring tightly bound electrons and systematically deriving the low-energy continuum theories, that describe two copies of linear and biquadratic band touching, respectively. Retaining the Fourier components in the vicinity of the band touching points, we arrive at analogous descriptions for MLG and BLG, featuring the same number of degrees of freedom. Incorporating the contact interactions then involves constructing all quartic terms allowed by symmetry. As the point group describing the symmetry transformations of the lattice models is the same $D_{3d}$ group for MLG and BLG, the interacting Lagrangian in Eq.~(\ref{eq:L_int}) is \emph{identical} in the two systems. We address the effect of short range electronic interactions in a perturbative fashion, via Wilsonian momentum-shell RG analysis and the $\epsilon$ expansion scheme up to the one-loop order.

The central distinction in the continuum theories lies in the band structure, and the fact that the band touching points are linear (quadratic) in MLG (BLG). The purely linear (quadratic) dispersions result in the dynamic scaling exponent $z=1\ (2)$ for Dirac (Luttinger) fermions at the non-interacting fixed point, and hence $\epsilon=d-z=1\ (0)$ in the $\epsilon$ expansion. Since the coupling constants of contact interactions scale as
\begin{equation}
[g]=z-d=-\epsilon,
\end{equation}
such interactions are irrelevant (marginal) in MLG (BLG). However, this only sets the physical value of $\epsilon$ in the RG scheme and therefore the location of the phase boundaries. More importantly, the $k$-linear functions ($k_x$ and $k_y$) describing Dirac fermions in MLG are odd under inversion and transform under the $E_u$ representation of the $D_{3d}$ point group. In comparison, the $d$-wave harmonics ($k_x^2-k_y^2$ and $2k_xk_y$) are even under inversion and thus transform under the $E_g$ representation. To obtain an $A_{1g}$ quantity, they are multiplied by matrices from the same irreducible representation in the Hamiltonian. This algebraic difference is the origin of order differentiation between MLG and BLG, as in the case for the purely on-site Hubbard repulsion, which brings us to the main objective of this paper.

In Sec.~\ref{sec:extendedsum:selection} we outline a set of selection rules proposed in Ref.~\cite{PhysRevB.103.165139}, and argue that while the selection rules (Ia) and (Ib) in Eq.~(\ref{eq:rule_I}) allow for two distinct ways for an ordered phase to be promoted via a certain contact interaction term, the selection rule (II) in Eq.~(\ref{eq:rule_II}) is a generalized energy-entropy argument and organizes these phases along the temperature axis. The example of MLG and BLG is specifically suited to demonstrate these principles, since only (II) relies on the underlying band structure, while (I) is oblivious to it. Therefore, the ultimate ordered phases for a given interaction can in principle be different in MLG and BLG. We demonstrate the validity of the selection rules and organizing principle by increasing the strength of interactions in the individual channels and identifying the nature of symmetry breaking, see Tables~\ref{tab:ptransitions_sgl} and \ref{tab:ptransitions_tr}. By adding chemical doping and forming an extended Fermi surface, we also obtain superconductivity from repulsive electronic interactions. We show that the nucleating pairing phases also obey the selection rules, and that the adjacent excitonic and pairing phases consequently constitute composite order parameters, that form an \emph{enlarged} O(N) algebra.

Besides the individual interaction channels, we also consider a microscopic description of interacting electrons on the honeycomb lattice, the extended Hubbard model, containing the on-site, the NN and the NNN components of the Coulomb repulsion. See Figs.~\ref{fig:Hubbard_U}, \ref{fig:Hubbard_V1} and \ref{fig:Hubbard_V2} for the corresponding phase diagrams. Even though in this case we increase the strength of interactions in multiple channels at once, the dominantly diverging channel plays the role of $M$ in the selection rules (Ia) and (Ib). Then, our findings for the extended Hubbard model are in complete agreement with the selection rules. Namely, for repulsive Hubbard $U$ at half filling ($\mu=0$) we find antiferromagnetic orderings in both systems, while chemical doping ($\mu>0$) gives rise to $E_g$ nematic pairing in MLG and $A_{1k}$ singlet Kekul\' e superconductor in BLG. Either of these two pairings forms O(5) composite order parameter with the antiferromagnet [see Eq.~(\ref{eq:SOP_HubbardU})]. In the presence of the NN repulsion ($V_1>0$), we find charge density wave ordering when $\mu=0$ and simultaneous nucleation of both $s$-wave and $f$-wave pairing for $\mu>0$ in both systems. The $s$-wave and $f$-wave pairings respectively form O(3) and O(4) composite order parameters with charge density wave [see Eq.~(\ref{eq:SOP_HubbardV})]. On the other hand, with the NNN repulsion ($V_2>0$), we observe a quantum spin Hall insulator phase at $\mu=0$ and $s$-wave pairing for $\mu>0$ in both systems, where these two phases maximally anticommute and form an O(5) vector [see Eq.~(\ref{eq:SOP_HubbardV2})].

This observation is in agreement with Ref.~\cite{PhysRevB.98.045142}, where the authors carried out a non-perturbative functional RG analysis for the NN repulsive interaction. Note that only the $f$-wave pairing was found in this study, which, however, considered spinless fermions in MLG that does not permit $s$-wave pairing, due to the requisite antisymmetry property of the electronic wave function. On the other hand, a quantum Monte Carlo study in Ref.~\cite{wang2020dopinginduced} found quantum spin Hall insulator and $s$-wave superconductor phases, respectively at zero and finite doping. The dominant four-fermion interaction in this case is expected to be in the QSHI channel after integrating out the bosonic degrees of freedom. See Fig.~\ref{fig:QSHI}.

In light of these qualitative agreements with non-perturbative numerical results, and considering their purely algebraic nature (based on (anti)commutation among matrices), we believe that the validity of our proposed selection rules goes beyond one-loop RG calculations. In Appendix~\ref{sec:App_diagrams} we show how they manifest in a one-loop RG calculation at zero and finite temperature and zero chemical doping. Possible future investigations could target systems with existing experimental and/or numerical results, e.g. twisted bilayer graphene~\cite{2018Natur.556...43C, 2019Natur.574..653L, Codecidoeaaw9770, Yankowitz1059}, Weyl semimetals~\cite{PhysRevB.90.035126, PhysRevB.95.201102, PhysRevB.103.125132}, nodal-loop semimetals~\cite{Sur_2016, PhysRevB.96.041113}, to name a few.

%%%%%%%%%%%%%%%%%%%%%%%%%%%%%%%%%%%%%%%%%%%%%%%%%%%%%%%%%%%%%%%%%%%%%%%%%
%%%%%%%%%%%%%%%%%%%%%%%%%%%%%%%%%%%%%%%%%%%%%%%%%%%%%%%%%%%%%%%%%%%%%%%%%
%%%%%%%%%%%%%%%%%%%%%%% FEYMNANN - SUSCEPTIBILITY %%%%%%%%%%%%%%%%%%%%%%%
%%%%%%%%%%%%%%%%%%%%%%%%%%%%%%%%%%%%%%%%%%%%%%%%%%%%%%%%%%%%%%%%%%%%%%%%%
%%%%%%%%%%%%%%%%%%%%%%%%%%%%%%%%%%%%%%%%%%%%%%%%%%%%%%%%%%%%%%%%%%%%%%%%%
\begin{figure}[t!]
\includegraphics[width=0.95\linewidth]{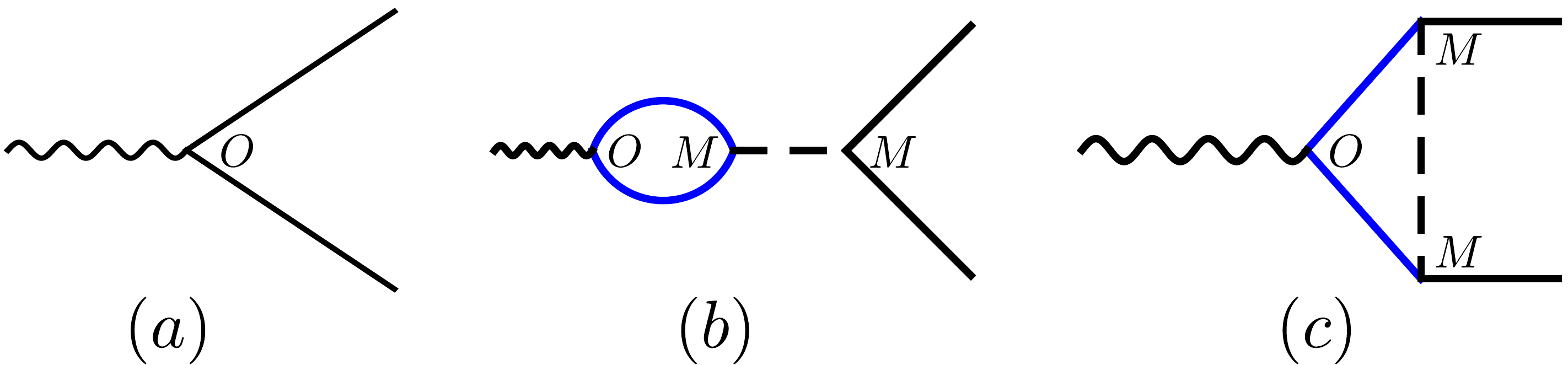}
\caption{(a) The bare vertex associated with the source term $\Psi^\dagger O \Psi$. The leading order renormalization of such vertices arises from Feynman diagrams (b) and (c), yielding the RG flow of the source terms or conjugate fields. Here, wavy lines stand for the conjugate field, while solid lines for fermions, and the dashed lines for the interaction vertex. The black (blue) solid lines represent slow (fast) modes. Respectively the diagrams (b) and (c) yield $B(t)$ and $V(t)$ in Appendix~\ref{sec:App_diagrams}.}
~\label{fig:Feynmann_susc}
\end{figure}
%%%%%%%%%%%%%%%%%%%%%%%%%%%%%%%%%%%%%%%%%%%%%%%%%%%%%%%%%%%%%%%%%%%%%%%%%
%%%%%%%%%%%%%%%%%%%%%%%%%%%%%%%%%%%%%%%%%%%%%%%%%%%%%%%%%%%%%%%%%%%%%%%%%
%%%%%%%%%%%%%%%%%%%%%%%%%%%%%%%%%%%%%%%%%%%%%%%%%%%%%%%%%%%%%%%%%%%%%%%%%
%%%%%%%%%%%%%%%%%%%%%%%%%%%%%%%%%%%%%%%%%%%%%%%%%%%%%%%%%%%%%%%%%%%%%%%%%
%%%%%%%%%%%%%%%%%%%%%%%%%%%%%%%%%%%%%%%%%%%%%%%%%%%%%%%%%%%%%%%%%%%%%%%%%

We close the discussion by qualitatively addressing the role of the trigonal warping [stemming from $t_{BA}$ in Eq.~(\ref{eq:BLG_tb})] on the phase diagrams of interacting BLG. Note that trigonal warping splits the biquadratic band touching near each valley into four Dirac points, yielding linearly vanishing DOS, namely $\rho(E) \sim |E|$ at the lowest energy~\cite{PhysRevB.82.201408}. Consequently, all orderings in BLG strictly take place at finite coupling, which is, however, much smaller than those in MLG, as $t_{BA} \ll t$. Recently, it has been shown that even when one begins with purely biquadratic band touchings in BLG, quantum fluctuations via self-energy corrections can split them into Dirac points (analogous to the trigonal warping), thereby deferring onset of any ordered phase to finite coupling~\cite{PhysRevLett.117.086404, PhysRevB.98.245128}. Even then our conclusions remain qualitatively valid for BLG. For example, at zero doping when the interaction strength is stronger (weaker) than the scale of the trigonal warping, the ordered phases are determined by the selection rules for Luttinger (Dirac) fermions. At finite doping (setting an infrared cutoff for the RG flow), when the chemical doping is above (below) the scale of trigonal warping, the emergent superconducting phases at the lowest temperature are qualitatively similar to the ones we show for Luttinger (Dirac) fermions. As such for repulsive on-site Hubbard model, a doped BLG is expected to support singlet $E_g$ nematic (Kekul\'e) pairing for small (large) doping at the lowest temperature~\cite{PhysRevLett.112.147002}.       

This anticipation can be germane (at least qualitatively) in twisted BLG near the magic angle, where a cascade of ordered states (including both excitonic and superconducting) have been observed by tuning the filling of the valley and spin degenerate nearly flat bands~\cite{2018Natur.556...43C, 2019Natur.574..653L, Codecidoeaaw9770, Yankowitz1059}. In particular, as the system approaches the magic angle, the Fermi velocity of slow Dirac fermions in twisted bilayer graphene gradually vanishes~\cite{PhysRevLett.99.256802, Bistritzer12233, PhysRevX.8.031089, PhysRevX.8.031087, PhysRevB.99.035111, bernevig2020tbg}, that in turn enhances the importance of the higher gradient terms, similar to the ones we considered here for Bernal BLG~\cite{PhysRevB.99.035111, bernevig2020tbg}, which may therefore play crucial roles in determining the ultimate nature of the ordered phases. In the future, our formalism can be extended to shed light on this enigmatic correlated system.

%%%%%%%%%%%%%%%%%%%%%%%%%%%%%%%%%%%%%%%%%%%%%%%%%%%%%%%%%%%%%%%%%%%%%%%%%%%%%%%%%%%%%%%
%%%%%%%%%%%%%%%%%%%%%%%%%%%%%%%%%%%%%%%%%%%%%%%%%%%%%%%%%%%%%%%%%%%%%%%%%%%%%%%%%%%%%%%
%%%%%%%%%%%%%%%%%%%%%%%%%%%%%%%%%%%%%%%%%%%%%%%%%%%%%%%%%%%%%%%%%%%%%%%%%%%%%%%%%%%%%%%

\acknowledgments

B.R. was supported by a start-up grant from Lehigh University and acknowledges hospitality of Max Planck Institute for the Physics of Complex Systems (MPIPKS), Dresden, Germany.

%%%%%%%%%%%%%%%%%%%%%%%%%%%%%%%%%%%%%%%%%%%%%%%%%%%%%%%%%%%%%%%%%%%%%%%%%%%
%%%%%%%%%%%%%%%%%%%%%%%%%%%%%%%%%%%%%%%%%%%%%%%%%%%%%%%%%%%%%%%%%%%%%%%%%%%
%%%%%%%%%%%%%%%%%%%%%%%%%%% Energy entropy figure %%%%%%%%%%%%%%%%%%%%%%%%%
%%%%%%%%%%%%%%%%%%%%%%%%%%%%%%%%%%%%%%%%%%%%%%%%%%%%%%%%%%%%%%%%%%%%%%%%%%%
%%%%%%%%%%%%%%%%%%%%%%%%%%%%%%%%%%%%%%%%%%%%%%%%%%%%%%%%%%%%%%%%%%%%%%%%%%%
\begin{figure}[t!]
\includegraphics[width=0.95\linewidth]{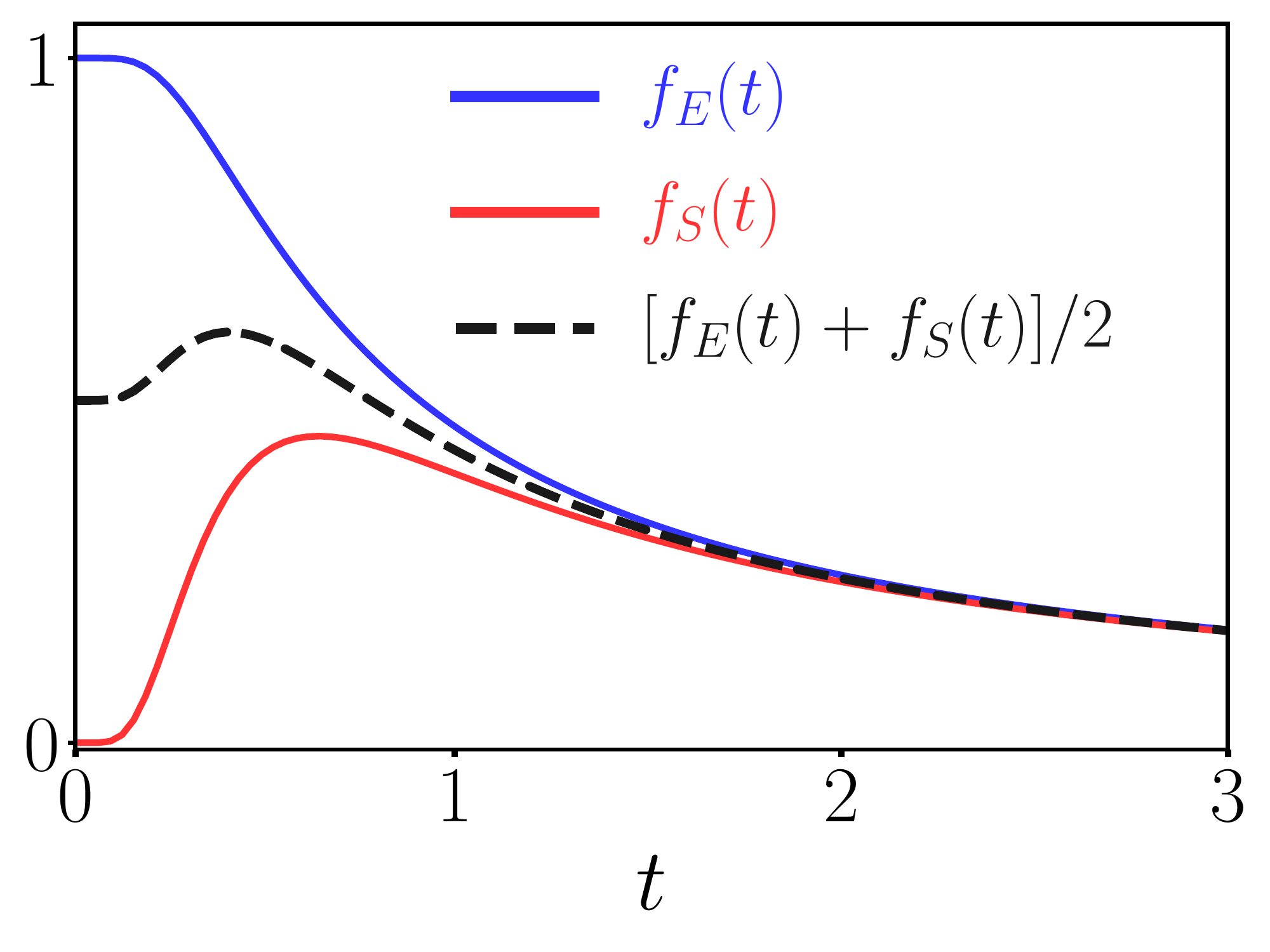}
\caption{Functions of dimensionless temperature $t$ from Eq.~(\ref{eq:f_t}) resulting from the loop integrals at zero chemical doping. At low temperatures $f_E(t)$ dominates, while $f_S(t)$ is small. Consequently, order parameters that fully commute with the Hamiltonian only get renormalized through their engineering dimensions at zero temperature. Partially anticommuting order parameters acquire a one-loop correction proportional to the arithmetic mean $[f_E(t)+f_S(t)]/2$.}
~\label{fig:loop_functions}
\end{figure}
%%%%%%%%%%%%%%%%%%%%%%%%%%%%%%%%%%%%%%%%%%%%%%%%%%%%%%%%%%%%%%%%%%%%%%%%%%%
%%%%%%%%%%%%%%%%%%%%%%%%%%%%%%%%%%%%%%%%%%%%%%%%%%%%%%%%%%%%%%%%%%%%%%%%%%%
%%%%%%%%%%%%%%%%%%%%%%%%%%%%%%%%%%%%%%%%%%%%%%%%%%%%%%%%%%%%%%%%%%%%%%%%%%%
%%%%%%%%%%%%%%%%%%%%%%%%%%%%%%%%%%%%%%%%%%%%%%%%%%%%%%%%%%%%%%%%%%%%%%%%%%%
%%%%%%%%%%%%%%%%%%%%%%%%%%%%%%%%%%%%%%%%%%%%%%%%%%%%%%%%%%%%%%%%%%%%%%%%%%%

%%%%%%%%%%%%%%%%%%%%%%%%%%%%%%%%%%%%%%%%%%%%%%%%%%%%%%%%%%%%%%%%%%%%%%%%%%%%%%%%%%%%%%%
%%%%%%%%%%%%%%%%%%%%%%%%%%%%%%%%%%%%%%%%%%%%%%%%%%%%%%%%%%%%%%%%%%%%%%%%%%%%%%%%%%%%%%%
%%%%%%%%%%%%%%%%%%%%%%%%%%%%%%%%%%%%%%%%%%%%%%%%%%%%%%%%%%%%%%%%%%%%%%%%%%%%%%%%%%%%%%%
%%%%%%%%%%%%%%%%%%%%%%%%%%%%%%%%%%%%%%%%%%%%%%%%%%%%%%%%%%%%%%%%%%%%%%%%%%%%%%%%%%%%%%%
%%%%%%%%%%%%%%%%%%%%%%%%%%%%%%%%%%%%%%%%%%%%%%%%%%%%%%%%%%%%%%%%%%%%%%%%%%%%%%%%%%%%%%%

\appendix

%%%%%%%%%%%%%%%%%%%%%%%%%%%%%%%%%%%%%%%%%%%%%%%%%%%%%%%%%%%%%%%%%%%%%%%%%%%%%%%%%%%%%%%
%%%%%%%%%%%%%%%%%%%%%%%%%%%%%%%%%%%%%%%%%%%%%%%%%%%%%%%%%%%%%%%%%%%%%%%%%%%%%%%%%%%%%%%
%%%%%%%%%%%%%%%%%%%%%%%%%%%%%%%%%%%%%%%%%%%%%%%%%%%%%%%%%%%%%%%%%%%%%%%%%%%%%%%%%%%%%%%
%%%%%%%%%%%%%%%%%%%%%%%%%%%%%%%%%%%%%%%%%%%%%%%%%%%%%%%%%%%%%%%%%%%%%%%%%%%%%%%%%%%%%%%
%%%%%%%%%%%%%%%%%%%%%%%%%%%%%%%%%%%%%%%%%%%%%%%%%%%%%%%%%%%%%%%%%%%%%%%%%%%%%%%%%%%%%%%

\section{Diagramatic contributions to order parameter fields}~\label{sec:App_diagrams}

In this appendix, we review the possible one-loop contributions to the anomalous dimension of various order parameter or conjugate fields that couple to some fermion bilinear as $\Delta_O(\Psi^\dag O \Psi)$, in an interacting theory containing local four fermion interactions of the schematic form $g_{_M} (\Psi^\dag M \Psi)^2$. We here consider one-component order parameter and quartic interaction terms. In general, we can apply the following to multicomponent interactions and order parameters by adding the contributions from individual pairs. In order to demonstrate the organizing principle, in this appendix we incorporate finite temperature. However, we set chemical potential to zero. Note that the effect of chemical doping is an increased propensity toward superconducting orders.

In the above outlined scenario there are two sources of one-loop correction to an order parameter field, the bubble and vertex diagrams [see Fig.~\ref{fig:Feynmann_susc}], denoted by $B(t)$ and $V(t)$, respectively, where $t$ is the dimensionless temperature [see Eq.~(\ref{eq:dimless_pars})]. The contribution from the bubble diagram reads
\begin{align}
B(t)&=-2 T \Tr \sum_{n=-\infty}^\infty \int \frac{\D^d \vec{k}}{(2 \pi)^d} \left[ MG(i\omega_n,\vec{k}) OG(i\omega_n,\vec{k}) \right] \nonumber \\
&= 2T\Tr \sum_{n=-\infty}^\infty \int \frac{\D^d \vec{k}}{(2\pi)^d} \frac{\omega^2_n + g(\vec{k})}{(\omega^2_n+\epsilon_{\vec{k}}^2)^2}MO,
\end{align}
where
\begin{align}
g(\vec{k})=\begin{cases}
   +\epsilon_{\vec{k}}^2,& \text{if }\{O,H\}=0, \\
   -\epsilon_{\vec{k}}^2,& \text{if }\left[ O,H \right]=0,\\
   0,& \text{if } A_H=C_H,
\end{cases}\label{eq:OH_cases}
\end{align}
where $A_H$ ($C_H$) is the number of anticommuting (commuting) matrix pairs between the order parameter and the Hamiltonian ($H$), $\omega_n=(2n+1)\pi T$ are the fermionic Matsubara frequencies, $G(i\omega_n,\vec{k})=\left[ i \omega_n - H\right]^{-1}$ is the fermionic Greens function and $\epsilon_{\vec{k}}$ is the positive eigenvalue of $H$ (see Appendix~\ref{app:beta_func}). Due to the trace (Tr), $B(t)$ is nonzero only if $O=M$ [selection rule (Ia)].

%%%%%%%%%%%%%%%%%%%%%%%%%%%%%%%%%%%%%%%%%%%%%%%%%%%%%%%%%%%%%%%%%%%%%%%%%%%%%%%%
%%%%%%%%%%%%%%%%%%%%%%%%%%%%%%%%%%%%%%%%%%%%%%%%%%%%%%%%%%%%%%%%%%%%%%%%%%%%%%%%
%%%%%%%%%%%%%%%%%% DIAGRAM CONTRIBUTIONS SUMMARY TABLE %%%%%%%%%%%%%%%%%%%%%%%%%
%%%%%%%%%%%%%%%%%%%%%%%%%%%%%%%%%%%%%%%%%%%%%%%%%%%%%%%%%%%%%%%%%%%%%%%%%%%%%%%%
%%%%%%%%%%%%%%%%%%%%%%%%%%%%%%%%%%%%%%%%%%%%%%%%%%%%%%%%%%%%%%%%%%%%%%%%%%%%%%%%
\begin{table}[t!]
\renewcommand{\arraystretch}{1.5}
\begin{tabular}{|>{\centering}m{0.7cm}|c|c|c|c|}
\hline
\multicolumn{2}{|c|}{} & $[O,H]=0$ & $\{ O,H \}=0$ & $A_H=C_H=1$ \\

\rowcolor{RowColor}
\multicolumn{2}{|c|}{$O=M$} & 0 & $\dim(M)-1$ & $\frac{1}{2}[\dim(M)-1]$ \\

\parbox[t]{2mm}{\multirow{2}{*}{\rotatebox[origin=c]{90}{$O \neq M$}}} & $\{ O, M \}=0$ & 0 & $1$ & $1/2$ \\
 & $[O,M]=0$ & 0 & $-1$ & $-1/2$ \\
\hline
\end{tabular}
\caption{Anomalous dimension to the one-loop order (see Fig.~\ref{fig:Feynmann_susc}) in units of $2\pi \Lambda^d/ \epsilon_{\Lambda}$ acquired by the order parameter $\Delta_O(\Psi^\dag O \Psi)$ in the presence of a four fermion interaction term $g_{_M} (\Psi^\dag M \Psi)^2$. Here $\dim(M)$ is the dimension of the Hermitian matrix $M$ and the normal state band structure is described by the single-particle Hamiltonian $H$.}\label{tab:diagram_contribution}
\end{table}
%%%%%%%%%%%%%%%%%%%%%%%%%%%%%%%%%%%%%%%%%%%%%%%%%%%%%%%%%%%%%%%%%%%%%%%%%%%%%%%%
%%%%%%%%%%%%%%%%%%%%%%%%%%%%%%%%%%%%%%%%%%%%%%%%%%%%%%%%%%%%%%%%%%%%%%%%%%%%%%%%
%%%%%%%%%%%%%%%%%%%%%%%%%%%%%%%%%%%%%%%%%%%%%%%%%%%%%%%%%%%%%%%%%%%%%%%%%%%%%%%%
%%%%%%%%%%%%%%%%%%%%%%%%%%%%%%%%%%%%%%%%%%%%%%%%%%%%%%%%%%%%%%%%%%%%%%%%%%%%%%%%
%%%%%%%%%%%%%%%%%%%%%%%%%%%%%%%%%%%%%%%%%%%%%%%%%%%%%%%%%%%%%%%%%%%%%%%%%%%%%%%%

Performing the Matsubara sum and integrating over the momentum shell $\Lambda e^{-\ell} < |\vec{k}| < \Lambda$, we obtain 
\begin{align}
B(t) = \dim(M)  \frac{\Lambda^d}{2 \pi \epsilon_\Lambda}
\begin{cases}
   f_E(t), \text{ if }\{O,H\}=0,& \\
   f_S(t), \text{ if }\left[ O,H \right]=0,&   \\
   [f_E(t)+f_S(t)]/2, \text{ if } A_H=C_H,&
\end{cases} \label{eq:B_contribution}
\end{align}
where $\dim(M)$ is the dimension of $M$, and
\begin{align}
f_E(t)=\tanh\left( \frac{1}{2t} \right), \hspace{0.7cm} f_S(t)= \frac{1}{2t}{\rm sech}^2\left(\frac{1}{2t}\right).\label{eq:f_t}
\end{align}
The functions $f_E(t)$ and $f_S(t)$, as well as their arithmetic average are shown in Fig.~\ref{fig:loop_functions}. Note that $f_E(0)=1$ and $f_S(0)=0$, and thus the energetically favorable fully gapped phases dominate at low temperatures. On the other hand, at higher temperatures $f_E(t)$ decreases and $f_S(t)$ increases, hence giving way to phases with higher entropy gain, a manifestation of the organizing principle (II). In this context recall that while $\{ O, H \}=0$ leads to fully and isotropically gapped spectra in the ordered phase, $[O, H]=0$ yields gapless quasiparticles therein. A partially anticommuting ($A_H=C_H$) order parameter acquires a contribution that is the average of the former two cases.

The vertex correction yields the following contribution
\begin{align}
V(t)&=2 T \sum_{n=-\infty}^\infty \int \frac{\D^d \vec{k}}{(2\pi)^d} \left[ MG(i\omega_n,\vec{k}) OG(i\omega_n,\vec{k})M \right] \nonumber \\
 &= -2 \sum_{n=-\infty}^\infty \int \frac{\D^d \vec{k}}{(2\pi)^d} \frac{\omega_n^2 +g(\vec{k})}{(\omega_n^2+\epsilon_{\vec{k}}^2)^2} MOM,
\end{align}
where $g(\vec{k})$ is defined in Eq.~(\ref{eq:OH_cases}). After performing the Matsubara sum and the shell integral this contribution reads
\begin{align}
V(t)=\pm \frac{\Lambda^d}{2 \pi \epsilon_\Lambda}
\begin{cases}
   f_E(t), \text{ if }\{O,H\}=0,& \\
   f_S(t), \text{ if }\left[ O,H \right]=0,&   \\
   [f_E(t)+f_S(t)]/2, \text{ if } A_H=C_H,&
\end{cases} \label{eq:V_contribution}
\end{align}
where the positive (negative) sign follows if $O$ and $M$ mutually anticommute (commute).

The above results at \emph{zero temperature} can be summarized as follows.
\vspace{0.2cm}

\noindent (a) An order parameter $O$ that fully commutes with the Hamiltonian acquires identically zero anomalous dimension, irrespective of the interaction channel.
\vspace{0.2cm}

\noindent (b) An order parameter that fully anticommutes with the Hamiltonian acquires \emph{positive} anomalous dimension if it fulfills $O=M$ or $\{O,M\}=0$ (selection rules (Ia) and (Ib), respectively), and \emph{negative} anomalous dimension if it fulfills $[O,M]=0$ but $O \neq M$.
\vspace{0.2cm}

\noindent (c) An order parameter that partially anticommutes with the Hamiltonian (here $A_H=C_H=1$) acquires anomalous dimension that is the \emph{average} of cases (a) and (b).

\vspace{0.2cm}
\noindent These findings are also summarized in Table~\ref{tab:diagram_contribution}. Note since $\epsilon_{\bf k}\sim |\vec{k}|^z$ the quantity $\Lambda^d/\epsilon_\Lambda\sim \Lambda^{d-z}=\Lambda^\epsilon$. Indeed at the lower critical dimension where $\epsilon=0$ the cutoff dependence drops out from $B(t)$ and $V(t)$.

\widetext

\section{Fierz reduction of local interaction terms}~\label{app:Fierz}

In this appendix we present the Fierz reduction of 18 momentum-independent quartic terms, that describe all the local quartic interactions on monolayer and bilayer graphene. We report the existence of 9 linearly independent four fermion terms, which we choose, without the loss of generality, to be the spin singlet interactions. Since we write all quartic terms in the particle-hole basis, we here refrain from using the Nambu basis, and $\Psi$ and $\Psi^\dag$ are eight-component spinors structured according to Eq.~(\ref{eq:spinor}).

We start by organizing the interaction terms into a vector as $\vec{X}=\left[ \vec{X}^s, \vec{X}^t \right]^\top$, where
\allowdisplaybreaks[4]
\begin{align}
&\vec{X}^s= \bigg\{
(\Psi^\dag \Gamma_{000} \Psi)^2, 
(\Psi^\dag \Gamma_{033} \Psi)^2,
(\Psi^\dag \Gamma_{001} \Psi)^2 + (\Psi^\dag \Gamma_{032} \Psi)^2,
(\Psi^\dag \Gamma_{030} \Psi)^2,
(\Psi^\dag \Gamma_{003} \Psi)^2,
(\Psi^\dag \Gamma_{031} \Psi)^2 + (\Psi^\dag \Gamma_{002} \Psi)^2, \nonumber\\
&\hspace{1.2cm} \sum_{j=1}^2 (\Psi^\dag \Gamma_{0j1} \Psi)^2,
 \sum_{j=1}^2 (\Psi^\dag \Gamma_{0j2} \Psi)^2,
 \sum_{j=1}^2\Big[ (\Psi^\dag \Gamma_{0j0} \Psi)^2 + (\Psi^\dag \Gamma_{0j3} \Psi)^2 \Big]\bigg\}^\top,\\
&\vec{X}^t= \bigg\{\sum_{s=1}^3(\Psi^\dag \Gamma_{s00} \Psi)^2, 
\sum_{s=1}^3(\Psi^\dag \Gamma_{s33} \Psi)^2,
\sum_{s=1}^3\Big[(\Psi^\dag \Gamma_{s01} \Psi)^2 + (\Psi^\dag \Gamma_{0s32} \Psi)^2\Big],
\sum_{s=1}^3(\Psi^\dag \Gamma_{s30} \Psi)^2,
\sum_{s=1}^3(\Psi^\dag \Gamma_{s03} \Psi)^2, \nonumber \\
&\sum_{s=1}^3(\Psi^\dag \Gamma_{s31} \Psi)^2 + (\Psi^\dag \Gamma_{s02} \Psi)^2,
\sum_{s=1}^3\sum_{j=1}^2(\Psi^\dag \Gamma_{sj1} \Psi)^2,
\sum_{s=1}^3\sum_{j=1}^2(\Psi^\dag \Gamma_{sj2} \Psi)^2,
\sum_{s=1}^3\sum_{j=1}^2\Big[(\Psi^\dag \Gamma_{sj0} \Psi)^2 + (\Psi^\dag \Gamma_{sj3} \Psi)^2 \Big] \bigg\}^\top,
\end{align}
and $\vec{X}^s$ and $\vec{X}^t$ contain the spin singlet and triplet quartic terms, respectively.

The Fierz identity for quartic terms of eight-component Grassmann variables reads~\cite{PhysRevB.79.085116, PhysRevB.82.205106}
\begin{align}
[\Psi^\dag (x) M \Psi(x)][\Psi^\dag (y) N \Psi(y)]=-\frac{1}{64}\Tr(M\Gamma^a N \Gamma^b) [\Psi^\dag (x) \Gamma^b \Psi(y)][\Psi^\dag (y) \Gamma^a \Psi(x)],
\end{align}
where $M$ and $N$ are eight-dimensional Hermitian matrices, $\Gamma^a$ span a basis in the space of such matrices, and the summation over repeated indices is assumed. Here we apply the above identity for contact interactions, for which $x=y$. Using this identity we write each element in $\vec{X}$ as a linear combination of the remaining ones. We can summarize these linear connections with the matrix equation $F\vec{X}=0$, with the eighteen-dimensional Fierz matrix
\begin{align}
F=
\left[
\begin{array}{*{20}c}
 -9 & -1 & -1 & -1 & -1 & -1 & -1 & -1 & -1 & -1 & -1 & -1 & -1 & -1 & -1 & -1 & -1 & -1 \\
 -1 & -9 & 1  & -1 & -1 & 1  & -1 & -1 & 1  & -1 & -1 & 1  & -1 & -1 & 1  & -1 & -1 & 1  \\
 -2 & 2  & -8 & -2 & 2  & 0  & -2 & 2  & 0  & -2 & 2  & 0  & -2 & 2  & 0  & -2 & 2  & 0  \\
 -1 & -1 & -1 & -9 & -1 & -1 & 1  & 1  & 1  & -1 & -1 & -1 & -1 & -1 & -1 & 1  & 1  & 1  \\
 -1 & -1 & 1  & -1 & -9 & 1  & 1  & 1  & -1 & -1 & -1 & 1  & -1 & -1 & 1  & 1  & 1  & -1 \\
 -2 & 2  & 0  & -2 & 2  & -8 & 2  & -2 & 0  & -2 & 2  & 0  & -2 & 2  & 0  & 2  & -2 & 0  \\
 -2 & -2 & -2 & 2  & 2  & 2  & -8 & 0  & 0  & -2 & -2 & -2 & 2  & 2  &  2 & 0  & 0  & 0  \\
 -2 & -2 & 2  & 2  & 2  & -2 & 0  & -8 & 0  & -2 & -2 & 2  & 2  & 2  & -2 & 0  & 0  & 0  \\
 -4 & 4  & 0  & 4  & -4 & 0  & 0  & 0  & -8 & -4 & 4  & 0  & 4  & -4 & 0  & 0  & 0  & 0  \\
 -3 & -3 & -3 & -3 & -3 & -3 & -3 & -3 & -3 & -7 & 1  & 1  & 1  & 1  & 1  & 1  & 1  & 1  \\
 -3 & -3 & 3  & -3 & -3 & 3  & -3 & -3 & 3  & 1  & -7 & -1 & 1  & 1  & -1 & 1  & 1  & -1 \\
 -6 & 6  & 0  & -6 & 6  & 0  & -6 & 6  & 0  & 2  & -2 & -8 & 2  & -2 & 0  & 2  & -2 & 0  \\
 -3 & -3 & -3 & -3 & -3 & -3 & 3  & 3  & 3  & 1  & 1  & 1  & -7 & 1  & 1  & -1 & -1 & -1 \\
 -3 & -3 & 3  & -3 & -3 & 3  & 3  & 3  & -3 & 1  & 1  & -1 & 1  & -7 & -1 & -1 & -1 & 1  \\
 -6 & 6  & 0  & -6 & 6  & 0  & 6  & -6 & 0  & 2  & -2 & 0  & 2  & -2 & -8 & -2 & 2  & 0  \\
 -6 & -6 & -6 & 6  & 6  & 6  & 0  & 0  & 0  & 2  & 2  & 2  & -2 & -2 & -2 & -8 & 0  & 0  \\
 -6 & -6 & 6  & 6  & 6  & -6 & 0  & 0  & 0  & 2  & 2  & -2 & -2 & -2 & 2  & 0  & -8 & 0  \\
 -12& 12 & 0  & 12 &-12 & 0  & 0  & 0  & 0  & 4  & -4 & 0  & -4 & 4  & 0  & 0  & 0  & -8 \\
\end{array}
\right].
\end{align}

The number of independent quartic terms is given by 18 (dimension of $F$) - 9 (rank of $F$)=9. We can extract the relevant equations by reordering the columns of $F$ and performing row reduction. Choosing the spin singlet interactions as the independent ones we obtain nine equations which we can write compactly in matrix form as
\begin{align}
\vec{X}^t=\frac{1}{2}
\left(
\begin{array}{ccccccccc}
 3 & 1 & 1 & 1 & 1 & 1 & 1 & 1 & 1 \\
 1 & 3 & -1 & 1 & 1 & -1 & 1 & 1 & -1 \\
 2 & -2 & 2 & 2 & -2 & 0 & 2 & -2 & 0 \\
 1 & 1 & 1 & 3 & 1 & 1 & -1 & -1 & -1 \\
 1 & 1 & -1 & 1 & 3 & -1 & -1 & -1 & 1 \\
 2 & -2 & 0 & 2 & -2 & 2 & -2 & 2 & 0 \\
 2 & 2 & 2 & -2 & -2 & -2 & 2 & 0 & 0 \\
 2 & 2 & -2 & -2 & -2 & 2 & 0 & 2 & 0 \\
 4 & -4 & 0 & -4 & 4 & 0 & 0 & 0 & 2
\end{array}
\right)
\vec{X}^s.\label{eq:Fierz_contraint}
\end{align}
However, when 9 linearly independent couplings are chosen in the spin singlet channels, we do not generate any quartic term in the spin triplet channel via coarse grain. Nevertheless, the above linear relations allow us to set the initial condition for any quartic interaction in the spin triplet channel in the RG equations, even though they are expressed in terms of the quartic interactions in the spin singlet channel, and that way construct various cuts of the phase diagrams shown in Figs.~\ref{fig:mass_PD}-\ref{fig:Kekule_PD}.

%%%%%%%%%%%%%%%%%%%%%%%%%%%%%%%%%%%%%%%%%%%%%%%%%%%%%%%%%%%%%%%%%%%%%%%%%
%%%%%%%%%%%%%%%%%%%%%%%%%%%%%%%%%%%%%%%%%%%%%%%%%%%%%%%%%%%%%%%%%%%%%%%%%
%%%%%%%%%%%%%%%%%%%%%%%%% FEYMNANN - INTERACTION %%%%%%%%%%%%%%%%%%%%%%%%
%%%%%%%%%%%%%%%%%%%%%%%%%%%%%%%%%%%%%%%%%%%%%%%%%%%%%%%%%%%%%%%%%%%%%%%%%
%%%%%%%%%%%%%%%%%%%%%%%%%%%%%%%%%%%%%%%%%%%%%%%%%%%%%%%%%%%%%%%%%%%%%%%%%
\begin{figure}[t!]
\includegraphics[width=0.95\linewidth]{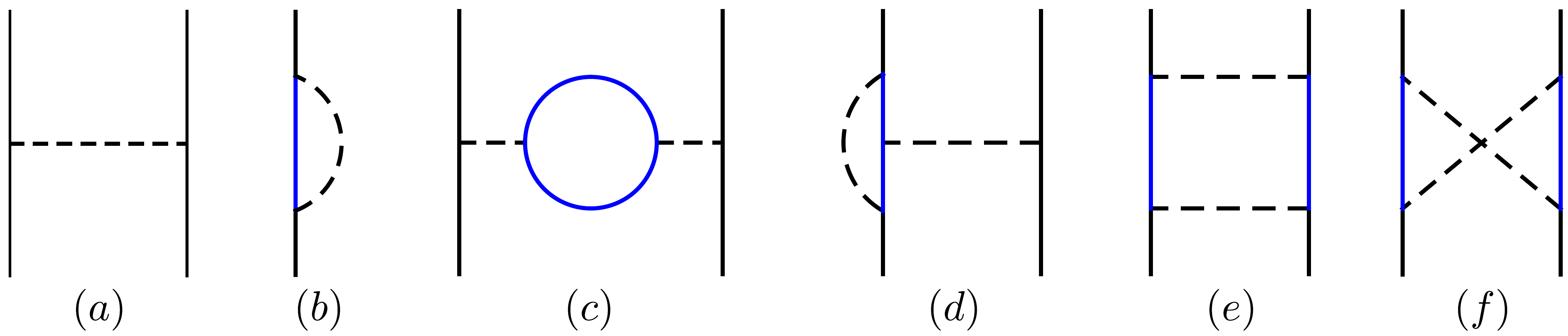}
\caption{(a) Bare four-fermion interaction vertex and (b) the self-energy correction due to four-fermion interaction. The contribution from the Feynman diagram (b) is finite only when the chemical potential ($\mu$) is finite, and it renormalizes $\mu$. Feynman diagrams (c)-(f) yield corrections to the bare interaction vertex to the leading order in the $\epsilon$ expansion, where $\epsilon=d-z$. Here, solid lines represent fermions. While the blue lines in (b)-(f) correspond to the fast modes, living within a thin Wilsonian momentum shell $\Lambda \; e^{-\ell} <|{\bf k}|<\Lambda$, where $\Lambda$ is the ultraviolet momentum cut-off and $\ell$ is the logarithm of the renormalization group scale, the black lines are the slow modes with $|{\bf k}|<\Lambda \; e^{-\ell}$.}
~\label{fig:Feynmann_int}
\end{figure}
%%%%%%%%%%%%%%%%%%%%%%%%%%%%%%%%%%%%%%%%%%%%%%%%%%%%%%%%%%%%%%%%%%%%%%%%%
%%%%%%%%%%%%%%%%%%%%%%%%%%%%%%%%%%%%%%%%%%%%%%%%%%%%%%%%%%%%%%%%%%%%%%%%%
%%%%%%%%%%%%%%%%%%%%%%%%%%%%%%%%%%%%%%%%%%%%%%%%%%%%%%%%%%%%%%%%%%%%%%%%%
%%%%%%%%%%%%%%%%%%%%%%%%%%%%%%%%%%%%%%%%%%%%%%%%%%%%%%%%%%%%%%%%%%%%%%%%%
%%%%%%%%%%%%%%%%%%%%%%%%%%%%%%%%%%%%%%%%%%%%%%%%%%%%%%%%%%%%%%%%%%%%%%%%%

%%%%%%%%%%%%%%%%%%%%%%%%%%%%%%%%%%%%%%%%%%%%%%%%%%%%%%%%%%%
%%%%%%%%%%%%%%%%%%%%%%%%%%%%%%%%%%%%%%%%%%%%%%%%%%%%%%%%%%%
%%%%%%%%%%%%%%%%%%%%%%%%%%%%%%%%%%%%%%%%%%%%%%%%%%%%%%%%%%%
%%%%%%%%%%%%%%%%%%%%%%%%%%%%%%%%%%%%%%%%%%%%%%%%%%%%%%%%%%%
%%%%%%%%%%%%%%%%%%%%%%%%%%%%%%%%%%%%%%%%%%%%%%%%%%%%%%%%%%%
\section{Renormalization group (RG) flow equations}~\label{app:beta_func}

Here we provide further details to the RG analysis at finite temperature and chemical doping. Finite temperature causes the previously continuous Matsubara frequency to take on discrete values, $\omega\to\omega_n=(2n+1)\pi T$, with $n\in \mathbb{Z}$ (integers) for fermions, where $T$ is the dimensionfull temperature. Using the compact notations $\Omega_\pm=i\omega\pm \mu$, the four distinct loop integrals read
\begin{align}
I_\omega &= T\sum_{\omega_n} \int \frac{\D \vec{k}}{(2 \pi)^d} \frac{\Omega_+^2}{(\Omega_+^2-\epsilon_{\vec{k}}^2)^2}=
-\frac{1}{4 \pi} \frac{\Lambda^d}{\epsilon_\Lambda} f_\omega(t,\mu) l,\:\:\:
\tilde{I}_\omega =T\sum_{\omega_n} \int \frac{\D \vec{k}}{(2 \pi)^d} \frac{\Omega_+ \Omega_-}{(\Omega_+^2-\epsilon_{\vec{k}}^2)(\Omega_-^2-\epsilon_{\vec{k}}^2)}=
-\frac{1}{4 \pi} \frac{\Lambda^d}{\epsilon_\Lambda} \tilde{f}_\omega (t,\mu)l,\nonumber \\
I_k &= T\sum_{\omega_n} \int \frac{\D \vec{k}}{(2 \pi)^d} \frac{v_i^2(\vec{k})}{(\Omega_+^2-\epsilon_{\vec{k}}^2)^2}=
\frac{1}{4 \pi} \frac{\Lambda^d}{\epsilon_\Lambda} f_k(t,\mu) l,\:\:\:
\tilde{I}_k =T\sum_{\omega_n} \int \frac{\D \vec{k}}{(2 \pi)^d} \frac{v_i^2(\vec{k})}{(\Omega_+^2-\epsilon_{\vec{k}}^2)(\Omega_-^2-\epsilon_{\vec{k}}^2)} =
\frac{1}{4 \pi} \frac{\Lambda^d}{\epsilon_\Lambda} \tilde{f}_k (t,\mu)l,
\end{align}
where $\epsilon_{\vec{k}}^2=v^2_1(\vec{k})+v^2_2(\vec{k})$, with $v_i(\vec{k})=p_i(\vec{k})$ and $d_i(\vec{k})$ for Dirac and Luttinger fermions, respectively. See the definitions of these cubic harmonics in Sec.~\ref{sec:lowenergy_noninteract} of the main text, and $\epsilon_{\Lambda}=\epsilon_{|\vec{k}|=\Lambda}$ is the given quasiparticle spectrum evaluated at the ultraviolet momentum cutoff $\Lambda$. The functions of dimensionless temperature and chemical potential [see Eq.~(\ref{eq:dimless_pars})] are of the form
\begin{align}~\label{eq:loop_functions}
f_\omega(t,\mu)&=\sum_{\tau=\pm}\Big[ \frac{\sech^2(\frac{1+\tau \mu}{2t})}{4 t} + \frac{\tanh(\frac{1+\tau \mu}{2t})}{2} \Big], \quad
\tilde{f}_\omega(t,\mu)=\sum_{\tau=\pm} \frac{\tau \tanh(\frac{1+\tau \mu}{2t})(1+\tau \mu -2 \mu^2)}{2 \mu (1-\mu^2)}, \nonumber \\
f_k(t,\mu)&=\frac{1}{2}\sum_{\tau=\pm} \Big[ -\frac{\sech^2(\frac{1+\tau \mu}{2t})}{4 t} + \frac{\tanh(\frac{1+\tau \mu}{2t})}{2} \Big], \quad
\tilde{f}_k(t,\mu)=\frac{1}{2} \sum_{\tau=\pm} \frac{-\tau \tanh(\frac{1+\tau \mu}{2t})(1-\tau \mu)}{2 \mu (1-\mu^2)}.
\end{align}
Note, in the limit of zero chemical doping (but at finite temperature) they become
\begin{align}
\frac{1}{2}\Big[ f_\omega(t,\mu\to 0)\pm 2 f_k(t,\mu\to 0)\Big]=
\begin{cases}
   +: f_E(t),  \\
   -: f_S(t).             
\end{cases}
\end{align}
On the other hand, in the limit of both zero temperature and zero chemical doping the functions take on the values
\begin{align}
f_\omega(t\to 0,\mu\to0)&=1, &
f_k(t\to 0,\mu\to 0)&=\frac{1}{2}, &
\tilde{f}_\omega(t\to 0, \mu \to 0) &= 1, &
\tilde{f}_k(t\to 0, \mu \to 0) &= \frac{1}{2}. 
\end{align}

In the rest of this appendix we display the flow equations of various coupling constants and order parameter conjugate fields. Let us recall the schematic form of them from Eqs.~(\ref{eq:beta_g}) and (\ref{eq:beta_D}) in the main text,
\begin{align}
\frac{\D g^s_{_i}}{\D \ell}&= -\epsilon g^s_{_i} + \sum_{j,k} g^s_{_j} g^s_{_k} H^i_{jk}(t,\mu), \\
\bar{\beta}^j_{\Delta_k}&\equiv\frac{\D \Delta_k}{\D \ell}-z=\sum_{l} F_{k,l}^j (t,\mu) g^s_{_l}.
\end{align}
To differentiate between the two systems, we use the notation $H^i_{jk}(t,\mu)=M^i_{jk}(t,\mu)$ and $B^i_{jk}(t,\mu)$ for MLG and BLG respectively. These functions are obtained by computing the Feynman diagrams (c)-(f) in Fig.~\ref{fig:Feynmann_int}. All the terms that are not explicitly shown are zero. In case of the conjugate fields we write out the full $\beta$ functions, $\bar{\beta}^D_{\Delta_k}$ ($\bar{\beta}^L_{\Delta_k}$) for MLG (BLG), which are obtained by computing the Feynman diagrams from Fig.~\ref{fig:Feynmann_susc}. For brevity, we suppress the $t$ and $\mu$ dependence of the functions from Eq.~(\ref{eq:loop_functions}).

\subsection{Monolayer graphene}

The functions appearing in the RG flow equation for $g^{s}_{_1}$ in MLG are
\allowdisplaybreaks
\begin{flalign}
&M^{1}_{11} = -24 f_{k} - 10 f_{\omega} - 2 \tilde{f}_{\omega},\  
M^{1}_{22} = 2 f_{\omega} - 2 \tilde{f}_{\omega},\  
M^{1}_{33} = 4 f_{\omega} - 4 \tilde{f}_{\omega},\  
M^{1}_{44} = 2 f_{\omega} - 2 \tilde{f}_{\omega},\  
M^{1}_{55} = 2 f_{\omega} - 2 \tilde{f}_{\omega},&\nonumber\\&
M^{1}_{66} = 4 f_{\omega} - 4 \tilde{f}_{\omega},\  
M^{1}_{77} = 4 f_{\omega} - 4 \tilde{f}_{\omega},\  
M^{1}_{88} = 4 f_{\omega} - 4 \tilde{f}_{\omega},\  
M^{1}_{99} = 8 f_{\omega} - 8 \tilde{f}_{\omega},\  
M^{1}_{12} = 8 f_{k} + 4 f_{\omega},&\nonumber\\&  
M^{1}_{13} = 16 f_{k} + 8 f_{\omega},\ 
M^{1}_{14} = 8 f_{k} + 4 f_{\omega},\  
M^{1}_{15} = 8 f_{k} + 4 f_{\omega},\  
M^{1}_{16} = 24 f_{k} + 8 f_{\omega} - 8 \tilde{f}_{k},\  
M^{1}_{17} = 16 f_{k} + 8 f_{\omega},&\nonumber\\&    
M^{1}_{18} = 16 f_{k} + 8 f_{\omega},\  
M^{1}_{19} = 32 f_{k} + 16 f_{\omega},\
M^{1}_{26} = 8 f_{k} + 8 \tilde{f}_{k},\  
M^{1}_{34} = 8 f_{k} - 8 \tilde{f}_{k},\  
M^{1}_{35} = 8 f_{k} + 8 \tilde{f}_{k},&\nonumber \\&  
M^{1}_{79} = 16 f_{k} + 16 \tilde{f}_{k},\  
M^{1}_{89} = 16 f_{k} - 16 \tilde{f}_{k}.&
\end{flalign}
The functions appearing in the RG flow equation for $g^{s}_{_2}$ in MLG are
\begin{flalign}
&M^{2}_{22} = 24 f_{k} - 12 f_{\omega},\  
M^{2}_{33} = 4 f_{\omega} + 4 \tilde{f}_{\omega},\  
M^{2}_{66} = 4 f_{\omega} + 4 \tilde{f}_{\omega},\  
M^{2}_{99} = 8 f_{\omega} + 8 \tilde{f}_{\omega},\  
M^{2}_{12} = -8 f_{k} + 8 f_{\omega} - 4 \tilde{f}_{\omega},&\nonumber\\&
M^{2}_{16} = 8 f_{k} + 8 \tilde{f}_{k},\
M^{2}_{23} = 16 f_{k} - 8 f_{\omega},\  
M^{2}_{24} = -8 f_{k} + 4 f_{\omega},\  
M^{2}_{25} = -8 f_{k} + 4 f_{\omega},\  
M^{2}_{26} = 24 f_{k} - 8 f_{\omega} - 8 \tilde{f}_{k},&\nonumber\\&
M^{2}_{27} = -16 f_{k} + 8 f_{\omega},\
M^{2}_{28} = -16 f_{k} + 8 f_{\omega},\  
M^{2}_{29} = 32 f_{k} - 16 f_{\omega},\  
M^{2}_{34} = 8 f_{k} + 8 \tilde{f}_{k},\  
M^{2}_{35} = 8 f_{k} - 8 \tilde{f}_{k},&\nonumber\\&
M^{2}_{45} = 4 f_{\omega} - 4 \tilde{f}_{\omega},\  
M^{2}_{78} = 8 f_{\omega} - 8 \tilde{f}_{\omega},\  
M^{2}_{79} = 16 f_{k} - 16 \tilde{f}_{k},\  
M^{2}_{89} = 16 f_{k} + 16 \tilde{f}_{k}.&
\end{flalign}
The functions appearing in the RG flow equation for $g^{s}_{_3}$ in MLG are
\begin{flalign}
&M^{3}_{33} = -16 f_{\omega},\  
M^{3}_{77} = 4 f_{k} + 4 \tilde{f}_{k},\  
M^{3}_{88} = 4 f_{k} + 4 \tilde{f}_{k},\  
M^{3}_{99} = 16 f_{k} + 16 \tilde{f}_{k},\  
M^{3}_{13} = 8 f_{\omega} - 4 \tilde{f}_{\omega},\
M^{3}_{14} = 4 f_{k} - 4 \tilde{f}_{k},&\nonumber\\&
M^{3}_{15} = 4 f_{k} + 4 \tilde{f}_{k},\  
M^{3}_{23} = 4 \tilde{f}_{\omega},\  
M^{3}_{24} = 4 f_{k} + 4 \tilde{f}_{k},\  
M^{3}_{25} = 4 f_{k} - 4 \tilde{f}_{k},\  
M^{3}_{34} = 4 f_{\omega},\  
M^{3}_{35} = -4 f_{\omega},&\nonumber\\&
M^{3}_{36} = 16 f_{k} - 16 \tilde{f}_{k},\ 
M^{3}_{37} = 8 f_{\omega},\  
M^{3}_{38} = -8 f_{\omega},\
M^{3}_{46} = 4 f_{\omega} - 4 \tilde{f}_{\omega},\  
M^{3}_{56} = 4 f_{\omega} + 4 \tilde{f}_{\omega},&\nonumber\\&
M^{3}_{78} = 8 f_{k} - 8 \tilde{f}_{k},\  
M^{3}_{79} = 8 f_{\omega} - 8 \tilde{f}_{\omega},\  
M^{3}_{89} = 8 f_{\omega} + 8 \tilde{f}_{\omega}.
\end{flalign}
The functions appearing in the RG flow equation for $g^{s}_{_4}$ in MLG are
\begin{flalign}
&M^{4}_{44} = -24 f_{k} - 12 f_{\omega},\  
M^{4}_{77} = 4 f_{\omega} + 4 \tilde{f}_{\omega},\  
M^{4}_{88} = 4 f_{\omega} + 4 \tilde{f}_{\omega},\  
M^{4}_{99} = 8 f_{\omega} + 8 \tilde{f}_{\omega},\  
M^{4}_{13} = 8 f_{k} - 8 \tilde{f}_{k},&\nonumber\\&
M^{4}_{14} = 8 f_{k} + 8 f_{\omega} - 4 \tilde{f}_{\omega},\ 
M^{4}_{23} = 8 f_{k} + 8 \tilde{f}_{k},\  
M^{4}_{24} = 8 f_{k} + 4 f_{\omega},\  
M^{4}_{25} = 4 f_{\omega} - 4 \tilde{f}_{\omega},\  
M^{4}_{34} = 16 f_{k} + 8 f_{\omega},&\nonumber\\&
M^{4}_{36} = 8 f_{\omega} - 8 \tilde{f}_{\omega},\  
M^{4}_{45} = 8 f_{k} + 4 f_{\omega},\  
M^{4}_{46} = 24 f_{k} + 8 f_{\omega} - 8 \tilde{f}_{k},\  
M^{4}_{47} = -16 f_{k} - 8 f_{\omega},\  
M^{4}_{48} = -16 f_{k} - 8 f_{\omega},&\nonumber\\& 
M^{4}_{49} = -32 f_{k} - 16 f_{\omega},\  
M^{4}_{56} = 8 f_{k} + 8 \tilde{f}_{k},\  
M^{4}_{79} = 16 f_{k} - 16 \tilde{f}_{k},\  
M^{4}_{89} = 16 f_{k} + 16 \tilde{f}_{k}.&
\end{flalign}
The functions appearing in the RG flow equation for $g^{s}_{_5}$ in MLG are
\begin{flalign}
&M^{5}_{55} = 24 f_{k} - 12 f_{\omega},\  
M^{5}_{99} = 8 f_{\omega} - 8 \tilde{f}_{\omega},\  
M^{5}_{13} = 8 f_{k} + 8 \tilde{f}_{k},\  
M^{5}_{15} = -8 f_{k} + 8 f_{\omega} - 4 \tilde{f}_{\omega},\  
M^{5}_{23} = 8 f_{k} - 8 \tilde{f}_{k},&\nonumber\\&
M^{5}_{24} = 4 f_{\omega} - 4 \tilde{f}_{\omega},\  
M^{5}_{25} = -8 f_{k} + 4 f_{\omega},\  
M^{5}_{35} = 16 f_{k} - 8 f_{\omega},\  
M^{5}_{36} = 8 f_{\omega} + 8 \tilde{f}_{\omega},\  
M^{5}_{45} = -8 f_{k} + 4 f_{\omega},&\nonumber\\&
M^{5}_{46} = 8 f_{k} + 8 \tilde{f}_{k},\   
M^{5}_{56} = 24 f_{k} - 8 f_{\omega} - 8 \tilde{f}_{k},\  
M^{5}_{57} = 16 f_{k} - 8 f_{\omega},\  
M^{5}_{58} = 16 f_{k} - 8 f_{\omega},\  
M^{5}_{59} = -32 f_{k} + 16 f_{\omega},&\nonumber\\&  
M^{5}_{78} = 8 f_{\omega} + 8 \tilde{f}_{\omega},\  
M^{5}_{79} = 16 f_{k} + 16 \tilde{f}_{k},\  
M^{5}_{89} = 16 f_{k} - 16 \tilde{f}_{k}.
\end{flalign}
The functions appearing in the RG flow equation for $g^{s}_{_6}$ in MLG are
\begin{flalign}
&M^{6}_{11} = 2 f_{k} - 2 \tilde{f}_{k},\  
M^{6}_{22} = 2 f_{k} - 2 \tilde{f}_{k},\  
M^{6}_{33} = 8 f_{k} - 8 \tilde{f}_{k},\  
M^{6}_{44} = 2 f_{k} - 2 \tilde{f}_{k},\  
M^{6}_{55} = 2 f_{k} - 2 \tilde{f}_{k},&\nonumber\\&
M^{6}_{66} = 8 f_{k} - 16 f_{\omega} - 8 \tilde{f}_{k},\  
M^{6}_{77} = 4 f_{k} - 4 \tilde{f}_{k},\  
M^{6}_{88} = 4 f_{k} - 4 \tilde{f}_{k},\  
M^{6}_{99} = 16 f_{k} - 16 \tilde{f}_{k},\  
M^{6}_{12} = 4 f_{k} + 4 \tilde{f}_{k},&\nonumber\\&
M^{6}_{16} = 8 f_{\omega} - 4 \tilde{f}_{\omega},\  
M^{6}_{26} = 4 \tilde{f}_{\omega},\  
M^{6}_{34} = 4 f_{\omega} - 4 \tilde{f}_{\omega},\  
M^{6}_{35} = 4 f_{\omega} + 4 \tilde{f}_{\omega},\  
M^{6}_{45} = 4 f_{k} + 4 \tilde{f}_{k},\ 
M^{6}_{46} = 4 f_{\omega},&\nonumber\\&
M^{6}_{56} = -4 f_{\omega},\  
M^{6}_{67} = -8 f_{\omega},\  
M^{6}_{68} = 8 f_{\omega},\  
M^{6}_{78} = 8 f_{k} + 8 \tilde{f}_{k},\
M^{6}_{79} = 8 f_{\omega} + 8 \tilde{f}_{\omega},\  
M^{6}_{89} = 8 f_{\omega} - 8 \tilde{f}_{\omega}.&
\end{flalign}
The functions appearing in the RG flow equation for $g^{s}_{_7}$ in MLG are
\begin{flalign}
&M^{7}_{77} = 32 f_{k} - 16 f_{\omega},\ 
M^{7}_{17} = -8 f_{k} + 8 f_{\omega} - 4 \tilde{f}_{\omega},\  
M^{7}_{19} = 8 f_{k} + 8 \tilde{f}_{k},\  
M^{7}_{27} = -8 f_{k} + 4 f_{\omega},\  
M^{7}_{28} = 4 f_{\omega} - 4 \tilde{f}_{\omega},&\nonumber\\&
M^{7}_{29} = 8 f_{k} - 8 \tilde{f}_{k},\  
M^{7}_{37} = -8 f_{k} + 8 f_{\omega} + 8 \tilde{f}_{k},\  
M^{7}_{38} = 8 f_{k} - 8 \tilde{f}_{k},\  
M^{7}_{39} = 8 f_{\omega} - 8 \tilde{f}_{\omega},\  
M^{7}_{47} = 8 f_{k} + 4 \tilde{f}_{\omega},&\nonumber\\&
M^{7}_{49} = 8 f_{k} - 8 \tilde{f}_{k},\  
M^{7}_{57} = 8 f_{k} - 4 f_{\omega},\  
M^{7}_{58} = 4 f_{\omega} + 4 \tilde{f}_{\omega},\  
M^{7}_{59} = 8 f_{k} + 8 \tilde{f}_{k},\  
M^{7}_{67} = 24 f_{k} - 8 f_{\omega} - 8 \tilde{f}_{k},&\nonumber\\&
M^{7}_{68} = 8 f_{k} + 8 \tilde{f}_{k},\  
M^{7}_{69} = 8 f_{\omega} + 8 \tilde{f}_{\omega}.
\end{flalign}
The functions appearing in the RG flow equation for $g^{s}_{_8}$ in MLG are
\begin{flalign}
&M^{8}_{88} = -32 f_{k} - 16 f_{\omega},\  
M^{8}_{18} = 8 f_{k} + 8 f_{\omega} - 4 \tilde{f}_{\omega},\  
M^{8}_{19} = 8 f_{k} - 8 \tilde{f}_{k},\  
M^{8}_{27} = 4 f_{\omega} - 4 \tilde{f}_{\omega},\  
M^{8}_{28} = 8 f_{k} + 4 f_{\omega},&\nonumber\\&
M^{8}_{29} = 8 f_{k} + 8 \tilde{f}_{k},\  
M^{8}_{37} = 8 f_{k} - 8 \tilde{f}_{k},\  
M^{8}_{38} = -8 f_{k} - 8 f_{\omega} + 8 \tilde{f}_{k},\  
M^{8}_{39} = 8 f_{\omega} + 8 \tilde{f}_{\omega},\  
M^{8}_{48} = -8 f_{k} + 4 \tilde{f}_{\omega},&\nonumber\\&
M^{8}_{49} = 8 f_{k} + 8 \tilde{f}_{k},\  
M^{8}_{57} = 4 f_{\omega} + 4 \tilde{f}_{\omega},\  
M^{8}_{58} = -8 f_{k} - 4 f_{\omega},\  
M^{8}_{59} = 8 f_{k} - 8 \tilde{f}_{k},\  
M^{8}_{67} = 8 f_{k} + 8 \tilde{f}_{k},&\nonumber\\&
M^{8}_{68} = 24 f_{k} + 8 f_{\omega} - 8 \tilde{f}_{k},\  
M^{8}_{69} = 8 f_{\omega} - 8 \tilde{f}_{\omega}.
\end{flalign}
The functions appearing in the RG flow equation for $g^{s}_{_9}$ in MLG are
\begin{flalign}
&M^{9}_{99} = -16 f_{\omega},\  
M^{9}_{17} = 4 f_{k} + 4 \tilde{f}_{k},\  
M^{9}_{18} = 4 f_{k} - 4 \tilde{f}_{k},\
M^{9}_{19} = 8 f_{\omega} - 4 \tilde{f}_{\omega},\  
M^{9}_{27} = 4 f_{k} - 4 \tilde{f}_{k},\
M^{9}_{28} = 4 f_{k} + 4 \tilde{f}_{k},&\nonumber\\&  
M^{9}_{29} = 4 \tilde{f}_{\omega},\  
M^{9}_{37} = 4 f_{\omega} - 4 \tilde{f}_{\omega},\  
M^{9}_{38} = 4 f_{\omega} + 4 \tilde{f}_{\omega},\  
M^{9}_{39} = 16 f_{k} + 16 \tilde{f}_{k},\  
M^{9}_{47} = 4 f_{k} - 4 \tilde{f}_{k},\ 
M^{9}_{48} = 4 f_{k} + 4 \tilde{f}_{k},&\nonumber\\&   
M^{9}_{49} = 4 \tilde{f}_{\omega},\  
M^{9}_{57} = 4 f_{k} + 4 \tilde{f}_{k},\  
M^{9}_{58} = 4 f_{k} - 4 \tilde{f}_{k},\ 
M^{9}_{59} = 8 f_{\omega} - 4 \tilde{f}_{\omega},\ 
M^{9}_{67} = 4 f_{\omega} + 4 \tilde{f}_{\omega},\  
M^{9}_{68} = 4 f_{\omega} - 4 \tilde{f}_{\omega},&\nonumber\\&   
M^{9}_{69} = 16 f_{k} - 16 \tilde{f}_{k}.  
\end{flalign}

The RG flow equations for the conjugate fields associated with all the spin singlet excitonic orders in MLG are
\begin{align}
\bar{\beta}_{\Delta^s_1}^{\rm D}& = -2 (7 g^s_{_1}-g^s_{_2}-2 g^s_{_3}-g^s_{_4}-g^s_{_5}-2 g^s_{_6}-2 g^s_{_7}-2 g^s_{_8}-4 g^s_{_9}) (2 f_{k}+f_{\omega}),\nonumber\\
\bar{\beta}_{\Delta^s_2}^{\rm D}& = -2 (g^s_{_1}-7 g^s_{_2}-2 g^s_{_3}+g^s_{_4}+g^s_{_5}-2 g^s_{_6}+2 g^s_{_7}+2 g^s_{_8}-4 g^s_{_9}) (2 f_{k}-f_{\omega}),\nonumber\\
\bar{\beta}_{\Delta^s_3}^{\rm D}& = 2 (g^s_{_1}-g^s_{_2}-8 g^s_{_3}+g^s_{_4}-g^s_{_5}+2 g^s_{_7}-2 g^s_{_8}) f_{\omega},\nonumber\\
\bar{\beta}_{\Delta^s_4}^{\rm D}& = 2 (g^s_{_1}+g^s_{_2}+2 g^s_{_3}-7 g^s_{_4}+g^s_{_5}+2 g^s_{_6}-2 g^s_{_7}-2 g^s_{_8}-4 g^s_{_9}) (2 f_{k}+f_{\omega}),\nonumber\\
\bar{\beta}_{\Delta^s_5}^{\rm D}& = -2 (g^s_{_1}+g^s_{_2}-2 g^s_{_3}+g^s_{_4}-7 g^s_{_5}-2 g^s_{_6}-2 g^s_{_7}-2 g^s_{_8}+4 g^s_{_9}) (2 f_{k}-f_{\omega}),\nonumber\\
\bar{\beta}_{\Delta^s_6}^{\rm D}& = 2 (g^s_{_1}-g^s_{_2}+g^s_{_4}-g^s_{_5}-8 g^s_{_6}-2 g^s_{_7}+2 g^s_{_8}) f_{\omega},\nonumber\\
\bar{\beta}_{\Delta^s_7}^{\rm D}& = -2 (g^s_{_1}+g^s_{_2}+2 g^s_{_3}-g^s_{_4}-g^s_{_5}-2 g^s_{_6}-8 g^s_{_7}) (2 f_{k}-f_{\omega}),\nonumber\\
\bar{\beta}_{\Delta^s_8}^{\rm D}& = 2 (g^s_{_1}+g^s_{_2}-2 g^s_{_3}-g^s_{_4}-g^s_{_5}+2 g^s_{_6}-8 g^s_{_8}) (2 f_{k}+f_{\omega}),\nonumber\\
\bar{\beta}_{\Delta^s_9}^{\rm D}& = 2 (g^s_{_1}-g^s_{_2}-g^s_{_4}+g^s_{_5}-8 g^s_{_9}) f_{\omega}.
\end{align}
The RG flow equations for the conjugate fields associated with all the spin triplet excitonic orders in MLG are
\begin{align}
\bar{\beta}_{\Delta^t_1}^{\rm D}& = 2 (g^s_{_1}+g^s_{_2}+2 g^s_{_3}+g^s_{_4}+g^s_{_5}+2 g^s_{_6}+2 g^s_{_7}+2 g^s_{_8}+4 g^s_{_9}) (2 f_{k}+f_{\omega}),\nonumber\\
\bar{\beta}_{\Delta^t_2}^{\rm D}& = -2 (g^s_{_1}+g^s_{_2}-2 g^s_{_3}+g^s_{_4}+g^s_{_5}-2 g^s_{_6}+2 g^s_{_7}+2 g^s_{_8}-4 g^s_{_9}) (2 f_{k}-f_{\omega}),\nonumber\\
\bar{\beta}_{\Delta^t_3}^{\rm D}& = 2 (g^s_{_1}-g^s_{_2}+g^s_{_4}-g^s_{_5}+2 g^s_{_7}-2 g^s_{_8}) f_{\omega},\nonumber\\
\bar{\beta}_{\Delta^t_4}^{\rm D}& = 2 (g^s_{_1}+g^s_{_2}+2 g^s_{_3}+g^s_{_4}+g^s_{_5}+2 g^s_{_6}-2 g^s_{_7}-2 g^s_{_8}-4 g^s_{_9}) (2 f_{k}+f_{\omega}),\nonumber\\
\bar{\beta}_{\Delta^t_5}^{\rm D}& = -2 (g^s_{_1}+g^s_{_2}-2 g^s_{_3}+g^s_{_4}+g^s_{_5}-2 g^s_{_6}-2 g^s_{_7}-2 g^s_{_8}+4 g^s_{_9}) (2 f_{k}-f_{\omega}),\nonumber\\
\bar{\beta}_{\Delta^t_6}^{\rm D}& = 2 (g^s_{_1}-g^s_{_2}+g^s_{_4}-g^s_{_5}-2 g^s_{_7}+2 g^s_{_8}) f_{\omega},\nonumber\\
\bar{\beta}_{\Delta^t_7}^{\rm D}& = -2 (g^s_{_1}+g^s_{_2}+2 g^s_{_3}-g^s_{_4}-g^s_{_5}-2 g^s_{_6}) (2 f_{k}-f_{\omega}),\nonumber\\
\bar{\beta}_{\Delta^t_8}^{\rm D}& = 2 (g^s_{_1}+g^s_{_2}-2 g^s_{_3}-g^s_{_4}-g^s_{_5}+2 g^s_{_6}) (2 f_{k}+f_{\omega}),\nonumber\\
\bar{\beta}_{\Delta^t_9}^{\rm D}& = 2 (g^s_{_1}-g^s_{_2}-g^s_{_4}+g^s_{_5}) f_{\omega}.
\end{align}
The RG flow equations for the conjugate fields associated with all the local pairing orders in MLG are 
\begin{align}
\bar{\beta}_{\Delta^p_1}^{\rm D}& = 2 (g^s_{_1}-g^s_{_2}+2 g^s_{_3}-g^s_{_4}+g^s_{_5}-2 g^s_{_6}+2 g^s_{_7}-2 g^s_{_8}+4 g^s_{_9}) (2 \tilde{f}_{k}-\tilde{f}_{\omega}),\nonumber\\
\bar{\beta}_{\Delta^p_2}^{\rm D}& = -2 (g^s_{_1}-g^s_{_2}-2 g^s_{_3}-g^s_{_4}+g^s_{_5}+2 g^s_{_6}+2 g^s_{_7}-2 g^s_{_8}-4 g^s_{_9}) (2 \tilde{f}_{k}+\tilde{f}_{\omega}),\nonumber\\
\bar{\beta}_{\Delta^p_3}^{\rm D}& = -2 (g^s_{_1}+g^s_{_2}-g^s_{_4}-g^s_{_5}+2 g^s_{_7}+2 g^s_{_8}) \tilde{f}_{\omega},\nonumber\\
\bar{\beta}_{\Delta^p_4}^{\rm D}& = 2 (g^s_{_1}-g^s_{_2}+2 g^s_{_3}-g^s_{_4}+g^s_{_5}-2 g^s_{_6}-2 g^s_{_7}+2 g^s_{_8}-4 g^s_{_9}) (2 \tilde{f}_{k}-\tilde{f}_{\omega}),\nonumber\\
\bar{\beta}_{\Delta^p_5}^{\rm D}& = -2 (g^s_{_1}-g^s_{_2}-2 g^s_{_3}-g^s_{_4}+g^s_{_5}+2 g^s_{_6}-2 g^s_{_7}+2 g^s_{_8}+4 g^s_{_9}) (2 \tilde{f}_{k}+\tilde{f}_{\omega}),\nonumber\\
\bar{\beta}_{\Delta^p_6}^{\rm D}& = 2 (-g^s_{_1}-g^s_{_2}+g^s_{_4}+g^s_{_5}+2 (g^s_{_7}+g^s_{_8})) \tilde{f}_{\omega},\nonumber\\
\bar{\beta}_{\Delta^p_7}^{\rm D}& = -2 (g^s_{_1}-g^s_{_2}+2 g^s_{_3}+g^s_{_4}-g^s_{_5}+2 g^s_{_6}) (2 \tilde{f}_{k}+\tilde{f}_{\omega}),\nonumber\\
\bar{\beta}_{\Delta^p_8}^{\rm D}& = 2 (g^s_{_1}-g^s_{_2}-2 g^s_{_3}+g^s_{_4}-g^s_{_5}-2 g^s_{_6}) (2 \tilde{f}_{k}-\tilde{f}_{\omega}),\nonumber\\
\bar{\beta}_{\Delta^p_9}^{\rm D}& = -2 (g^s_{_1}+g^s_{_2}+g^s_{_4}+g^s_{_5}) \tilde{f}_{\omega}.
\end{align}

\subsection{Bilayer graphene}

The functions appearing in the RG flow equation for $g^{s}_{_1}$ in BLG are 
\begin{flalign}
&B^1_{11} = -24 f_{k} - 10 f_{\omega} - 2 \tilde{f}_{\omega},\ 
B^1_{22} = 2 f_{\omega} - 2 \tilde{f}_{\omega},\ 
B^1_{33} = 4 f_{\omega} - 4 \tilde{f}_{\omega},\ 
B^1_{44} = 2 f_{\omega} - 2 \tilde{f}_{\omega},\ 
B^1_{55} = 2 f_{\omega} - 2 \tilde{f}_{\omega},&\nonumber\\&
B^1_{66} = 4 f_{\omega} - 4 \tilde{f}_{\omega},\ 
B^1_{77} = 4 f_{\omega} - 4 \tilde{f}_{\omega},\ 
B^1_{88} = 4 f_{\omega} - 4 \tilde{f}_{\omega},\ 
B^1_{99} = 8 f_{\omega} - 8 \tilde{f}_{\omega},\ 
B^1_{12} = 8 f_{k} + 4 f_{\omega},&\nonumber\\&
B^1_{13} = 24 f_{k} + 8 f_{\omega} + 8 \tilde{f}_{k},\ 
B^1_{14} = 8 f_{k} + 4 f_{\omega},\ 
B^1_{15} = 8 f_{k} + 4 f_{\omega},\ 
B^1_{16} = 16 f_{k} + 8 f_{\omega},\ 
B^1_{17} = 16 f_{k} + 8 f_{\omega},&\nonumber\\&
B^1_{18} = 16 f_{k} + 8 f_{\omega},\ 
B^1_{19} = 32 f_{k} + 16 f_{\omega},\ 
B^1_{23} = 8 f_{k} - 8 \tilde{f}_{k},\ 
B^1_{46} = 8 f_{k} + 8 \tilde{f}_{k},\ 
B^1_{56} = 8 f_{k} - 8 \tilde{f}_{k},&\nonumber\\& 
B^1_{79} = 16 f_{k} + 16 \tilde{f}_{k},\ 
B^1_{89} = 16 f_{k} - 16 \tilde{f}_{k}.
\end{flalign}
The functions appearing in the RG flow equation for $g^{s}_{_2}$ in BLG are
\begin{flalign}
&B^2_{22} = 24 f_{k} - 12 f_{\omega},\ 
B^2_{33} = 4 f_{\omega} + 4 \tilde{f}_{\omega},\ 
B^2_{66} = 4 f_{\omega} + 4 \tilde{f}_{\omega},\ 
B^2_{99} = 8 f_{\omega} + 8 \tilde{f}_{\omega},\ 
B^2_{12} = -8 f_{k} + 8 f_{\omega} - 4 \tilde{f}_{\omega},&\nonumber\\&
B^2_{13} = 8 f_{k} - 8 \tilde{f}_{k},\ 
B^2_{23} = 24 f_{k} - 8 f_{\omega} + 8 \tilde{f}_{k},\ 
B^2_{24} = -8 f_{k} + 4 f_{\omega},\ 
B^2_{25} = -8 f_{k} + 4 f_{\omega},\ 
B^2_{26} = 16 f_{k} - 8 f_{\omega},&\nonumber\\&
B^2_{27} = -16 f_{k} + 8 f_{\omega},\ 
B^2_{28} = -16 f_{k} + 8 f_{\omega},\ 
B^2_{29} = 32 f_{k} - 16 f_{\omega},\ 
B^2_{45} = 4 f_{\omega} - 4 \tilde{f}_{\omega},\ 
B^2_{46} = 8 f_{k} - 8 \tilde{f}_{k},&\nonumber\\& 
B^2_{56} = 8 f_{k} + 8 \tilde{f}_{k},\ 
B^2_{78} = 8 f_{\omega} - 8 \tilde{f}_{\omega},\ 
B^2_{79} = 16 f_{k} - 16 \tilde{f}_{k},\ 
B^2_{89} = 16 f_{k} + 16 \tilde{f}_{k}.
\end{flalign}
The functions appearing in the RG flow equation for $g^{s}_{_3}$ in BLG are
\begin{flalign}
&B^3_{11} = 2 f_{k} + 2 \tilde{f}_{k},\ 
B^3_{22} = 2 f_{k} + 2 \tilde{f}_{k},\ 
B^3_{33} = 8 f_{k} - 16 f_{\omega} + 8 \tilde{f}_{k},\ 
B^3_{44} = 2 f_{k} + 2 \tilde{f}_{k},\ 
B^3_{55} = 2 f_{k} + 2 \tilde{f}_{k},\ 
B^3_{66} = 8 f_{k} + 8 \tilde{f}_{k},&\nonumber\\&  
B^3_{77} = 4 f_{k} + 4 \tilde{f}_{k},\ 
B^3_{88} = 4 f_{k} + 4 \tilde{f}_{k},\ 
B^3_{99} = 16 f_{k} + 16 \tilde{f}_{k},\ 
B^3_{12} = 4 f_{k} - 4 \tilde{f}_{k},\ 
B^3_{13} = 8 f_{\omega} - 4 \tilde{f}_{\omega},\ 
B^3_{23} = 4 \tilde{f}_{\omega},&\nonumber\\& 
B^3_{34} = 4 f_{\omega},\ 
B^3_{35} = -4 f_{\omega},\ 
B^3_{37} = 8 f_{\omega},\ 
B^3_{38} = -8 f_{\omega},\ 
B^3_{45} = 4 f_{k} - 4 \tilde{f}_{k},\ 
B^3_{46} = 4 f_{\omega} - 4 \tilde{f}_{\omega},\ 
B^3_{56} = 4 f_{\omega} + 4 \tilde{f}_{\omega},&\nonumber\\& 
B^3_{78} = 8 f_{k} - 8 \tilde{f}_{k},\ 
B^3_{79} = 8 f_{\omega} - 8 \tilde{f}_{\omega},\ 
B^3_{89} = 8 f_{\omega} + 8 \tilde{f}_{\omega}.
\end{flalign}
The functions appearing in the RG flow equation for $g^{s}_{_4}$ in BLG are
\begin{flalign}
&B^4_{44} = -24 f_{k} - 12 f_{\omega},\ 
B^4_{77} = 4 f_{\omega} + 4 \tilde{f}_{\omega},\ 
B^4_{88} = 4 f_{\omega} + 4 \tilde{f}_{\omega},\ 
B^4_{99} = 8 f_{\omega} + 8 \tilde{f}_{\omega},\ 
B^4_{14} = 8 f_{k} + 8 f_{\omega} - 4 \tilde{f}_{\omega},&\nonumber\\&
B^4_{16} = 8 f_{k} + 8 \tilde{f}_{k},
B^4_{24} = 8 f_{k} + 4 f_{\omega},\ 
B^4_{25} = 4 f_{\omega} - 4 \tilde{f}_{\omega},\ 
B^4_{26} = 8 f_{k} - 8 \tilde{f}_{k},\ 
B^4_{34} = 24 f_{k} + 8 f_{\omega} + 8 \tilde{f}_{k},\ 
B^4_{35} = 8 f_{k} - 8 \tilde{f}_{k},&\nonumber\\&	
B^4_{36} = 8 f_{\omega} - 8 \tilde{f}_{\omega},\ 
B^4_{45} = 8 f_{k} + 4 f_{\omega},\ 
B^4_{46} = 16 f_{k} + 8 f_{\omega},\ 
B^4_{47} = -16 f_{k} - 8 f_{\omega},\ 
B^4_{48} = -16 f_{k} - 8 f_{\omega},&\nonumber\\&
B^4_{49} = -32 f_{k} - 16 f_{\omega},\ 
B^4_{79} = 16 f_{k} - 16 \tilde{f}_{k},\ 
B^4_{89} = 16 f_{k} + 16 \tilde{f}_{k}. 
\end{flalign}
The functions appearing in the RG flow equation for $g^{s}_{_5}$ in BLG are
\begin{flalign}
&B^5_{55} = 24 f_{k} - 12 f_{\omega},\ 
B^5_{99} = 8 f_{\omega} - 8 \tilde{f}_{\omega},\ 
B^5_{15} = -8 f_{k} + 8 f_{\omega} - 4 \tilde{f}_{\omega},\ 
B^5_{16} = 8 f_{k} - 8 \tilde{f}_{k},\ 
B^5_{24} = 4 f_{\omega} - 4 \tilde{f}_{\omega},&\nonumber\\&
B^5_{25} = -8 f_{k} + 4 f_{\omega},\ 
B^5_{26} = 8 f_{k} + 8 \tilde{f}_{k},\ 
B^5_{34} = 8 f_{k} - 8 \tilde{f}_{k},\ 
B^5_{35} = 24 f_{k} - 8 f_{\omega} + 8 \tilde{f}_{k},\ 
B^5_{36} = 8 f_{\omega} + 8 \tilde{f}_{\omega},&\nonumber\\&
B^5_{45} = -8 f_{k} + 4 f_{\omega},\ 
B^5_{56} = 16 f_{k} - 8 f_{\omega},\ 
B^5_{57} = 16 f_{k} - 8 f_{\omega},\ 
B^5_{58} = 16 f_{k} - 8 f_{\omega},\ 
B^5_{59} = -32 f_{k} + 16 f_{\omega},&\nonumber\\&
B^5_{78} = 8 f_{\omega} + 8 \tilde{f}_{\omega},\ 
B^5_{79} = 16 f_{k} + 16 \tilde{f}_{k},\ 
B^5_{89} = 16 f_{k} - 16 \tilde{f}_{k}.&
\end{flalign}
The functions appearing in the RG flow equation for $g^{s}_{_6}$ in BLG are
\begin{flalign}
&B^6_{66} = -16 f_{\omega},\ 
B^6_{77} = 4 f_{k} - 4 \tilde{f}_{k},\ 
B^6_{88} = 4 f_{k} - 4 \tilde{f}_{k},\ 
B^6_{99} = 16 f_{k} - 16 \tilde{f}_{k},\ 
B^6_{14} = 4 f_{k} + 4 \tilde{f}_{k},\ 
B^6_{15} = 4 f_{k} - 4 \tilde{f}_{k},&\nonumber\\&
B^6_{16} = 8 f_{\omega} - 4 \tilde{f}_{\omega},\ 
B^6_{24} = 4 f_{k} - 4 \tilde{f}_{k},\ 
B^6_{25} = 4 f_{k} + 4 \tilde{f}_{k},\ 
B^6_{26} = 4 \tilde{f}_{\omega},\ 
B^6_{34} = 4 f_{\omega} - 4 \tilde{f}_{\omega},\ 
B^6_{35} = 4 f_{\omega} + 4 \tilde{f}_{\omega},&\nonumber\\&
B^6_{36} = 16 f_{k} + 16 \tilde{f}_{k},\ 
B^6_{46} = 4 f_{\omega},\ 
B^6_{56} = -4 f_{\omega},\ 
B^6_{67} = -8 f_{\omega},\ 
B^6_{68} = 8 f_{\omega},\ 
B^6_{78} = 8 f_{k} + 8 \tilde{f}_{k},\ 
B^6_{79} = 8 f_{\omega} + 8 \tilde{f}_{\omega},&\nonumber\\&
B^6_{89} = 8 f_{\omega} - 8 \tilde{f}_{\omega}.
\end{flalign}
The functions appearing in the RG flow equation for $g^{s}_{_7}$ in BLG are
\begin{flalign}
&B^7_{77} = -32 f_{k} - 16 f_{\omega},\ 
B^7_{17} = 8 f_{k} + 8 f_{\omega} - 4 \tilde{f}_{\omega},\ 
B^7_{19} = 8 f_{k} + 8 \tilde{f}_{k},\ 
B^7_{27} = 8 f_{k} + 4 f_{\omega},\ 
B^7_{28} = 4 f_{\omega} - 4 \tilde{f}_{\omega},&\nonumber\\&
B^7_{29} = 8 f_{k} - 8 \tilde{f}_{k},\ 
B^7_{37} = 24 f_{k} + 8 f_{\omega} + 8 \tilde{f}_{k},\ 
B^7_{38} = 8 f_{k} - 8 \tilde{f}_{k},\ 
B^7_{39} = 8 f_{\omega} - 8 \tilde{f}_{\omega},\ 
B^7_{47} = -8 f_{k} + 4 \tilde{f}_{\omega},&\nonumber\\&
B^7_{49} = 8 f_{k} - 8 \tilde{f}_{k},\ 
B^7_{57} = -8 f_{k} - 4 f_{\omega},\ 
B^7_{58} = 4 f_{\omega} + 4 \tilde{f}_{\omega},\ 
B^7_{59} = 8 f_{k} + 8 \tilde{f}_{k},\ 
B^7_{67} = -8 f_{k} - 8 f_{\omega} - 8 \tilde{f}_{k},&\nonumber\\&
B^7_{68} = 8 f_{k} + 8 \tilde{f}_{k},\ 
B^7_{69} = 8 f_{\omega} + 8 \tilde{f}_{\omega}.
\end{flalign}
The functions appearing in the RG flow equation for $g^{s}_{_8}$ in BLG are
\begin{flalign}
&B^8_{88} = 32 f_{k} - 16 f_{\omega},\ 
B^8_{18} = -8 f_{k} + 8 f_{\omega} - 4 \tilde{f}_{\omega},\ 
B^8_{19} = 8 f_{k} - 8 \tilde{f}_{k},\ 
B^8_{27} = 4 f_{\omega} - 4 \tilde{f}_{\omega},\ 
B^8_{28} = -8 f_{k} + 4 f_{\omega},&\nonumber\\&
B^8_{29} = 8 f_{k} + 8 \tilde{f}_{k},\ 
B^8_{37} = 8 f_{k} - 8 \tilde{f}_{k},\ 
B^8_{38} = 24 f_{k} - 8 f_{\omega} + 8 \tilde{f}_{k},\ 
B^8_{39} = 8 f_{\omega} + 8 \tilde{f}_{\omega},\ 
B^8_{48} = 8 f_{k} + 4 \tilde{f}_{\omega},&\nonumber\\&
B^8_{49} = 8 f_{k} + 8 \tilde{f}_{k},\ 
B^8_{57} = 4 f_{\omega} + 4 \tilde{f}_{\omega},\ 
B^8_{58} = 8 f_{k} - 4 f_{\omega},\ 
B^8_{59} = 8 f_{k} - 8 \tilde{f}_{k},\ 
B^8_{67} = 8 f_{k} + 8 \tilde{f}_{k},&\nonumber\\&
B^8_{68} = -8 f_{k} + 8 f_{\omega} - 8 \tilde{f}_{k},\ 
B^8_{69} = 8 f_{\omega} - 8 \tilde{f}_{\omega}.
\end{flalign}
The functions appearing in the RG flow equation for $g^{s}_{_9}$ in BLG are
\begin{flalign}
&B^9_{99} = -16 f_{\omega},\ 
B^9_{17} = 4 f_{k} + 4 \tilde{f}_{k},\ 
B^9_{18} = 4 f_{k} - 4 \tilde{f}_{k},\ 
B^9_{19} = 8 f_{\omega} - 4 \tilde{f}_{\omega},\ 
B^9_{27} = 4 f_{k} - 4 \tilde{f}_{k},\ 
B^9_{28} = 4 f_{k} + 4 \tilde{f}_{k},&\nonumber\\&
B^9_{29} = 4 \tilde{f}_{\omega},\ 
B^9_{37} = 4 f_{\omega} - 4 \tilde{f}_{\omega},\ 
B^9_{38} = 4 f_{\omega} + 4 \tilde{f}_{\omega},\ 
B^9_{39} = 16 f_{k} + 16 \tilde{f}_{k},\ 
B^9_{47} = 4 f_{k} - 4 \tilde{f}_{k},\ 
B^9_{48} = 4 f_{k} + 4 \tilde{f}_{k},&\nonumber\\& 
B^9_{49} = 4 \tilde{f}_{\omega},\ 
B^9_{57} = 4 f_{k} + 4 \tilde{f}_{k},\ 
B^9_{58} = 4 f_{k} - 4 \tilde{f}_{k},\ 
B^9_{59} = 8 f_{\omega} - 4 \tilde{f}_{\omega},\ 
B^9_{67} = 4 f_{\omega} + 4 \tilde{f}_{\omega},\ 
B^9_{68} = 4 f_{\omega} - 4 \tilde{f}_{\omega},&\nonumber\\&
B^9_{69} = 16 f_{k} - 16 \tilde{f}_{k}.
\end{flalign}

The RG flow equations for the conjugate fields associated with all the spin singlet excitonic orders in BLG are
\begin{align}
\bar{\beta}_{\Delta^s_1}^{\rm L}&=-2 (2 f_{k}+f_{\omega}) (7 g^s_{_1}-g^s_{_2}-2 g^s_{_3}-g^s_{_4}-g^s_{_5}-2 g^s_{_6}-2 g^s_{_7}-2 g^s_{_8}-4 g^s_{_9}), \nonumber \\
\bar{\beta}_{\Delta^s_2}^{\rm L}&=-2 (2 f_{k}-f_{\omega}) (g^s_{_1}-7 g^s_{_2}-2 g^s_{_3}+g^s_{_4}+g^s_{_5}-2 g^s_{_6}+2 g^s_{_7}+2 g^s_{_8}-4 g^s_{_9}), \nonumber \\
\bar{\beta}_{\Delta^s_3}^{\rm L}&=2 f_{\omega} (g^s_{_1}-g^s_{_2}-8 g^s_{_3}+g^s_{_4}-g^s_{_5}+2 g^s_{_7}-2 g^s_{_8}), \nonumber \\
\bar{\beta}_{\Delta^s_4}^{\rm L}&=2 (2 f_{k}+f_{\omega}) (g^s_{_1}+g^s_{_2}+2 g^s_{_3}-7 g^s_{_4}+g^s_{_5}+2 g^s_{_6}-2 g^s_{_7}-2 g^s_{_8}-4 g^s_{_9}), \nonumber \\
\bar{\beta}_{\Delta^s_5}^{\rm L}&=-2 (2 f_{k}-f_{\omega}) (g^s_{_1}+g^s_{_2}-2 g^s_{_3}+g^s_{_4}-7 g^s_{_5}-2 g^s_{_6}-2 g^s_{_7}-2 g^s_{_8}+4 g^s_{_9}), \nonumber \\ 
\bar{\beta}_{\Delta^s_6}^{\rm L}&=2 f_{\omega} (g^s_{_1}-g^s_{_2}+g^s_{_4}-g^s_{_5}-8 g^s_{_6}-2 g^s_{_7}+2 g^s_{_8}), \nonumber \\
\bar{\beta}_{\Delta^s_7}^{\rm L}&=2 (2 f_{k}+f_{\omega}) (g^s_{_1}+g^s_{_2}+2 g^s_{_3}-g^s_{_4}-g^s_{_5}-2 g^s_{_6}-8 g^s_{_7}), \nonumber \\
\bar{\beta}_{\Delta^s_8}^{\rm L}&=-2 (2 f_{k}-f_{\omega}) (g^s_{_1}+g^s_{_2}-2 g^s_{_3}-g^s_{_4}-g^s_{_5}+2 g^s_{_6}-8 g^s_{_8}), \nonumber \\
\bar{\beta}_{\Delta^s_9}^{\rm L}&=2 f_{\omega} (g^s_{_1}-g^s_{_2}-g^s_{_4}+g^s_{_5}-8 g^s_{_9}).
\end{align}
The RG flow equations for the conjugate fields associated with all the spin triplet excitonic orders in BLG are
\begin{align}
\bar{\beta}_{\Delta^t_1}^{\rm L}&=2 (2 f_{k}+f_{\omega}) (g^s_{_1}+g^s_{_2}+2 g^s_{_3}+g^s_{_4}+g^s_{_5}+2 g^s_{_6}+2 g^s_{_7}+2 g^s_{_8}+4 g^s_{_9}), \nonumber \\
\bar{\beta}_{\Delta^t_2}^{\rm L}&=-2 (2 f_{k}-f_{\omega}) (g^s_{_1}+g^s_{_2}-2 g^s_{_3}+g^s_{_4}+g^s_{_5}-2 g^s_{_6}+2 g^s_{_7}+2 g^s_{_8}-4 g^s_{_9}), \nonumber \\
\bar{\beta}_{\Delta^t_3}^{\rm L}&=2 f_{\omega} (g^s_{_1}-g^s_{_2}+g^s_{_4}-g^s_{_5}+2 g^s_{_7}-2 g^s_{_8}), \nonumber \\
\bar{\beta}_{\Delta^t_4}^{\rm L}&=2 (2 f_{k}+f_{\omega}) (g^s_{_1}+g^s_{_2}+2 g^s_{_3}+g^s_{_4}+g^s_{_5}+2 g^s_{_6}-2 g^s_{_7}-2 g^s_{_8}-4 g^s_{_9}), \nonumber \\
\bar{\beta}_{\Delta^t_5}^{\rm L}&=-2 (2 f_{k}-f_{\omega}) (g^s_{_1}+g^s_{_2}-2 g^s_{_3}+g^s_{_4}+g^s_{_5}-2 g^s_{_6}-2 g^s_{_7}-2 g^s_{_8}+4 g^s_{_9}), \nonumber \\
\bar{\beta}_{\Delta^t_6}^{\rm L}&=2 f_{\omega} (g^s_{_1}-g^s_{_2}+g^s_{_4}-g^s_{_5}-2 g^s_{_7}+2 g^s_{_8}), \nonumber \\
\bar{\beta}_{\Delta^t_7}^{\rm L}&=2 (2 f_{k}+f_{\omega}) (g^s_{_1}+g^s_{_2}+2 g^s_{_3}-g^s_{_4}-g^s_{_5}-2 g^s_{_6}), \nonumber \\
\bar{\beta}_{\Delta^t_8}^{\rm L}&=-2 (2 f_{k}-f_{\omega}) (g^s_{_1}+g^s_{_2}-2 g^s_{_3}-g^s_{_4}-g^s_{_5}+2 g^s_{_6}), \nonumber \\
\bar{\beta}_{\Delta^t_9}^{\rm L}&=2 f_{\omega} (g^s_{_1}-g^s_{_2}-g^s_{_4}+g^s_{_5}).
\end{align}
The RG flow equations for the conjugate fields associated with all the local pairing orders in BLG are 
\begin{align}
\bar{\beta}_{\Delta^p_1}^{\rm L}&=2 (2 \tilde{f}_{k}-\tilde{f}_{\omega}) (g^s_{_1}-g^s_{_2}+2 g^s_{_3}-g^s_{_4}+g^s_{_5}-2 g^s_{_6}+2 g^s_{_7}-2 g^s_{_8}+4 g^s_{_9}), \nonumber \\
\bar{\beta}_{\Delta^p_2}^{\rm L}&=-2 (2 \tilde{f}_{k}+\tilde{f}_{\omega}) (g^s_{_1}-g^s_{_2}-2 g^s_{_3}-g^s_{_4}+g^s_{_5}+2 g^s_{_6}+2 g^s_{_7}-2 g^s_{_8}-4 g^s_{_9}), \nonumber \\
\bar{\beta}_{\Delta^p_3}^{\rm L}&=-2 \tilde{f}_{\omega} (g^s_{_1}+g^s_{_2}-g^s_{_4}-g^s_{_5}+2 g^s_{_7}+2 g^s_{_8}), \nonumber \\
\bar{\beta}_{\Delta^p_4}^{\rm L}&=2 (2 \tilde{f}_{k}-\tilde{f}_{\omega}) (g^s_{_1}-g^s_{_2}+2 g^s_{_3}-g^s_{_4}+g^s_{_5}-2 g^s_{_6}-2 g^s_{_7}+2 g^s_{_8}-4 g^s_{_9}), \nonumber \\
\bar{\beta}_{\Delta^p_5}^{\rm L}&=-2 (2 \tilde{f}_{k}+\tilde{f}_{\omega}) (g^s_{_1}-g^s_{_2}-2 g^s_{_3}-g^s_{_4}+g^s_{_5}+2 g^s_{_6}-2 g^s_{_7}+2 g^s_{_8}+4 g^s_{_9}), \nonumber \\
\bar{\beta}_{\Delta^p_6}^{\rm L}&=2 \tilde{f}_{\omega}(-g^s_{_1}-g^s_{_2}+g^s_{_4}+g^s_{_5}+2 g^s_{_7}+2 g^s_{_8}), \nonumber \\
\bar{\beta}_{\Delta^p_7}^{\rm L}&=2 (2 \tilde{f}_{k}-\tilde{f}_{\omega}) (g^s_{_1}-g^s_{_2}+2 g^s_{_3}+g^s_{_4}-g^s_{_5}+2 g^s_{_6}), \nonumber \\
\bar{\beta}_{\Delta^p_8}^{\rm L}&=-2 (2 \tilde{f}_{k}+\tilde{f}_{\omega}) (g^s_{_1}-g^s_{_2}-2 g^s_{_3}+g^s_{_4}-g^s_{_5}-2 g^s_{_6}), \nonumber\\
\bar{\beta}_{\Delta^p_9}^{\rm L}&=-2 \tilde{f}_{\omega} (g^s_{_1}+g^s_{_2}+g^s_{_4}+g^s_{_5}).
\end{align}

\twocolumngrid

%\nocite{*} % Comment out to only cite referred publications

%\bibliographystyle{apsrev4-1}
\bibliography{library_honeycomb}

\end{document}